\documentclass[10pt,a4paper,twoside]{report}

\usepackage{macros}

\begin{document}
% Title page________________________________________________________________
%
% THIS IS NOT THE COVER FOR THE PRINTED VERSION, BUT THE FIRST PAGE AFTER THE COVER, I GENERATE THE COVER SEPARATELY WITH ANOTHER FONT SUITABLE FOR HARD COVERS AND ADD IT TO THE DOCUMENT AS A PDF
\SetWatermarkText{}
\begin{titlepage}

\begin{flushright}
% OUTLINE  \\ 
% DRAFT \\ 
% NOTES \\ 
% arXiv:2507.12849 \\ 
September 2026
\end{flushright}
\begin{centering}
\vfill

{\Large{\bf
    Warm-Inflation: a software for first-principles  \\[3mm]
    ~integration of warm inflation power spectra 
}} 

\vspace{0.8cm}

A.~Rogelj\hspace*{0.4mm}$^\text{a}_{ }$

\vspace{0.8cm}

$^\text{a}_{ }${\em
AEC, 
Institute for Theoretical Physics, 
University of Bern, \\ 
Sidlerstrasse 5, CH-3012 Bern, Switzerland \\}

\vspace*{0.8cm}

\mbox{\bf Abstract}
 
\end{centering}

\vspace*{0.3cm}
 
\noindent
As part of a PhD thesis entitled ``Scalar perturbations in warm inflation and during reheating'' we present a software for computing warm inflation observables. In particular, our \texttt{Mathematica} implementation computes the power spectrum, spectral tilt, and tensor-to-scalar ratio for a given momentum scale. Compared with other implementations, we start in a Bunch-Davies state deep inside the horizon, ideally at a time when thermal noise is exponentially suppressed. We integrate gauge-invariant equations for three curvature perturbations across horizon exit. We continue deep enough outside of horizon so that all curvature perturbations agree. Results are illustrated for Standard Model embedded warm inflation, and we also provide notebooks for several other popular potentials and friction coefficients. The theoretical foundations for the code can be found in the main part of the thesis or in earlier papers \cite{Alica1,Alica2}. Specifically, part \ref{part:analyt} derives gauge-invariant, model-agnostic evolution equations, as well as the noise autocorrelator interpolating between quantum and classical domains. Part \ref{part:num} presents the general ideas behind the numerical implementation, while part \ref{part:resultsdisc} presents selected results. The code is described, line-by-line, in Appendix \ref{app:CodeOverview}. The published PhD thesis is available at \texttt{https://doi.org/10.48620/101166} while the notebooks can be downloaded from \texttt{https://github.com/alicarogeljblackhole/Warm-Inflation}.

\vspace*{0.5cm}

\noindent
{\small
{\em Email}: 
{\tt alica.rogelj@unibe.ch}
}

\vfill

%% %\noindent
%% %PACS numbers: 
%% %11.10.Wx, %        Finite temperature field theory
%% %11.15.Ha, %        Lattice gauge theory  
%% %12.38.Bx, %        Perturbative calculations in QCD
%% %12.38.Mh, %        Quark--gluon plasma
%% %14.40.Nd, %        Bottom mesons
%% %\\
%% %Keywords: cosmological perturbation theory [100] 
%%            inflation [100]
%%            particle physics - cosmology connection [50] 
%%            axions [50]

\end{titlepage}

\SetWatermarkText{}
%%%%%%%%%%%%%%%%%%%%%%%%%%%%%%%%%%%%%%%%%%%%%%%%%%%%%%%%%%%%%%%%%%%%%%%%%%%%%%%%%%%%%%%%%%%
%
% FOR THE PRINTED VERSION AN EMPTY PAGE BETWEEN THE TITLE AND THE DEDICATION PAGE IS NEEDED
%\newpage\null\thispagestyle{empty}\newpage
%
%%%%%%%%%%%%%%%%%%%%%%%%%%%%%%%%%%%%%%%%%%%%%%%%%%%%%%%%%%%%%%%%%%%%%%%%%%%%%%%%%%%%%%%%%%%
%

% Preamble______________________________________________________________________

%%%%%%%%%%%%%%%%%%%%%%%%%%%%%%%%%%%%%%%%%%%%%%%%%%%%%%%%%%%%%%%%%%%%%%%%%%%%%%%%%%%%%%%%%%%

\pagestyle{plain} 
\pagenumbering{roman} 				% roman page numbering (i,ii,...)
%\thispagestyle{empty}

%\cleardoublepage

%---------------------------------------------------------------------------
% Table of contents

\setcounter{tocdepth}{1}
\tableofcontents

%---------------------------------------------------------------------------
\clearpage
\section*{Notation and conventions}
\addcontentsline{toc}{chapter}{Notation}
Indices denoted by Greek letters run from $0$ to $3$, describing the components on the four-dimensional space-time. Roman letters are used for spatial indices and run from $1$ to $3$. Repeated indices are summed over. 
Analytical computations are performed using natural units, where 
\begin{equation*}\label{eq_units}
c = \hbar = k_\text{\tiny B} = 1 \ ,
\end{equation*}
but retaining the Planck mass $m_{Pl}$. 
\\\\
Conformal ($\tau$) and physical time ($t$) differ by a scale factor: $dt=a\;d\tau$. Derivative with respect to conformal time is denoted by a prime,
$$\frac{d}{d\tau}f(\tau)\equiv f'(\tau)\;\;,$$
while the physical time derivative is denoted by a dot,
$$\frac{d}{dt}f(t)\equiv \dot{f}(t)\;\;.$$
A partial derivative can be denoted by a comma after the function,
$$\partial_{\varphi}V\equiv V_{,\varphi}\;. $$
Similarly, a covariant derivative can be written as a semicolon,
$$\nabla_\mu \;T^{\mu\nu}
\equiv\; T^{\mu\nu}_{;\mu}\;.$$
We use the ``mostly plus'' convention for the metric tensor.
\newpage
%---------------------------------------------------------------------------
\vfill

\section*{List of globally defined symbols}
\addcontentsline{toc}{chapter}{List of symbols}
The symbols introduced in this thesis, together with some relevant reference values, are listed below.
\vspace{.2cm}
%%%%%%%%%%%%%%%%%%%%%%%%%%%%%%%%%%%%%%%%%%%%%%%%%%%%%%%%%%%%%%%%%%%%%%%%%%%%%%%%%%%%%%%%%%%
\begin{center}
\begin{tabular*}{\textwidth}{@{\extracolsep{\fill}}l l l }
\hline
\hline
Symbol                      & Value (SI units) and/or definition                                      
	                                                                  					\\
\hline
\hline
$c$ 					       & $\approx 3.00 \times 10^8$ m\,s$^{-1}$, velocity of light	
										  								\\
$\hbar$ 			           & $\approx 1.05 \times 10^{-34}$ kg\,m$^2$\,s$^{-1}$, reduced Planck's constant				 		                 	\\	
$k_\text{\tiny B}$ 		       & $\approx 1.38 \times 10^{-23}$ kg\,m$^2$\,s$^{-2}$\,K$^{-1}$, Boltzmann constant    		  				                \\
$m_{Pl}$                 	   & $\approx 2.18 \times 10^{-8}$ kg $\approx 1.22\times 10^{19}$ GeV\,c$^{-2}$, Planck mass      \\
\hline
$t_{Pl}$					& $\approx 5.39 \times 10^{-44}$ s, Planck time
\\
$t_\text{dec}$          & $\sim 10^{12}$ s $\sim 3\times 10^5$ y, time at decoupling              \\
$t_\text{0}$          & $\sim 4.4\times 10^{17}$ s $\sim 1.4\times 10^{10}$ y, time today&              \\
$T_\text{ref}$          & $\sim 10^{27}$ K, temperature estimate during inflation
\\
$T_\text{dec}$          & $\sim 3000$ K, temperature at decoupling                                                   	                                                   \\
$T_0$                   & $\approx 2.73$ K, CMB temperature today                          	                                           \\
$d_\text{dec}$    		& $\sim 3\times 10^{23}$ m $\sim 100$ Mpc, size of visible universe at decoupling                                       \\
$d_0$                   & $\sim 4\times 10^{25}$ m $\sim 14$ Gpc, size of visible universe today                                                   										        \\
\hline
$t_\text{ref}$ & reference time (when we start the background evolution)
\\
$t_1$ & time at which we start the evolution of first-order perturbations
\\
$t_\star$ & time of the horizon exit, $k=aH$
\\
\hline
$t$                     & physical time coordinate
										       \\		
                                          $\tau$                  & conformal time coordinate                                   
	                                         \\
$a$                     & scale factor                                     
	                                           \\
$H$                     & Hubble parameter               
	                                           \\
$k/a$ & physical momentum
\\
$\Upsilon$ & thermal friction
                                               \\
                                          $\HH$                   & conformal Hubble parameter                      	
		                                   \\
$\eta_{\m\n}$           & Minkowski metric         
										   	 \\
                                          $\kappa$                & spatial curvature                              
                                         \\
$\hat{H}$     & Hamiltonian (operator)
                                         \\
%\end{tabular*}
%\newpage
%%%%%%%%%%%%%%%%%%%%%%%%%%%%%% CONTINUE ON NEW PAGE %%%%%%%%%%%%%%%%%%%%%%%%%%%%%%%%%%%%%%%%%%%%%%%%%
%\begin{tabular*}{\textwidth}{@{\extracolsep{\fill}}l l l l l}
\hline
$\mathcal{R}_\varphi$ & inflaton curvature perturbation in real space\\
$\mathcal{R}_v,\,\mathcal{R}_T$ & plasma curvature perturbations in real space\\
$\mathcal{S}_v,\,\mathcal{S}_T$ & isocurvature perturbations in real space\\
$\hat{\mathcal{Q}}_\varphi\equiv a\mathcal{Q}_\varphi$ & rescaled inflaton curvature perturbations in real space\\
$\mathcal{R}_{\varphi k}$ & inflaton curvature perturbation in Fourier space\\
$\mathcal{R}_{vk},\,\mathcal{R}_{Tk}$ & plasma curvature perturbations in Fourier space\\
$\mathcal{S}_{vk},\,\mathcal{S}_{Tk}$ & isocurvature perturbations in Fourier space\\
$\hat{\mathcal{Q}}_{\varphi k}\equiv a\mathcal{Q}_{\varphi k}$ & rescaled inflaton curvature perturbations in Fourier space\\
\hline
\hline
\end{tabular*}
\end{center}
%
%%%%%%%%%%%%%%%%%%%%%%%%%%%%%%%%%%%%%%%%%%%%%%%%%%%%%%%%%%%%%%%%%%%%%%%%%%%%%%%%%%%%%%%%%%%%%%%%%%%

% Chapters______________________________________________________________________

%%%%%%%%%%%%%%%%%%%%%%%%%%%%%%%%%%%%%%%%%%%%%%%%%%%%%%%%%%%%%%%%%%%%%%%%%%%%%%%%%%%%%%%%%%%
%
% FOR THE PRINTED VERSION AN EMPTY PAGE BETWEEN PREAMPLE AND BEGINNING OF THE CHAPTERS IS NEEDED
%\newpage 
%\thispagestyle{empty}
%\ 
%\newpage
%%%%%%%%%%%%%%%%%%%%%%%%%%%%%%%%%%%%%%%%%%%%%%%%%%%%%%%%%%%%%%%%%%%%%%%%%%%%%%%%%%%%%%%%%%%

\pagestyle{fancy}               	% headings
\pagenumbering{arabic}			% arabic page numbering (1,2,...)

\part{Introduction}\label{part:intro}
\newpage
\chapter{General context}
\begin{figure}[!b]
    \centering
    \begin{subfigure}[t]{0.32\linewidth}
        \centering
        \adjustbox{valign=c}{\includegraphics[width=\linewidth]{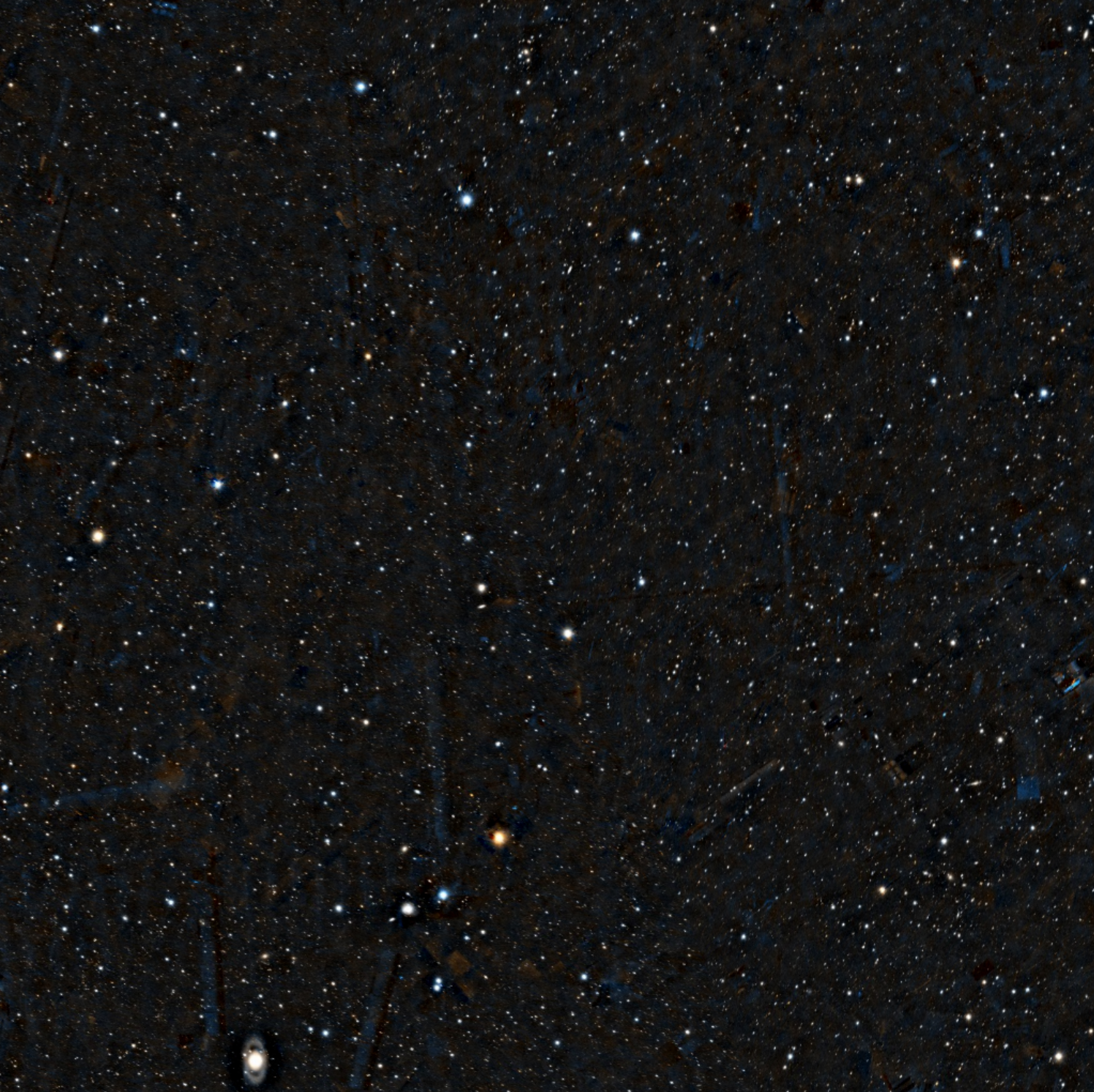}}
        \caption{visible spectrum, ref.~\cite{chambers16}}\label{esaa}
    \end{subfigure}
    \hfill
    \begin{subfigure}[t]{0.32\linewidth}
        \centering
        \adjustbox{valign=c}{\includegraphics[width=\linewidth]{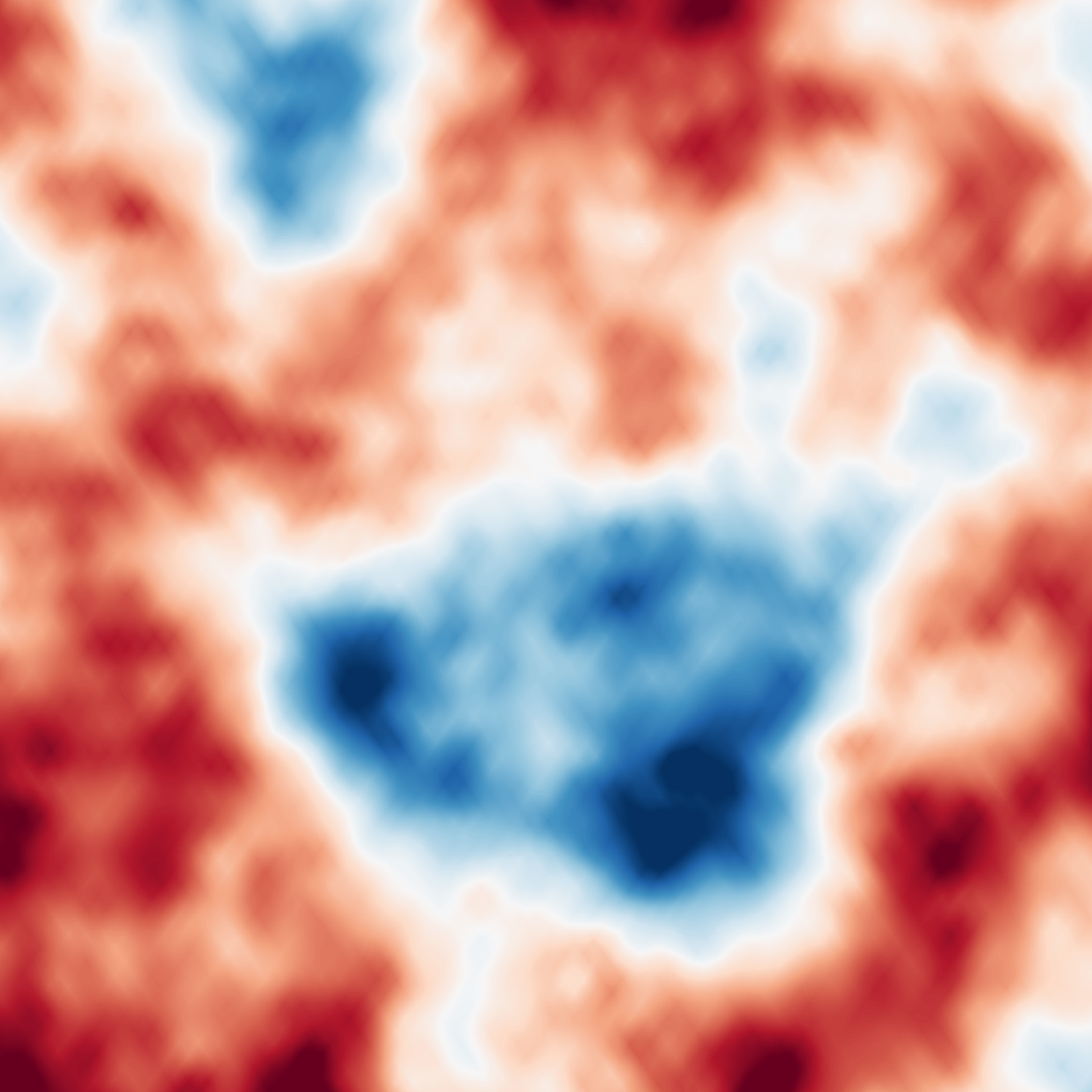}}
        \caption{temperature map, ref.~\cite{Planck:2015mis}}\label{esab}
    \end{subfigure}
    \hfill
    \begin{subfigure}[t]{0.32\linewidth}
        \centering
        \adjustbox{valign=c}{\includegraphics[width=\linewidth]{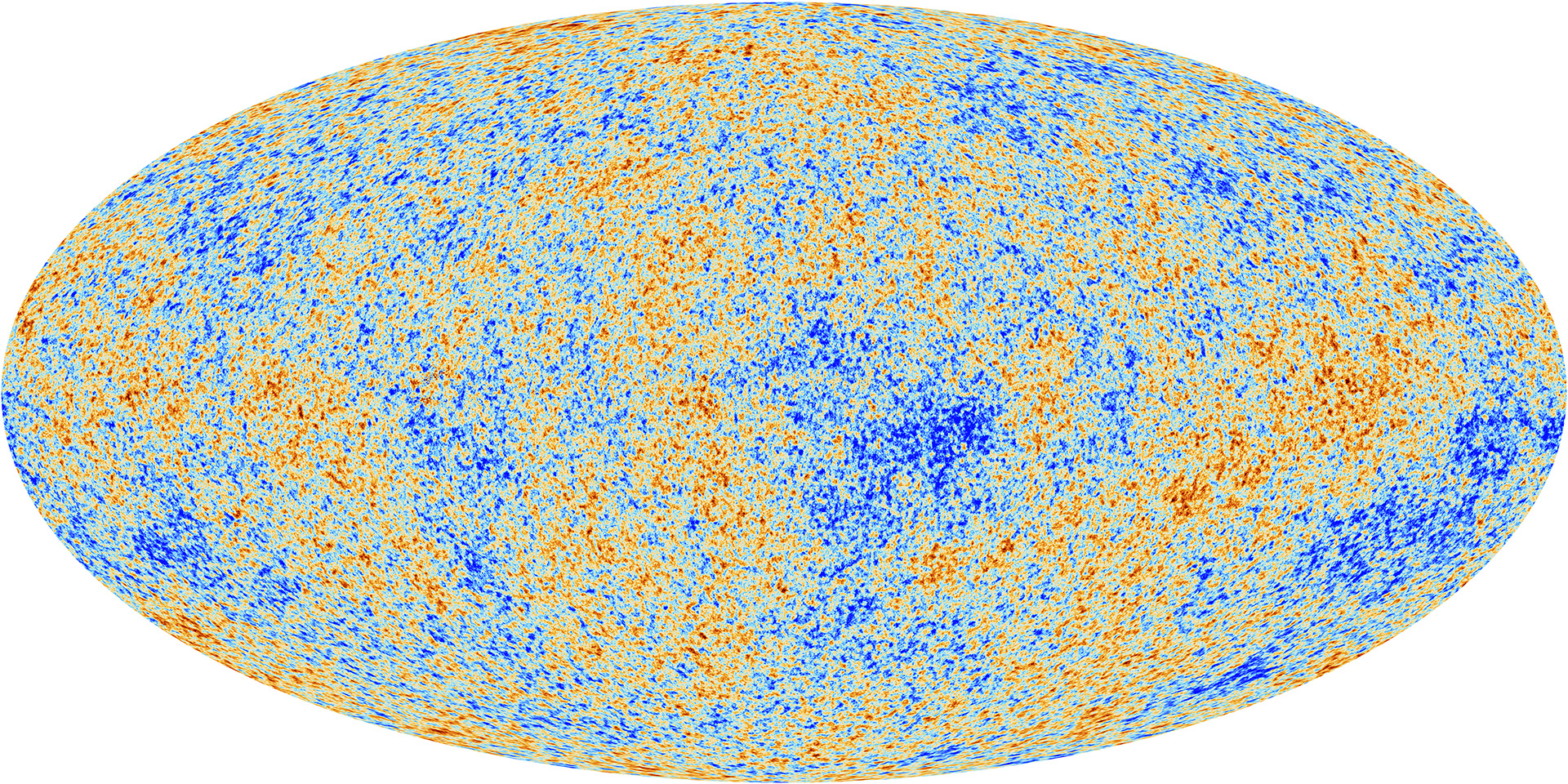}}
        \vspace{1.15cm}
        \caption{CMB (full sky), ref.~\cite{planckcmb2013}}\label{esac}
    \end{subfigure}
    \caption[Perturbations across different length scales.]{Perturbations across different length scales. Fig.~\ref{esaa} shows a map of the sky of the size $2^\circ \times 2^\circ$ in the visible spectrum (centered at $l=180^\circ$, $b=60^\circ$ in galactic coordinates); a number of small scale structures can be seen. On the other hand, fig.~\ref{esab} shows the same patch of the sky as a temperature map; in comparison with fig.~\ref{esaa}, we already see the perturbations span larger length scales. For reference, fig.~\ref{esac} shows the full sky version of the temperature map, which also contains large scale perturbations. The difference between the hot (red) and the cold (blue) parts of the universe is $\mathcal{O}(10^{-5}) \,K$.\\ Figures \ref{esaa} and \ref{esab} were obtained using \texttt{hips2fits},\footnotemark a service provided by CDS.}
    \label{fig:differentscalesofperturbations}
\end{figure}

Our universe looks largely the same in all directions, which we call isotropy. In addition, we assume the universe looks the same from every point in space, which we refer to as homogeneity. Those two statements are together known as the cosmological principle. Nonetheless, our universe is not perfectly homogeneous and isotropic. For example, on cosmologically ``small'' distance scales, objects such as galaxies and stars act as perturbations to the homogeneity and isotropy. Perturbations can be also observed on larger distance scales. This can manifest through small fluctuations in the background temperature of our universe (in particular, temperature fluctuations in the cosmic microwave background, CMB). The fluctuations evidently span many different scales; see fig.~\ref{fig:differentscalesofperturbations} for an illustration. A central question in cosmology is where these fluctuations came from, and how they evolved after they were created. 
\\
\\
The leading theory is that the fluctuations were created roughly 13.8 billion years ago. At that time, the universe was undergoing an exponential expansion known as cosmic inflation \cite{Starobinsky:1980te}. An essential prediction of inflation is the generation of quantum fluctuations. These serve as seeds of the inhomogeneities that we observe today \cite{Mukhanov:1981xt}. In particular, in a microscopic picture, the inflation-driving inflaton field eventually decays, stopping the inflation and filling the universe with Standard Model (SM) particles. This, perhaps less well-understood process of energy transfer, is called reheating. By the time it is completed, the universe is estimated to be $10^{-10}$ seconds old. Afterwards, an empirically better established thermal history of the universe begins.  While we will not work directly on the subsequent universe evolution in this thesis, we briefly outline it in the paragraph below for completeness.
\\\\
Shortly after inflation and reheating, the existing particles formed a hot quark-gluon plasma, which hadronized as the universe kept expanding and simultaneously cooling down. About a microsecond after inflation, the universe had cooled down enough for quarks to bind into protons and neutrons, which bonded into the first nuclei some three minutes later. Roughly 380 000 years afterwards, electrons started becoming bound to protons, drastically reducing photon scatterings off free electrons, and consequently making the universe transparent. Then the photons could travel freely for the first time, which we can observe as CMB radiation. This radiation already exhibited fluctuations across different length scales. After the subsequent non-linear evolution through the dark ages, the fluctuations on smaller length scales formed the first stars, and later galaxies, as well as the large scale structure.
\\\\
In this thesis, we will explore the primordial fluctuations, which can have both quantum and thermal origins. We further describe their evolution through the late stage of inflation and the subsequent period of reheating. We will largely work on the level of an effective description, being as model-independent as possible.
\\\\
The structure of the thesis is as follows. In Ch.~\ref{WIsmoothreheating}, we briefly touch on various inflation and reheating mechanisms, eventually zooming into warm inflation (WI) and smooth reheating. We identify pre-existing gaps in the literature and explain how our work contributes to filling them. In Ch.~\ref{ch:essentials}, we didactically present the basics of the perturbation theory in warm inflation, which serves as a starting point for our analytical work. The latter is then described in Part \ref{part:analyt}, see Chs.~\ref{ch:gaugeinv}-\ref{ch:noise}. There we develop a gauge-invariant framework for computing the evolution of scalar perturbations. Two main analytical results are a closed set of perturbation equations \eqref{result}, as well as a noise autocorrelator interpolating between the initial quantum and subsequent classical regimes, eqs.~\eqref{correlator} and \eqref{omega}. In Part \ref{part:num}, see Chs.~\ref{sec:NumMethodIntegration}-\ref{sec:ModelBuilding}, we present a numerical implementation of our equations, including a summary of the model on which we test our framework. We proceed to Part \ref{part:resultsdisc}, see Chs.~\ref{sec:numres}-\ref{chapter:conclusion}, where we show and discuss some selected results of our numerical implementation. The thesis concludes with several appendices. A detailed overview of our code can be found in Appendix \ref{app:CodeOverview}, while the code is also available online: \texttt{https://github.com/alicarogeljblackhole/Warm-Inflation}. A detailed derivation of the gauge-invariant evolution equations is found in Appendix \ref{sec:curvature}.
\\
\footnotetext{\url{https://alasky.cds.unistra.fr/hips-image-services/hips2fits}}

\newpage
\SetWatermarkText{}
\chapter{Warm inflation and smooth reheating}\label{WIsmoothreheating}

\section{Motivation}
Inflation was initially introduced to solve some ``background-level'' puzzles. One of them is the horizon problem, with inflation explaining why the temperature of the universe is uniform across such large distances. Other puzzles solved by inflation are e.g.~the flatness and the relic problems, see refs.~\cite{inflGuth,Albrecht:1982wi,Linde:1982uu,Linde:1983gd,wilczek,InflationLinde:1981}. Once we extend our analysis beyond the background level, by e.g. perturbing the inflaton and other relevant fields, inflation also offers predictions, e.g. explaining the origin of primordial scalar perturbations and nearly scale-invariant power spectrum \cite{Brandenberger:1993zc,Kodama:1984ziu}.\footnote{The start of cosmological perturbation theory can be traced all the way back to ref.~\cite{Lifshitz:1963ps}.} Specific observational signatures will be discussed in Ch.~\ref{ch:observables}. For a clear theoretical overview, see e.g.~refs.~\cite{Baumann:2022mni,LaineProcacci2026InflationHotBigBang}. 
\\\\
Eventually, however, inflation needs to end. It should, at some point, be succeeded by the state of the ``hot big bang''. Such a hot, dense universe, filled with SM plasma, can then be seen as a start of an empirically better established thermal history of the universe. This transition between inflation and the hot big bang, dubbed ``reheating'', is not yet well established, with many directions explored (see e.g.~\cite{reheating_summary} for a review). The simplest realization assumes that during inflation, the universe supercooled, while instantaneously transferring all of its energy density to radiation afterwards (``instantaneous reheating''). A somewhat more realistic model is provided by parametric reheating, see refs.~\cite{Starob2,starob&comp}. In particular, after a period of inflation, the universe might experience a short ``preheating'' period, during which an inflaton field oscillates around the minimum of its potential, rapidly sourcing particles in an out-of-equilibrium process. This is followed by perturbative reheating which establishes a thermal bath. 
\\\\
This and similar approaches, however, call for a separate, poorly constrained epoch between inflation and the hot big bang. In the following, we explore a different approach, in which the hot big bang naturally follows from inflation, without the need for a separate reheating period.

\section{Warm inflation and smooth reheating}
An alternative to the conventional picture of emptying and supercooling the universe during inflation, followed by the generation of plasma and the onset of the hot big bang, is a model of smooth reheating. In this minimal framework, the inflaton couplings to the plasma are non-vanishing during inflation, allowing for a smooth transition to a radiation-dominated universe. If the couplings between the inflaton and the plasma are not minimal, but rather large, we normally speak of warm inflation, ref.~\cite{Warminflation_berera} (yet the model in principle also covers the minimal case of smooth reheating). An effective description of the low-energy degrees of freedom of the process is provided by a \textit{Langevin equation}, see eq.~\eqref{langevin}. On a microscopic level, the inflaton interacts with a plasma. These interactions are, on a macroscopic level, encapsulated by the effective friction coefficient $\Upsilon$. By the fluctuation-dissipation theorem, we then also expect the presence of thermal fluctuations, sourced by the noise $\varrho$ in eq.~\eqref{langevin}; see refs.~\cite{Berera_thermal_pert} and \cite{Berera_1994_noise} for a warm-inflation-specific and an earlier cosmological reference respectively. 
\\\\
The dissipative term, $\Upsilon$, provides a source of friction in addition to the Hubble rate, $H$, see eqs.~\eqref{hubble} and \eqref{LangevinBackground}. This is a source of many differences/benefits, as opposed to the standard, ``cold inflation'' picture. Aside from the conceptual benefit of having the reheating phase smoothly follow the inflationary one, there are other, practical benefits of the extra friction, too. For example, for the models of cold inflation, very flat inflaton potentials, $V$, are needed in order to keep the universe expanding exponentially for a sufficiently long  time to reproduce the observed amplitude of perturbations. However, the additional thermal friction slows down the inflaton field, allowing for a wider range of steeper potentials to realize inflation (this can be mathematically seen through modifications of ``slow-roll'' parameters\footnote{Slow-roll is a common approximation used during inflation, assuming the inflaton dynamics is mostly determined by its potential. The inflaton field changes very slowly.}). Hence, a number of potentials, e.g.~$\varphi^4$, which were excluded in the cold inflation scenario due to insufficient flatness, see ref.~\cite{Planck_inflation_2020}, are still allowed in the warm inflation setting. Additionally, the extra source of friction decreases the tensor-to-scalar ratio, $r$, which only has an upper bound experimentally (since inflation can now happen at a lower scale $H$, while tensor perturbations grow as $H^2$ --- see Ch.~\ref{ch:observables}). Thus, warm inflation could help save models predicting too high $r$.
\\\\
The early ideas of warm inflation can perhaps be traced back to ref.~\cite{Moss}, which discusses the existence of thermal friction during inflation; though the idea was also entertained in e.g.~refs.~\cite{Yokoyama,Zimdahl_1996}. Nonetheless, the modern warm inflation framework started with ref.~\cite{Warminflation_berera}. The subject was further developed in several subsequent works, see e.g.~refs.~\cite{Berera_1996,Chris_Graham_2009,Bastero_Gil_2011,Ramos_2013,Bastero_Gil_2014,Mirbabayi_2023,Bastero_Gil_2024,ballesteros2024monomialwarminflationrevisited}, with more precise quantum-field-theoretic treatment e.g.~in refs.~\cite{Simona1,Klose_2022,kcube2,UpsilonIR}. For reviews, see e.g.~refs.~\cite{overview_warminflation_rangarajan,Kamali_2023,overview_Berera,Yokoyama_1999}.

\section{Challenges of warm inflation }\label{challenges}
There are two types of challenges associated with describing the late stage of inflation and reheating. One is describing the interactions between a plasma and an inflaton field microscopically, i.e.~model-building. The other is describing the reheating process on an effective level, without specifying the microscopic details. In this thesis we focus on the latter; we elaborate on specific issues in Chs.~\ref{ch:gaugeintro}-\ref{ch:numintro}. We end the introduction by referring the reader to some model-building literature, see Ch.~\ref{ch:modelbuildintro}.

\subsection{Perturbation equations: gauge invariance, slow-roll}\label{ch:gaugeintro}
When consulting warm inflation literature, one might observe that no single set of perturbation equations is universally agreed upon. Perhaps the most important reason for this is gauge freedom, a topic introduced in Ch.~\ref{ch:gaugeinv}. In a theory with gauge freedom, there are two ways to proceed; one can either fix the gauge, or proceed in a gauge-invariant way. One might be tempted to fix the gauge, hoping to simplify equations of motion. This is in principle fine, as long as it is done consistently. For example, an early ref.~\cite{de_Oliveira_2001} derives the perturbation equations in the longitudinal gauge, while making use of the slow-roll approximation.
\\\\
However, a fully gauge-invariant approach has its benefits. For example, it ensures that no gauge artifacts are accidentally interpreted as physical, and that there are no other gauge inconsistencies. To illustrate the potential problem, let us look at a simple estimate for the power spectrum sometimes used in the literature (e.g.~ref.~\cite{Bastero_Gil_2019}). In such an estimate, the perturbation equations are solved in a particular gauge (e.g.~Newtonian gauge, $h=\vartheta=0,\;h_D=h_0$ in eq.~\eqref{pert_metric}), while the power spectrum is estimated from merely a rescaling of the inflaton field ($\mathcal{R}_\varphi\approx (H/\dot{\bar{\varphi}})\,\delta\varphi $, see eq.~\eqref{curvature0}). The latter would require a different gauge choice. This also has practical implications, see Ch.~\ref{ch;importanceofgaugeinvariance}. Essentially, even in the absence of plasma, the evolution of the gauge-invariant variable is different from the inflaton rescaling estimate, as soon as we depart from pure de Sitter (dS) expansion. Therefore, in precision computations, or computations at higher momentum scales (which exit the horizon later), the difference might be relevant.
\\\\
A good alternative to the gauge-fixed approach is to define gauge-invariant variables. This was first well-established in the cold inflation literature (see e.g.~\cite{Sasaki,Mukhanov,Bardeen:1980kt,Mukhanov:1990me,Hwang_2002}). In warm inflation, some works have already derived equations in a gauge-invariant way, for example ref.~\cite{Bastero_Gil_2011}, however, they still used the slow-roll approximation. In many other cases, the equations were solved in a particular gauge only, see e.g.~refs.~\cite{Bastero_Gil_2014,WarmSPy,ballesteros2024monomialwarminflationrevisited,WI2Easy}. We thus aim to derive and solve gauge-invariant equations, without making use of the slow-roll approximation.
\\\\
Deriving gauge-invariant equations requires defining gauge-invariant variables. The latter may include definitions of ``Bardeen potentials'' (ref.~\cite{Bardeen:1980kt}), or various definitions of ``curvature perturbations'', see e.g.~eq.~\eqref{curvature0}. While the former only include metric perturbations, the latter also include other, e.g.~inflaton perturbations, as already defined in cold inflation refs.~\cite{Sasaki,Mukhanov}. In warm inflation, one gets two extra degrees of freedom associated with the plasma, which need to be accounted for. Therefore, we construct two additional gauge-invariant curvature perturbations describing plasma dynamics, see eqs.~\eqref{curvature1}. We describe the evolution of these variables, and define the power spectra as the moduli squared of the individual curvature perturbations (this is not always the case in the literature; see e.g.~ref.~\cite{Bastero_Gil_2011}, where different ``curvature perturbations'' add up to a single power spectrum). While the inflaton curvature perturbation is expected to dominate at very early times, we show that it eventually decouples from the plasma. Then, in radiation domination, the evolution is governed by the plasma curvature perturbations instead.

\subsection{Noise treatment}\label{ch:noiseintro}
Another non-trivial aspect of warm inflation, and a source of differences between different approaches, is the handling of the noise. Properties of a Gaussian noise are given through its autocorrelator, see eq.~\eqref{correlator}. Its explicit expression is a subject of discussion. 
\\\\
In the high temperature limit, a text-book computation leads to the thermal noise autocorrelator being $\langle\varrho(t)\varrho(t')\rangle=2\Upsilon T \delta(t-t')$. While widely used, it only captures thermal fluctuations and is not valid at low temperatures, so further work is called for. On top of thermal fluctuations, one can also expect to have vacuum fluctuations, even in cold inflation. There are different ideas on how to describe that. A theoretically straightforward way to account for the quantum contributions is to start the evolution very early, when the universe can be well described by a Bunch-Davies vacuum. Provided that the physical momentum of perturbations is much larger than the Hubble scale, one can define quantum initial conditions. An alternative approach is inspired by the framework of ``stochastic inflation''. In this case, the inflaton field is split into long and short wavelength modes; the long modes are tracked by the Langevin equation, while the short ones influence the former through an effective ``quantum noise'' term. This is generally understood to capture quantum effects, and allows one to only start the evolution at a later time (when the physical momentum is comparable to the Hubble rate).
\\\\ 
Let us analyse the concept of ``quantum noise'' for a moment. The noise autocorrelator depends on both, the scale at which one does the splitting into short and long wavelength modes (dubbed ``coarse-graining scale''), as well as on the ``window function'' describing how the transition happens. The quantum noise autocorrelator sometimes quoted in WI literature is $\langle\varrho(t)\varrho(t')\rangle=\frac{H^2\sqrt{9+12\pi Q}}{\pi a^3}(1+2n)\delta(t-t')$, see ref.~\cite{Ramos_2013} with explicit expression in e.g.~ref.~\cite{Kamali_2023}. In the equation, the occupation number of short-wavelength inflaton modes $n$ can either be zero (vacuum state) or the Einstein-Boltzmann distribution (thermal population). In a practical implementation, the noise autocorrelator is sometimes represented as a sum of its ``quantum'' and its ``thermal'' contributions, with the latter being the high temperature classical noise autocorrelator (see e.g.~ref.~\cite{swim}).
The problem with this approach is, however, that it overcounts the thermal contribution (as explained in e.g.~ref.~\cite{ballesteros2024monomialwarminflationrevisited}). A simple way to see this is to note that the thermal part of the autocorrelator enters both the classical autocorrelator (through $T$) as well as the quantum one (through $n$).
\\\\
It is clear that the noise treatment, and with it the treatment of thermal and vacuum fluctuations, requires more nuance. Hence, in our work, we present an alternative, first-principles approach (see Ch.~\ref{ch:noise}). Instead of treating quantum effects on the level of the short-frequency noise, we set quantum-initial conditions in the Bunch-Davies vacuum at early times. Therefore we have a single noise term in our evolution equations. Using a matching computation, we make sure that the noise smoothly interpolates between quantum and classical domains, while avoiding the double counting problem. Furthermore, we show that the noise autocorrelator obtains its classical value in the high temperature limit, while it exponentially vanishes in the zero-temperature limit. On the level of perturbations, this means that in the regime of vanishing dissipation, the perturbations should be entirely of quantum origin, while where the dissipation is very large, they should be entirely of thermal origin, sourced by the noise term.

\subsection{Numerical implementation}\label{ch:numintro}
Warm inflation equations are in general too complicated to be solved analytically. Several authors have published their own numerical implementations, e.g.~\texttt{WarmSPy}, see ref.~\cite{WarmSPy}. The \texttt{Python} code successfully obtains a semi-analytical stochastic power spectrum. Yet, it also has some theoretical limitations, for example, it assumes that the dissipation is proportional to the Hubble rate, $\Upsilon/H=\text{const}$. Recently, a new \texttt{Mathematica} implementation was released, named \texttt{WI2easy},  see ref.~\cite{WI2Easy}. It is directly based on methods developed in refs.~\cite{ballesteros2024monomialwarminflationrevisited} and \cite{Ballesteros:2022hjk}, and obtains a semi-analytical, deterministic power spectrum, based on gauge-fixed equations and the ``quantum $+$ thermal noise'' treatment. The most recent publicly available code is (\texttt{D})\texttt{SWIM} \cite{swim,dswim}. The implementation is done either stochastically (\texttt{SWIM}), or using (the deterministic) matrix formalism (\texttt{DSWIM}), see Ch.~\ref{sec:matrixformalism}. In terms of evolution equations, (\texttt{D})\texttt{SWIM} is very similar to \texttt{WI2easy}; it also implements Newtonian-gauge perturbation equations with a power spectrum as defined in ref.~\cite{Bastero_Gil_2011}, and it implements the ``quantum $+$ thermal noise'' treatment. The implementation is done in \texttt{C++} and \texttt{Python}. The output is the full power spectrum, meant mainly for application in parameter estimation. 
\\\\
In our work, we provide our own numerical implementation. That is because the theoretical basis of our work is different than those used in previous implementations (gauge invariance, noise treatment, no slow-roll). We use \texttt{Mathematica} to predict several observables for a chosen momentum scale (see Ch.~\ref{ch:observables}), provided the necessary model parameters.

\subsection{Model-building}\label{ch:modelbuildintro}
While our work will be mostly agnostic on the microscopic details, with model-dependence captured in coefficients $\Upsilon,\; V,\;p_r$ (radiation pressure), and $e_r$ (radiation energy density), the explicit expressions of those coefficients are needed for a numerical implementation. Here we do not aim to develop a particular model, but rather we test our model-agnostic framework on the proposal developed by Berghaus, Drewes and Zell in ref.~\cite{ZellSM}, described in Ch.~\ref{sec:ModelBuilding}. We additionally note that our framework can be used to explore the model-independence of reheating predictions, see Ch.~\ref{chapter:conclusion}.
\\\\
We now mention an incomplete list of references on the model building. In general, they work around the problems of preventing too large quantum (see refs.~\cite{vaccor1,vaccor2,vaccor3,vaccor4}) and/or thermal corrections to the inflaton potential; see also Ch.~\ref{ch:SM} for more explanation. Some earlier investigations include, for example, ref.\cite{Yokoyama_1999}, which investigates a number of models, among other things suggesting that $T$-dependent potentials tend to lead to large temperature corrections possibly spoiling inflation. Some work on ``axion inflation'' can be found in e.g.~refs.~\cite{Zell_2025,ZellSM} (see Ch.~\ref{sec:ModelBuilding} and also references therein). Selected references on the friction term include, e.g.,~refs.~\cite{Fu_2022,B_deker_2022,ZellSM}. For the inflationary potential, see e.g.~refs.~\cite{Linde:1983gd,broadberry2026warminflationpseudoscalarcouplings,Freese:1990rb}. An overview of inflationary models is presented in ref.~\cite{encyclopedia}.

%Other, specific challenges of model building. Also, maintaining sufficient friction while not wanting quantum perurbative corrections is hard (so use sphaleron heating instead).  

%parameters determination (how do we measure/choose them (Upsilon comes from micro theory, fa $\approx$ Mpl might avoid swampland, hierarchies allowing our potential and constants)
%Different scales: inflaton mass, decay constant, confinement scale of YM plasma, T, initial inflaton value.

%problem outline: cosine potential solves quantum corrections (eta problem). For $\Upsilon$ you need coupling that breaks axion protection?

%\hl{But you automatically still get some quantum (loop corrections, eta problem, can be avoided with e.g.~ shift symmtery of the potential) and thermal corrections, no? And that is on top of the cosine potential you do discuss?}

%CMB ruled out simple monomial potentials (?) \cite{Planck:2018jri,BICEP:2021xfz}. There exist many (other) potentials, an overview is given in . Inflation numerical simulations (nonpert) -\cite{CaravanoCode}.

\chapter{The perturbed universe: essentials}\label{ch:essentials}
We start by introducing our setup: an isotropic and homogeneous metric tensor, a thermal plasma, and an inflaton field driving the expansion of the universe. We treat each of these three components on a perturbative level, focusing only on scalar perturbations.

\section{Gravitational perturbations}
Gravitational perturbations can be described by a perturbed metric tensor. We can decompose the perturbations into tensor, vector, and scalar parts. In this thesis, we only focus on the scalar components. One might wonder whether we are allowed to ignore the rest, and why we would focus on scalar components in particular. Since we are working at the linear level in perturbations, we may simplify our task by only considering one type of perturbations at a time. The scalar ones are particularly interesting, as they are responsible for creating the temperature fluctuations in the CMB and the large scale structure formation. Furthermore, scalar perturbations may lead to primordial black holes, which could partially represent dark matter content of our universe, see refs.~\cite{PBHHawking,PBHHawkingComp,PBHCarr:1975qj}, or ref.~\cite{PBSreviewCarr:2020xqk} for a review. Vector perturbations are generally thought of as negligible, decaying rapidly with the expansion of the universe. %\footnote{They would, however, be responsible for potential vortices, and could contribute to B-mode polarizations in CMB.} 
On the other hand, tensor modes (the transverse traceless parts of the metric tensor), represent propagating degrees of freedom, observable as gravitational waves. They are very interesting to compute as well, and contribute to an important observational constraint, tensor-to-scalar ratio (see Ch.~\ref{ch:observables}). While it is true that the different types of perturbations do not mix on the linear level, this is no longer the case when one goes to the second order. There, scalar perturbations could source gravitational waves, potentially providing an extra probe of early universe physics. In this thesis, however, we restrict ourselves to the linear regime.
\\\\
The specific metric we are using is the \textit{Friedmann-Lemaître-Robertson-Walker} (FLRW) metric. It describes an expanding, isotropic and homogeneous universe. The expansion is described by a time-dependent scale factor, $a(t)$. In four dimensions, the metric can be written as\footnote{One could also include a curvature parameter, $\kappa$. However, observations indicate $\kappa$ to be approximately zero, i.e.~the universe to be fairly flat.}
\begin{equation}
    ds^2=-dt^2+a(t)^2[dx^2+dy^2+dz^2]\;.
\end{equation}
From now on, the time dependence of $a(t)$ will be suppressed for better readability of the equations. This form of the metric, however, ignores the local density perturbations, such as galaxies. If one wishes to eventually describe those, one needs to perturb the FLRW metric. Over time, these perturbations grow and evolve into structures such as galaxy clusters or stars. Hence, our task is to describe those metric perturbations at early times.
\\\\
Since we are working in 4 dimensions, the metric tensor consists of 16 degrees of freedom, yet only 10 of them are independent, due to the symmetry of the metric tensor, $g_{\mu\nu}=g_{\nu\mu}$. Among those, 4 are of the scalar type ($h$, $h_D$, $h_0$, $\vartheta$), parameterizing the metric as
\begin{subequations}\label{pert_metric}
    \begin{align}
g^{ }_{\mu\nu} & \equiv  
 a^2_{ }
 \left(
  \begin{array}{cc} 
   -1 - 2 h^{ }_0 & \partial^{ }_i h \\ 
   \partial^{ }_i h & 
   ( 1 - 2 h^{ }_{D}) \delta^{ }_{ij} + 
   2 \bigl( 
     \partial^{ }_i \partial^{ }_j - \delta^{ }_{ij} \frac{\nabla^2}{3}
     \bigr) \vartheta
  \end{array} 
 \right) %% + \rmO(\delta^2) 
 \;, \label{pert_metric_a} \\[2mm]
%%%%%
 g^{\mu\nu}_{ } & =  
 a^{-2}_{ }
 \left(
  \begin{array}{cc} 
   -1 + 2 h^{ }_0 & \partial^{ }_i h \\ 
   \partial^{ }_i h & 
   ( 1 + 2 h^{ }_{D}) \delta^{ }_{ij} - 
   2 \bigl( 
     \partial^{ }_i \partial^{ }_j - \delta^{ }_{ij} \frac{\nabla^2}{3}
     \bigr) \vartheta
  \end{array} 
 \right) + \mathcal{O}(\delta^2) 
 \;.\label{pert_metric_b}
    \end{align}
\end{subequations}
This metric is written in conformal time $\tau$, which differs from the physical time $t$ by a scale factor, $a\; d\tau=dt$. Physical time derivatives are denoted by a dot, while a prime denotes a derivative with respect to conformal time. The above metric parametrization is of course not unique, and can be related to other parameterizations in the literature by simple redefinitions. The inverse metric $g^{\mu\nu}$ is obtained from the standard identity from General Relativity (GR), $g_{\mu\nu}g^{\mu\kappa}=\delta^\kappa_\nu$.
\\\\
Having introduced the metric perturbations, one can go further and describe how they evolve over time. That information is provided by the Einstein Field Equations (EFE),
\begin{equation}\label{efe}
    G^\mu_{\;\nu}=8\pi G \; T^\mu_{\;\nu}\;.
\end{equation}
We now provide a brief summary of GR essentials needed to make use of this equation. The left-hand side is the Einstein tensor, describing the curvature of spacetime. It is given by a combination of metric tensors and their derivatives,
\begin{equation}\label{GRbasics1}
  G_{\mu\nu}=R_{\mu\nu}-\frac{1}{2} g^{\kappa\sigma}R_{\kappa\sigma} g_{\mu\nu} \;, 
\end{equation}
      where        \begin{subequations}\label{GRbasics}
\begin{align}
R_{\mu\nu}&=\partial_{\kappa}\G^\kappa_{\mu\nu}-\partial_\mu \G^\sigma_{\nu\sigma}+\G^\tau_{\tau\alpha}\G^{\alpha}_{\mu\nu}-\G^\beta_{\mu\chi}\G^\chi_{\beta\nu}\;, \\
        \G^\alpha_{\mu\nu}&=\frac{1}{2}g^{\alpha\kappa}\left(\partial_\mu g_{\nu\kappa}+\partial_\nu g_{\mu\kappa}-\partial_\kappa g_{\nu\mu}\right)\;.
\end{align}
\end{subequations}
Here $R_{\mu\nu}$ is the Ricci tensor and $\G^\mu_{\alpha\beta}$ is the Christoffel symbol (also called Levi-Civita connection). When the latter is added to a partial derivative, it forms a \textit{covariant derivative} $\nabla_\mu$, i.e.~a derivative operator that transforms as a tensor. When acting on a rank 1 tensor $f^\nu$, the covariant derivative reads\footnote{A pedagogical comment: as is the convention in GR, an index repeated upstairs and downstairs (such as $\beta$ in eq.~\eqref{cov}) implies a summation over all the values this index can take (for example, $\beta=t,x,y$ or $z$, for all the spacetime dimensions). A lonely index just tells how many equations there are. For example, $\mu$ and $\nu$ can each take 4 values, which means eq.~\eqref{cov} is really 16 equations. The right-hand side of each eq.~\eqref{cov} has 5 terms (one derivative term, and 4 coming from the dummy index $\beta$).}
\begin{equation}\label{cov}
    \nabla_\mu f^\nu=\partial_\mu f^\nu +\G^\nu_{\beta\mu}f^\beta\;.
\end{equation}
We now return to eq.~\eqref{efe}; its right-hand side is given by the energy-momentum tensor, which is discussed in Ch.~\ref{mandrpert}. The tensor describes the matter content of the universe (in our case, a thermal plasma and an inflaton field). Together with the left-hand side, the EFE provide a simple message: matter tells spacetime how to curve, and spacetime curvature tells matter how to move.
\\\\
Since we are working in the perturbative picture, the EFE from eq.~\eqref{efe} are further split into two sets of equations: background equations,
\begin{equation}\label{efebck}
      \bar{ G}^\mu_{\;\nu}=8\pi G \; \bar{ T}^\mu_{\;\nu}\;,
\end{equation}
and first-order perturbations,
\begin{equation}\label{efepert}
     \delta G^\mu_{\;\nu}=8\pi G \; \delta T^\mu_{\;\nu}\;.
\end{equation}
Here we used $G^\mu_{\;\nu}=\bar{ G}^\mu_{\;\nu}+\delta G^\mu_{\;\nu}$. The tensors $T^\mu_{\;\nu}$ and $g_{\mu\nu}$ are expanded in exactly the same way. In principle, we could already write out the left-hand side of the perturbed EFE in terms of the perturbations of the metric tensor from eq.~\eqref{pert_metric}. We will do that in eqs.~\eqref{Gperturbed}, but first let us introduce the matter and the radiation content of the universe, together with their perturbations, appearing on the right-hand sides of eqs.~\eqref{efebck} and \eqref{efepert}.

\section{Matter and radiation perturbations}\label{mandrpert}
The energy-momentum tensor describing the matter and radiation content of the universe is assumed to take the form
\begin{equation}\label{stressenergy}
     T^{\mu}_{\;\nu}
   \; \equiv \; 
 T^{\mu}_{\;\nu\;\text{ideal}}
   \; \equiv \;  ( e + p ) \,  u^{\mu}_{ }u_{\nu}+ p \, g^{\mu}_{\;\nu } +
 \varphi^{;\mu}_{ }\varphi_{;\nu}
 - \frac{ g^{\mu}_{\;\nu }  \varphi^{ }_{;\alpha} \varphi^{;\alpha}_{ } }{2} 
 \;.
\end{equation}
Partial derivatives are written as $\partial_\mu(...)=(...)_{,\mu}$, while covariant derivatives are written as $\nabla_\mu(...)=(...)_{;\mu}$. Since $\varphi$ is a scalar, its covariant derivative is just a partial derivative. If a covariant derivative is applied to $T^{\mu\nu}$ (imposing the stress-energy conservation), the combination $\varphi^{;\mu}$ should be treated as a tensor and the overall $\nabla^{\mu}$ should not be replaced by $\partial^\mu$ alone.
\\\\
Eq.~\eqref{stressenergy} includes several terms: $\varphi(x)\equiv \varphi$ is the inflaton field, $e$ is the energy density including the radiation and inflaton components, $p$ is the pressure, and $u^\mu$ is the four-velocity of the plasma. The inflaton field has an associated potential $V(\varphi,T)\equiv V$, which enters the expressions $e$ and $p$. These do not include derivatives of the inflaton field (``kinetic energy'') --- rather, as the inflaton is not just a part of the plasma, the derivative terms are kept separate (the last two terms in eq.~\eqref{stressenergy}). The scalar quantities are now expanded to linear order in terms of temperature and inflaton perturbations, which are themselves expanded around their background values,
\begin{equation}\label{expansionTPhi}
    T\equiv \bar{T}+\delta T \;,\;\;\;\;\;\;\;\;\;\;\;\;\;\;\;\;\;\;\;\;\;\; \varphi\equiv\Bar{\varphi}+\delta \varphi\;.
\end{equation}
In particular, we make the expansions
\begin{subequations}
\begin{align}
    p&=\bar p + \delta p=\bar p + \left( \bar p_{,T}\;\delta T+\bar p_{,\varphi} \; \delta \varphi \right)\;,\\
    e&=\bar e + \delta e=\bar e + \left( \bar e_{,T}\;\delta T+ \bar e_{,\varphi} \; \delta \varphi \right)\;,\\
    V&=\bar V + \delta V=\bar V + \left(\bar V_{,T}\;\delta T+ \bar V_{,\varphi} \; \delta \varphi \right)\;.
\end{align}
\end{subequations}
We suppressed the dependencies on $T$ and $\varphi$ for better readability. The average pressure can be written as a sum of its radiation $\bar p_r$ and inflaton components $\bar V$,
\begin{equation}\label{pressureaverage}
    \bar p(T,\varphi)=\bar{p}_r(T)-\bar{V}(T,\varphi) \;.
\end{equation}
Its perturbation becomes
\begin{equation}\label{p}
    \delta p =\partial_T \left(\bar p_r-\bar V\right)\delta T-\partial_{\varphi} \bar V \delta   \varphi= \bar s\; \delta T-\bar{V}_{,\varphi} \delta \varphi\;.
\end{equation}
Here we introduced the entropy density ($\bar s\equiv \bar p_{,T}$). From $\bar e+\bar p=T\bar s$ we can deduce the background energy density,
\begin{equation}\label{energyaverage}
    \bar e=T \, \bar p_{r,T}- T\,  \bar V_{,T} -\bar p_r + \bar V =\bar e_r + \bar V -T\, \bar V_{,T}\;.
\end{equation}
The associated perturbation to the energy density reads
\begin{equation}\label{e}
    \delta e=\bar c\, \delta T+\left(\bar V_{,\varphi} -T\, \bar V_{,T\varphi}\right)\delta\varphi=\left(\bar e_{r ,T} -T\, \bar V_{,TT}\right)\delta T+\left(\bar V_{,\varphi} -T\, \bar V_{,T\varphi}\right)\delta\varphi\;.
\end{equation}
Here, the heat capacity was defined as $\bar c\equiv\bar e_{,T}$. 
\\\\
One more quantity needs to be expanded before we can write the energy-momentum tensor in the desired form $T_{\mu\nu}=\bar T_{\mu\nu}+\delta T_{\mu\nu}$. That is the four-velocity, $u^\mu=\bar u^\mu+\delta u^\mu$, which we expand around a homogeneous and isotropic background, i.e.~$\bar u^\mu = a^{-1} (1,0,0,0)$. Contracting this with the metric tensor gives $\bar u_\mu = a(-1,0,0,0)$. The perturbation we initially parameterize as $\delta u^\mu=(\delta u^0,-\partial_iv)$. From $u_\mu u^\mu=-1$ it follows $\delta u^t=-h_0 a^{-1}$, and hence 
\begin{equation}\label{4velocity}
    u^\mu=a^{-1} (1-h_0,\;-\partial_i v)\;\;\;\;\;,\;\;\;\;\;\;\;\;\;\;\;\;\;\;\;\;\;\;\;\;u_\mu=a (-1-h_0,\;\partial_i (h-v))\;.
\end{equation}
With this in hand, we write the background stress-energy tensor as
\begin{equation}\label{plasmabck}
 {\bar T^{0}_{ }}{}^{ }_0  
 = 
 - \, 
 \biggl[
  \frac{(\bar\varphi ')^2_{ }}{2 a^2} + \bar e 
 \biggr] 
 \;, \quad
 {\bar T^{i}_{ }}{}^{ }_j
 = 
 \delta^{ }_{ij}
 \,  
 \biggl[
  \frac{(\bar\varphi ')^2_{ }}{2 a^2} + \bar p 
 \biggr] 
 \;. 
\end{equation}
For the perturbations,\footnote{One might notice that $T^{0}_{\;\;i}\neq T^{i}_{\;\;0}$. This is not a problem; while it indeed holds that $T_{\mu\nu}=T_{\nu\mu}$, the same cannot be said for its mixed (not co-/contra-variant) version, for which $T^{0}_{\;\;i}=-T^{i}_{\;\;0}+(\bar T^i_{\;\;i}-\bar T^0_{\;\;0})\partial_i h$. %Wrongly requiring $T^{0}_{\;\;i}= T^{i}_{\;\;0}$ would lead one to incorrectly conclude the isocurvature perturbation $S_v$ from \eqref{isocurvature} vanishes when the universe undergoes a perfect exponential expansion.
} the inflaton kinetic terms give
\begin{equation}
  \delta {T^{\mu}_{ }}^{ }_\nu \, |^{ }_{\varphi} 
 = 
 \frac{ \bar\varphi' }{a^2_{ }} \, 
 \left(\, 
   \begin{array}{cc}
     h^{ }_0\, \bar\varphi' - \delta \varphi'
     & 
    -\delta\varphi^{ }_{,j} \\  
     h^{ }_{,i}\, \bar\varphi'  + \delta\varphi^{ }_{,i}    &
     \quad \delta^{ }_{ij}\,
     \bigl[ 
     - h^{ }_0\, \bar\varphi' + \delta \varphi'
     \bigr]
   \end{array}
 \,\right)
 \;, \label{delta_Tmunu_varphi}
\end{equation}
while the radiation part becomes
\begin{equation}
  \delta {T^{\mu}_{ }}^{ }_\nu \, |^{ }_{r} 
 = 
 \left(\, 
   \begin{array}{cc}
    -\delta e & 
   (\bar{e} + \bar{p})\, (h-v)^{ }_{,j} \\ 
   (\bar{e} + \bar{p})\, v^{ }_{,i} & 
   \delta^{ }_{ij} \, \delta p 
   \end{array}
 \,\right)
 \;. \label{delta_Tmunu_r}
\end{equation}
Note that $\delta p$ and $\delta e$ also contain contributions from $\bar{V}$, as given by \eqref{p} and \eqref{e}.
\\\\
Apart from the Einstein field equations \eqref{efebck} and \eqref{efepert}, we impose two more equations at the perturbative level.\footnote{While stress-energy conservation equations are not independent of the EFE, they will help us to simplify evolution equations in the next chapters.} One follows from the Bianchi identity and is given by the stress-energy conservation requirement,
\begin{equation}\label{conserv}
    \nabla_\mu  T^{\mu}_{\;\nu}=0\;.
\end{equation}
The other is given by a Langevin equation,
\begin{equation}\label{langevin}
     {\varphi^{;\mu}_{ }}^{ }_{;\mu} - V^{ }_{,\varphi}  - \Upsilon \, u^{\mu}_{ } \varphi^{ }_{,\mu}
  + \varrho 
 \; = \; 0
 \;. 
\end{equation}
The Langevin equation is an effective description of a system with slow degrees of freedom/long wavelengths (the inflaton field) that couple weakly to fast degrees of freedom (a strongly interacting plasma). It is a rather universal equation applied in many areas of physics, from simple Brownian motion to, for example, heavy quark diffusion, nucleation dynamics in first-order phase transitions, or warm inflation. Due to its assumed strong self-interactions, the plasma can be thought of as being in thermal equilibrium and thus having a temperature $T$. On the other hand, the slow degree of freedom $\varphi$ needs not be in thermal equilibrium. It is the power of this approach that one does not need to assume whether $\varphi$ is thermalized or not, since its thermalization is dealt with dynamically.
\\\\
The Langevin equation as written in eq.~\eqref{langevin} is quite generic. The specific physical context enters through the choice of the inflaton potential, $V(\varphi,T)$, and the coupling coefficient between the fast and slow degrees of freedom, $\Upsilon(\varphi,T)$. The former could be introduced in an effective theory sense, i.e.~by writing down the allowed terms in the effective Lagrangian. Alternatively, if an ultraviolet (UV) theory is known, the effective potential could be derived from it by integrating out the UV degrees of freedom. For now we shall be agnostic about the specific forms of $V(\varphi,T)$ and $\Upsilon(\varphi,T)$, and thus keep a certain level of model-independence in our analysis. 
\\\\
Aside from the Klein-Gordon terms and the dissipative term proportional to $\Upsilon$, the Langevin equation \eqref{langevin} also includes a stochastic noise term, $\varrho$. It is responsible for sourcing thermal fluctuations. We will study it in detail in Ch.~\ref{ch:noise}.
\\\\
All in all, the Langevin equation describes an initially dominant inflaton field dissipating its energy density to the surrounding plasma, while the plasma back-reacts by fluctuations.
The two effects are related by a fluctuation-dissipation theorem: if the dissipation ($\Upsilon$) is large, the noise ($\varrho$) should be non-negligible as well. In the context of an expanding universe, large is defined as relative to the expansion of the universe. The latter is described by the Hubble rate,
\begin{equation}\label{hubble}
    H\equiv\frac{\partial_t a}{a}\;.
\end{equation}
The time derivative is defined in terms of the physical time. When we wish to express the Hubble parameter in terms of conformal time, as given in the metric \eqref{pert_metric}, we denote it by $\mathcal{H}=a'/a$.  Now, if the dissipation is large, that is $\Upsilon \gg H$, we talk about the ``strong regime''. On the other hand, if the dissipation is small, $\Upsilon \ll H$, we are in the ``weak regime''.
\\\\
In the following, the three sets of equations \eqref{efe}, \eqref{conserv}, and \eqref{langevin}, will be expanded to linear order (Ch.~\ref{ch:evol_eq_bck_pert_basic}) and combined into a convenient set of model-agnostic perturbation equations (Ch.~\ref{ch:gaugeinv}, eqs.~\ref{result}).

\section{Evolution equations: background and perturbations}\label{ch:evol_eq_bck_pert_basic}
Prior to deriving the perturbation equations, we should derive the background equations. For that purpose, we use the background Einstein field equations \eqref{efebck}, the stress-energy tensor conservation requirement \eqref{conserv} together with the background stress-energy tensor itself \eqref{plasmabck}, and the Langevin equation \eqref{langevin}. Starting by expanding the contractions in the Langevin equation, taking into account the metric \eqref{pert_metric} and the four-velocity \eqref{4velocity} perturbations, we find
\begin{subequations}
    \begin{align}  
    \nabla^\mu\nabla_\mu\,\varphi&= \frac{1 - 2 h^{ }_0}{a^2_{ }}\left[
  -\varphi'' + \nabla^2_{ }\varphi
 + \left( h_0' + 3 h_{D}' + \nabla^2_{ } h - 2\mathcal{H} \right) \varphi' 
 + \mathcal{O}(\delta^2_{ })\right]\;,
    \\
 u^\mu\partial_\mu\,\varphi&=\frac{ 1 - h^{ }_0}{a}\left[ 
   \varphi' + \mathcal{O}(\delta^2_{ })
 \,\right]\;.
    \end{align}
\end{subequations}
We now expand the inflaton field, as well as the potential and the friction coefficient, in terms of inflaton and temperature perturbations as
\begin{equation}
    V(\varphi,T)=\bar V+\bar V_{,\varphi}\, \delta\varphi+\bar V_{,T}\,\delta T \;\;,\;\;\;\;\;\;\Upsilon(\varphi,T)=\bar\Upsilon+\bar\Upsilon_{,\varphi}\,\delta\varphi+\bar\Upsilon_{,T}\,\delta T\;.
\end{equation}
With this in hand, the Langevin equation \eqref{langevin} becomes
\begin{equation}
\begin{split}
     &-\left(  \frac{1}{a^2_{ }}\left[
  \bar\varphi'' + 2\mathcal{H}  \bar\varphi'
 \right]+\bar V_{,\varphi}+\bar\Upsilon\frac{ \bar\varphi' }{a}\right)-\left(\bar V_{,\varphi T}+\bar\Upsilon_{,T} \frac{ \bar\varphi' }{a}\right)
\delta T\\   
    &+ \frac{1}{a^2_{ }}\left[
   \nabla^2_{ }\delta\varphi -\delta\varphi''
 - 2\mathcal{H}  \delta\varphi' 
\right]-\bar V_{,\varphi\varphi}\,\delta\varphi-\bar\Upsilon \frac{ \delta\varphi'  }{a}-\bar\Upsilon_{,\varphi} \frac{ \bar\varphi' }{a}
\delta\varphi\\
 &+\frac{1}{a^2_{ }}\left[
  h_0' + 3 h_{D}' + \nabla^2_{ } h    
 \right]\,\bar\varphi'+\frac{ 2 h^{ }_0}{a^2_{ }}\left[
  \bar\varphi'' 
  +2\mathcal{H} \bar\varphi' 
 \right]+\bar\Upsilon\, \frac{  h_0\bar\varphi' }{a}+\varrho +\mathcal{O}(\delta^2_{ })=0    \;.
\end{split}
\end{equation}
We can now split it into the background and first-order equations. We assume the inflation works within a minimal set-up by treating the noise $\varrho$ on the level of a perturbation. This gives the background Langevin equation,
\begin{equation}\label{LangevinBackground}
  \bar\varphi''
 + \left(   2\mathcal{H}+a\Upsilon \right)\, \bar\varphi' 
+a^2V_{,\varphi}
   =0  \;.
\end{equation}
From here onwards, we shall omit the bar on top of the potential, temperature and the friction when discussing background quantities, whenever the distinction can be clearly inferred from the context.
\\\\
We also explicitly write the first-order Langevin equation, using eq.~\eqref{LangevinBackground} to simplify it,
\begin{equation}\label{LangevinPerturbation}
\begin{split}
\delta\varphi''
 &+ \left(2\mathcal{H} + a \Upsilon\right)  \delta\varphi'
 - \nabla^2 \delta\varphi 
 + \left( a^2 V^{ }_{,\varphi\varphi}  
           + a \bar\varphi' \Upsilon^{ }_{,\varphi} \right) 
  \, \delta\varphi  \\
 &+ \,
  \left(
   2 a^2 V^{ }_{,\varphi} 
 + a \Upsilon \bar\varphi'
 \right)\, h^{ }_0
 - \bar\varphi' \,
 \left( h_0' + 3 h_{D}' + \nabla^2 h \right)  
 + 
 \left( a^2 V^{ }_{,\varphi T} + a \bar\varphi' \Upsilon^{ }_{,T} \right)
 \, \delta T
 = 
 a^2_{ }\varrho
 \end{split}
\end{equation}
As the next step, we derive the background and the first-order EFE. For that purpose, we first compute the Einstein tensor components, using  eqs.~\eqref{pert_metric}-\eqref{efebck},
\begin{equation}
    \bar G^0_{\;0}=-3\left(\frac{a'}{a^2}\right)^2\;\;,\;\;\;\;\;\;\;\; \bar G^1_{\;1}=\bar G^2_{\;2}=\bar G^3_{\;3}=\left(\frac{a'}{a^2}\right)^2-2\left(\frac{a''}{a^3}\right)\;.
\end{equation}
After a long but straightforward computation, we obtain the components of the perturbed Einstein tensor,
\begin{subequations}\label{Gperturbed}
    \begin{align}
        \delta G_{\;0}^{0}&=\frac{2}{a^2}\left(3\mathcal{H}^2h_0+\mathcal{H}\left(3h_D'+\nabla^2h\right)-\nabla^2\left(h_D+\frac{\nabla^2\vartheta}{3}\right)\right)\;,\label{Gperturbed_a}
        \\ 
        \delta G_{\;i}^{0}&=-\frac{2}{a^2}\partial_i\left(\mathcal{H}h_0+\left(h_D+\frac{\nabla^2\vartheta}{3}\right)'\right)\;,\label{Gperturbed_c} 
        \\
        \delta G_{\;j}^{i}&=-\frac{1}{a^2}\partial_i\partial_j\left(h_0+\left(2\mathcal{H}+\partial_\tau\right)(h-\vartheta')-\left(h_D+\frac{\nabla^2\vartheta}{3}\right)\right)\;,\label{Gperturbed_d}
        \\
        \delta G_{\;1}^{1}&=\frac{2}{a^2}\left(\mathcal{H}h_0'+\left(2\mathcal{H}'+\mathcal{H}^2\right)h_0+\left(\partial_\tau+2\mathcal{H}\right)\left(h_D+\frac{\nabla^2\vartheta}{3}\right)'\right)\;.\label{Gperturbed_b}
    \end{align}
\end{subequations}
We only explicitly wrote down the distinct components (note that $ \delta G_{\;1}^{1}= \delta G_{\;2}^{2}=\delta G_{\;3}^{3}$). Additionally, in eq.~\eqref{Gperturbed_d}, $i\neq j$.\\\\ Fortunately, we have already obtained the stress-energy tensor in its perturbed form in eqs.~\eqref{plasmabck}-\eqref{delta_Tmunu_r}. We can now additionally use eq.~\eqref{efebck} to obtain the background equations for $\mathcal{H}$ and its derivative,
\begin{equation}\label{EFEbckexpanded}
    \mathcal{H}^2_{ } = \frac{4\pi G}{3}
 \bigl[ (\bar\varphi')^2_{ }  + 2 a^2 \bar{e}
 \bigr]
 \;, \quad 
 \mathcal{H}' = \frac{4\pi G}{3}
 \bigl[ -  2 (\bar\varphi')^2_{ } - a^2 (\bar{e} + 3 \bar{p})
 \bigr]\;.
\end{equation}
In Ch.~\ref{ch:gaugeinv} we make use of the linear combinations of the background identities from \eqref{EFEbckexpanded}, in particular,
\begin{subequations}\label{bck1}
\begin{align}
    \HH^2 - \HH' &= 4 \pi G \Bigl[ (\bar\varphi')^2 + a^2(\bar{e} + \bar{p}) \Bigr] \;, \label{bck1_a} \\
    2 \HH^2 + \HH' &= 4 \pi G \, a^2(\bar{e} - \bar{p}) \;. \label{bck1_b}
\end{align}
\end{subequations}
Additionally, it will be useful to take a derivative of eqs.~\eqref{bck1},
\begin{subequations}\label{bck2}
    \begin{align}
        2\HH\HH' - \HH'' \; &= \; 4\pi G  \bigl\{ 
   2 \bar\varphi' \bar\varphi'' + 
  a^2 \bigl[ \bar{e}' + \bar{p}' + 2\HH(\bar{e} + \bar{p}) \bigr] \;,\label{bck2_a} \\
  \HH'' + 2 \HH \HH' - 4 \HH^3 \; &= \; 4 \pi G  a^2(\bar{e}' - \bar{p}')\;.\label{bck2_b}
    \end{align}
\end{subequations}
We now use eq.~\eqref{efepert} to obtain the perturbed equations. The equations below correspond to components $\delta G_{\;0}^{0}$, $\delta G_{\;i}^{0}$, $\delta G_{\;j}^{i}$, and $\delta G_{\;1}^{1}$, respectively (where, just as before, $i\neq j$, and $\delta G_{\;1}^{1}= \delta G_{\;2}^{2}=\delta G_{\;3}^{3}$),\footnote{This gives a full set of Einstein field equations. For example, upon using the background EFE, $G^t_{\;x}=8\pi G T^t_{\;x}$ is equivalent to $G^x_{\;t}=8\pi G T^x_{\;t}$.}
\begin{subequations}\label{EFEpertexpanded}
\begin{align}
    3\mathcal{H}^2h_0+\mathcal{H}\left(3h_D'+\nabla^2h\right)-\nabla^2\left(h_D+\frac{\nabla^2\vartheta}{3}\right)&=\label{EFEpertexpanded1}\\4\pi G\,\bigl[\,h^{ }_0\, &\bar\varphi'^2 - \bar\varphi'\delta \varphi'-a^2(\bar e_{,T}\,\delta T+\bar e_{,\varphi}\,\delta \varphi)\bigr]\nonumber\,,
    \\
    \mathcal{H}h_0+\left(h_D+\frac{\nabla^2\vartheta}{3}\right)'&=4\pi G\,\left[\,\bar\varphi'\delta\varphi-a^2(\bar e+\bar p)(h-v)\right]\label{EFEpertexpanded2}\,,
    \\
    h_0+\left(2\mathcal{H}+\partial_\tau\right)(h-\vartheta')-\left(h_D+\frac{\nabla^2\vartheta}{3}\right)&=0\label{EFEpertexpanded3}\,,
    \\
    \mathcal{H}h_0'+\left(2\mathcal{H}'+\mathcal{H}^2\right)h_0+\left(\partial_\tau+2\mathcal{H}\right)\left(h_D+\frac{\nabla^2\vartheta}{3}\right)'&=\label{EFEpertexpanded4}\\
    4\pi G\,\bigl[\,\bar\varphi'\delta&\varphi'-h_0\bar\varphi'^2+a^2\left(\bar p_{,T}\,\delta T+\bar p_{,\varphi}\,\delta\varphi\right)\bigr] \,.\nonumber
\end{align}    
\end{subequations}
In the process, we have assumed the perturbations vanish at spatial infinity, such that $\partial_if=0$ implies $ f=0$.\footnote{To see that, consider for example equation $\delta G^t_{\;x}-8\pi G \delta T^t_{\;x}=0$. Expanding it out gives an equation of the form $\der_x(...)=0$. This means that $(...)=f(t,y,z)$, and not vanishing a priori. However, equations $\delta G^t_{\;y}-8\pi G \delta T^t_{\;y}=0$ and $\delta G^t_{\;z}-8\pi G \delta T^t_{\;z}=0$ give identical requirements that need to be satisfied at the same time, except that $x$ is replaced by $y, z$, respectively. Consequently, one can really only allow a time-dependent constant $f(t)$, without any spatial dependence: $(...)=f(t)$. We can then, however, require perturbations to vanish at spatial infinity, fixing $f(t)=0$ and leading to \eqref{EFEpertexpanded2}.}  
\\\\
Lastly, let us write down the stress-energy conservation equations. This will lead to more equations than there are variables (metric, inflaton and plasma perturbations), hence they are not all independent. Nonetheless, writing them out explicitly turns out to be useful. On the background level, only the zeroth component contributes. We expand $\nabla_\mu \bar T^\mu_{\;\;0}=(\partial_0+\Gamma^\mu_{\mu0})\,\bar T^0_{\;\;0}-\Gamma^\nu_{\mu0}\bar T^\mu_{\;\;\nu}=0$, and after explicitly writing out the Christoffel symbols and the energy-momentum tensor, we obtain
\begin{equation}\label{plasmaconservbck}
    \bar\varphi'\left(\bar\varphi''+2\mathcal{H}\bar\varphi'\right)+a^2\left[\,3\mathcal{H}(\bar e+\bar p)+ \bar e'\,\right]=0\;.
\end{equation}
On the level of first-order perturbations, we get two equations describing energy-momentum conservation. The $\nabla_\mu \bar T^\mu_{\;\;i}=0$ component gives
\begin{equation}\label{plasmaconservP1}
     \bigl( \bar\varphi'' + 2 \HH \bar\varphi'\bigr) \delta \varphi
 - 
 a^2_{ } \bigl\{ 
     \bar{p}^{ }_{, T} \delta T
   + \bar{p}^{ }_{,\varphi} \delta\varphi
   + [(\bar{e} + \bar{p})(h-v)]'
   + (\bar{e} + \bar{p}) [ h^{ }_0 + 4 \HH  (h-v) ]
 \bigr\}
 = 
 0\;.
\end{equation}
On the other hand, $\nabla_\mu \bar T^\mu_{\;\;0}=0$ results in
\begin{equation}\label{plasmaconservP2}
\begin{split}
   & \bar\varphi' \, 
 \bigl(\, \delta\varphi'' - \nabla^2 \delta\varphi \,\bigr)
 + 
 \bigl(\, \bar\varphi'' + 4\HH \bar\varphi' \,\bigr)
 \delta\varphi'
 - 
 \bigl(\,
  h_0' + 3 h_{D}' + \nabla^2 h 
 \,\bigr) 
 (\bar\varphi' )^2_{ }+\\
 &- 2 h^{ }_0\, \bar\varphi' 
 \bigl(\, \bar\varphi'' + 2 \HH \bar\varphi' \,\bigr)
+
 a^2\bigl[\, 
 \delta e' + 3\HH \bigl( \delta e + \delta p \bigr)
 - \bigl( \bar{e} + \bar{p} \bigr)
   \bigl(\,
   \nabla^2 v + 3 h_{D}'  
   \,\bigr)
 \,\bigr]
 \; = \; 
 0 \;.
\end{split}
\end{equation}
The background equations are hence given by eqs.~\eqref{LangevinBackground}, \eqref{EFEbckexpanded}, and \eqref{plasmaconservbck}. The perturbation equations are given by eqs.~\eqref{LangevinPerturbation}, \eqref{EFEpertexpanded}, and \eqref{plasmaconservP1}-\eqref{plasmaconservP2}. These will serve as a starting point for Part \ref{part:analyt}. Before rewriting them in a more useful form, we discuss one of the central issues of our analysis, namely gauge invariance.
\newpage
\part{Analytical derivation}\label{part:analyt}

\chapter{Gauge invariance}\label{ch:gaugeinv}
After a brief discussion of gauge invariance in GR (Ch.~\ref{ch:gaugetransforms}), we present a convenient set of gauge-invariant variables (Ch.~\ref{ch:ginvpar}). We then rewrite the perturbation equations from Ch.~\ref{ch:essentials} in terms of these gauge-invariant variables in Ch.~\ref{sec:gaugeinveq}, with the final set of equations given in eqs.~\eqref{result}. \\\\
The present chapter is supplemented by two more chapters, together forming Part \ref{part:analyt} of the thesis. Specifically, Ch.~\ref{ch:extraremarks} discusses eqs.~\eqref{result} in different regimes and highlights the importance of the gauge-invariant treatment. We finish by deriving the analytic form of the noise autocorrelator in Ch.~\ref{ch:noise}.

\section{Gauge transformations}\label{ch:gaugetransforms}
%\textbf{ $https://arxiv.org/pdf/hep-ph/0210162$ page 55 equation 213.}
We have so far obtained first-order equations in terms of 7 degrees of freedom: $\delta \varphi,\,\delta T,\,v,\,h,\,h_D,\,h_0,$ and $\vartheta$. Yet, they are not all independent. The reason is the gauge freedom in GR, in other words, the freedom of choosing coordinates. Starting with coordinates $x^\mu$, one can go to a different coordinate system, $x^\mu \rightarrow \tilde{x}^\mu= x^\mu+\xi^\mu$. The small parameter $\xi^\mu \equiv \xi^\mu(x)$  introduces two scalar gauge parameters. To see that, we decompose the 4-vector $\xi^\mu$, written as
\begin{equation}
    \xi^\mu=\begin{pmatrix}
        \xi^0\\
        \boldsymbol{\xi}
    \end{pmatrix}\;,
\end{equation}
into its scalar and vector components. While $\xi^0$ is already a scalar, the vector components $\boldsymbol{\xi}$ can be decomposed using Helmholtz decomposition. This splits the vector in an irrotational gradient part (which, in momentum space, is parallel to $\vec{k}$), and in a divergence-free curl contribution (which, in momentum space, is perpendicular to $\vec{k}$), 
\begin{equation}
    \begin{split}
        \boldsymbol{\xi}&=\underbrace{-\nabla \xi\,}_{ \boldsymbol{\xi}_\parallel}+\underbrace{\nabla \times \Vec{A}\,}_{ \,\boldsymbol{\xi}_\perp}\;.
    \end{split}
\end{equation}
Hence, we have two scalar gauge parameters, $\xi^0$ and $\xi$. %Appendix \ref{app:gauges} derives the corresponding gauge transformations 
For the derivation of their coordinate transformations, see e.g.~ref.~\cite{Riotto:2018pcx}, Ch.~7.5, while here we just cite the end results%from eqs.~\eqref{scalars}, \eqref{vectors}, and \eqref{tensors}
, 
\begin{subequations}\label{gauges}
\begin{align}
     \tilde h^{ }_0 & =  h^{ }_0 - {\xi^0_{ }}' - \mathcal{H} \xi^0_{ }
 \;,  \\[2.5mm]
%%%
 \tilde{h} & =  h + \xi^0_{ } + \xi'
 \;, \\[0.5mm] 
%%%
 \tilde h^{ }_{D} & =  h^{ }_D + 
 \mathcal{H} \xi^0_{ }  - \frac{\nabla^2\xi}{3}
 \;,   \\[2mm]
%%%
 \tilde\vartheta & =  \vartheta + \xi 
 \;, \\[2mm] 
%%%
 \delta\tilde\varphi & =  \delta\varphi - \bar{\varphi}' \xi^0_{ } 
 \;, \\[2mm] 
%%%
 \delta\tilde T & =  \delta T - T' \xi^0_{ } 
 \;, \\[2mm]
%%% 
 \tilde v & =  v + \xi'
 \;.
\end{align}
\end{subequations}
\section{Gauge-invariant parameters}\label{ch:ginvpar}
\begin{figure}[!b]
    \centering
    \includegraphics[width=0.78\linewidth]{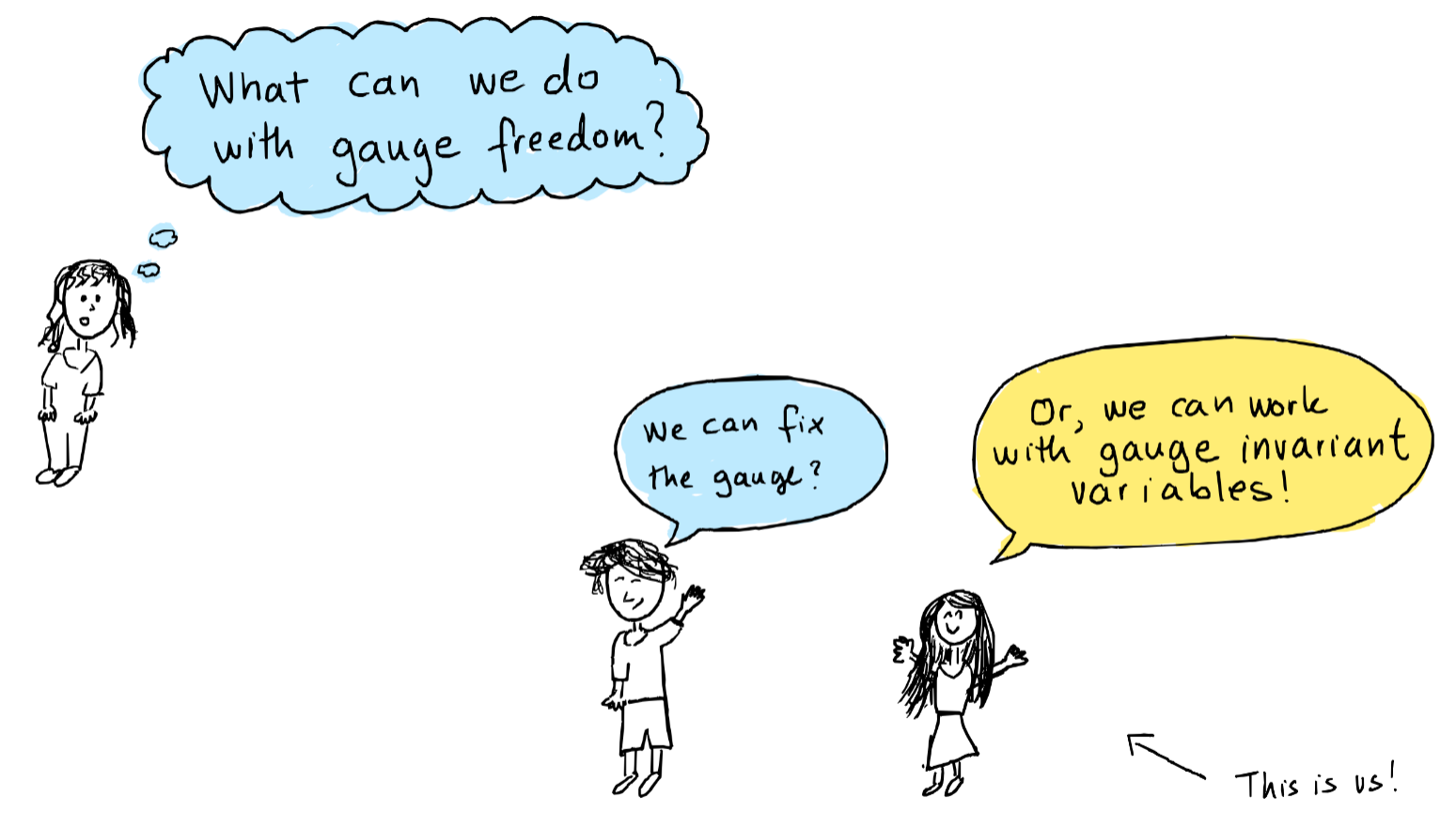}
    \caption{A comic on gauge freedom.
    }
    \label{fig:comic}
\end{figure}

As we have seen in Ch.~\ref{ch:gaugetransforms}, different coordinate systems, corresponding to different gauge choices, give rise to different perturbations. That means that working directly with the quantities in eq.~\eqref{gauges} will lead to gauge-dependent results. At this point, there are two ways to proceed. The first, usually technically simpler way, is to choose a particular gauge and accept that the results may include gauge artifacts. It also makes the comparison to results obtained in different gauges complicated. One may even run into a risk of simultaneously using results from different computations, which were performed using different gauge choices, as explained in the introduction. Hence, while performing a computation in a particular gauge might initially be quicker, it is still worthwhile to explore another option that avoids the drawbacks just described. As fig.~\ref{fig:comic} puts it, the goal is to derive equations satisfied by \textit{gauge-invariant} linear combinations of our degrees of freedom in eqs.~\eqref{gauges}. 
\\
\\
This prompts a question: which gauge-invariant objects should we use? One of the most well-known examples are Bardeen potentials, combinations of metric perturbations defined as
\begin{subequations}\label{bardeen}
\begin{align}
        \phi&\equiv h_0+(\partial_\tau +\mathcal{H})(h-\vartheta')\;,\label{bardeen1}\\
    \psi&\equiv h_D+\frac{\nabla^2\vartheta}{3}-\mathcal{H}(h-\vartheta')\;.\label{bardeen2}
\end{align}
\end{subequations}
Since these are gauge-invariant variables, performing the gauge transformations in eqs.~\eqref{gauges} on them renders the variables unchanged, and by extension, the equations satisfied by them remain unchanged also. For didactic purposes, let us verify gauge invariance for one variable explicitly,
\begin{equation}
    \begin{split}
        \phi\rightarrow\tilde{\phi}&\overset{\textcolor{white}{(4.3)}}{=}\tilde h_0+(\partial_\tau +\mathcal{H})(\tilde h-\tilde\vartheta')\\
        &\overset{\text{\eqref{gauges}}}{=}\;h^{ }_0 - {\xi^0_{ }}' - \mathcal{H} \xi^0_{ } +(\partial_\tau +\mathcal{H})(h + \xi^0_{ } + \xi'-\vartheta' - \xi')\\
        &\overset{\textcolor{white}{(4.3)}}{=} \phi- {\xi^0_{ }}' - \mathcal{H} \xi^0_{ } +(\partial_\tau +\mathcal{H})(\xi^0_{ } )\\
        &\overset{\textcolor{white}{(4.3)}}{=}\phi\;.
    \end{split}
\end{equation}
Bardeen potentials, however, capture only the metric perturbations and give us no information about the perturbations of the inflaton or the plasma. Furthermore, Bardeen potentials are only two variables, which is insufficient. In our setup, we have 7 perturbations, and given two gauge degrees of freedom, we require $7-2=5$ gauge-invariant degrees of freedom. So, we need to find three more. Preferably, they would carry some physical interpretation as well. To this end, we start by considering a so-called curvature perturbation. \\\\
In GR, one can compute the spatial trace of the Ricci tensor, which is itself a measure of curvature (in particular, it is intrinsic spatial curvature on hypersurfaces of constant conformal time). In our perturbed FLRW universe, the spatial trace becomes
\begin{equation}
    g^{ij}R_{ij}=\frac{4}{a^2}\nabla^2\left(h_D+\frac{\nabla^2\vartheta}{3}\right)\;.
\end{equation}
The quantity in the brackets is sometimes referred to as curvature perturbation. However, the combination $\left(h_D+\frac{\nabla^2\vartheta}{3}\right)$ alone is not gauge-invariant. One could add to it a rescaled inflaton perturbation, to obtain a gauge-invariant combination,
\begin{equation}\label{curvature0}
    \mathcal{R}_\varphi\equiv -\left(h_D+\frac{\nabla^2\vartheta}{3}\right)-\mathcal{H}\frac{\delta \varphi}{\bar\varphi'}\;.
\end{equation}
We shall refer to this as \textit{inflaton curvature perturbation}. It is related to the \textit{Mukhanov-Sasaki} variable. In particular, its rescaled version, $-\frac{\bar \varphi'}{H}a \mathcal{R}_\varphi$, is the central gauge-invariant variable proposed by Mukhanov, see ref. \cite{Mukhanov}, eq.~(22). Meanwhile, $\mathcal{R}_\varphi$ is exactly the physically important gauge-invariant variable $\mathcal{R}_m$ proposed by Sasaki in ref. \cite{Sasaki}, eq.~(2.33) with eq.~(2.9b). Mukhanov and Sasaki were both studying a cold system including inflaton and metric perturbations, without any plasma couplings.
\\\\
Taking this as an inspiration, we form two other \textit{plasma curvature perturbations}, by adding rescaled plasma perturbations $\delta T$ and $v$ to $\left(h_D+\frac{\nabla^2\vartheta}{3}\right)$, in order to ensure gauge invariance. In this way we obtain
\begin{subequations}\label{curvature1}
    \begin{align}
        \mathcal{R}_v&\equiv -\left(h_D+\frac{\nabla^2\vartheta}{3}\right)+\mathcal{H}(h-v)\;,\\
        \mathcal{R}_T&\equiv -\left(h_D+\frac{\nabla^2\vartheta}{3}\right)-\mathcal{H}\frac{\delta T}{ T'}\;.
    \end{align}
\end{subequations}
It is important to note that the gauge-invariant variables presented so far represent just one possible choice. For numerical implementation, instead of using $\mathcal{R}_v$ and $\mathcal{R}_T$, a different choice of variables will be made, though the end result will still be presented in terms of curvature perturbations. We call the new variables \textit{isocurvature} perturbations, owing to the fact that the \textit{curvature} part drops out,
\begin{subequations}\label{isocurvature}
    \begin{align}
        S_v&\equiv (\bar e+\bar p)\left(\mathcal{R}_v-\mathcal{R}_\varphi\right)\;,\label{isocurvature_a}\\
        S_T&\equiv \bar e_{,T}\dot{ T}\left(\mathcal{R}_T-\mathcal{R}_\varphi\right)\;.\label{isocurvature_b}
    \end{align}
\end{subequations}
On the formal level, going to gauge-invariant variables can be described as a projection,
\begin{equation}
            \begin{bmatrix}
        \textcolor{black}{\mathcal{R}_\varphi} \\
        \textcolor{black}{\mathcal{S}_v} \\
        \textcolor{black}{\mathcal{S}_T} \\
        \textcolor{black}{\phi} \\
        \textcolor{black}{\psi} 
        \end{bmatrix}
        =
        \begin{bmatrix}
        0 & 0 & -1 & -\frac{\nabla^2}{3} & -\frac{\mathcal{H}}{\bar\varphi'} & 0 & 0 \\
        0 & (\bar e+\bar p)\,\mathcal{H} & 0 & 0 & \frac{(\bar e+\bar p)\,\mathcal{H}}{\bar\varphi'} & 0 & -(\bar e+\bar p)\,\mathcal{H} \\
        0 & 0 & 0 & 0 & \frac{\partial_T \bar e \, \dot T}{\bar\varphi'}\mathcal{H} & -H\, \partial_T \bar e  & 0 \\
        1 & (\mathcal{H}+\partial_\tau) & 0 & -(\mathcal{H}+\partial_\tau)\partial_\tau & 0 & 0 & 0 \\
        0 & -\mathcal{H} & 1 & (\frac{\nabla^2}{3}+\mathcal{H}\partial_\tau)& 0 & 0 & 0 \\
        \end{bmatrix}
        \begin{bmatrix}
        h_0 \\
        h \\
        h_D \\
        \vartheta \\
        \delta \varphi \\
        \delta T \\
        v
        \end{bmatrix}\;.
\end{equation}
Note that we only included 5 gauge-invariant perturbations, as only 5 are linearly independent.

\subsubsection{Other gauge-invariant variables}
Another choice of variables may be encountered while browsing the literature, namely that of (non)adiabatic perturbations. Adiabatic perturbations all have the same value; for us, this happens for curvature perturbations, eqs.~\eqref{curvature0} and \eqref{curvature1}, after they cross the horizon; then we can also refer to them as \textit{adiabatic} perturbations. On the other hand, the isocurvature perturbations, eqs.~\eqref{isocurvature}, can also be referred to as \textit{non-adiabatic} perturbations. In particular, in single-field models, isocurvature perturbations are expected to exponentially vanish while the modes are outside of the Hubble horizon \cite{Weinberg_2004}. We will later see this explicitly (analytically in Ch.~\ref{ch:freezeout}, numerically in fig.~\ref{fig:isocurvature}). 
\\\\
The isocurvature perturbations are a measure of the mismatch between different components (inflaton and plasma), and are as such also a measure of entropy. Entropy perturbation is conventionally defined as
\begin{equation}
    \mathcal{S}_{\text{tot}}\equiv \mathcal{H}\left(\frac{\delta p}{\bar p'}-\frac{\delta e}{\bar e'}\right)=\frac{(\bar e,_T\bar p,_\varphi-\bar p,_T\bar e,_\varphi)\dot{\bar\varphi}\dot{ T}}{\dot{\bar e}\,\dot{\bar p}}\left(\mathcal{R}_T-\mathcal{R}_\varphi\right)\;,
\end{equation}
which is just a rescaled isocurvature perturbation, and not an independent degree of freedom. 
\\\\
Another alternative variable to consider is the ``energy density'' perturbation. We may define it as
\begin{equation}\label{energydensityperturbation}
     \Delta^{ }_v
 \equiv 
 \frac{ 
% \, \bigl[ \, 
  \delta e + \bar e ' 
 \, (h-v) 
% \,   \bigr]
 }{ - \overline{T}{}^{0}_{ }{}^{ }_{0} }
 \; = \; 
 \frac{
   \bar{e}^{ }_{,T}\, \dot{T}\,
   (\mathcal{R}^{ }_v - \mathcal{R}^{ }_T) 
 + \bar{e}^{ }_{,\varphi}\, \dot{\bar\varphi}\, 
   (\mathcal{R}^{ }_v - \mathcal{R}^{ }_\varphi) 
 }{\bigl( \bar{e} + \frac{1}{2} \dot{\bar\varphi}^2_{ } \bigr)\, H}\;.
\end{equation}
This is again not a new degree of freedom, just a rewriting of the existing curvature perturbations. One should keep in mind that the energy density definition in eq.~\eqref{stressenergy} does not include the derivatives of the inflaton field. Only when the latter are very small can we try to assign the physical interpretation of the energy density to eq.~\eqref{energydensityperturbation}. The object in eq.~\eqref{energydensityperturbation} is, however, a valid variable at all times. 
\\\\
Let us close this section by pointing out that the three gauge-invariant curvature perturbations will be central in our analysis in the next chapter. As it will turn out in the next section, the reheating dynamics can be described using curvature perturbations alone; the Bardeen potentials decouple (they are sourced by the curvature perturbations). This is similar to the findings of refs. \cite{Sasaki} and \cite{Mukhanov}.

\section{Gauge-invariant perturbation equations}\label{sec:gaugeinveq}
Let us return to the first-order equations: the Langevin equation \eqref{LangevinPerturbation}, four Einstein field equations \eqref{EFEpertexpanded}, and the two stress-energy tensor conservation equations \eqref{plasmaconservP1}-\eqref{plasmaconservP2}. Ch.~\ref{sec:gaugeinvequ} rewrites them in terms of the gauge-invariant variables presented in Ch.~\ref{ch:ginvpar}, eventually leading to three coupled differential equations for curvature perturbations, see eqs.~\eqref{es}. In the subsequent Ch.~\ref{ch:finalset} we show that eqs.~\eqref{es} can be significantly simplified by switching to physical time and by using variables $\mathcal{S}_v$ and $\mathcal{S}_T$ in lieu of $\mathcal{R}_v$ and $\mathcal{R}_T$. This gives the final set of eqs.~\eqref{result}, which we eventually solve numerically (see Ch.~\ref{sec:numres}). We finish this section by deriving the remaining two equations for Bardeen potentials, see Ch.~\ref{ch:Bardeens}, thus completing the set-up.

\subsection{Evolution equations for curvature perturbations}\label{sec:gaugeinvequ}
Since the derivation of all three equations is lengthy, we describe it in detail in Appendix \ref{sec:curvature}. Here, we explain the logic behind the derivation and cite the results. \\\\
We start by picking one of the first-order equations. The equation is written in terms of (a subset of) gauge-dependent variables $\delta \varphi,\,\delta T,\,v,\,h,\,h_D,\,h_0,$ and $\vartheta$. The goal is to turn this into an equation with variables $\mathcal{R}_\varphi$, $\mathcal{R}_v$, and $\mathcal{R}_T$. Those gauge-invariant variables contain specific combinations of gauge-dependent variables (see eqs.~\eqref{curvature0}-\eqref{curvature1}), namely $$\left(h_D+\frac{\nabla^2\vartheta}{3}\right),\;\quad \HH (h-v),\;\quad\HH \frac{\delta T}{T'}\;\quad,\quad\text{and }\quad\HH \frac{\delta \varphi}{\bar\varphi'}\;.$$ In order to rewrite our starting equation exclusively in terms of these combinations, we add to it linear combinations of other first-order equations. We then recognize curvature perturbations, possibly at the expense of adding and subtracting additional terms proportional to $\HH \frac{\delta \varphi}{\bar\varphi'}$. Those terms either add up to $\mathcal{R}_\varphi$ or cancel in the end. We use background identities to simplify expressions throughout. After explicitly performing this procedure in Appendix \ref{sec:curvature}, we find that the set of curvature perturbations satisfies
\begin{subequations}\label{es}
\begin{align}
     a^2\frac{\HH}{\bar\varphi'}\varrho&=-\mathcal{M}_{1\varphi}\mathcal{R}_\varphi-\mathcal{M}_{1v}(\mathcal{R}_v-\mathcal{R}_\varphi)-\mathcal{M}_{1T}(\mathcal{R}_T-\mathcal{R}_\varphi) \;,\label{e1}\\
      0&= \mathcal{M}^{ }_{2\varphi}\, \mathcal{R}^{ }_\varphi 
 + \mathcal{M}^{ }_{2v} 
   \bigl(\mathcal{R}^{ }_v - \mathcal{R}^{ }_\varphi\bigr)  
 + \mathcal{M}^{ }_{2T} 
   \bigl(\mathcal{R}^{ }_T - \mathcal{R}^{ }_\varphi\bigr)\;,\label{e2}\\
   0&=\mathcal{M}_{3\varphi}\mathcal{R}_\varphi+\mathcal{M}_{3v}\left(\mathcal{R}_v-\mathcal{R}_\varphi\right)+\mathcal{M}_{3T} \left(\mathcal{R}_T-\mathcal{R}_\varphi\right)\;.\label{e3}
\end{align}
\end{subequations}
The coefficients $\mathcal{M}_{i\varphi},\;\mathcal{M}_{iv},\text{ and }\mathcal{M}_{i T}$ are defined through, for $i=1$,
\begin{subequations}
    \begin{align}
 \mathcal{M}^{ }_{1\varphi}
 & =  
   \partial_\tau^2 - \nabla^2_{ }
   + 
   \biggl[\, 
    a \Upsilon + 2 \HH + 
    \frac{2\HH}{\bar\varphi'}
    \biggl( \frac{\bar\varphi'}{\HH} \biggr)'_{ }
   \,\biggr]\,\partial^{ }_\tau
  \;, 
  \\
%%%%%
 \mathcal{M}^{ }_{1v}
 & =   -\, 
 \biggl[\, 
   a \Upsilon
   +
   \frac{2 \HH}{\bar\varphi'}
   \biggl( \frac{\bar\varphi'}{\HH} \biggr)'   
 \,\biggr]
    \frac{ 4\pi G a^2_{ }(\bar{e} + \bar{p}) }{\HH }
  \;, 
  \\
%%%%%
 \mathcal{M}^{ }_{1T}
 & =  
 \biggl[\, 
   \frac{4\pi G a^2_{ } ( \bar{e}^{ }_{,T}  - \bar{p}^{ }_{,T} ) }{ \HH}
  +
  \frac{  
    a^2_{ } V^{ }_{,\varphi T}  
  +  a \bar\varphi' \Upsilon^{ }_{,T}  }
  { \bar\varphi' }
 \,\biggr] \, T' 
 \;.
    \end{align}
\end{subequations} 
For $i=2$, the coefficients are given by
\begin{equation}
    \mathcal{M}^{ }_{2\varphi}
  =
 \partial^{ }_\tau
 \;, \quad
%%%%
 \mathcal{M}^{ }_{2v}
 = 
 \partial^{ }_\tau
 + 
    \frac{3 \HH^2_{ }+ 4\pi G (\bar\varphi')^2_{ } }{\HH }
    + \frac{(\bar e + \bar p)'}{\bar e + \bar p}
 \;, \quad
%%%%
 \mathcal{M}^{ }_{2T}
 = 
 - 
 \frac{\bar{p}^{ }_{,T} \, T'}{\bar e + \bar p} 
 \;,
\end{equation}
while for $i=3$ we have
\begin{subequations}\label{M3}
    \begin{align}
         \mathcal{M}^{ }_{3\varphi}
  =&  -
 \frac{(\bar\varphi')^2_{ }}{a^2_{ }}
 \biggl\{\, 
   \partial_\tau^2 - \nabla^2_{ }
   + 
   2 \biggl[\, 
    \HH + 
    \frac{\HH}{\bar\varphi'}
    \biggl( \frac{\bar\varphi'}{\HH} \biggr)'_{ }
   \,\biggr]\,\partial^{ }_\tau
  \,\biggr\}
\\
 &+ 
 \bigl( \bar{e} + \bar{p} \bigr) 
 \,\biggl[\,
   \biggl( 2\HH + \frac{\HH'}{\HH} \biggr)
   \,\partial^{ }_\tau + \nabla^2_{ }
 \,\biggr]
 \;,\nonumber \\
 \mathcal{M}^{ }_{3v}
 =&  
 + \frac{(\bar\varphi')^2_{ }}{a^2_{ }}
 \biggl\{\, 
    \frac{2 \HH}{\bar\varphi'}
    \biggl( \frac{\bar\varphi'}{\HH} \biggr)'_{ }
    \frac{4\pi G a^2(\bar{e} + \bar{p})}{\HH}
 \biggr\}
\\
  &-
 \bigl( \bar{e} + \bar{p} \bigr) 
 \biggl\{\, 
    \frac{4\pi G a^2(\bar{e} + \bar{p})}{\HH}
 \biggl[\,
    \partial^{ }_\tau
    + 6 \HH 
    + \frac{(\bar e + \bar p)'}{\bar e + \bar p}   
 \,\biggr]
 -\nabla^2_{ }
 \,\biggr\} 
 \;, \nonumber
\\
 \mathcal{M}^{ }_{3T}
  =&
 - \bar{e}^{ }_{,T}  \, T'
 \biggl\{\,
   \partial^{ }_\tau 
  + 2 \, \biggl(2\HH - \frac{\HH'}{\HH} \biggr)
  + \frac{(\bar{e}^{ }_{,T} T')'}
         {\bar{e}^{ }_{,T} T'}
  + \biggl(2\HH + \frac{\HH'}{\HH} \biggr)
    \frac{\bar{p}^{ }_{,T}}{\bar{e}^{ }_{,T}}
 \,\biggr\}
 \;.
\end{align}
\end{subequations}
While we refer to different $\mathcal{M}$ as ``coefficients'', it should be noted that some terms inside $\mathcal{M}$ are really operators (temporal and spatial derivatives).

\subsection{Final set of equations in Fourier space and physical time}\label{ch:finalset}
We will now simplify eqs.~\eqref{es}. We begin by going from conformal time, $\tau$, to physical time, $t$, at the background level, 
\begin{subequations}\label{physical}
    \begin{align}
       \bar\varphi''
 + \left(   2\mathcal{H}+a\Upsilon \right)\, \bar\varphi' 
+a^2V_{,\varphi}
   =0   \quad\quad &\rightarrow\quad\quad  \ddot{\bar\varphi}+(3H+\Upsilon)\dot{\bar\varphi}+V_{,\varphi}
   =0 \label{physical_a}\;,
   \\
    \mathcal{H}^2_{ } = \frac{4\pi G}{3}
 \bigl[ (\bar\varphi')^2_{ }  + 2 a^2 \bar{e}
 \bigr]
 \; \quad\quad  &\rightarrow\quad\quad H^2_{ } = \frac{4\pi G}{3}
 \bigl[ (\dot{\bar\varphi})^2_{ }  + 2 \bar{e}
 \bigr] \label{physical_b}\;,
 \\ 
 \mathcal{H}' = \frac{4\pi G}{3}
 \bigl[ -  2 (\bar\varphi')^2_{ } - a^2 (\bar{e} + 3 \bar{p})
 \bigr]\quad\quad &\rightarrow\quad\quad \dot{H}=-4\pi G(\dot{\bar\varphi}^2+\bar e+\bar p) \label{physical_c}\;,
 \\
 \bar\varphi'\left(\bar\varphi''+2\mathcal{H}\bar\varphi'\right)+a^2\left[3\mathcal{H}(\bar e+\bar p)+ \bar e'\right]=0\quad\quad &\rightarrow\quad\quad \dot{\bar\varphi}\left(\ddot{\bar\varphi}+3H\dot{\bar\varphi}\right)+3H(\bar e+\bar p)+\dot{\bar e}=0\label{physical_d}\;.
    \end{align}
\end{subequations}
At the first order in perturbations, we simplify each of our equations \eqref{es} through three steps: (i) switching to isocurvature perturbations, (ii) going to physical time, and (iii) going to momentum space. Starting with eq.~\eqref{e1}, we switch from $\mathcal{R}_v$ and $\mathcal{R}_T$ to $\mathcal{S}_v$ and $\mathcal{S}_T$ by using eqs.~\eqref{isocurvature}. After dividing by $a^2$, defining 
\begin{equation}
    \mathcal{F}\equiv \frac{H}{\dot{\bar\varphi}}\partial_t \left(\frac{\dot{\bar\varphi}}{H}\right),
\end{equation}
and going to physical time, the first perturbation equation becomes
\begin{equation}\label{ee1}
    \begin{split}
        \ddot{\mathcal{R}}^{ }_\varphi
  = &
 - \frac{\varrho H}{\dot{\bar\varphi}}
 - \dot{\mathcal{R}}^{ }_\varphi 
    \, \bigl[\, \Upsilon + 3 H + 2 \mathcal{F} \,\bigr]
 + \frac{\nabla^2}{a^2}  \mathcal{R}^{ }_\varphi 
\\
  &+  
   \mathcal{S}^{ }_v\, 
   \biggl[\,
       \frac{4\pi G ( \Upsilon  +  2 \mathcal{F} )}{H}
   \,\biggr] 
 - \mathcal{S}_T \, 
   \biggl[\,
   \frac{4\pi G}{H}
 \biggl( 1 - \frac{\bar{p}^{ }_{,T}}{\bar{e}^{ }_{,T}}\biggr)
   + \frac{V^{ }_{,\varphi T} 
   + \Upsilon^{ }_{,T} \dot{\bar\varphi} }
     {\dot{\bar\varphi} \, \bar{e}^{ }_{,T}}
   \,\biggr]
 \;.
    \end{split}
\end{equation}
The second perturbation equation follows from eq.~\eqref{e2}. Multiplying by $(\bar e+\bar p)/a$, switching to isocurvature perturbations, and once again going to physical time $t$, we arrive at
\begin{equation}\begin{split}\label{ee2}
\dot{\mathcal{S}}^{ }_v =-(\bar e+\bar p)\,\dot{ \mathcal{R}}^{ }_\varphi 
 - \left[ \frac{3 H^2_{ }+ 4\pi G (\dot{\bar\varphi})^2_{ } }{H }\right] 
   \mathcal{S}^{ }_v 
  + \frac{\bar{p}^{ }_{,T} \, }{\bar e_{,T}}  \mathcal{S}^{ }_T\;.
\end{split}
\end{equation}
The simplification of the third perturbation, eq.~\eqref{e3}, is slightly more tedious. We switch to isocurvature perturbations and physical time, as well as divide the expression by $a^2$, to obtain
\begin{equation}\begin{split}\label{ee3a}
     \left\{\bigl( \bar{e} + \bar{p} \bigr) 
 \,\biggl[\,
   \biggl( 3H + \frac{\dot{H}}{H} \biggr)
   \,\partial^{ }_t + \frac{\nabla^2_{ }}{a^2}
 \,\biggr]-
\dot{\bar\varphi}^2
 \biggl[\,
   \partial_t^2 - \frac{\nabla^2_{ }}{a^2}
   + 
    \biggl(\, 
   3 H + 2
    \mathcal{F}
   \,\biggr)\,\partial^{ }_t
  \,\biggr]\right\}\;&\mathcal{R}_\varphi
  \\
    +\left\{\frac{4\pi G }{H}\Biggl[ \dot{\bar\varphi}^2 
    2\mathcal{F}-  
     (\bar{e} + \bar{p})
 \bigl(\,
    \partial^{ }_t
    + 6 H   
 \,\bigr)\Biggr]
 +\frac{\nabla^2_{ }}{a^2}
 \right\}\;&\mathcal{S}^{ }_v
 \\
    -
 \biggl\{\,
   \partial^{ }_t 
  +  \,3H -2\frac{\dot{H}}{H}
  + \biggl(3H + \frac{\dot{H}}{H} \biggr)
    \frac{\bar{p}^{ }_{,T}}{\bar{e}^{ }_{,T}}
 \,\biggr\}\; & \mathcal{S}^{ }_T
 \; = \; 
 0 \;.
 \end{split}
\end{equation}
We notice that this equation includes a second time derivative of $\mathcal{R}_{\varphi}$. Using eq.~\eqref{ee1} we can turn eq.~\eqref{ee3a} into a first-order equation. We can further simplify the expression by using eq.~\eqref{ee2} to remove the derivative from the variable $ \mathcal{S}^{ }_v$. Using the background identity \eqref{physical_c}, we additionally cancel 3 terms proportional to $\frac{\bar{p}^{ }_{,T} \, }{\bar e_{,T}}$,
\begin{equation}\begin{split}\label{ee3c}
    \dot{\mathcal{S}}_T=& \;\dot{\bar\varphi}\varrho H+\left[\frac{4\pi G }{H}
     (\bar{e} + \bar{p})^2 +\bigl( \bar{e} + \bar{p} \bigr) 
   \biggl( 3H + \frac{\dot{H}}{H} \biggr)
 +
\dot{\bar\varphi}^2
   \Upsilon 
   \,
  \,  \right]\dot{\mathcal{R}}_\varphi+\bigl( \bar{e} + \bar{p} \bigr) \frac{\nabla^2_{ }}{a^2}\mathcal{R}_\varphi \\
    &+\left\{\frac{4\pi G }{H}\left[(\bar{e} + \bar{p}) \frac{ 4\pi G (\dot{\bar\varphi})^2_{ } }{H } -\dot{\bar\varphi}^2 \Upsilon-  
     (\bar{e} + \bar{p})
 3 H   
\right]
 +\frac{\nabla^2_{ }}{a^2}
 \right\}\, \mathcal{S}^{ }_v
 \\
    &+
 \biggl\{\,\frac{4\pi G}{H}
\dot{\bar\varphi}^2
   + \frac{V^{ }_{,\varphi T} 
   + \Upsilon^{ }_{,T} \dot{\bar\varphi} }
     {\bar{e}^{ }_{,T}}\dot{\bar\varphi}
  +2\frac{\dot{H}}{H}
  - 3H \left(1+
    \frac{\bar{p}^{ }_{,T}}{\bar{e}^{ }_{,T}}\right)
 \,\biggr\}\;\mathcal{S}^{ }_T\;.
 \end{split}
\end{equation}
Note that using eq.~\eqref{ee1} introduced the noise term, $\varrho$, in this equation as well. We can still slightly simplify eq.~\eqref{ee3c}. In the first line, we use identities \eqref{physical_b} and \eqref{physical_c}. In the second line, we use eq.~\eqref{physical_b}, while in the third line we use eq.~\eqref{physical_c}. This leads us to
\begin{equation}
    \begin{split}
        \dot{\mathcal{S}}^{ }_T
  =&  
 \;\varrho \, \dot{\bar\varphi}  H
 + \dot{\mathcal{R}}^{ }_\varphi\,
  \biggl[\,
   \Upsilon\dot{\bar\varphi}^2_{ }
  + \frac{8\pi G \bar{e}(\bar{e}+\bar{p})}{H}
  \,\biggr]
 +\biggl[\, 
      \frac{(\bar{e} + \bar{p})}{a^2_{ }}
   \,\biggr] \nabla^2 \mathcal{R}^{ }_\varphi \, 
   \\
&- 
 \mathcal{S}^{ }_v \,
 \biggl[\,
  \frac{k^2_{ }}{a^2_{ }} 
%% + \frac{4\pi G \Upsilon \dot{\bar\varphi}^2_{ }}{H}
%% + \frac{32 \pi^2_{ } G^2_{ }\bar{e}(\bar{e} + \bar{p})}{H^2_{ }}
+ \frac{4\pi G }{H}
   \biggl(\, \Upsilon \dot{\bar\varphi}^2_{ } 
       + \frac{8 \pi G \bar{e}(\bar{e} + \bar{p})}{H}
   \,\biggr)
 \,\biggr]
\\
  &+ 
 \mathcal{S}^{ }_T \, 
 \biggl[\,
   \frac{\dot{H} - 4\pi G(\bar{e} + \bar{p})}{H}
  - 3 H \,\biggl( 1 
  + \frac{\bar{p}^{ }_{,T}}{\bar{e}^{ }_{,T}}\,\biggr)
   + \frac{( V^{ }_{,\varphi T} 
   + \Upsilon^{ }_{, T} \dot{\bar\varphi})\,\dot{\bar\varphi}  }
     {\bar{e}^{ }_{, T}}
 \,\biggr]\;.
    \end{split}
\end{equation}
As a final step, we go from the physical to the 3-dimensional Fourier space for all three equations. Using a Fourier decomposition analogous to eq.~\eqref{fourier}, we notice $-\nabla^2\rightarrow k^2$ when going from position to momentum space. The resulting equations for the \textit{mode functions}, $\mathcal{R}_{\varphi k}$, $\mathcal{S}_{vk}$, and $\mathcal{S}_{Tk}$, are thus
\begin{subequations}\label{result}
    \begin{align}
         \ddot{\mathcal{R}}^{ }_{\varphi k}
  =&  
 - \frac{\varrho_k H}{\dot{\bar\varphi}}
 - \dot{\mathcal{R}}^{ }_{\varphi k} 
    \, \bigl[\, \Upsilon + 3 H + 2 \mathcal{F} \,\bigr]
 -  \mathcal{R}^{ }_{\varphi k}
    \, \biggl[\, \frac{k^2_{ }}{a^2_{ }} \,\biggr]\label{result_a}
\\
  &+  
   \mathcal{S}^{ }_{v k}\, 
   \biggl[\,
       \frac{4\pi G ( \Upsilon  +  2 \mathcal{F} )}{H}
   \,\biggr] 
 - \mathcal{S}^{ }_{T k} \, 
   \biggl[\,
   \frac{4\pi G}{H}
 \biggl( 1 - \frac{\bar{p}^{ }_{,T}}{\bar{e}^{ }_{, T}}\biggr)
   + \frac{V^{ }_{,\varphi T} 
   + \Upsilon^{ }_{, T} \dot{\bar\varphi} }
     {\dot{\bar\varphi} \, \bar{e}^{ }_{, T}}
   \,\biggr]
 \;,\nonumber
 \\
 \dot{\mathcal{S}}^{ }_{v k}
 =& 
  - \dot{\mathcal{R}}^{ }_{\varphi k}
    \, \bigl[ \bar{e} + \bar{p} \bigr]
  - \mathcal{S}^{ }_{v k} 
 \biggl[\, 
 3 H + 
 \frac{4 \pi G \dot{\bar\varphi}^2_{ }}{H}
 \,\biggr]
 + 
 \mathcal{S}^{ }_{T k}
 \,
 \biggl[\, 
   \frac{\bar{p}^{ }_{, T}}{\bar{e}^{ }_{, T}}
 \,\biggr]
 \;,\label{result_b}
 \\
 \dot{\mathcal{S}}^{ }_{T k}
  =&\,  
 \varrho_k \, \dot{\bar\varphi}  H
 + \dot{\mathcal{R}}^{ }_{\varphi k}\,
  \biggl[\,
   \Upsilon\dot{\bar\varphi}^2_{ }
  + \frac{8\pi G \bar{e}(\bar{e}+\bar{p})}{H}
  \,\biggr]
 - \mathcal{R}^{ }_{\varphi k} \, 
   \biggl[\, 
     (\bar{e} + \bar{p}) \frac{k^2_{ }}{a^2_{ }}
   \,\biggr]\nonumber\\
&- 
 \mathcal{S}^{ }_{v k} \,
 \biggl[\,
  \frac{k^2_{ }}{a^2_{ }} 
%% + \frac{4\pi G \Upsilon \dot{\bar\varphi}^2_{ }}{H}
%% + \frac{32 \pi^2_{ } G^2_{ }\bar{e}(\bar{e} + \bar{p})}{H^2_{ }}
+ \frac{4\pi G }{H}
   \biggl(\, \Upsilon \dot{\bar\varphi}^2_{ } 
       + \frac{8 \pi G \bar{e}(\bar{e} + \bar{p})}{H}
   \,\biggr)
 \,\biggr]\label{result_c}
\\
  &+ 
 \mathcal{S}^{ }_{T k} \, 
 \biggl[\,
   \frac{\dot{H} - 4\pi G(\bar{e} + \bar{p})}{H}
  - 3 H \,\biggl( 1 
  + \frac{\bar{p}^{ }_{,T}}{\bar{e}^{ }_{,T}}\,\biggr)
   + \frac{( V^{ }_{,\varphi T} 
   + \Upsilon^{ }_{, T} \dot{\bar\varphi})\,\dot{\bar\varphi}  }
     {\bar{e}^{ }_{, T}}
 \,\biggr]\;.\nonumber
    \end{align}
\end{subequations}
This is our final set of gauge-invariant evolution equations, which is also used in the numerical implementation of our framework, see Part \ref{part:num}. Let us stress that since the variables are now mode functions, they can be complex. We will later use this fact while removing the singularities introduced by $1/\dot{\bar\varphi}$, see Ch.~\ref{ch:singularities}.

\subsection{Completing the set-up: Bardeen potentials}\label{ch:Bardeens}
In order to complete the setup, two more gauge-invariant perturbation equations have to be provided. Those can be written for the Bardeen potentials in eqs.~\eqref{bardeen}. Starting with eq.~\eqref{EFEpertexpanded3}, we easily recognize the equation of motion
\begin{equation}\label{BardeenEq}
    \begin{split}
         \phi-\psi=0\;.
    \end{split}
\end{equation}
To obtain the fifth equation, we start with eq.~\eqref{EFEpertexpanded2}. We rewrite $\delta\varphi$ in terms of $\mathcal{R}_{\varphi}$, and $(h-v)$ in terms of $\mathcal{R}_v$, using eqs.~\eqref{curvature0} and \eqref{curvature1} respectively, 
\begin{equation}
    \begin{split}
        \mathcal{H}h_0+&\left(h_D+\frac{\nabla^2\vartheta}{3}\right)'=\\
        4\pi G&\left[-\frac{\bar\varphi'^2+a^2(\bar e+\bar p)}{\HH}\left(h_D+\frac{\nabla^2\vartheta}{3}\right)-\frac{\bar\varphi'^2}{\HH}\mathcal{R}_\varphi-\frac{a^2(\bar e+\bar p)}{\HH}\mathcal{R}_v\right]\;.
    \end{split}
\end{equation}
After using the background identity \eqref{bck1_a} to simplify the prefactor of $\left(h_D+\frac{\nabla^2\vartheta}{3}\right)$, we further add and subtract $\HH(\partial_\tau+\HH)(h-\vartheta')$ and $\HH'(h-\vartheta')$, in order to recognize $\phi$ and $\psi$,
\begin{equation}
    \begin{split}
        \mathcal{H}\phi+\left[\partial_\tau+\HH-\frac{\HH'}{\HH}\right]\psi&=4\pi G\left[-\frac{\bar\varphi'^2}{\HH}\mathcal{R}_\varphi-\frac{a^2(\bar e+\bar p)}{\HH}\mathcal{R}_v\right]\;.
    \end{split}
\end{equation}
Lastly, we go to physical time, momentum space, and divide by $a$,
\begin{equation}
    H \phi_k+\left[\partial_t-\frac{\dot{H}}{H}\right]\psi_k=4\pi G\left[-\frac{\dot{\bar\varphi}^2}{H}\mathcal{R}_{\varphi k}-\frac{(\bar e+\bar p)}{H}\mathcal{R}_{vk}\right]\;.
\end{equation}
Using eq.~\eqref{BardeenEq}, we can further simplify this,
\begin{equation}\label{bardeen3}
  \dot{\psi}_k+ \left[H-\frac{\dot{H}}{H}\right]\psi_k=-4\pi G\left[\frac{\dot{\bar\varphi}^2}{H}\mathcal{R}_{\varphi k}+\frac{(\bar e+\bar p)}{H}\mathcal{R}_{vk}\right]\;.
\end{equation}
In practice, we would first solve for eqs.~\eqref{result} numerically. We can then use the results as a source term for eq.~\eqref{bardeen3}. This gives us $\psi$ and, by extension, also $\phi$, see eq.~\eqref{BardeenEq}. Then our system is complete.

\newpage
\chapter{Perturbations: regime analysis}\label{ch:extraremarks}
\begin{figure}[!b]
    \centering
    \includegraphics[width=0.78\linewidth]{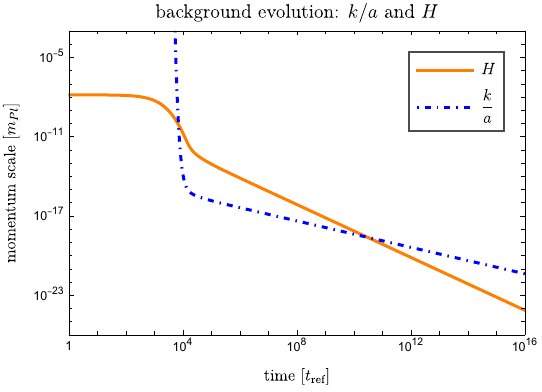}
    \caption{Illustration of the exit and re-entering of a particular momentum mode, $k/a$, from a Hubble horizon. 
    }
    \label{fig:regimes}
\end{figure}
The main result of Ch.~\ref{ch:gaugeinv}, eqs.~\eqref{result}, may seem somewhat hard to interpret at first sight. Hence, this Chapter simplifies the equations in several regimes. Firstly, we simplify them during early times, where we set \textit{initial conditions} (Ch.~\ref{ch:initialconditions}). Further, in the overdamped regime we observe the perturbations \textit{freeze-out}, and obtain an analytic solution (Ch.~\ref{ch:freezeout}). Finally, in \textit{radiation domination}, the simplified equations offer a nice theoretical insight (Ch.~\ref{ch:radiationdomination}). With the relevant equations in hand, we return to the topic of gauge invariance. Namely, we demonstrate the importance of gauge-invariant treatment even in the absence of a plasma (Ch.~\ref{ch;importanceofgaugeinvariance}).
\\\\
Let us clarify how the different regimes are defined before proceeding with the detailed analysis. Figure \ref{fig:regimes} shows the time evolution of two quantities: the physical momentum, $k/a$, and the Hubble rate, $H$. At early times, a particular momentum mode is deep inside the horizon, $k/a\gg H$. This means that, to a good approximation, we find ourselves in a Bunch-Davies vacuum, and we can use the standard procedure to fix initial conditions. Once the mode exits the Hubble horizon, $k/a< H$, we enter the overdamped regime. The name comes from the fact that the coefficient in front of the term with one time derivative of the inflaton curvature perturbation is relatively large --- see the simplified eq.~\eqref{justRphiequation}. Eventually the mode re-enters the horizon, and then $k/a> H$ again.
\\\\
This, however, is not the only way one can define different regimes. An alternative partition would be: inflation, matter domination, radiation domination. During inflation, the inflaton curvature perturbation is the most relevant degree of freedom. At the end of inflation, the inflaton field rolls down its potential and starts oscillating around zero. This period of rapid oscillations is a characteristic of matter domination. On the other hand, at some point the energy density of the radiation overtakes the energy density of the inflaton; when the radiation's energy density is orders of magnitude larger than that of the inflaton field, we speak of radiation domination. The mode can re-enter the horizon either during matter or radiation domination. In some models, defined by a particular choice of inflaton potential, the  early period of matter domination may not even exist.

\section{Initial conditions}\label{ch:initialconditions}
In order to set initial conditions, it is easiest to find a regime where one component of the solution vector dominates, and find the precise value of that component alone. With this in mind, we study eqs.~\eqref{result} at early times, when the mode is deep inside the horizon, where we perform a number of approximations. We note that those are used to set initial conditions, while to find the numerical power spectrum, the entire eqs.~\eqref{result} are integrated.
\\\\
At early times, the energy density of the inflaton is much larger than that of the radiation, $\dot{\bar{\varphi}}^2/2+V-T V_T \gg e_r$. Using the background eq.~\eqref{physical_b} together with eq.~\eqref{energyaverage}, we note that $ 3 H^2 \gg 8\pi G\, e_r$. Similarly, assuming the potential depends negligibly on the temperature, and that $\mathcal{O}(e_r)\approx\mathcal{O}(p_r)$, we note that $\bar{e}+\bar{p}\approx e_r$, and hence $ 3 H^2 \gg 8\pi G\, (\bar{e}+\bar{p})$. 
\\\\
We now look at how this affects the isocurvature terms in eq.~\eqref{result_a}. Focusing on $S_{vk}$, eq.~\eqref{isocurvature_a}, we use the approximations from the previous paragraph to conclude
\begin{equation}\label{SvDiscard}
    \left|S_{vk}\;\frac{4\pi G}{H}\;(\Upsilon+2\mathcal{F})\right|\ll H \; \left|\Upsilon+2\mathcal{F}\right|\; \left|\mathcal{R}_{vk}-\mathcal{R}_{\varphi k}\right|\;.
\end{equation}
At the moment, we further assume curvature perturbations are of roughly the same size.\footnote{This may be a model-dependent statement; however, the final conclusion (setting isocurvature $\mathcal{S}_{v k}=\mathcal{S}_{T k}=0$ at the beginning) turned out to be numerically an excellent approximation even in a model where $\mathcal{R}_{T k}$ was initially significantly larger than $\mathcal{R}_{v k}$ and $\mathcal{R}_{\varphi k}$; see Ch.~\ref{ch:results:perturb}.} Then, the right-hand side of ineq.~\eqref{SvDiscard} is already smaller than $\mathcal{R}_{\varphi k}$ terms in eq.~\eqref{result_a}  (except perhaps very deep in the strong regime), since $k/a$ is the largest scale at early times, while the derivative terms are then expected to be non-negligible (power spectra can be observed to decrease fast at early times in fig.~\ref{fig:perturb_strong}). 
\\\\
We similarly inspect the $\mathcal{S}_{Tk}$ term from eq.~\eqref{result_a}. Assuming that during inflation, the temperature of the bath changes only very slowly, such that $\dot{e}_r$ in eq.~\eqref{energyinserted_b} is negligible, we have $|\dot{e}_r|\ll 3H(\bar{e}+\bar{p})$; where we again assumed $V_{,T}\approx 0$. If we further assume $\Upsilon_{,T}/\bar{e}_{,T}\approx \Upsilon/e_r$, we may estimate the $\mathcal{S}_{Tk}$ (see eq.~\eqref{isocurvature_b}) contribution to eq.~\eqref{result_a} as
\begin{equation}
    \left|\mathcal{S}_{Tk}\; \frac{4\pi G}{H}\; \right| \ll H^2\,\left|\mathcal{R}_{Tk}-\mathcal{R}_{\varphi k}\right|\;,\quad\quad  \left|\mathcal{S}_{Tk}\; \frac{\Upsilon_{,T}}{\bar{e}_{,T}} \right| \ll H\,\Upsilon\,\left|\mathcal{R}_{Tk}-\mathcal{R}_{\varphi k}\right|\;.
\end{equation}
As before, we estimate these terms to be less important than $\mathcal{R}_{\varphi k}$, which gives the approximate equation valid at very early times (in the weak and moderately strong regime),
%\footnote{\hl{Argument why} $S_v$ and $S_T$ can be omitted from \eqref{result}. The differences between curvature perturbations are not necessarily small (power spectra are orders of magnitude apart). $V_T$ does not in general vanish. Why would $\dot{\bar\varphi}^2\Upsilon \sim 3H(\bar e+\bar p)$ in general? Can p in general be $\leq \bar e$?}
\begin{equation}\label{justRphiequation}
    \ddot{\mathcal{R}}_{\varphi k} + (\Upsilon+2\mathcal{F}+3H)\, \dot{\mathcal{R}}_{\varphi k} + \frac{k^2}{a^2}\,\mathcal{R}_{\varphi k}\,\approx
    -\varrho_k \frac{H}{\dot{\bar\varphi}}\,.
\end{equation}
It was checked numerically that $S_{vk}$ and $S_{Tk}$ are many orders of magnitude smaller than $\mathcal{R}_{\varphi k}$, giving further credibility to approximately dropping them in eq.~\eqref{justRphiequation} at early times (see fig.~\ref{fig:isocurvature}).
\\\\
We now make a series of redefinitions aimed at massaging some of the friction terms into a time-dependent ``mass term''. We start by defining $\mathcal{Q}_{\varphi k}\equiv \frac{\dot{\bar\varphi}}{H} \mathcal{R}_{\varphi k} $, which eliminates $\mathcal{F}$ from the derivative term,
\begin{equation}
\begin{split}
     -\varrho_k=\;\ddot Q_{\varphi k}+\left(\Upsilon+3H\right)\dot Q_{\varphi k}+\left(\frac{\ddot H}{H}-\frac{\dddot{\bar{\varphi}}}{\dot{\bar\varphi}}-\mathcal{F}\Upsilon-3H\mathcal{F}-2\frac{\dot{H}^2}{H^2}+2\frac{\ddot{\bar{\varphi}}}{\dot{\bar{\varphi}}}\frac{\dot H}{H}+\frac{k^2}{a^2}\right) Q_{\varphi k}\,.
\end{split}
\end{equation}
We make another redefinition, $Q_{\varphi k}=\frac{\hat{Q}_{\varphi k}}{a}$, which eliminates $2H$ from the derivative term,
\begin{equation}
\begin{split}
     -a\,\varrho_k\;=\;&\ddot{\hat{Q}}_{\varphi k}+\left(\Upsilon+H\right)\dot{\hat{Q}}_{\varphi k}\\
     &+\left(    H^2-\dot{H}-3H^2-\Upsilon H+\frac{\ddot H}{H}-\frac{\dddot{\bar{\varphi}}}{\dot{\bar\varphi}}-\mathcal{F}\Upsilon-3H\mathcal{F}-2\frac{\dot{H}^2}{H^2}+2\frac{\ddot{\bar{\varphi}}}{\dot{\bar{\varphi}}}\frac{\dot H}{H}+\frac{k^2}{a^2}\right) \hat{Q}_{\varphi k}\,.
\end{split}
\end{equation}
We succeed at removing the final $H$ from the derivative term by going to conformal time,
\begin{equation}
\begin{split}
     -a^3\,\varrho_k\;=\;&\hat{Q}_{\varphi k}''+a \Upsilon\,\hat{Q}_{\varphi k}'\\&+a^2\left(    -2H^2-\dot{H}-\Upsilon H+\frac{\ddot H}{H}-\frac{\dddot{\bar{\varphi}}}{\dot{\bar\varphi}}-\mathcal{F}\Upsilon-3H\mathcal{F}-2\frac{\dot{H}^2}{H^2}+2\frac{\ddot{\bar{\varphi}}}{\dot{\bar{\varphi}}}\frac{\dot H}{H}+\frac{k^2}{a^2}\right) \hat{Q}_{\varphi k}\,.
\end{split}
\end{equation}
Lastly, we wish to tidy up this expression by defining
\begin{equation}\label{rescaled}
    \begin{split}
    a^2\epsilon_{k\Upsilon}^2\equiv\hat{\epsilon}_{k\Upsilon}^2\equiv k^2\underbrace{-a^2(\partial_t+\mathcal{F}+2H)(\mathcal{F}+H)}_{\equiv \hat{m}^2}\underbrace{-a(\mathcal{F}+H)}_{\equiv \hat{\mu}} \underbrace{(a\Upsilon)}_{\equiv \hat{\Upsilon}}\,,\;\;\;\;\;\;\;\;\;\;\;\;\hat{\varrho}_k=-a^3\varrho_k\,,
    \end{split}
\end{equation}
where $\hat{m}^2$ could be written more compactly as
\begin{equation}
    \hat{m}^2\equiv -\frac{\HH}{a\bar\varphi'}\left(\frac{a\bar\varphi'}{\HH}\right)''\;.
\end{equation}
With this in hand, we arrive at a simple equation,
\begin{equation}\label{mink}
     \hat{Q}_{\varphi k}''+\hat{\Upsilon}\,\hat{Q}_{\varphi k}'+\underbrace{(k^2+\hat{\mu}\hat{\Upsilon}+\hat{m}^2)}_{\equiv \hat{\epsilon}_{k\Upsilon}^2}\hat{Q}_{\varphi k}=\hat{\varrho}_k\,.
\end{equation}
A comment on the interpretation of $\hat{\epsilon}_{k\Upsilon}^2$ is in order. First, consider the limit of constant $\Upsilon,a$ in $\epsilon_{k\Upsilon}^2 \equiv \hat{\epsilon}_{k\Upsilon}^2/a^2$,
\begin{equation}
\begin{split}
     \epsilon_{k\Upsilon}^2|_{a,\Upsilon=\text{const.}}&=\frac{k^2}{a^2}-\frac{\dddot{\bar{\varphi}}}{\dot{\bar{\varphi}}}-\Upsilon\frac{\ddot{\bar{\varphi}}}{\dot{\bar{\varphi}}}=\frac{k^2}{a^2}+V_{,\varphi\varphi}\;.
\end{split}
\end{equation}
For the last equality we have used the time derivative of eq.~\eqref{physical_a} under the assumption $\dot{a}=0$, $-\frac{\dddot{\bar{\varphi}}}{\dot{\bar{\varphi}}}-\Upsilon\frac{\ddot{\bar{\varphi}}}{\dot{\bar{\varphi}}}=V_{,\varphi\varphi}+\dot{\Upsilon}$. The result is the standard Minkowskian energy, suggesting that $ \epsilon_{k\Upsilon}$ is the energy. Nonetheless, as $k/a$ is decreasing, it is possible for $ \epsilon_{k\Upsilon}^2$ to become negative. To see this one can go to the early times where $\dot{H}\ll H^2$ and $\ddot{\bar{\varphi}}\ll H \dot{\bar{\varphi}}$. Using the latter to introduce the first inequality, and the former in the underbrace approximation, one can deduce that 
\begin{equation}\begin{split}
     \mathcal{F}=\frac{1}{ \dot{\bar{\varphi}}}\left(\ddot{\bar{\varphi}}-\dot{\bar{\varphi}}\frac{\dot{H}}{H}\right) \ll \underbrace{\left(H-\frac{\dot{H}}{H}\right)}_{\approx H} \;\;\;\;\;\rightarrow\;\;\;\;\;
      \mathcal{F} \ll  H\;.
\end{split}
\end{equation}
Now, with $\mathcal{F} \ll  H$ and $\dot{H}\ll H^2$ in mind, we can simplify the energy squared as
\begin{equation}\begin{split}
     \epsilon_{k\Upsilon}^2&=\frac{k^2}{a^2}-(\partial_t+\mathcal{F}+2H)(\mathcal{F}+H)-(\mathcal{F}+H) \Upsilon\\
     &\approx \frac{k^2}{a^2}-(2H+\Upsilon)H\;.
\end{split}
\end{equation}
As $k/a$ initially decreases much faster than $H$ or $\Upsilon$, it is not guaranteed that the energy squared remains positive. Negative energy squared would prevent us from constructing an instantaneous Hilbert space. Luckily, at early times this is not an issue, and in practice the energy becomes negative only well after the initial conditions are set.\footnote{The energy squared would become negative around the time of the freeze-out, and then the mode in question, $\hat{Q}_{\varphi k}$, would become tachyonic. This is related to the freeze-out of the inflaton curvature perturbation, since the two are related via the scale factor, $\hat{Q}_{\varphi k} =a (\dot{\bar{\varphi}}/H) \mathcal{R}_{\varphi k}$.}
\\\\
It is now finally the time to use eq.~\eqref{mink} to determine initial conditions. Suppose we investigate eq.~\eqref{mink} in the weak regime, where $\Upsilon$ is negligible. By the fluctuation-dissipation theorem, the noise should then also be negligible. Then the damping term and the noise vanish, and we are left with
\begin{equation}\label{init}
    \hat{Q}_{\varphi k}''+\underbrace{(k^2+\hat{m}^2)}_{\equiv \hat{\epsilon}_{k}^2}\hat{Q}_{\varphi k}=0\;.
\end{equation}
Since the scale $k/a$ is in practice much larger than any other scale at early times, dropping $\Upsilon$ and $\varrho$ is expected to be acceptable even when one moves away from the weak regime. Furthermore, even if the dissipation would be very large, the harmonic oscillator, eq.~\eqref{mink}, would be strongly damped, and hence the initial conditions would be quickly attenuated, while new fluctuations would be generated by the noise term. Hence, in practice, we use eq.~\eqref{init} to determine initial conditions everywhere.
\\\\
Equation \eqref{init} is a second-order classical equation. To determine initial conditions, we view it as an equation for a quantum-mechanical mode function. In the quantum-mechanical picture, the field and its conjugate momentum are promoted to operators, and the canonical commutation relations are imposed. This allows us to define a vacuum state at very early times, the Bunch-Davies vacuum, see ref.~\cite{Bunch:1978yq}. After choosing the positive-energy solution,\footnote{We define a set of mode functions that behave like positive-energy modes deep inside the Hubble horizon. The vacuum state is then defined as the state annihilated by the corresponding operators. There are no excitations of the positive-energy modes in the asymptotic past.} one finds initial conditions at time $t_1$,\footnote{Defining an instantaneous Hilbert space and not making the $k^2\gg \hat{m}^2$ approximation would replace $k$ in the initial conditions by $\sqrt{k^2+\hat{m}^2}$.}
\begin{equation}
    \hat{Q}_{\varphi k}(t_1)\approx \frac{1}{\sqrt{2k}}\;,\;\;\;\; \hat{Q}_{\varphi k}'(t_1)\approx -ik  \;\hat{Q}_{\varphi k}(t_1)\;.
\end{equation}
One should translate this into the initial conditions for our variables of interest, $\mathcal{R}_{\varphi k},\; \mathcal{S}_{T k}, $ and $ \mathcal{S}_{v k}$. Aside from recalling $\hat{Q}_{\varphi k}\equiv a\frac{\dot{\bar\varphi}}{H}\mathcal{R}_{\varphi k}$, we carry out an additional rescaling. The rescaled variables differ from the previous ones as $\mathcal{R}_{\varphi k}(t_1)_{\text{rescaled}}=\frac{k^{3/2}}{\pi\sqrt{2}}\mathcal{R}_{\varphi k}(t_1)$, to directly absorb the angular integration constants into the power spectrum, thus obtaining the dimensionless power spectrum, to be defined in eq.~\eqref{dimlessPS}.  Our initial conditions become
\begin{subequations}\label{ourinitialconditions}
\begin{align}
    &\mathcal{R}_{\varphi k}(t_1)_{\text{rescaled}}=-\frac{k}{a(t_1)}\frac{H(t_1)}{2\pi\dot{\bar\varphi}(t_1)} \;,\label{ourinitialconditions_r}  \\
    &\dot{\mathcal{R}}_{\varphi k}(t_1)_{\text{rescaled}}=i\frac{k^2}{a(t_1)^2}\frac{H(t_1)}{2\pi\dot{\bar\varphi}(t_1)}\;,\label{ourinitialconditions_rdot} \\
    &\mathcal{S}_{v k}(t_1) =\mathcal{S}_{T k}(t_1)=0\;.\label{ourinitialconditions_ss}
\end{align}
\end{subequations}

\section{Freeze-out}\label{ch:freezeout}
A quantum fluctuation is produced with a certain momentum. For the scales we are considering, the initial physical momentum will be much larger than the Hubble rate, i.e.~$k/a \gg H$. During inflation, $H$ is approximately constant, while $k/a$ rapidly decreases due to the expansion of the universe. At some point, $k/a$ becomes smaller than $H$, or in other words, the physical wavelength of a perturbation is greater than the Hubble radius (see fig.~\ref{fig:regimes}). Here we show that in this regime the curvature perturbations freeze-out, meaning that they freeze to a constant value. 
\\\\
Starting from eqs.~\eqref{result}, if one goes in the regime where $k/a$ is negligible compared to the other scales in question, and if in that regime one searches for a stationary solution, one is left with three independent equations describing two variables,
\begin{subequations}\label{crossedresult}
\begin{align}
        \textcolor{black}{\cancel{\ddot{\mathcal{R}}^{ }_{\varphi k}}} &= \textcolor{black}{- \frac{\varrho H}{\dot{\bar\varphi}}
           - \cancel{\dot{\mathcal{R}}^{ }_{\varphi k}} \left[ \Upsilon + 3 H + 2 \mathcal{F} \right]
           - \mathcal{R}^{ }_{\varphi k} \cancel{\left[ \frac{k^2}{a^2} \right]}} \\
        &\quad + \mathcal{S}^{ }_{v k} \left[\frac{4\pi G ( \Upsilon  +  2 \mathcal{F} )}{H} \right] 
           -  \mathcal{S}^{ }_{T k} \left[\frac{4\pi G}{H} \left( 1 - \frac{\bar{p}_{,T}}{\bar{e}_{,T}} \right)
           + \frac{V_{,\varphi T} + \Upsilon_{,T} \dot{\bar\varphi}}{\dot{\bar\varphi} \bar{e}_{,T}}\right]
           \;,\nonumber\\[0.5em]
%%%%%%%%
        \cancel{\dot{ \mathcal{S}}^{ }_{v k}} &= -\cancel{ \dot{\mathcal{R}}^{ }_{\varphi k}} \left[ \bar{e} + \bar{p} \right]
           -  \mathcal{S}^{ }_{v k} \left[3 H + \frac{4 \pi G \dot{\bar\varphi}^2}{H} \right]
           +  \mathcal{S}^{ }_{T k} \left[ \frac{\bar{p}_{,T}}{\bar{e}_{,T}} \right]
           \;, \\[0.5em]
%%%%%%%%
        \cancel{\dot{ \mathcal{S}}^{ }_{T k}} &= \textcolor{black}{\varrho \dot{\bar\varphi} H
           +\cancel{ \dot{\mathcal{R}}^{ }_{\varphi k}} \left[\Upsilon \dot{\bar\varphi}^2 + \frac{8\pi G \bar{e}(\bar{e}+\bar{p})}{H} \right]
           - \mathcal{R}^{ }_{\varphi k} \left[(\bar{e} + \bar{p}) \cancel{\frac{k^2}{a^2}} \right]}\nonumber \\
        &\quad -  \mathcal{S}^{ }_{v k} \left[\cancel{\frac{k^2}{a^2}} + \frac{4\pi G }{H} \left( \Upsilon \dot{\bar\varphi}^2 + \frac{8 \pi G \bar{e}(\bar{e} + \bar{p})}{H} \right) \right] \\
        &\quad +  \mathcal{S}_{T k} \left[\frac{\dot{H} - 4\pi G(\bar{e} + \bar{p})}{H}
           - 3 H \left( 1 + \frac{\bar{p}_{,T}}{\bar{e}_{,T}} \right)
           + \frac{( V_{,\varphi T} + \Upsilon_{,T} \dot{\bar\varphi}) \dot{\bar\varphi}}{\bar{e}_{,T}} \right]
           \;.\nonumber
\end{align}    
\end{subequations}
The equations above describe an overconstrained system, which is solved by vanishing isocurvature perturbations. This in turn implies the curvature perturbations obtain the same value,
\begin{equation}\label{overdampedconclusion}
    \begin{split}
        \mathcal{S}^{ }_{v k}=\mathcal{S}^{ }_{T k}=0 \;\;\overset{\text{\eqref{isocurvature}}}{\rightarrow}\;\;\;\mathcal{R}^{ }_{v k}=\mathcal{R}^{ }_{T k}=\mathcal{R}^{ }_{\varphi k}=\text{constant}\;.
    \end{split}
\end{equation}
This analytical result will be further confirmed numerically (see figs.~\ref{fig:perturb_strong} and \ref{fig:isocurvature} for the conclusion on the curvature and isocurvature freeze-out behaviour respectively).

\section{Radiation domination}\label{ch:radiationdomination}
While the inflaton field is expected to be the dominant component of the energy density in the beginning, this role is at some point taken over by the plasma. In this section we simplify the evolution equations \eqref{result} by assuming the inflaton contributions to be negligibly small,
\begin{equation}\label{limits}
\Dot{\bar{\varphi}} \rightarrow 0 \;,\;\;\;\;\; \;\; \Ddot{\bar{\varphi}} \rightarrow 0 \;,\;\;\;\;\; \;\; V\rightarrow 0\;.
\end{equation}
Starting with eq.~\eqref{result_b}, and omitting the $\dot{\bar\varphi}^2$ term, we find
\begin{equation}
    \mathcal{\dot{S}}_{v k}=-\mathcal{\dot{R}}_{\varphi k} (\bar{e}+\bar{p})-\mathcal{S}_{v k} \; 3H +\mathcal{S}_{T k} \;\frac{\bar{p}_{,T}}{\bar{e}_{,T}}\;.
\end{equation}
Rewriting $\mathcal{S}_{vk},\;\mathcal{S}_{Tk}  $ in terms of the curvature perturbations, and using the background identity $\frac{\bar{e}_{,T}\dot{T}}{\bar{e}+\bar{p}}=-3H$, which follows from eq.~\eqref{physical_d} in the absence of the inflaton field, leads to
\begin{equation}\label{radiation1}
    \dot{\mathcal{R}}_{v k}=\frac{\dot{\bar{p}}}{\bar{e}+\bar{p}}\left(\mathcal{R}_{Tk}-\mathcal{R}_{vk}\right)\;.
\end{equation}
Note that in radiation domination, $\dot{\bar{p}}=\bar{p}_{,T}\dot{T}$. 
\\\\
Next, we turn to eq.~\eqref{result_c}, where we again take the limits in eq.~\eqref{limits}, and replace the variables $\mathcal{S}_{vk},\;\mathcal{S}_{Tk}  $ with their curvature counterparts,
\begin{equation}
\begin{split}
    \dot{\mathcal{R}}_{Tk}=&\mathcal{R}_{Tk}\left[\frac{\dot H}{H}\left(1+\frac{3H}{\dot{\bar e}}(\bar e+\bar p)\right) -\frac{4\pi G}{H}(\bar e+\bar p) \right]\\&-\frac{(\bar e+\bar p)}{\dot{\bar e}}\left[ \frac{k^2}{a^2}+\frac{4\pi G}{H}\frac{8\pi G \bar e}{H}(\bar e+\bar p) \right]\mathcal{R}_{vk}.\\
\end{split}
\end{equation}
In the process we used an identity derived from eq.~\eqref{physical_d} by taking its time derivative in the absence of $\varphi$,
\begin{equation}
    -\frac{3\dot{H}}{\dot{\bar e}}(\bar e+\bar p)-3H\left(1+\frac{\bar p_{,T}}{\bar e_{,T}}\right)=\frac{\dot{\bar e}_{,T}}{\bar e_{,T}}+\frac{\ddot T}{\dot T}\;.
\end{equation}
We can further use eq.~\eqref{physical_b} to remove the first term multiplying $\mathcal{R}_{Tk}$ and express $8\pi G \bar e$ in terms of $3H^2$, while also simplifying the $\mathcal{R}_{vk}$ term using eq.~\eqref{physical_d},
\begin{equation}\label{radiation2}
    \dot{\mathcal{R}}_{Tk}=\frac{\mathcal{R}_{vk}}{3H}\frac{k^2}{a^2}-\frac{4\pi G(\bar{e}+\bar{p})}{H}\left( \mathcal{R}_{Tk}-\mathcal{R}_{vk}\right)\;.
\end{equation}
Our final equations in radiation domination are thus given by \eqref{radiation1} and \eqref{radiation2}. Note that they offer a nice theoretical insight: in radiation domination the inflaton curvature perturbation, $\mathcal{R}_{\varphi k}$, decouples from $\mathcal{R}_{vk}$ and $\mathcal{R}_{Tk}$. While $\mathcal{R}_{\varphi k}$ still exists and evolves, it no longer influences the plasma.
\\\\
We can also verify that the freeze-out conclusion in eq.~\eqref{overdampedconclusion} still applies if the overdamped regime overlaps with radiation domination. Focusing first on eq.~\eqref{radiation1}, and looking for a stationary solution, we notice $\mathcal{R}_{Tk}=\mathcal{R}_{vk}=$ constant. Similarly, looking for a stationary solution in eq.~\eqref{radiation2} in the regime where $k/a$ is negligible, $\mathcal{R}_{Tk}=\mathcal{R}_{vk}=$ constant. Hence, the conclusion in eq.~\eqref{overdampedconclusion} is recovered.
\\\\
It was numerically verified that equations \eqref{radiation1} and \eqref{radiation2} perfectly describe the dynamics of the plasma curvature perturbations deep in the radiation domination. Needless to say, it is numerically cheaper to solve eqs.~\eqref{radiation1} and \eqref{radiation2} than the full set of eqs.~\eqref{result}.

\section{Importance of gauge invariance}\label{ch;importanceofgaugeinvariance}
In the introduction, Ch.~\ref{ch:gaugeintro}, we have already illustrated the importance of gauge freedom by describing a rough estimate for the power spectrum (sometimes used in the literature), which essentially sets all metric perturbations to zero. The precise equations were then omitted, but now we are in a good position to present the problem explicitly.
\\\\
We can illustrate the importance of gauge freedom already in a simplified setting, where $T=\Upsilon=\varrho=\delta T=0$, i.e.~in the absence of plasma. If we in addition omit the metric perturbations, in momentum space eq.~\eqref{LangevinPerturbation} becomes
\begin{equation}\label{haha}
\begin{split}
\delta\varphi''
 &+ 2\mathcal{H}   \delta\varphi'
 +k^2 \delta\varphi 
 + a^2 V^{ }_{,\varphi\varphi}    
  \, \delta\varphi  =0\;.
 \end{split}
\end{equation}
We now perform a change of variables, $\delta\hat{\varphi}\equiv a\delta\varphi$, and multiply eq.~\eqref{haha} by the scale factor, yielding
\begin{equation}\label{gaugedeop}
    \delta\hat{\varphi}''+\left[k^2+a^2 V_{,\varphi\varphi}-\frac{a''}{a}\right]\delta\hat{\varphi}=0\;.
\end{equation}
We proceed to derive an equation for the gauge-invariant version of $\delta\hat{\varphi}$, namely $\hat{Q}_{\varphi k}=-\delta\hat{\varphi}-a\dot{\bar{\varphi}}(h_D+\nabla^2\vartheta/3)/H$. As a starting point, let us note that if $T=\Upsilon=\varrho=0$, the full equation \eqref{result_a} reduces to eq.~\eqref{justRphiequation}, with $\Upsilon$ and $\varrho$ set to zero. As we have shown in Ch.~\ref{ch:initialconditions}, this can then be rewritten into eq.~\eqref{init},
\begin{equation}\label{jaja}
    \hat{Q}_{\varphi k}''+\left[k^2-\frac{\HH}{a\bar\varphi'}\left(\frac{a\bar\varphi'}{\HH}\right)''\right]\hat{Q}_{\varphi k}=0\;.
\end{equation}
We write out the derivatives of the background quantities explicitly. We further take a conformal time derivative of eq.~\eqref{LangevinBackground} to eliminate $\bar{\varphi}'''$ from eq.~\eqref{jaja}. Finally, we remove the remaining $\bar{\varphi}''$ term through eq.~\eqref{LangevinBackground}, resulting in
\begin{equation}\label{gaugenondeop}
    \hat{Q}_{\varphi k}''+\left[k^2+a^2 V_{,\varphi\varphi}-\frac{a''}{a}+\left\{\frac{\HH''}{\HH}-2\frac{\HH'^2}{\HH^2}+\frac{2a^2\,V_{,\varphi}}{\bar{\varphi}'}\left(\HH-\frac{\HH'}{\HH}\right)\right\}\right]\hat{Q}_{\varphi k}=0\;.
\end{equation}
Comparing eqs.~\eqref{gaugedeop} and \eqref{gaugenondeop}, we notice that they differ by the factors in the curly brackets of eq.~\eqref{gaugenondeop}. In pure dS space, $\HH=-1/\tau$,\footnote{In dS, $H=\text{constant}$ with $a\propto e^{H t}$. Combining this with the integrated relation $ad\tau=dt$, we get $\tau=-1/(aH)$, but since $\HH=H a$, we get $\HH=-1/\tau$. } and hence the terms in the curly brackets vanish.\footnote{The curly brackets can also be written as proportional to the slow roll parameter $\epsilon_H\equiv-\dot{H}/H^2$.} However, this is not true in general. We see that even in vacuum, the gauge-invariant prescription differs from the rough estimate as soon as the expansion is not exactly exponential; this should in practice be of more importance for higher momentum modes, as they exit the horizon later (when the expansion may be further away from pure dS).

\chapter{Noise autocorrelator}\label{ch:noise}

\section{The idea}
In order to solve the equations, certain properties of the noise need to be assumed or derived. First, our system exhibits a separation of time scales: the plasma is assumed to equilibrate much faster than the inflaton. In this picture, the fast plasma fluctuations are perceived by the inflaton as random, memory-less ``kicks''.\footnote{The white noise assumption is an idealized setting, where plasma self-interactions are strong enough that the influence of a previous ``noise-kick'' is forgotten by the time the next one arrives (the correlation time approaches zero). An analogy could be throwing stones in water: each stone throw is a ``kick'', after which the water surface settles down (in a correlation time), before the next stone (``kick'') arrives. If the ``stones'' would be thrown faster than the water had a chance to settle down, the white noise assumption would no longer be valid.} In momentum space, the noise autocorrelator is then given by
\begin{equation}\label{correlator}
    \langle \varrho_k(t)\varrho_k(t') \rangle = \Omega_k(t) \,\delta(t-t')\;,
\end{equation}
or, using rescaled variables from eq.~\eqref{rescaled},
\begin{equation}\label{correlator2}
    \langle\hat{\varrho}_k(\tau)\hat{\varrho}_k(\tau')\rangle=a^6\Omega_k\underbrace{\delta(\tau-\tau')\frac{1}{a}}_{\delta(t-t')}\equiv \hat{\Omega}_k \,\delta(\tau-\tau')\;.
\end{equation}
We suppressed the time dependence of $\Omega_k$ for brevity. The brackets $\langle\rangle$ denote thermal average. %In a practical implementation, we treat the noise as gaussian. However, in this thesis we are only computing the power spectrum, i.e.~a two-point correlator, hence the gaussianity assumption has no formal influence. 
Evidently, the amplitude of the noise autocorrelator depends on time. Finding its analytical (implicit) time dependence will be the focus of this chapter.
\\\\
The coefficient $\Omega_k$ can be determined by a matching computation. In essence, we derive the analytical expression for the power spectrum in two ways: by using quantum-statistical and classical methods. In the regime where both the inflaton and the plasma are in equilibrium, the two results can be equated, fixing the form of $\Omega_k$. Important to note is that in practice, the inflaton does not need to reach the equilibrium. The hypothetical scenario is just used to fix the form of $\Omega_k$, as is normally the case with matching computations. The resulting dynamics should then smoothly interpolate between the quantum and classical domains, with the latter having the benefit of allowing out-of-equilibrium dynamics of the inflaton field. In our derivation we assume that the values of the coefficients in the equations do not change significantly in the equilibration process.
\\\\
The organization of this subsection is as follows. First, we use the simplified, early time evolution equation for $\hat{Q}_{\varphi k}$ as a starting point to derive the quantum-statistical power spectrum. Then, we compute the power spectrum using classical-statistical physics. Lastly, we use those results to read off what $\Omega_k$ needs to be and check that it has the correct quantum and classical limits.

\section{Quantum-statistical equilibrium power spectrum }\label{sec:noisequantum}
We return to eq.~\eqref{mink}, which upon omitting classical matching coefficients $\Omega_k$ and $\Upsilon$ becomes
\begin{equation}\label{minkquantum}
     \hat{Q}_{\varphi k}''(\tau)+ \hat{\epsilon}_{k}^2\;\hat{Q}_{\varphi k}(\tau)=0\;.
\end{equation}
This is the equation we use in the quantum regime, which also defined $\hat{\epsilon}_{k}^2$, see eq.~\eqref{init}. Let us briefly note that this is exactly Mukhanov's result (eq. (31) from ref. \cite{Mukhanov}, with our $\hat{Q}_{\varphi k}$ exactly matching Mukhanov's $v_k$: $\hat{Q}_{\varphi k}=v_k$), so that in the absence of the plasma, known gauge-invariant results are reproduced. This equation has two canonical initial conditions defined at time $t_1$ (at this time we define an instantaneous Hilbert space). The first one is
\begin{equation}\label{commie1}
    \hat{Q}_{\varphi k}'(\tau_1)=-i \hat{\epsilon}_k (\tau_1)\, \hat{Q}_{\varphi k}(\tau_1)\;,% \;\;\;\;\rightarrow \;\;\;\; \dot{\hat{Q}}_{\varphi k}(t_1)=-i  \epsilon_k (t_1)\, \hat{Q}_{\varphi k}(t_1)\;,
\end{equation}
which corresponds to a forward-propagating mode. The second initial condition is given by the canonical commutation relations,
\begin{equation}\label{commie}
    [\hat{Q}_\varphi(\tau,\textbf{x}),\hat{Q}_\varphi'(\tau,\textbf{y})]=i\delta^{(3)}(\textbf{x}-\textbf{y})\;. %\;\;\;\;\;\;\; \approx \;\;\;\;\;\;\;     [\delta\varphi(\tau,\textbf{x}),\delta\dot{\varphi}(\tau,\textbf{y})]=\frac{i}{a(\tau_1)^3}\delta^{(3)}(\textbf{x}-\textbf{y})
\end{equation}
%where the similarity sign comes from neglecting the metric perturbations in $\mathcal{R}_\varphi$. Both initial conditions will be used to compute the quantum-statistical power spectrum. 
The field $\hat{Q}_\varphi(\tau,\textbf{x})$ has a mode expansion,
\begin{equation}\label{fourier}
    \hat{Q}_\varphi(\tau,\textbf{x})\equiv \int \frac{d^3\textbf{k}}{\sqrt{(2\pi)^3}}\left[ w_\textbf{k}\hat{Q}_{\varphi k}(\tau)e^{i\textbf{k}\cdot\textbf{x}}+w_\textbf{k}^\dagger\hat{Q}^*_{\varphi k}(\tau)e^{-i\textbf{k}\cdot\textbf{x}} \right]\;.
\end{equation}
In the spirit of the paragraph above eq.~\eqref{ourinitialconditions}, we can rescale mode functions such that the power spectrum $|\hat{Q}_k|^2$ comes out dimensionless,  $\hat{Q}_k\equiv\frac{k^{3/2}}{\sqrt{2\pi^2}}\hat{Q}_{\varphi k}$, leading to
\begin{equation}\label{dimlessPS}
    \begin{split}
        \langle0||\hat{Q}_\varphi(\tau,\textbf{x})|^2|0\rangle&=\Bigl\langle0\Bigr|\int \frac{d^3\textbf{k}}{\sqrt{(2\pi)^3}}\int \frac{d^3\textbf{h}}{\sqrt{(2\pi)^3}}\left[ w_\textbf{k}\hat{Q}_{\varphi k}(\tau)e^{i\textbf{k}\cdot\textbf{x}}+w_\textbf{k}^\dagger\hat{Q}^*_{\varphi k}(\tau)e^{-i\textbf{k}\cdot\textbf{x}} \right]\\
        &\quad \quad \quad\quad\hspace{3cm} \times\quad\left[ w_\textbf{h}\hat{Q}_{\varphi h}(\tau)e^{i\textbf{h}\cdot\textbf{x}}+w_\textbf{h}^\dagger\hat{Q}^*_{\varphi h}(\tau)e^{-i\textbf{h}\cdot\textbf{x}} \right]\Bigl|0\Bigr\rangle
        \\
        &=\Bigl\langle0\Bigr|\int\int \frac{d^3\textbf{k}\;d^3\textbf{h}}{(2\pi)^3} \left[\Bigl(w_\textbf{h}^\dagger w_\textbf{k}+\delta^{(3)}(\textbf{k}-\textbf{h})\Bigr) \hat{Q}_{\varphi k}(\tau)e^{i(\textbf{k}-\textbf{h})\cdot\textbf{x}}\hat{Q}^*_{\varphi h}(\tau)\right.\\
        &\left.\hspace{5.3cm}+\;w_\textbf{k}^\dagger w_\textbf{h} \hat{Q}^*_{\varphi k}(\tau)\hat{Q}_{\varphi h}(\tau)e^{i(\textbf{h}-\textbf{k})\cdot\textbf{x}} \right] \Bigl|0\Bigr\rangle\\
        &=\Bigl\langle 0\Bigr| \int \frac{d^3\textbf{k}}{(2\pi)^3} \left|\hat{Q}_{\varphi k}(\tau)\right|^2 \Bigl|0\Bigr\rangle =\int \text{d}\ln{k}\,|\hat{Q}_{ k}(\tau)|^2\;.
    \end{split}
\end{equation} 
To rewrite the second initial condition, we consider the following commutator, while keeping in mind that the result should equal eq.~\eqref{commie},
\begin{equation}\label{commie3}
\begin{split}
     [\hat{Q}_\varphi (\tau_1,\textbf{x}),\hat{Q}_\varphi'(\tau_1,\textbf{y})]&\overset{\textcolor{white}{(5.4)}}{=}\int \frac{d^3\textbf{k}}{(2\pi)^3}\left[ \hat{Q}_{\varphi k}(\tau_1)\hat{Q}'^*_{\varphi k}(\tau_1)\, e^{i\textbf{k}(\textbf{x}-\textbf{y})} -\hat{Q}'_{\varphi k}(\tau_1)\hat{Q}^*_{\varphi k}(\tau_1)\, e^{-i\textbf{k}(\textbf{x}-\textbf{y})} \right]\\
     &\overset{\text{\eqref{commie1}}}{=}2i \hat{\epsilon}_k(\tau_1) \delta^{(3)}(\textbf{x}-\textbf{y}) | \hat{Q}_{\varphi k}(\tau_1)|^2\\
     &\overset{\text{\eqref{commie}}}{=}i\delta^{(3)}(\textbf{x}-\textbf{y})\\
     & \;\;\Rightarrow\; |\hat{Q}_{\varphi k}(\tau)|^2=\frac{1}{2\hat{\epsilon}_k(\tau)}\;.
\end{split}
\end{equation}
In the first line we used $[w_\textbf{k},w^\dagger_\textbf{h}]=\delta^{(3)}(\textbf{k}-\textbf{h})$, while in the second line we recognized a delta function and used eq.~\eqref{commie1}. With this in hand, the vacuum expectation value, eq.~\eqref{dimlessPS}, becomes
\begin{equation}\label{quantumvac}
    \langle0||\hat{Q}_\varphi(\tau,\textbf{x})|^2|0\rangle=\int \text{d}\ln{k}\, \frac{k^3}{2\pi^2}\frac{1}{2\hat{\epsilon}_k}\;.
\end{equation}
Next we turn to computing the thermal expectation value of $\langle|\hat{Q}_\varphi(\tau,\textbf{x})|^2\rangle$,
\begin{equation}\label{quantumnonvac}
    \begin{split}
        \langle|\hat{Q}_\varphi(\tau,\textbf{x})|^2 \rangle&=\left\langle\int \frac{d^3\textbf{k}}{\sqrt{(2\pi)^3}}\int \frac{d^3\textbf{h}}{\sqrt{(2\pi)^3}}\left[ w_\textbf{k}\hat{Q}_{\varphi k}(\tau)e^{i\textbf{k}\cdot\textbf{x}}+w_\textbf{k}^\dagger\hat{Q}^*_{\varphi k}(\tau)e^{-i\textbf{k}\cdot\textbf{x}} \right] \right.\\ 
        &\hspace{5cm}\left. \times \left[ w_\textbf{h}\hat{Q}_{\varphi h}(\tau)e^{i\textbf{h}\cdot\textbf{x}}+w_\textbf{h}^\dagger\hat{Q}^*_{\varphi h}(\tau)e^{-i\textbf{h}\cdot\textbf{x}} \right]\right\rangle\\
        &=\int \frac{d^3\textbf{k}}{\sqrt{(2\pi)^3}}\int \frac{d^3\textbf{h}}{\sqrt{(2\pi)^3}} \left[\left\langle\left(w_\textbf{h}^\dagger w_\textbf{k} +\delta^{(3)}(\textbf{k}-\textbf{h})\right)\right\rangle\; \hat{Q}_{\varphi k}(\tau)e^{i(\textbf{k}-\textbf{h})\cdot\textbf{x}}\hat{Q}^*_{\varphi h}(\tau)\right.\\
        &\hspace{5cm}\left.+\left\langle w_\textbf{k}^\dagger w_\textbf{h}\right\rangle \;\hat{Q}^*_{\varphi k}(\tau)\hat{Q}_{\varphi h}(\tau)e^{i(\textbf{h}-\textbf{k})\cdot\textbf{x}}\right]  \\
        &=  \int \frac{d^3\textbf{k}}{(2\pi)^3}\left|\hat{Q}_{\varphi k}(\tau)\right|^2 \left(2\langle n_\textbf{k}\rangle +1\right)\;.
    \end{split}
\end{equation}
Going to the last line, we recognized the number density operator $\langle w_\textbf{k}^\dagger w_\textbf{h}\rangle=\delta^{(3)}(\textbf{k}-\textbf{h}) \langle n_\textbf{k}\rangle$. The expectation value $\langle n_\textbf{k}\rangle$ is just the Bose-Einstein distribution, denoted by $n_\text{B}$,
\begin{equation}
\begin{split}
    \langle n_\textbf{k}\rangle &=\frac{\underset{n_\textbf{k}=0}{\sum^\infty} \bra{n_\textbf{k}}n_\textbf{k} \exp{-\frac{\hat{H}}{T}}\ket{n_\textbf{k}}}{\underset{n_\textbf{k}=0}{\sum^\infty} \bra{n_\textbf{k}}\exp{-\frac{\hat{H}}{T}}\ket{n_\textbf{k}}}=\frac{\underset{n_\textbf{k}=0}{\sum^\infty} n_\textbf{k} \exp{-\frac{E_{n_\textbf{k}}}{T}}}{\underset{n_\textbf{k}=0}{\sum^\infty} \exp{-\frac{E_{n_\textbf{k}}}{T}}}=\frac{\frac{\partial}{\partial (-\frac{\hat{\epsilon}_\textbf{k}}{aT})}\underset{n_\textbf{k}=0}{\sum^\infty}  \exp{-\frac{\hat{\epsilon}_\textbf{k}n_\textbf{k}}{aT}}}{\underset{n_\textbf{k}=0}{\sum^\infty} \exp{-\frac{\hat{\epsilon}_\textbf{k}n_\textbf{k}}{aT}}}
    \\&=\frac{\partial}{\partial (-\frac{\hat{\epsilon}_\textbf{k}}{aT})} \ln{\left(\frac{1}{1-\exp{-\frac{\hat{\epsilon}_\textbf{k}}{aT}}}\right)}=\frac{1}{\exp{\frac{\hat{\epsilon}_\textbf{k}}{aT}}-1}\equiv n_\text{B}\left(\frac{\hat{\epsilon}_\textbf{k}}{a}\right)\;.
    \end{split}
\end{equation}
The first fraction still includes the number density operator and the Hamiltonian $\hat{H}$, while in the second fraction we have replaced them with their expectation values. The second line follows from the geometric series expansion. The thermal expectation value thus becomes
\begin{equation}\label{quantumnonvac}
    \begin{split}
        \langle|\hat{Q}_\varphi(\tau,\textbf{x})|^2 \rangle=\int \text{d}\ln{k}\, \frac{k^3}{2\pi^2}\frac{1}{2\hat{\epsilon}_k}\left[2\,n_\text{B}\left(\frac{\hat{\epsilon}_\textbf{k}}{a}\right) +1\right]\;.
    \end{split}
\end{equation}
Before moving on to the classical-statistical power spectrum, we compare eqs.~\eqref{quantumvac} and \eqref{quantumnonvac}, and briefly note that the non-vacuum (thermal) contribution to the power spectrum is given by 
\begin{equation}\label{nonhomo}
\langle|\hat{Q}_\varphi(\tau,\textbf{x})|^2\rangle_{\text{therm.}}=\int \text{d}\ln{k}\, \frac{k^3}{2\pi^2}\frac{1}{\hat{\epsilon}_k}n_\text{B}\left(\frac{\hat{\epsilon}_\textbf{k}}{a}\right)\;\;\rightarrow\;\;\;\langle |\hat{Q}_{\varphi k}(\tau)|^2\rangle_\text{therm.}= \frac{k^3}{2\pi^2}\frac{1}{\hat{\epsilon}_k}n_\text{B}\left(\frac{\hat{\epsilon}_\textbf{k}}{a}\right)\;.
\end{equation}
While the equation being solved, eq.~\eqref{minkquantum}, is a homogeneous equation, the thermal background makes the contribution \eqref{nonhomo} behave as if it solved a non-homogeneous equation. Therefore, we treat it on the level of a particular solution, when comparing to the particular solution obtained by classical statistical methods, eq.~\eqref{nonhomoC}. 

\section{Classical-statistical equilibrium power spectrum}
We now return to eq.~\eqref{mink}. We obtain the particular solution via the Green's function method. Noting that $ G_\text{k}(\tau,\tau_1)=\int_{-\infty}^\infty \frac{d\omega}{2\pi} e^{-i\omega (\tau-\tau_1)}G_\text{k}(\omega)$, we obtain $G_\text{k}(\omega)$ from
\begin{equation}\label{greens}
    \begin{split}
        (\partial_\tau^2+\hat{\Upsilon}\partial_\tau+\hat{\epsilon}_\text{k}^2)\,G_\text{k}(\tau)&=\delta(\tau)\\
        \Leftrightarrow\quad\int_{-\infty}^\infty \frac{d\omega}{2\pi} (-\omega^2-i\omega \hat{\Upsilon}+\hat{\epsilon}_\text{k}^2)\,G_\text{k}(\omega) \,e^{-i\omega \tau}&=\int_{-\infty}^\infty \frac{d\omega}{2\pi}  \,e^{-i\omega \tau}\\
        \Rightarrow \;\; G_\text{k}(\omega)&=\frac{1}{-\omega^2-i\omega \hat{\Upsilon}+\hat{\epsilon}_\text{k}^2}\;. 
    \end{split}
\end{equation}
The particular solution is then given by
\begin{equation}
\begin{split}
    \hat{Q}_{\varphi \text{k}, \,\text{part.}}(\tau)&=\int_{-\infty}^\infty d\tau_1 \, G_\text{k}(\tau,\tau_1)\,\hat{\varrho}_\text{k}(\tau_1)=\int_{-\infty}^\infty d\tau_1 \, \int_{-\infty}^\infty \frac{d\omega}{2\pi} e^{-i\omega (\tau-\tau_1)}G_\textbf{k}(\omega)\,\hat{\varrho}_\text{k}(\tau_1)\;.
\end{split}
\end{equation}
The power spectrum becomes
\begin{equation}\label{nonhomoC}
    \begin{split}
        \langle|\hat{Q}_{\varphi \text{k}}(\tau)|^2\rangle_{\text{part.}}&\overset{\textcolor{white}{(6.14)}}{=}\int_{-\infty}^\infty d\tau_1 \, \int_{-\infty}^\infty \frac{d\omega}{2\pi} \int_{-\infty}^\infty d\tau_2 \, \int_{-\infty}^\infty \frac{d\omega_1}{2\pi}\Bigl[ e^{i\omega_1 (\tau-\tau_2)}e^{-i\omega (\tau-\tau_1)}\;\times\\
        &\hspace{6cm}\times\;G^*_\text{k}(\omega_1)G_\text{k}(\omega)\,\langle\hat{\varrho}^*_\text{k}(\tau_2)\,\hat{\varrho}_\text{k}(\tau_1)\rangle \Bigr]\\
        &\overset{\textcolor{white}{.}\text{\eqref{correlator}\textcolor{white}{.}}}{=}\hat{\Omega}_{\text{k}}\int_{-\infty}^\infty d\tau_1 \, \int_{-\infty}^\infty \frac{d\omega}{2\pi} \, \int_{-\infty}^\infty \frac{d\omega_1}{2\pi} e^{i\omega_1 (\tau-\tau_1)}e^{-i\omega (\tau-\tau_1)}G^*_\text{k}(\omega_1)G_\text{k}(\omega)\\
        &\overset{\textcolor{white}{(6.14)}}{=}\hat{\Omega}_{\text{k}} \, \int_{-\infty}^\infty d\omega \, \int_{-\infty}^\infty \frac{d\omega_1}{2\pi} e^{i(\omega_1-\omega) \tau}G^*_\text{k}(\omega_1)G_\text{k}(\omega)\delta(\omega-\omega_1)\\
        &\overset{\textcolor{white}{(6.14)}}{=}\hat{\Omega}_{\text{k}} \, \int_{-\infty}^\infty \frac{d\omega_1}{2\pi} |G_\text{k}(\omega_1)|^2\\
        &\overset{\eqref{greens}}{=}\hat{\Omega}_{\text{k}} \, \int_{-\infty}^\infty \frac{d\omega_1}{2\pi}\left[ \frac{1}{(\omega_1-i\frac{\hat{\Upsilon}}{2}+\sqrt{\hat{\epsilon}^2_\text{k}-\frac{\hat{\Upsilon}^2}{4}})}\frac{1}{(\omega_1+i\frac{\hat{\Upsilon}}{2}+\sqrt{\hat{\epsilon}^2_\text{k}-\frac{\hat{\Upsilon}^2}{4}})}\;\times \right. \\
        &\hspace{3cm}\times \;\left. \frac{1}{(\omega_1-i\frac{\hat{\Upsilon}}{2}-\sqrt{\hat{\epsilon}^2_\text{k}-\frac{\hat{\Upsilon}^2}{4}})}\frac{1}{(\omega_1+i\frac{\hat{\Upsilon}}{2}-\sqrt{\hat{\epsilon}^2_\text{k}-\frac{\hat{\Upsilon}^2}{4}})}\right]\\
        &\overset{\textcolor{white}{(6.14)}}{=}\frac{\hat{\Omega}_\text{k}}{2\hat{\Upsilon}\sqrt{\hat{\epsilon}_\text{k}^2-\frac{\hat{\Upsilon}^2}{4}}}\left[\frac{1}{i\hat{\Upsilon}+2\sqrt{\hat{\epsilon}_\text{k}^2-\frac{\hat{\Upsilon}^2}{4}}}+\frac{1}{2\sqrt{\hat{\epsilon}_\text{k}^2-\frac{\hat{\Upsilon}^2}{4}}-i\hat{\Upsilon}}\right]\\
        &\overset{\textcolor{white}{(6.14)}}{=}\frac{\hat{\Omega}_\text{k}}{2\hat{\Upsilon}\hat{\epsilon}_\text{k}^2}\;.
    \end{split}
\end{equation}
We used the residue theorem for carrying out the last integral. For positive $\hat{\Upsilon}$ and $\hat{\epsilon}_\text{k}^2$, the poles of $G_\text{k}(\omega)$ lie in the lower half-plane, while $G_\text{k}^*(\omega)$ contributes two poles in the upper half-plane. Since the integral along the large semicircle vanishes in both half-planes, we can close the contour in either, leading to the result in the last line. \\\\
Before deriving the form of $\Omega_k$, let us comment on the Green's function. Closing the contour of $G_\text{k}(\tau,\tau_1)=\int_{-\infty}^\infty \frac{d\omega}{2\pi} e^{-i\omega (\tau-\tau_1)}G_\text{k}(\omega)$ in the lower half-plane, the Green's function is non-zero for $\tau>\tau_1$. On the other hand, closing the contour in the upper half-plane corresponds to the Green's function where $\tau<\tau_1$. Since there are no poles in the upper half-plane, time dependent Green's function must vanish for $\tau<\tau_1$. In other words, we are dealing with retarded Green's function, describing causal phenomena.

\section{Matching}
We now equate $\langle |\hat{Q}_{\varphi k}(\tau)|^2\rangle_\text{therm.}$ from \eqref{nonhomo} and $ \langle|\hat{Q}_{\varphi k}(\tau)|^2\rangle_{\text{part.}}$ from \eqref{nonhomoC}, as motivated by the concluding paragraph of Ch.~\ref{sec:noisequantum}. Henceforth, the $\hat{\Omega}_\text{k}$ from the two-point noise correlator \eqref{correlator} becomes
\begin{equation}
    \frac{k^3}{2\pi^2}\frac{1}{\hat{\epsilon}_k}n_\text{B}\left(\frac{\hat{\epsilon}_\textbf{k}}{a}\right)=\frac{\hat{\Omega}_\text{k}}{2\hat{\Upsilon}\hat{\epsilon}_\text{k}^2} \;\;\;\;\;\;\;\;\Rightarrow \;\;\;\;\;\;\;\;\hat{\Omega}_\text{k}=2\,\hat{\Upsilon}\,\hat{\epsilon}_\text{k}\,n_\text{B}\hspace{-0.05cm}\left(\frac{\hat{\epsilon}_\text{k}}{a}\right)\, \frac{k^3}{2\pi^2}\;.
\end{equation}
This autocorrelator forms one of the central analytical results of the thesis.\footnote{The other key analytical result are eqs.~\eqref{result}.} While the form of $\hat{\Omega}_\textbf{k}$ was derived in the regime where the inflaton has had time to equilibrate faster than the quantities $\hat{\epsilon}_\textbf{k},\,\hat{\Upsilon},\,aT$ changed in time, it is expected to hold also for the cases where the inflaton field reaches equilibration much later.
\\\\
As a final step, let us return to the physical units,
\begin{equation}\label{omega}
    \Omega_\textbf{k}=2\,\Upsilon\,\epsilon_\textbf{k}\,n_\text{B}\hspace{-0.05cm}\left(\epsilon_\text{k}\right)\, \frac{(k/a)^3}{2\pi^2}\;,\quad\quad \epsilon_\text{k}\equiv\frac{\hat{\epsilon}_\text{k}}{a}\overset{\text{\eqref{init}}}{=} \sqrt{\frac{k^2}{a^2}+\frac{\hat{m}^2}{a^2}}\;.
\end{equation}
It is illuminating to consider some limiting cases. In the quantum regime, where $\epsilon_\textbf{k}\gg T$, the Bose distribution exponentially suppresses the autocorrelator by $\exp{-\epsilon_\textbf{k}/T}$. On the other hand, in the classical regime where $T\gg \epsilon_\textbf{k}$, or $T\approx \epsilon_\textbf{k}$, the autocorrelator approximately becomes
\begin{equation}
    \Omega_\textbf{k} \approx 2\Upsilon T\;\frac{(k/a)^3}{2\pi^2}\;.
\end{equation}
The ``textbook'' low frequency fluctuation-dissipation result, $\Omega_\textbf{k}=2\Upsilon T$, is here supplemented by the volume element $1/a^3$ which we decided to absorb into $\Omega_\textbf{k}$, as well as the rescaling factor $k^3/(2\pi^2)$. Hence, we have shown that the derived autocorrelator \eqref{omega} has the correct quantum and classical limits, where it respectively vanishes and is proportional to $2\Upsilon T$.
\\\\
To summarize, in \textit{Part II} (Chapters \ref{ch:essentials} -- \ref{ch:noise}) we have derived gauge-invariant stochastic differential equations, their initial conditions, as well as the corresponding noise autocorrelator. These equations describe inflation and reheating for a perturbed FLRW universe containing one scalar inflaton field surrounded by a plasma. In \textit{Part III}, we will describe the numerical method used to solve those equations.

\part{Numerical implementation}\label{part:num}

\chapter{Numerical method of integration}\label{sec:NumMethodIntegration}
The plan of the next chapters is the following. First, in Ch.~\ref{sec:NumMethodIntegration}, we describe the numerical method of integration for solving the full eqs.~\eqref{result}. In Ch.~\ref{sec:NumericalChallenges}, we present some numerical challenges, and show how we solved them. We furthermore apply our numerical implementation to the specific example of \textit{Warm Inflation with the Standard Model}, see ref.~\cite{ZellSM}, in Ch.~\ref{sec:ModelBuilding}.

\section{Discretization}
We now outline the key ideas needed for a practical implementation. An overview of the full Mathematica code can be found in Appendix \ref{app:CodeOverview}.
\\\\
The main theoretical ingredient needed to obtain the power spectrum is determining a suitable integration method. At early times, the curvature perturbations change their values very rapidly,\footnote{This is related to the presence of acoustic oscillations, to be discussed in Ch.~\ref{ch:inittimeofintegandacoustosc}.} in contrast to their slowly changing values later on, e.g.~during the freeze-out. In practice, this renders any kind of solver with a fixed time step futile. However, once the step size is determined dynamically at each moment in time, a simple Itô scheme implementation delivers good results.
\\\\
We start the numerical implementation by rewriting eqs.~\eqref{result} in a set of four first-order equations in a matrix form,
\begin{equation}\label{matrixeq}
   \dot{\mathcal{V}}\;=\;\mathcal{M}\;\mathcal{V}\;+\;\mathcal{N} \;.
\end{equation}
The explicit expressions for the solution vector $\mathcal{V}$ and the stochastic noise term $\mathcal{N}$ are
\begin{equation}\label{solutionvector}
\mathcal{V}=
    \begin{pmatrix}
    \mathcal{R}_{\varphi k}\\
    \dot{\mathcal{R}}_{\varphi k}\\
    \mathcal{S}_{v k}\\
    \mathcal{S}_{T k}
\end{pmatrix}
\;,\quad\quad\quad
\mathcal{N}=
\begin{pmatrix}
    0\\-\frac{H}{\dot{\bar{\varphi}}}\\0\\\dot{\bar{\varphi}}\,H
\end{pmatrix}\;\varrho\;.
\end{equation}
The deterministic drift term $\mathcal{M}$ takes the form
\begin{equation}\label{matrixM}
    \mathcal{M}=
    \begin{pmatrix}
    0&&1&&0&&0\\
    -\frac{k^2}{a^2}&&-\left(\Upsilon +3H+2\mathcal{F} \right) && \frac{4\pi G}{H}\left(\Upsilon+2\mathcal{F}\right) &&c_4\\
    0&&-(\bar{e}+\bar{p})&&-\left(3H+\frac{4\pi G}{H}\dot{\bar{\varphi}}^2\right)&&\frac{\bar{p}_{,T}}{\bar{e}_{,T}}\\
    -\frac{k^2}{a^2}\left(\bar{e}+\bar{p}\right)&&\Upsilon \dot{\bar{\varphi}}^2+8\pi G\,\bar{e}\,\frac{\bar{e}+\bar{p}}{H}&&-\left[\frac{k^2}{a^2} +\frac{4\pi G}{H}\left(\Upsilon \dot{\bar{\varphi}}^2+8\pi G\, \bar{e}\,\frac{\bar{e}+\bar{p}}{H}\right) \right]&&e_4
\end{pmatrix},
\end{equation}
where we have defined, for readability,
\begin{subequations}
    \begin{align}
        c_4&=-\frac{4\pi G}{H}\left(1-\frac{\bar{p}_{,T}}{\bar{e}_{,T}}\right)-\frac{V_{,\varphi T}+\dot{\bar{\varphi}}\Upsilon_{,T}}{\dot{\bar{\varphi}}\,\bar{e}_{,T}}\;,\\
        e_4&=\frac{\dot{H}-4\pi G(\bar{e}+\bar{p})}{H}-3H\left(1+\frac{\bar{p}_{,T}}{\bar{e}_{,T}}\right)+\frac{\left(V_{,\varphi T}+\dot{\bar{\varphi}}\Upsilon_{,T}\right)\dot{\bar{\varphi}}}{\bar{e}_{,T}}\;.
    \end{align}
\end{subequations}
We may now discretize eq.~\eqref{matrixeq} using the Itô scheme,
\begin{equation}\label{itofunny}
    \frac{\mathcal{V}_{i+1}-\mathcal{V}_{i}}{\varepsilon_i}=\mathcal{M}_i\mathcal{V}_i+
    \frac{
    \mathcal{N}_i}{\sqrt{\varepsilon}_i}\;,
\end{equation}
with the discrete noise autocorrelator being%\footnote{Switching from Dirac to Kroenecker delta in the discretization process introduced the unintuitive $1/\sqrt{\varepsilon_i}$, which was absorbed into $\mathcal{N}$ like $\mathcal{N}_i\rightarrow{\frac{\mathcal{N}_i}{\sqrt{\varepsilon}}}$ such that the discrete autocorrelator obtained the conventional form, 
%\begin{equation}
    %\begin{split}
       %% \langle  \;\underbrace{\mathcal{N}_i}_{\equiv \mathcal{N}(\varepsilon \cdot i)}\mathcal{N}_j\rangle&=W_i\;\frac{\delta_{ij}}{|\varepsilon|}\quad\quad\overset{\text{rescale  }\mathcal{N}_i\rightarrow{\frac{\mathcal{N}_i}{\sqrt{\varepsilon}}}}{\rightarrow}\\
        %% \end{split}
%\end{equation}
%In the second line, we have essentially used $\delta(t-t')= \delta((i-j)\varepsilon)\rightarrow
%\frac{\delta_{ij}}{|\varepsilon|}$
 %. \hl{this relies on a constant $\varepsilon$ - can you make an argument like this also for $\varepsilon$ that changes in time?}}
\begin{equation}\label{conventional}
    \langle \mathcal{N}_i\, \mathcal{N}_j^\dagger \rangle=W_i\;\delta_{ij}\;,
\end{equation}
where
\begin{equation}\label{noisematrix}
  W_i=
    2\Upsilon_i\,\epsilon_{\textbf{k}i}\;n_\text{B}\hspace{-0.05cm}\left(\epsilon_{\text{k}i}\right)\, \frac{(k/a_i)^3}{2\pi^2}\;
    \begin{pmatrix}
        0&0&0&0\\
        0& \frac{H_i^2}{\dot{\bar{\varphi}}^2_i}&0&-H^2_i\\
        0&0&0&0\\
        0&-H^2_i&0&H^2_i \dot{\bar{\varphi}}^2_i
    \end{pmatrix}\;.
\end{equation}
We adopted the definition $f_i\equiv f(t_i)$, and denoted a time step by $\varepsilon_i=t_{i+1}-t_i$.
\\\\
Switching from the Dirac to the Kronecker delta in the discretization process introduced a factor $\left|1/\varepsilon_i\right|$. This was then absorbed into noise through $\mathcal{N}_i\rightarrow \mathcal{N}_i/\sqrt{\varepsilon_i}$, such that the discrete autocorrelator obtained the conventional form, eq.~\eqref{conventional}, i.e.~without any factors of $\left|1/\varepsilon_i\right|$. Consequently, the discretized Ito equation contains $ 1/\sqrt{\varepsilon_i}$ in the noise term.\\\\
At this point, the solution can be recursively found as
\begin{equation}\label{curvaturediscreteevolution}
    \mathcal{V}_{i+1}=\bigl(1+\varepsilon_i\,\mathcal{M}_i\bigr)\mathcal{V}_i+
    \mathcal{N}_i \sqrt{\varepsilon}_i\;.
\end{equation}
As mentioned at the beginning of the section, the time step $\varepsilon_i$ is not constant in time. Here we describe how we chose its values. Theoretically, smaller time steps are closer to the continuum limit, however, too small steps become prohibitively expensive. In practice, we choose the value of $\varepsilon_i$ such that in a given time step, the drift term $\mathcal{M}$ does not change the solution too much,
\begin{equation}\label{tolerance}
    \varepsilon_i\equiv \frac{\texttt{tol}}{\text{max\{}|\mathcal{M}_i\;\text{eigenvalues|\}}}\;.
\end{equation}
The denominator makes sure that the time step is small enough to keep the drift evolution in eq.~\eqref{curvaturediscreteevolution} moderate. On the other hand, the tolerance, \texttt{tol}, determines how close to the continuum limit we wish to be. In a practical implementation, it typically\footnote{See discussion in Ch.~\ref{ch:inittimeofintegandacoustosc}.} appears to be sufficient to choose \texttt{tol}$=10^{-3}$.

\section{Observables}\label{ch:observables}
The discrete evolution eq.~\eqref{curvaturediscreteevolution} can be used to construct the power spectrum by rewriting the obtained solution vector $\mathcal{V}$ as moduli squared of curvature perturbations. An explicit example of this procedure is provided in Ch.~\ref{sec:matrixformalism}. In order to compare the results with the data, two more observational signatures can be easily computed: spectral tilt, and tensor-to-scalar ratio. Here we describe how to obtain them all.
\\\\
The power spectrum is defined as a square of the mode function related to curvature perturbations,
\begin{equation}
    \mathcal{P}_{\mathcal{R}_{\varphi k}}\equiv\left|\mathcal{R}_{\varphi k}\right|^2\;,\quad\quad \mathcal{P}_{\mathcal{R}_{vk}}\equiv\left|\mathcal{R}_{v k}\right|^2\;,\quad\quad \mathcal{P}_{\mathcal{R}_{Tk}}\equiv\left|\mathcal{R}_{T k}\right|^2\;.
\end{equation}
The power spectra are functions of the wave vector, $k$. The dependence on $k$ can be parameterized as a power-law,
\begin{equation}\label{amplitude}
    \mathcal{P}_{\mathcal{R}} (k)\equiv A_s\left(\frac{k}{k_\star}\right)^{n_s-1}\;,
\end{equation}
where $k_\star$ is a CMB reference scale, taken to be $0.05 \text{ Mpc}^{-1}$ (ref.~\cite{Planck_inflation_2020}), and $A_s$ is the amplitude at that scale. The spectral tilt, $n_s-1$ (where $n_s$ is the spectral index), is then just a logarithmic change of the frozen-out power spectrum with respect to momentum,
\begin{equation}\label{tilt}
    n_s-1\equiv \frac{d\ln{\langle \mathcal{P}_{\mathcal{R}} \rangle}}{d\ln{k}}\;.
\end{equation}
This quantity measures how far away we are from a \textit{scale invariant} power spectrum ($n_s-1=0$), i.e.~how far away we are from pure exponential expansion during inflation. 
\\\\
In practice, we start by running our numerical simulation for a specific choice of $k$, for example $k_\star$, and obtain the power spectrum for that $k$. To obtain the spectral tilt, we must additionally run the simulation for nearby $k$ values. Specifically, we approximate the nearby values as $10^{\delta_t} \;k$ and $10^{-\delta_t} \;k$, to denote a ``forward step'' and a ``backward step'' in momentum respectively. Then, we can use a numerical approximation to eq.~\eqref{tilt},
\begin{equation}\label{nsapprox}
    n_s-1\approx \frac{\log_{10}{\left\langle \mathcal{P}_{\mathcal{R}}\left(10^{\delta_t} \;k\right)  \right\rangle}-\log_{10}{\left\langle \mathcal{P}_{\mathcal{R}}\left(10^{-\delta_t} \;k\right)  \right\rangle}}{2\,\delta_t}\;.
\end{equation}
The numerical parameter $\delta_t$ needs to be small enough for the discretization approximation \eqref{nsapprox} to be reliable. However, since we expect the spectral index to be around $n_s\approx 1$, the two values of the numerator are expected to be very close. This can result in a loss of numerical precision, i.e.~significance loss. Hence, $\delta_t$ cannot be arbitrarily small either. In practice, we adopt the value of $\delta_t=0.1$. % (``0.1 decade''). Since we only compute observables at the pivot scale, it is not necessary to include the running of the spectral tilt in the computation.
\\\\
Aside from $A_s$ and $n_s$, we provide another observable, the tensor-to-scalar ratio. It is defined, as the name suggests, for the reference scale $k=k_\star$,
\begin{equation}\label{r}
    r=\frac{\mathcal{P}_t}{A_s},
\end{equation}
where $\mathcal{P}_t$ stands for tensor power spectrum. In vacuum, it takes the value
\begin{equation}
    \mathcal{P}_t = \frac{16 H_\star^2}{\pi\, m_{Pl}^2}\;.
\end{equation}
In principle, one should also include a thermal noise contribution to the tensor power spectrum. However, that is expected to grow as $k^3$ \cite{kcube1,kcube2,kcube3}. For CMB modes under consideration, this number is negligibly small. Hence, we adopt the vacuum value when computing $r$ of eq.~\eqref{r}.

\chapter{Numerical challenges}\label{sec:NumericalChallenges}
In this section we present some features of our equations hindering the numerics, and simple solutions that we implemented in our code.

\section{Singular coefficients}\label{ch:singularities}
\begin{figure}[!b]
    \centering
    \includegraphics[width=0.78\linewidth]{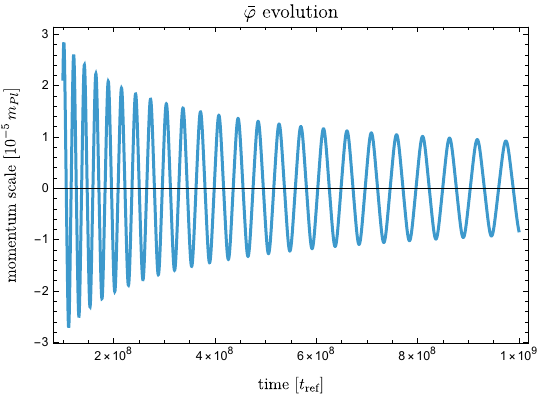}   \caption{The inflaton field typically experiences oscillations at late times, and thus there exist moments at which $\dot{\bar\varphi}$ vanishes.}
    \label{graph:baclgroundinflaton}
\end{figure}

After inflation has ended, the typical evolution of the background inflaton field  includes oscillations around zero, see fig.~\ref{graph:baclgroundinflaton}. 
Hence, there exist oscillation times $t_o$ at which the field's derivative vanishes, $\dot{\bar{\varphi}}(t_o)=0$. This can be difficult to handle, since the coefficients of our equations \eqref{result}, such as $\mathcal{F}$, contain terms proportional to $1/\dot{\bar{\varphi}}$, which at times $t_o$ become singular. We now investigate the behaviour of our equations around these oscillation times, illustrate the problem, and present a solution.
\\\\
Around a minimum, the background inflaton field can be approximately described as 
\begin{equation}\label{singulphi}
    \bar{\varphi}=-A\cos{[m(t-t_o)]}\;.
\end{equation}
The problematic coefficient $\mathcal{F}$ contains a term $\frac{\ddot{\bar{\varphi}}}{\dot{\bar{\varphi}}}$,  
\begin{equation}
    \begin{split}
        \frac{\ddot{\bar{\varphi}}}{\dot{\bar{\varphi}}}\;\overset{\eqref{singulphi}}{=}\; m\,\frac{\cos{[m(t-t_o)]}}{\sin{[m(t-t_o)]}}\;\overset{t\approx t_o}{\approx}\;\frac{1}{t-t_o}\;.
    \end{split}
\end{equation}
One could try to remedy the situation by replacing $\frac{1}{\dot{\bar{\varphi}}}$ with its principal value,
\begin{equation}
    \frac{\ddot{\bar{\varphi}}}{\dot{\bar{\varphi}}} \;\Rightarrow\; \ddot{\bar{\varphi}} \, \frac{\dot{\bar{\varphi}}}{\dot{\bar{\varphi}}^2+\delta^2} \;\overset{t\approx t_o}{\approx}\;\frac{t-t_o}{(t-t_o)^2+\underbrace{\left(\frac{\delta}{Am^2}\right)^2}_{\equiv\,\hat{\delta}^2}}\;.
\end{equation}
However, this turns out to be an unsatisfactory solution when one considers the limit $\hat{\delta}^2\rightarrow0$. Consider a problematic part of the inflaton perturbation equation,\footnote{This is not the only problematic term, however, other terms are of a similar type, $1/\dot{\bar{\varphi}}$. In this chapter we analytically discuss the solution of the simple case in eq.~\eqref{wrong}. It turns out this solution works numerically for the full set of eqs.~\eqref{result}.}
\begin{equation}\label{wrong}
    \begin{split}
         \ddot{\mathcal{R}}^{ }_{\varphi k}=  
 - \dot{\mathcal{R}}^{ }_{\varphi k} 
 \frac{2(t-t_o)}{(t-t_o)^2+\hat{\delta}^2}\;.
    \end{split}
\end{equation}
This has two solutions: $\mathcal{R}^{ }_{\varphi k}= \text{constant}$, but also
\begin{equation}
    \dot{\mathcal{R}}^{ }_{\varphi k}=\frac{c}{(t-t_o)^2+\hat{\delta}^2}\;\;\rightarrow \;\;\mathcal{R}^{ }_{\varphi k}=\frac{c}{\hat{\delta}}\,\arctan{\left[\frac{t-t_o}{\hat{\delta}}\right]}\;.
\end{equation}
Let us examine the behaviour of this solution across the minimum, where $t_+$ and $t_-$ are times shortly after and before $t_o$, respectively,
\begin{equation}
    \mathcal{R}^{ }_{\varphi k}(t_+)-\mathcal{R}^{ }_{\varphi k}(t_-)=\frac{c}{\hat{\delta}}\,\left(\arctan{\left[\frac{t_+-t_o}{\hat{\delta}}\right]}-\arctan{\left[\frac{t_--t_o}{\hat{\delta}}\right]}\right)\;\overset{|t_\pm-t_o|\gg\hat{\delta}}{\approx}\;\frac{c}{\hat{\delta}}\,\pi\;.
\end{equation}
In the limit $\hat{\delta} \rightarrow 0 $, the solution experiences an infinite jump. The physically relevant solution should thus be the constant. Nonetheless, in a practical numerical implementation, the second solution could eventually be excited, making the results after the jump meaningless. We should then search for another way of regularizing the system. We find solace in \textit{complexified} variables.
\\\\
Instead of using the principal value, consider moving the pole $1/\dot{\bar{\varphi}}$ off the real axis,
\begin{equation}\label{delta}
    \frac{1}{\dot{\bar{\varphi}}}\;\;\rightarrow\;\;\frac{1}{\dot{\bar{\varphi}}+i\delta}\;.
\end{equation}
Note that this is not what forces the curvature perturbations to be complex. The quantum initial conditions, eqs.~\eqref{ourinitialconditions}, already complexified our $\mathcal{R}_{\varphi k}$. 
\\\\
With this in mind, eq.~\eqref{wrong} changes into
\begin{equation}\label{right}
    \begin{split}
         \ddot{\mathcal{R}}^{ }_{\varphi k}=  
 - \dot{\mathcal{R}}^{ }_{\varphi k}\; 
 \frac{2}{t-t_o+i\hat{\delta}}\;,
    \end{split}
\end{equation}
where the non-constant solution is given by
\begin{equation}
    \begin{split}
        \dot{\mathcal{R}}^{ }_{\varphi k}=\frac{c}{(t-t_o+i\hat{\delta})^2}\;\;\rightarrow \;\;\mathcal{R}^{ }_{\varphi k}=-\frac{c}{t-t_o+i\hat{\delta}}\;.
    \end{split}
\end{equation}
Let us investigate the behaviour of the second solution when crossing the minimum,
\begin{equation}
    \mathcal{R}^{ }_{\varphi k}(t_+)-\mathcal{R}^{ }_{\varphi k}(t_-)=-c\,\left(\frac{1}{t_+-t_o+i\hat{\delta}}-\frac{1}{t_--t_o+i\hat{\delta}}\right)\;\overset{|t_\pm-t_o|\gg\hat{\delta}}{\approx}\;-\frac{2c}{|t_\pm-t_o|}\;.
\end{equation}
Hence, even if this second solution gets excited in the numerical implementation, it does not get infinitely amplified. For completeness, in a practical implementation (Appendix \ref{app:CodeOverview}), we set $\delta \approx10^{-15} \;m_{\text{pl}}$. We have verified that the results are independent of the precise value of $\delta$.

\section{Late time evolution}\label{averaging}
At late times, numerical integration may become expensive due to rapid inflaton oscillations. The problem, together with its solution, is outlined in Ch.~\ref{matterdom}. Such a discussion concerns models with potentials which drive harmonic oscillations of the inflaton field, with the oscillations preceding radiation domination. On the other hand, there also exist potentials without such harmonic oscillations, for which the treatment in Ch.~\ref{matterdom} is not a priori valid or even needed. Nonetheless, some late-time simplifications may still be made. An example is presented in Ch.~\ref{raddombck(matter)}.

\subsection{Harmonic oscillations}\label{matterdom}
Let us take a closer look at eq.~\eqref{physical_a} (also eq.~\eqref{energyinserted_a}). For a wide range of physically relevant potentials, after the end of inflation, the inflaton field would start oscillating around the minimum of its potential. Let us assume that those oscillations can be locally described by a potential $\frac{m^2}{2}\varphi^2$ (in a Taylor expansion, first derivatives of the potential vanish at the minimum, while the second derivatives give a mass term). Around the minimum, eq.~\eqref{physical_a} then behaves as
\begin{equation}\label{oscillator}
    \ddot{\bar\varphi}+(3H+\Upsilon)\,\dot{\bar\varphi}+m^2\bar{\varphi} \approx 0\;,
\end{equation}
where $m^2\equiv V_{,\varphi\varphi}|_{\bar{\varphi}=0}$. Eq.~\eqref{oscillator} is an oscillator equation with a damping coefficient and a mass term. If the damping becomes relatively small, $m\gg 3H+\Upsilon$, the inflaton field experiences harmonic oscillations. Those are characteristic of an intermediate period of matter domination, preceding radiation domination. If the oscillations continue over a significant period of time, they can pose a numerical, as well as a conceptual problem. 
\\\\
The conceptual problem concerns our assumption of plasma staying in thermal equilibrium throughout the process. This is an assumption on the separation of time scales. Generally, it is assumed that the plasma equilibration rate, $\Gamma_r$, is large enough so that the plasma can stay in equilibrium despite the expansion of the universe, $\Gamma_r\gg H$. Additionally, there are constraints on $\Gamma_r$ coming from the frequency of inflaton oscillations, set by a scale $m$. If the inflaton dynamics is much slower than that of the plasma, $\Gamma_r\gg m$, the plasma has time to adjust to the inflaton changes and thus it stays in thermal equilibrium. Similarly, if $\Gamma_r\ll m$, the plasma does not have time to react to the changes in the inflaton field; instead, it just ``sees'' the inflaton's average value, and still maintains equilibrium. The problem appears when $\Gamma_r\approx m$, when the plasma cannot quite adjust to the changes in the inflaton field, thus falling out of equilibrium. In practice, this is only expected to happen in a narrow temperature range, so we assume the plasma stays in equilibrium throughout.
\\\\
We now turn to the numerical problem. The oscillations may continue over a significant time period. Numerically integrating over all these oscillations is expensive. To mitigate this problem, we proceed to derive a set of equations that averages over the fast background oscillations. Firstly, we rewrite eqs.~\eqref{physical} using the background energy density and pressure equations, \eqref{energyaverage} and \eqref{pressureaverage},\footnote{For completeness, the remaining \eqref{physical_c}, with the average energy density inserted, becomes
\begin{equation}
	\dot{H}=-4\pi G \left(\dot{\bar{\varphi}}^2+e_r+p_r -T\;V_{,T} \right)\;.
\end{equation}
After averaging, see eqs.~\eqref{averagequantityy} and \eqref{virial}, this becomes
\begin{equation}
	\dot{H}=-4\pi G \left( e_r+p_r+\langle e_{\bar{\varphi}}\rangle \right)\;.
\end{equation}
However, this is not an independent equation, so we do not need to include it in our numerical implementation.}
\begin{subequations}\label{energyinserted}
    \begin{align}
    0&= \ddot{\bar\varphi}+(3H+\Upsilon)\dot{\bar\varphi}+V_{,\varphi}\;,\label{energyinserted_a}\\
    \Upsilon\,\dot{\bar{\varphi}}^2&=\dot{e}_r+3H\,(e_r+p_r-T\,V_{,T})-T\,\dot{V}_{,T}\;,\label{energyinserted_b}\\
        H&=\sqrt{\frac{8\pi G}{3}}\sqrt{\frac{1}{2}\dot{\bar{\varphi}}^2+e_r+V-T\,V_{,T}}\;.\label{energyinserted_c}
    \end{align}
\end{subequations}
In the following, let us simplify the equations above into a set of ``averaged'' equations, valid for a subset of models characterized by an intermediate period of matter domination (models which exhibit a period of fast, harmonic oscillations). We will also neglect the temperature dependence of the potential, see ref.~\cite{Yokoyama_1999}. Then, the inflaton energy density takes on a simplified form,
\begin{equation}\label{averagequantityy}
    e_{\bar{\varphi}}=\frac{1}{2}\dot{\bar{\varphi}}^2+V\;.
\end{equation}
Invoking the virial theorem, $\langle \dot{\bar{\varphi}}^2\rangle\approx n\;\langle V \rangle$, over the time period of one oscillation, with $n$ defined through $V\propto \bar{\varphi}^n$ in the minimum, we obtain
\begin{equation}\label{virial}
	\frac{1}{2}\left\langle \dot{\bar{\varphi}}^2 \right\rangle\approx\left\langle V\right\rangle\;.
\end{equation}
With this simplification in mind, and multiplying eq.~\eqref{energyinserted_a} with $\dot{\bar{\varphi}}$, the equations \eqref{energyinserted} become
\begin{subequations}\label{AVERAGED}
    \begin{align}
0&= \langle\dot{e}_{\bar{\varphi}}\rangle+(3H+\Upsilon)\, \langle e_{\bar{\varphi}}\rangle \;,\\
\Upsilon\,\left\langle e_{\bar{\varphi}}\right\rangle&=\dot{ e}_r  +3H\,(e_r+p_r )\;,\\
H&=\sqrt{\frac{8\pi G}{3}}\sqrt{ \langle e_{\bar{\varphi}} \rangle+e_r }\;.
    \end{align}
\end{subequations}
In practice, we switch to these equations after 20-40 oscillations. We can gauge the correctness of the procedure by e.g. changing the number of oscillations. For a smooth transition, the exact moment when we switch is such that eq.~\eqref{virial} holds at the time.

\subsection{Radiation domination}\label{raddombck(matter)}
Let us note that oscillations may also occur for potentials without a quadratic minimum. An example of this is in fact provided by fig.~\ref{graph:baclgroundinflaton}, obtained by using a quartic potential, eq.~\eqref{potential}. Those oscillations often have a time-dependent frequency. However, as is the case for the quartic potential, the oscillations primarily occur during radiation domination. This allows us to derive a vastly simplified set of background equations.
\\\\
During radiation domination, we can use the approximations $\{|e_r|,\,|p_r| \}\gg |e_{\bar{\varphi}}|$ on average,\footnote{This is fine for potentials of the type $\bar{\varphi}^p$ with $p\neq -2$ (using the virial theorem on eq.~\eqref{energyinserted_b}); alternatively, one can use $|\dot{\bar{\varphi}}|^2\ll |e_r|,\,|p_r|$ to obtain the same equations.} so if $V_{,T}\approx 0$, eqs.~\eqref{energyinserted} become
\begin{subequations}\label{raddombck}
\begin{align}
	\dot{e}_r&\approx -3H\left( e_r+p_r \right)\;,\label{raddombck_a}\\
	H&\approx \sqrt{\frac{8\pi G}{3}\; e_r}\;.\label{raddombck_b}
\end{align}
\end{subequations}
In practice, we switch to these equations when $e_r = 1000\, e_{\bar{\varphi}}$. 
\\\\
Depending on the type of dynamics, either eqs.~\eqref{AVERAGED} (intermediate matter domination) or eqs.~\eqref{raddombck} (no intermediate matter domination) can be used to simplify the background dynamics at late times.

\section{Matrix formalism}\label{sec:matrixformalism}
The equations we are dealing with are stochastic. Hence, to obtain a meaningful solution, one would need to solve eqs.~\eqref{result} as described in Ch.~\ref{sec:NumMethodIntegration} many (i.e.~hundred, thousand,... ) times. Then, one could plot the distribution of solutions, compute the average, etc. However, this is numerically expensive. If one is just interested in obtaining the average power spectrum, one can greatly simplify the task by using the \textit{matrix formalism} \cite{Ballesteros:2022hjk,ballesteros2024monomialwarminflationrevisited}.  
\\\\
In this approach, we compute the thermal average, $\langle\mathcal{V}_{i+1}\mathcal{V}_{i+1}^\dagger\rangle$. Since the solution vector $\mathcal{V}_i$ is only influenced by the noise at an earlier time step, $\mathcal{N}_{i-1}$, the correlations $\langle\mathcal{N}_i\mathcal{V}_i^\dagger\rangle$, $\langle\mathcal{V}_i\mathcal{N}_i^\dagger\rangle$ vanish, and we are left with
\begin{equation}\label{deterministicevolution}
    \begin{split}
        \langle\mathcal{V}_{i+1}\mathcal{V}_{i+1}^\dagger\rangle=\bigl(1+\varepsilon_i\,\mathcal{M}_i\bigr)\langle\mathcal{V}_i\mathcal{V}_i^\dagger\rangle\bigl(1+\varepsilon_i\,\mathcal{M}_i^\dagger\bigr)+\varepsilon_i\langle\mathcal{N}_i\mathcal{N}_i^\dagger\rangle\;.
    \end{split}
\end{equation}
The noise term, $\mathcal{N}_i$, now only appears in the form of a thermal average, see eqs.~\eqref{conventional} and \eqref{noisematrix}. As such, it is now deterministic, allowing us to solve for the time evolution, eq.~\eqref{deterministicevolution}, only once. Let us remark that eq.~\eqref{deterministicevolution} is a $4\times 4$ matrix equation, where the solution matrix takes the form
\begin{equation}
    \langle\mathcal{V}_i\mathcal{V}^\dagger_i \rangle\overset{\eqref{solutionvector}}{=}
    \begin{pmatrix}
        |\mathcal{R}_{\varphi k}|^2&\mathcal{R}_{\varphi k}\dot{\mathcal{R}}_{\varphi k}^*& \mathcal{R}_{\varphi k}\mathcal{S}_{v k}^*&\mathcal{R}_{\varphi k}\mathcal{S}_{T k}^*\\\
    \dot{\mathcal{R}}_{\varphi k}\mathcal{R}_{\varphi k}^*&|\dot{\mathcal{R}}_{\varphi k}|^2&  \dot{\mathcal{R}}_{\varphi k}\mathcal{S}_{v k}^*& \dot{\mathcal{R}}_{\varphi k}\mathcal{S}_{T k}^*\\\ 
    \mathcal{S}_{v k}\mathcal{R}_{\varphi k}^*& \mathcal{S}_{v k}\dot{\mathcal{R}}_{\varphi k}^*&|\mathcal{S}_{v k}|^2&\mathcal{S}_{v k}\mathcal{S}_{T k}^*\\
    \mathcal{S}_{T k}\mathcal{R}_{\varphi k}^*& \mathcal{S}_{T k}\dot{\mathcal{R}}_{\varphi k}^*& \mathcal{S}_{T k} \mathcal{S}_{v k}^*&| \mathcal{S}_{T k}|^2
    \end{pmatrix}_{t=t_1}\;.
\end{equation}
Therefore, the initial conditions are
\begin{equation}
    \langle\mathcal{V}_{1}\mathcal{V}_1^\dagger\rangle\overset{\eqref{ourinitialconditions}}{=}
    \begin{pmatrix}
        \frac{k^2}{a(t_1)^2}\frac{H(t_1)^2}{(2\pi)^2\dot{\bar{\varphi}}^2}&i\frac{k^3}{a(t_1)^3}\frac{H(t_1)^2}{(2\pi)^2\dot{\bar{\varphi}}^2}&0&0\\\
    -i\frac{k^3}{a(t_1)^3}\frac{H(t_1)^2}{(2\pi)^2\dot{\bar{\varphi}}^2}&\frac{k^4}{a(t_1)^4}\frac{H(t_1)^2}{(2\pi)^2\dot{\bar{\varphi}}^2}&0&0\\\ 
    0& 0&0&0\\
    0& 0&0&0
    \end{pmatrix}\;.
\end{equation}
In this way we can obtain an array of values of the thermal averages of power spectra at each time step. Denoting for convenience $\langle\mathcal{P}\rangle=\langle\mathcal{V}_i\mathcal{V}^\dagger_i\rangle$, the power spectra of curvature perturbations are given by
\begin{subequations}\label{matrixformalismeq}
\begin{align}
        \langle|\mathcal{R}_{\varphi k}|^2\rangle &=\langle\mathcal{P}\rangle_{(1,1)}\;,\\
         \langle|\mathcal{R}_{v k}|^2\rangle&=\left|\frac{\mathcal{S}_{v k}}{(\bar e+\bar p)}+\mathcal{R}_{\varphi k}\right|^2=\Bigl(\langle\mathcal{P}\rangle_{(1,1)}+\frac{\langle\mathcal{P}\rangle_{(3,3)}}{|\bar e+\bar p|^2}+\frac{\langle\mathcal{P}\rangle_{(3,1)}}{(\bar e+\bar p)}+\frac{\langle\mathcal{P}\rangle_{(1,3)}}{(\bar e+\bar p)^*}\Bigr)\;,\\
         \langle|\mathcal{R}_{T k}|^2\rangle&=\left|\frac{\mathcal{S}_{T k}}{\dot{T}\,\bar e_{,T}}+\mathcal{R}_{\varphi k}\right|^2=\Bigl(\langle\mathcal{P}\rangle_{(1,1)}+\frac{\langle\mathcal{P}\rangle_{(4,4)}}{|\dot{T}\,\bar e_{,T}|^2}+\frac{\langle\mathcal{P}\rangle_{(1,4)}}{(\dot{T}\,\bar e_{,T})^*}+\frac{\langle\mathcal{P}\rangle_{(4,1)}}{\dot{T}\,\bar e_{,T}}\Bigr)\;.
\end{align}
\end{subequations}

\chapter{Model building}\label{sec:ModelBuilding}
After some introduction on warm inflation model building, we succinctly summarize a minimal extension of the Standard Model, originally proposed in ref.~\cite{ZellSM}. It provides an embedding of warm inflation in a microscopic theory, yielding a concrete expression for the friction coefficient, $\Upsilon$; see Ch.~\ref{ch:SM}. The reason we discuss this particular model is that ref.~\cite{ZellSM}, aside from successfully generating friction, also solves the strong CP problem. Hence, this particular implementation answers two questions at once, making it a compelling example to test our general framework on.
\\\\
We proceed by defining background initial conditions, see Ch.~\ref{ch:initialvalues}. Those are, together with the model itself, used to obtain numerical results in Part \ref{part:resultsdisc}. There, the results are compared with the relevant experimental data, which we summarize in Ch.~\ref{ch:data}.

\section{Warm inflation with the Standard Model}\label{ch:SM}
While the evolution equations \eqref{result} are largely model-independent, for a concrete practical implementation, explicit expressions for several model-dependent quantities ($\Upsilon,\;V,\;e_r,\;p_r$) are needed. Provided the bath stays in thermal equilibrium, as is the assumption in this thesis, the energy density and pressure of the bath are given by their standard thermodynamic relations, eqs.~\eqref{radiationpressurenergy}. Determining the friction $\Upsilon$ and the potential $V$ is, however, less straightforward. It requires knowledge on the microscopic interactions, i.e.~the vertices in a Lagrangian describing the theory. The vertices can be both the source of friction and a source of corrections to the potential. A problem appears as it is typically difficult to create large friction without simultaneously introducing large corrections to the potential, as they have the same microscopic origin (the former being the imaginary, and the latter the real contribution to the self-energy) \cite{Yokoyama_1999}. Large corrections to the potential can be problematic, as they may spoil the flatness of the potential necessary to maintain inflation. There are, however, ways to remedy the situation. One is presented by ``sphaleron heating'', see ref.~\cite{sphalerons}, to be discussed in the paragraph below.
\\\\
Sphaleron heating is a non-perturbative approach, which has been successfully applied in cosmological contexts, see e.g.~refs. \cite{Simona1,Berghaus_2020}. It relies on the existence of non-Abelian gauge fields $F_{\mu\nu}$, which couple to an external ``axion'' field, $a(x)$, via a coupling like $a(x)F_{\mu\nu}\tilde{F}^{\mu\nu}$ (the dual tensor is defined with the Levi-Civita tensor, $\tilde{F}_{\mu\nu}=\varepsilon_{\mu\nu\rho\sigma}F^{\rho\sigma}$). After varying the action to obtain equations of motion for the expectation value of the axion field, one gets a term like $\langle F\tilde{F}\rangle$. After a lengthy computation, this term can be written as a quantity proportional to the derivative of the axion field (see e.g.~ref.~\cite{sphalerons}, eq.~(28)). In other words, the axion coupling induces friction.  Moreover, since $F\tilde{F}$ can be written as a total derivative, the vertex can be, after integration by parts, written as proportional to the derivative of axion field. Such a vertex does not generate perturbative corrections to the potential and thus this can be considered a good way to generate $\Upsilon$.
\\\\
On the microscopic level, this friction is really induced by the thermal transitions between topologically distinct vacua of the gauge theory. However, if the gauge theory is QCD, the transitions between the different vacua (labeled by the ``Chern--Simons number'') change the relative numbers of left- and right-handed quarks, in other words, they induce a quark axial chemical potential. On the level of the equations of motion, this acts as a suppression mechanism for the friction. So, sphaleron transitions induce friction, but they also build up a chemical potential, which typically makes it difficult to sustain the large friction necessary for the strong regime of warm inflation. 
\\\\
Nonetheless, there are ways to solve this problem. Historically, they were centered around postulating the existence of some new (dark) non-Abelian gauge fields, with no coupling to fermionic matter fields to induce this axial chemical potential; while it solved the problem, the question remained whether a solution can be found without the need to introduce a new Beyond-the-Standard-Model (BSM) sector. The proposal in ref.~\cite{ZellSM} instead demonstrates that $\Upsilon$ may arise solely from the axion-like interaction between the inflaton and the strong sector of the SM. They show this to be possible after accounting for Hubble dilution of the axial chemical potential, such that the sphaleron heating is not too suppressed by the quark's axial chemical potential (for details, see ref.~\cite{ZellSM}). In this way, the inflaton field needed for inflation is also the axion field required to solve the strong CP problem, solving two problems in physics by introducing a single BSM field.
\\\\
Interested readers may find reviews on QCD axion models and the strong CP problem in refs.~\cite{review1,review2,review3}, more background material on the sphaleron rate in refs.~\cite{lattice1,lattice2,lattice3,lattice4} (lattice) and ref.~\cite{semiclassical} (semiclassical), with the high temperature phase explored in e.g.~refs.~\cite{t1,t2,t3,t4,t5,t6,t7}.
\\\\
We now turn to the results of ref.~\cite{ZellSM}, which are directly relevant for our numerical implementation. In particular, the thermal friction derived in that paper has the form
\begin{equation}\label{upsilonSM}
    \Upsilon=\frac{T^2}{2\,f_a^2}\left(\frac{1}{\alpha^5\, N_c^5\, T}+\frac{2N_f}{N_c\, H}\right)^{-1}\;,
\end{equation}
    where $f_a$ is the inflaton decay constant (to be varied), $\alpha$ is the QCD gauge coupling, and $N_c$ is the number of colors ($N_c=3$). $N_f$ is the number of SM flavours that contribute ($N_f=5$; for the reason why the top quark is excluded, see the text between eqs.~(18) and (19) in ref.~\cite{ZellSM}).
\\\\
While eq.~\eqref{upsilonSM} was derived, the potential was postulated to be
\begin{equation}\label{potential}
    V=\lambda\;\varphi^4\;.
\end{equation}
Aside from the potential term $V=\lambda\;\varphi^4$, which explicitly appears in the Lagrangian of the theory, the QCD coupling to the inflaton employed in ref.~\cite{ZellSM} produces a cosine potential at energies just above $\Lambda_\text{QCD}$, see refs.~\cite{review1,review2}. Therefore, the full inflaton potential is 
\begin{equation}
    V_\text{total}=\lambda \varphi^4+\chi(T)\left[1-\cos{\left(\frac{\varphi}{f_a}-\theta\right)}\right]\;.
\end{equation}
The susceptibility, $\chi(T)$, is very heavily suppressed at temperatures much larger than $\Lambda_\text{QCD}$, hence in a practical implementation, only the $\lambda\varphi^4$ term is considered. 
\\\\
Lastly, we adopt the perfect fluid energy density and pressure,
\begin{equation}\label{radiationpressurenergy}
    e_r=\frac{g_\star \pi^2\,T^4}{30}\;,\;\;\;\;\;\;\;\;\;\;\;\;\;\;\; p_r=\frac{g_\star \pi^2\,T^4}{90}\;,
\end{equation}
where $g_\star$ is the number of effective degrees of freedom. Since our entire evolution happens above the electroweak scale, we use the value $g_\star=106.75$, which conservatively corresponds to the energies above $300$ GeV ($\approx 10^{-17} m_{Pl}$; we stop our evolution when $T=10^{-12}m_{Pl}$). During that epoch all quarks, leptons, and gauge bosons are effectively massless.

\section{Initial values}\label{ch:initialvalues}
\SetWatermarkText{}
\subsection{Background evolution}
We aim to set initial conditions for the background evolution during a period of quasi de Sitter expansion. We choose the initial value of $\varphi$ such that the inflationary period lasts sufficiently long. For the case considered here, it is sufficient to choose 
\begin{equation}
    \bar{\varphi}(t_\text{ref})=7\; m_{Pl}\;,
\end{equation}
where $t_\text{ref}$ is the initial time at which we start performing the computation. In practice, the evolution will be measured in the units of $t_\text{ref}$.\\\\
The initial value of $\dot{\bar\varphi}$ can be obtained from eq.~\eqref{physical_a} in the slow-roll regime,
\begin{equation}
    \dot{\bar\varphi}(t_\text{ref})=-\left.\frac{V_{,\varphi}}{3H+\Upsilon}\,\right|_{t_\text{ref}}\;.
\end{equation}
Similarly, assuming that during early times $\{\dot{\bar\varphi}^2\,,\;e_r\,,\;T\,V_{,T}\}\;\ll\; V$, eq.~\eqref{physical_b} becomes, in the units of Planck mass,
\begin{equation}\label{HubbleTref}
    H(t_\text{ref})\approx\left.\sqrt{\frac{8\pi \,V}{3}}\;m_{Pl}^{-1}\;\right|_{t_\text{ref}}\;.
\end{equation}
Even if the approximation in eq.~\eqref{HubbleTref} would be order(s) of magnitude different from its exact value (which does not happen in practice), the background solutions would quickly become that of de Sitter expansion, since the background seems to have an attractor solution.
\\\\
The initial value of $T$ can be determined by searching for an approximate constant temperature solution during inflation. Starting with eq.~\eqref{physical_d}, we drop $\dot{T}$ and $V_{,T}$ terms, as well as remove $\ddot{\bar\varphi}+3H\dot{\bar\varphi}$ using eq.~\eqref{physical_a}. This leads to
\begin{equation}
    e_r(T_\text{ref})+ p_r(T_\text{ref})=\frac{\Upsilon_\text{ref}}{3H_\text{ref}}\,\dot{\bar\varphi}^2_\text{ref}\;.
\end{equation}
To fully determine the initial values, we use a proxy
\begin{equation}\label{tref}
    t_\text{ref}\equiv \left.\sqrt{\frac{3}{8\pi \,V}}\;m_{Pl}\;\right|_{t_\text{ref}}\;.
\end{equation}
With eq.~\eqref{tref} in mind, $t_\text{ref}$ can be obtained with the help of eq.~\eqref{potential},
\begin{equation}
    t_\text{ref}=\sqrt{\frac{3}{8\pi \,\lambda}}\;\frac{m_{Pl}}{\bar{\varphi}_\text{ref}^2}\;.
\end{equation}

\subsection{Perturbation and momentum evolution}
A suggested initial time of the evolution of curvature perturbations can be defined by the numerical value of the ratio
\begin{equation}
    \frac{k}{a(t_1)H(t_1)} \in [10^2, 10^3] \;.
\end{equation}
The larger the value, the better the approximation we used to set the initial conditions, eq.~\eqref{ourinitialconditions}. At the same time, increasing the value renders the calculations numerically very expensive due to the initial phase rotation of $\mathcal{R}_{\varphi k}$. The topic is further discussed in Ch.~\ref{ch:inittimeofintegandacoustosc}.
\\\\
The final value of $\frac{k}{a(t_f)H(t_f)}$ should be small enough to allow for the curvature perturbations to have settled to a constant value. In practice, this turns out to be mostly achieved when the ratio is $10^{-5}$. However, there are scenarios where not all of the curvature perturbations have settled to exactly the same constant value when $k/(aH)=10^{-5}$. A more conservative and appropriate choice would thus be e.g.~$10^{-10}$, which comes with a little longer computation time. We discuss in the final paragraph in Ch.~\ref{ch:keyresults} how this affects the observables.
\\\\
In practice we compute the power spectrum for a particular mode of interest. We choose to track the pivot scale of CMB, which today has the value $k_\star/a_0=0.05\;\text{Mpc}^{-1}$. We need to evolve this mode backwards in time in order to set initial conditions for the momentum evolution. To do so, it would be helpful to estimate the number of $e$-folds between today and the time at which we stop our computation. A convenient choice for the latter is given by the time at which $T_e=10^{-12}\;m_{Pl}$, when the universe is conservatively radiation-dominated. To this end, we make use of the SM thermodynamics. The entropy is given by
\begin{equation}
    S=\frac{2\pi^2}{45}\,g_{\star s}(T)\; a^3\;T^3,
\end{equation}
where $g_{\star s}(T)$ is the effective number of relativistic degrees of freedom contributing to the entropy. Using entropy conservation, we obtain a helpful relation,
\begin{equation}
    \frac{a_0}{a_e}=\left(\frac{T_e^3}{T^3_0} \frac{g_{\star s,e}}{g_{\star s,0}}\right)^{1/3}\overset{\text{\cite{mikkodata}}}{\approx}e^{46.5}\;.
\end{equation}
We can now obtain the ``initial'' condition for the momentum evolution,
\begin{equation}
    \frac{k_\star}{a_e}= \frac{k_\star}{a_0} \frac{a_0}{a_e}\approx 4.10 \times 10^{-39}\;.
\end{equation}
In practice, we solve for the momentum evolution backwards in time, as explained in Appendix \ref{app:codeCOMPUTING}.

\section{Data-driven numerical values}\label{ch:data}
One application of our general framework is to compute precise reheating predictions for some model of interest, e.g.~the one presented in Ch.~\ref{ch:SM}, and compare those predictions to experimentally measured numbers. In practice, we compare the amplitude of the power spectrum, eq.~\eqref{amplitude}, the spectral tilt, eq.~\eqref{tilt}, and the tensor-to-scalar ratio, eq.~\eqref{r}. Here we cite the current experimental values of these observables. 
\\\\
The combined data analysis from Planck \cite{Planck_inflation_2020} (European Space Agency's satellite mission), Atacama Cosmology Telescope \cite{Planck_inflation_2020} (ground-based millimeter-wave telescope in Chile), and DESI  (ground-based Dark Energy Spectroscopic experiment in Arizona) data release 2 \cite{DESI_DR2}, ref.~\cite{combineddata}, gives the amplitude,
\begin{equation}\label{amplitudemeasured}
    \ln{10^{10} A_s}=3.062^{+0.010}_{-0.012}\quad\rightarrow\quad A_s=2.137^{+0.021}_{-0.025}\;\times 10^{-9}\;,
\end{equation}
and spectral index
\begin{equation}\label{nsexperiment}
    n_s=0.9752^{+0.0030}_{-0.0030}\;.
\end{equation}
The latter is higher than e.g.~Planck value (0.9651$\pm$0.0044), or any other (joint) analysis cited in ref.~\cite{combineddata}, Table 5.\\\\
The combined Planck and BICEP/Keck (cryogenic telescopes at the South Pole, for measuring B-mode polarization) analysis \cite{BICEP:2021xfz} sets the upper bound for the tensor-to scalar ratio to
\begin{equation}
    r<0.036\;.
\end{equation}
We note in passing that e.g.~the Planck data analysis, after cleaning the data (i.e.~foreground subtraction), produces 2-point correlators in temperature and polarization%, $\mathcal{C}_l^{TT}$, $\mathcal{C}_l^{TE}$, $\mathcal{C}_l^{EE}$, and $\mathcal{C}_l^{BB}$
. Afterwards, model-dependent assumptions are necessary to extract $A_s$, $n_s$, and $r$. The numbers we cited include several standard assumptions on cosmological evolution (e.g.~$\Lambda$-Cold-Dark-Matter model, power law shape of the power spectrum \eqref{amplitude}, etc). For non-standard evolution, one would need to use Boltzmann codes (e.g.~\texttt{CAMB}~\cite{Lewis:1999bs} or \texttt{CLASS}~\cite{Lesgourgues:2011re,Blas:2011rf}) to directly obtain theoretical correlators, and compare those with experimental ones (e.g.~with the help of Markov Chain Monte Carlo (MCMC) codes, such as \texttt{CosmoMC}~\cite{Lewis:2002ah}, \texttt{MontePython}~\cite{Audren:2012wb,Brinckmann:2018cvx}, or \texttt{Cobaya}~\cite{Torrado:2020dgo}, producing a region of allowed model parameter values). For cases considered here, however, such an analysis is not needed.

\part{Numerical results and discussion}\label{part:resultsdisc}

\chapter{Numerical results}\label{sec:numres}
Here we present examples of solutions, all for the model in Ch.~\ref{ch:SM}. We begin by examining the evolution of selected background quantities, followed by a discussion of the power spectrum and other observables. We run simulations for different choices of model parameters, see table \ref{table:sm}. We choose parameters such that they span both the weak and the strong regimes of warm inflation (by tuning $\lambda$, see eq.~\eqref{potential}), while producing the correct observational value for the amplitude of the power spectrum (by tuning the axion decay constant, $f_a$). The other observables, $n_s$ and $r$, can then be compared with experiments.

\section{Background}\label{ch:bckresults}
\begin{figure}[h]
    \centering
    \begin{subfigure}{0.49\linewidth}
        \centering
        \includegraphics[height=0.32\textheight]{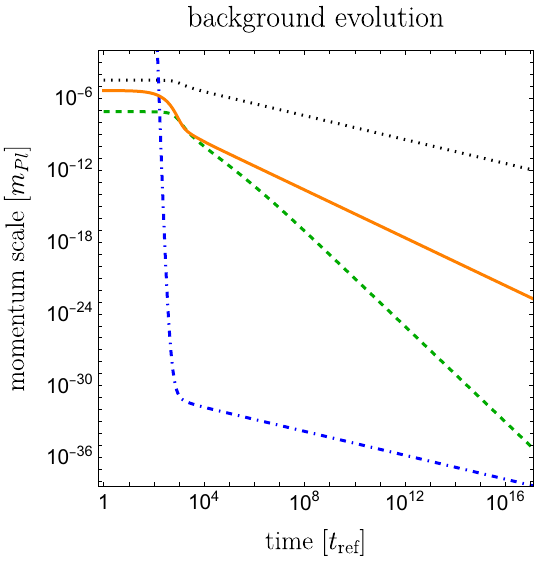}
        \caption{Weak regime: $Q_\star=0.015$,  $T_\star/H_\star=20.4$.}
    \end{subfigure}
     \hfill
    \begin{subfigure}{0.49\linewidth}
        \centering
        \includegraphics[height=0.32\textheight]{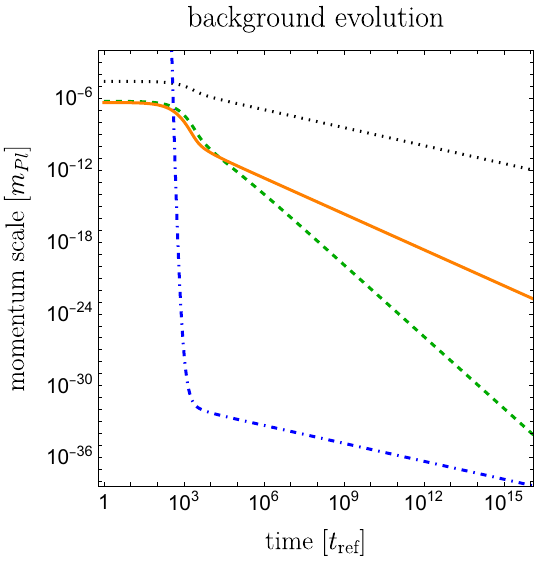}
        \caption{Weak regime:   $Q_\star=0.711$,  $T_\star/H_\star=213$.}
    \end{subfigure}
\par\vspace{1cm}
    \begin{subfigure}{0.49\linewidth}
        \centering
        \includegraphics[width=\linewidth]{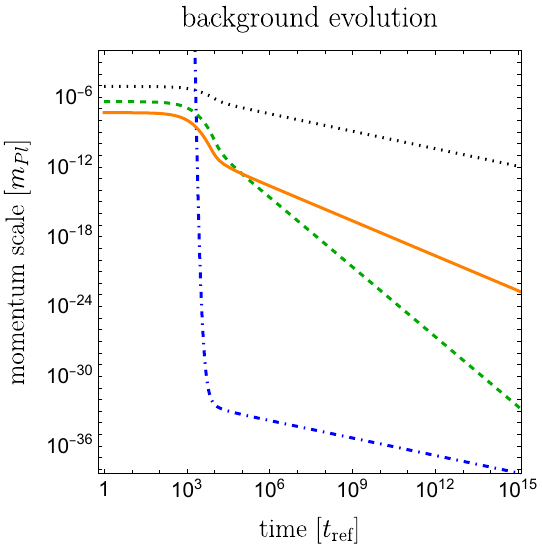}
        \caption{ Strong regime: $Q_\star=4.88$,  $T_\star/H_\star=1531$.}
    \end{subfigure}
    \hfill
    \begin{subfigure}{0.49\linewidth}
        \centering
        \includegraphics[width=\linewidth]{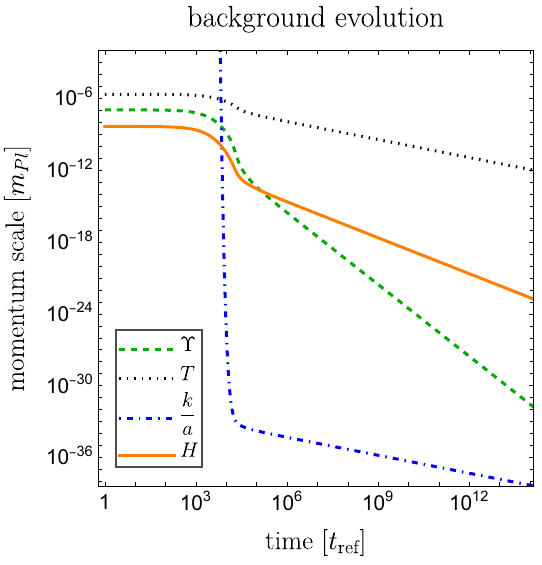}
        \caption{Strong regime: $Q_\star=13.95$,  $T_\star/H_\star=7804$.}
    \end{subfigure}
    \caption{Evolution of background quantities $H$, $T$, $\Upsilon$, and $k/a$, for a set of model parameters (see table \ref{table:sm}). We start the evolution during inflation, and end it when $T=10^{-12}m_{Pl}$, see Ch.~\ref{ch:initialvalues}. The legend from the strong regime is valid for all cases.}
    \label{fig:background_strong}
\end{figure}

\begin{figure}[!b]
    \begin{subfigure}{0.48\linewidth}
        \centering
        \includegraphics[width=\linewidth]{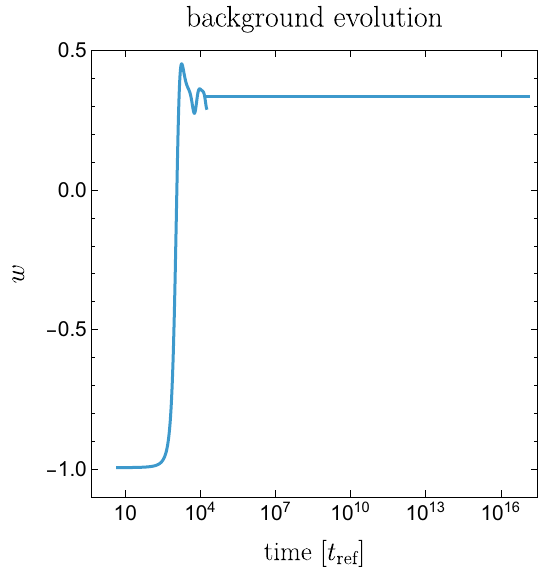}
        \caption{Weak regime: $Q_\star=0.015$,  $T_\star/H_\star=20.4$.}\label{w_weak}
    \end{subfigure}
        \begin{subfigure}{0.48\linewidth}
        \centering
        \includegraphics[width=\linewidth]{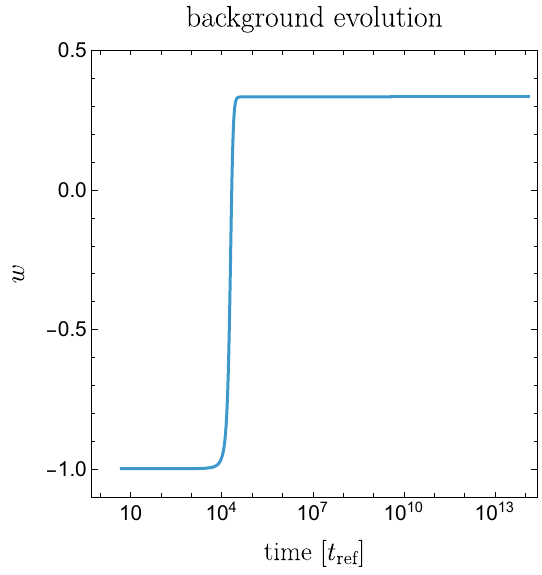}
        \caption{Strong regime: $Q_\star=13.95$,  $T_\star/H_\star=7804$.}\label{w_strong}
    \end{subfigure}
    \caption{Evolution of the equation of state parameter $w$, defined in eq.~\eqref{w}.}
    \label{fig:w}
\end{figure}

We start by presenting the general features of the background evolution, see fig.~\ref{fig:background_strong}. We notice that in all the graphs, the background solution starts as quasi de Sitter, as evidenced by the constant $H$ values. The universe continues to expand, but when $H\neq \text{const.}$, it ceases to do so exponentially fast. As the universe is expanding, the friction and temperature are decreasing. 
\\\\
Towards the end of inflation, the momentum mode we track crosses the horizon. At the moment of horizon exit, we can define the ratio $Q_\star=\Upsilon_\star/(3H_\star)$, and similarly $T_\star/H_\star$. Those can now define more precisely the weak and strong regimes of warm inflation: when $\{\Upsilon_\star,\,T_\star\}\gg H_\star$, thermal fluctuations are generated and strongly dissipated, while when $T_\star\gg H_\star\gg \Upsilon_\star$, thermal fluctuations are weakly dissipated. When $\{\Upsilon_\star,\,T_\star\}\ll H_\star$, there are essentially no thermal effects, though the plasma is still present. These starred quantities, along with other relevant numbers, are collected in table \ref{table:sm}, and will be explained in Ch.~\ref{ch:results:perturb}.
\\\\
The top two graphs in fig.~\ref{fig:background_strong}, left to right, describe the background evolution in the weak and marginally weak (intermediate) regime. The lower two graphs, left to right, describe the marginally strong (intermediate) and strong regimes.
\\\\
They all track the same momentum mode, $k_*=0.05\;\text{Mpc}^{-1}$, see Ch.~\ref{ch:initialvalues}, which has the amplitude of $A_s\approx 2.1 \times 10^{-9}$, see Ch.~\ref{ch:data}. We notice that deeper in the strong regime, the mode exits the horizon later, already towards the end of inflation, to produce the desired amplitude. 
\clearpage
\noindent
We can also discuss the scalings of $H,\;T,\;k/a$, and $\Upsilon$ in fig.~\ref{fig:background_strong} in radiation domination. The energy density scales as $a^{-4}$.\footnote{From the continuity eq.~\eqref{raddombck_a}, along with radiation equation of state, we have $\dot{\bar{e}}/\bar{e}=-4\,\dot{a}/a$, and hence $\bar{e}\propto a^{-4}$.} Then, according to eq.~\eqref{raddombck_b}, $H\propto a^{-2}$. But since $H=\dot{a}/a$, we have $a^2\propto t$, and hence $H\propto 1/t$. Again using eq.~\eqref{raddombck_b}, now with eq.~\eqref{radiationpressurenergy}, we have $H\propto T^2$, and hence $T\propto 1/\sqrt{t}$. Since $k/a\propto a^{-1}$, its scaling is the same as that of the temperature. For the dissipation coefficient, eq.~\eqref{upsilonSM}, we have, up to constants, $\Upsilon\propto(t^{3/2}+t^2)^{-1}$. Due to the much larger constant term in front of $t^{-3/2}$, this part prevails, so we have, approximately, $\Upsilon \propto t^{-3/2}$. Such scalings can be seen on all graphs in fig.~\ref{fig:background_strong}, after the transition into radiation domination.
\\\\
Another background quantity that might be of interest is $w$, the ratio between the background energy density and the pressure,
\begin{equation}\label{w}
    w=\frac{\bar{p}+\frac{1}{2}\dot{\bar{\varphi}}^2}{\bar{e}+\frac{1}{2}\dot{\bar{\varphi}}^2}\;.
\end{equation}
We plot this quantity in fig.~\ref{fig:w}.
We see $w$ transitioning from its inflationary value of $-1$, to $1/3$. We note in passing that in the weak regime, the quantity $w$ shortly fluctuates before settling to its constant value of 1/3, see fig.~\ref{w_weak}. This can be traced back to the potential, in particular, to the evolution of $\bar{\varphi}$. Specifically, dips in $w$ correspond to the first few extrema of $\bar{\varphi}$. This can be explained as $w=\frac{...-\lambda \bar{\varphi}^4}{...+\lambda \bar{\varphi}^4}$ (with $...$ being positive), hence any extrema in the background inflaton field will contribute to the decrease in $w$. The amplitude of the potential is decreasing with time, hence the fluctuations around $w=1/3$ are attenuating. In practice, we simplify the equations by setting $\bar{\varphi}=0$ deep in the radiation domination, see Ch.~\ref{raddombck(matter)}. This evidently sets $w$ to exactly $1/3$, which is clearly visible around $t=10^5\; t_\text{ref}$ in the weak regime, see fig.~\ref{w_weak}. The inflaton field also oscillates in the strong regime, yet as seen from fig.~\ref{w_strong}, no oscillations in $w$ are visible. That is because the inflaton potential is relatively small in the strong regime (note the $\lambda$ values in table \ref{table:sm}). Consequently, the amplitude of the oscillations is too small to be visible without significantly zooming in on the graph.
%\footnote{\hl{It is perhaps interesting to note that while radiation domination is in general achieved at smaller $t$ in the weak regime, the number of inflaton oscillations before radiation domination does not seem to either increase or decrease with varying $Q_\star$; it rather fluctuates. This points to the numerical origin of fluctuations. A suggestion would be that the moment at which we switch to late time evolution is imprecisely defined; as the numbers are very close to each other and the field oscillates, it could be that the same ratio is reached multiple times, but sometimes it is picked up on and sometimes not; one could plot the criterion to check}}

\section{Perturbations}\label{ch:results:perturb}

\subsection{Key results}\label{ch:keyresults}
Before describing the power spectrum evolution, let us focus on table \ref{table:sm}. For a given series of model-dependent inputs ($\lambda$, $f_a$, $\alpha_s$), we first produce outputs on the background level (columns from $t_1/t_{\text{ref}}$ to $|\dot{\bar\varphi}|^{ }_\star / H^2_\star$). The observables $A_s$, $n_s$, and $r$ are in the last three columns.
\\
\begin{table}[h!]
{\fontsize{8pt}{10pt}\selectfont

%%%%%%%%%%%%%%%%%%%%%%%%%%%%%%%%%%%%%%%%%%%%%%%%%%%%%%%%%%%%%%%%%%%
\vspace*{-3mm}

$$
\begin{array}{|ccc|ccccc|ccc|} % {llllllllll} %  {rrrrrrrrrr}
%%%%%%%%%%%%%%%
  \hline
  & & & & & & & & & & \\[-3mm]
\multicolumn{3}{|c|}{ \mbox{input} } 
  & 
  \multicolumn{5}{c|}{ \mbox{output: background} } 
  &
  \multicolumn{3}{c|}{ \mbox{output: perturbations} }
 \\[1mm]
%%%%%%%%%%%%%%%
  \lambda & 
  f^{ }_a \; [10^{12}_{ }\;\mbox{GeV}] &
  \alphas^{ } &
  t^{ }_1 / t^{ }_{ref} & 
  \bar\varphi^{ }_1/\mpl^{ } & 
  T^{ }_\star / H^{ }_\star & 
  Q^{ }_\star & % \Upsilon^{ }_* / 3H^{ }_* &
  |\dot{\bar\varphi}|^{ }_\star / H^2_\star &
  A^{ }_{s} &
  n^{ }_{s} &
  r 
 \\[1mm]
%%%%%%%%%%%%%%%
  \hline % 
  & & & & & & & & & & \\[-3mm]
%%%%
  10^{-21}_{ } & 
  0.153 & 
  0.0269 & 
  6896 & % t_1
  1.06 & % varphi*/mpl
  7804 & % T*/H*
  13.95 & % Ups*/H*
  1.11p8 & % phid*/HH*
  2.1m9 & % As
  0.967 & % ns 
  2.12m11   % r
   \\[1mm]
%%%%
  10^{-20}_{ } & 
  0.278 & 
  0.0263 & 
  3969 & % t_1
  1.31 & % varphi*/mpl
  3536 & % T*/H*
  8.59 & % Ups*/H*
  2.91p7 & % phid*/HH*
  2.1m9 & % As
  0.969 & % ns 
  4.99m10   % r
   \\[1mm]
%%%%
  10^{-19}_{ } & 
  0.516 & 
  0.0257 & 
  2136 & % t_1
  1.67 & % varphi*/mpl
  1531 & % T*/H*
  4.88 & % Ups*/H*
  7.24p6 & % phid*/HH*
  2.1m9 & % As
  0.969 & % ns 
  1.31m8   % r
   \\[1mm]
%%%%
  10^{-18}_{ } & 
  1.025 & 
  0.0251 & 
  995 & % t_1
  2.24 & % varphi*/mpl
  613 & % T*/H*
  2.30 & % Ups*/H*
  1.69p6 & % phid*/HH*
  2.1m9 & % As
  0.969 & % ns 
  4.27m7   % r
   \\[1mm]
%%%%
  10^{-17}_{ } & 
  2.27 & 
  0.0246 & 
  383 & % t_1
  3.15 & % varphi*/mpl
  213 & % T*/H*
  0.711 & % Ups*/H*
  3.68p5 & % phid*/HH*
  2.1m9 & % As
  0.968 & % ns 
  1.68m5  % r
   \\[1mm]
%%%%
  10^{-16}_{ } & 
  4.98 & 
  0.0243 & 
  179 & % t_1
  4.08 & % varphi*/mpl
  60.9 & % T*/H*
  0.0919 & % Ups*/H*
  8.38p4 & % phid*/HH*
  2.1m9 & % As
  0.971 & % ns 
  4.79m4   % r
   \\[1mm]
%%%%
  10^{-15}_{ } & 
  8.56 & 
  0.0242 & 
  150 & % t_1
  4.35 & % varphi*/mpl
  20.4 & % T*/H*
  0.0150 & % Ups*/H*
  2.33p4 & % phid*/HH*
  2.1m9 & % As
  0.944 & % ns 
  6.26m3   % r
   \\[1mm]
%%%%
  \hline
\end{array}
$$
}

\vspace*{-3mm}
\caption{Key results of our analysis. We use the notation $mX\equiv10^{-X}$ and $pX\equiv10^{X}$. %The errors are on less than a percent level.
}
\label{table:sm}
\end{table}
\\
It should be noted that ref.~\cite{ZellSM} provides their own predictions for observables $A_s$, $n_s$, and $r$. However, as the main contribution of ref.~\cite{ZellSM} is deriving the form of the thermal friction, their predictions for observables are somewhat quicker estimates based on analytical formulae available in their appendix. However, such analytical formulae all carry some level of model-dependence (or use significant approximations). Therefore, full numerical integration, such as the one provided here, is theoretically cleaner and more reliable.\footnote{In practice, the quicker predictions of ref.~\cite{ZellSM} (their Table 1) minimally differ from our more precise estimates (table \ref{table:sm}). The only significant difference is the spectral tilt for the lowest $Q_\star$ case, as discussed in the last paragraph in Ch.~\ref{ch:keyresults}. }
\\\\
Let us return to table \ref{table:sm}. On the level of the inputs, we choose the same values of the quartic inflaton coupling constant, $\lambda$, scanning over the weak and the strong regimes, as well as for the strong coupling constant, $\alpha_s$, as ref.~\cite{ZellSM}. Then we tune the axion decay constant, $f_a$, such that the scalar power spectrum amplitude, $A_s$, assumes its expected value, $2.1\times 10^{-9}$. The inflaton decay constant only needs to be changed minimally compared to ref.~\cite{ZellSM}, to predict the correct amplitude. Based on the provided inputs, one first obtains the background values, the first of which is $t_1/t_\text{ref}$. This is the time at which we start the evolution of the perturbations; see Ch.~\ref{ch:initialvalues}. The second background quantity is $\bar\varphi_1/m_{Pl}$, which is the value of the background inflaton field at $t_1/t_\text{ref}$. We continue by citing the already mentioned ratios between the temperature and the Hubble rate, as well as $Q_\star=\Upsilon_\star/H_\star$ at horizon crossing (defining weak and strong regimes of warm inflation). The last background quantity is the ratio of the inflaton field versus the Hubble rate squared at the horizon crossing. It is sometimes used in attempts at parameterizing the power spectrum in terms of the freeze-out quantities, as the vacuum power spectrum can be estimated as $\mathcal{P}_\mathcal{R}^\text{vac}\approx H^4/(2\pi\dot{\bar{\varphi}})^2$ \cite{Starobinsky:1982ee}.
\\\\
Let us discuss the results for the spectral tilt. We see that, except for the very weak regime, $Q_\star=0.015$, the model largely agrees with the data, eq.~\eqref{nsexperiment}, with $n_s$ having a maximal deviation of $2.5\sigma$. This is in fact the most pessimistic scenario, as the deviation is further reduced if one considers any other combination of measurements constraining $n_s$ that was considered in ref.~\cite{combineddata}. For example, comparing with the Atacama results alone, the maximum discrepancy is $0.6\sigma$, except for the case $Q_\star=0.015$, where it is $3\sigma$.
For the same case $Q_\star=0.015$, the deviation from the experimental result in eq.~\eqref{nsexperiment} is even much larger, above $10\sigma$. That would imply that the quartic potential, together with $\Upsilon$ from Ch.~\ref{ch:SM}, would be excluded deep in the weak regime, while allowed when thermal effects are important. 
\\\\
It should be pointed out that the errors discussed here are purely experimental. It is somewhat difficult to estimate theoretical errors. There are errors directly related to truncating our expansion at the linear level; to accurately estimate the correctness to that, one would need to go to the second order in the perturbation theory. %\footnote{\hl{This is related to non-Gaussianity. Earlier studies suggest that the impact of those on our observables would be minimal.}} 
The second type of theoretical errors concerns the inputs $\Upsilon$ and $V$, which we took at face value from ref.~\cite{ZellSM}. Since the potential was merely postulated, while the friction coefficient was derived using several approximations, we estimate that this is the largest source of theoretical errors. To make a quantitative statement, however, one would need to do a more precise computation predicting the model-dependent inputs, and compare it to the first estimates adopted here.
\\\\
It should be noted that the low spectral index for the lowest $Q_\star$ was not reported in ref.~\cite{Alica2}, where we first investigated the issue. There, a higher value was cited, which was in better agreement with ref.~\cite{ZellSM}.\footnote{Their $n_s$ estimate relied on quantities evaluated at the horizon crossing time.} Upon a subsequent closer inspection, it was discovered that not all curvature perturbations have completely frozen out by the time we previously computed the observables (when $k/a=10^{-5} H$). Even though the amplitude of the power spectrum changes minimally afterwards, the effect on the spectral tilt can be significant. Since the experimental results are obtained during the time of the freeze-out, and not during the time of the horizon crossing (which can be slightly earlier), one should be computing the observables only when all the power spectra obtain the same, constant value. Furthermore, since the spectral tilt is computed based on power spectra for $0.1$ decade higher and lower momentum, one should additionally make sure those power spectra are also completely frozen out. As the slightly larger momenta re-enter the Hubble horizon slightly later, the higher-decade power spectrum could need slightly more time to freeze-out. We checked that for all cases considered here, stopping the evolution at the time when $k/a=10^{-10}\,H$ is a very safe choice.

\subsection{Power spectrum evolution}
\begin{figure}[h!]
    \centering
    \begin{subfigure}{0.49\linewidth}
        \centering
        \includegraphics[height=0.32\textheight]{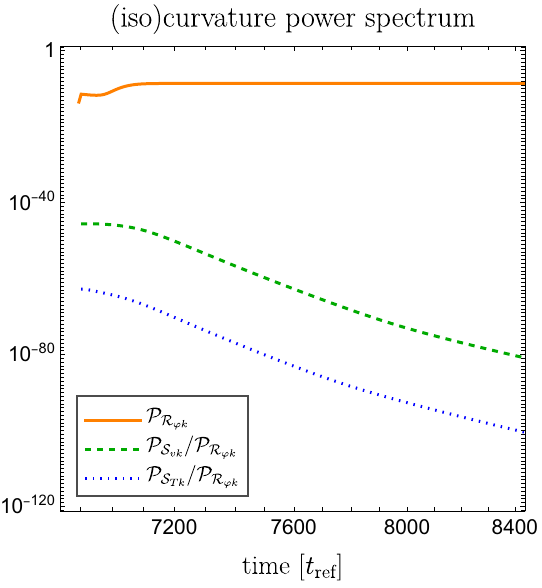}
        \caption{Strong regime: $Q_\star=13.95$,  $T_\star/H_\star=7804$.}\label{fig:isoa}
    \end{subfigure}
     \hfill
    \begin{subfigure}{0.49\linewidth}
        \centering
        \includegraphics[height=0.32\textheight]{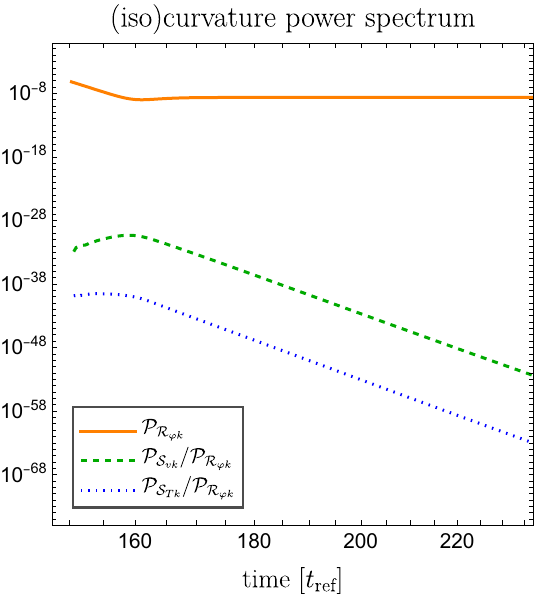}
        \caption{Weak regime: $Q_\star=0.015$,  $T_\star/H_\star=20.4$.}\label{fig:isob}
    \end{subfigure}
    \caption{Isocurvature perturbations are negligible compared to $\mathcal{R}_{\varphi k}$. We also notice isocurvature perturbations decay after the horizon exit, supporting the analytical conclusion from Ch.~\ref{ch:freezeout}. The notation is like in table \ref{table:sm}.}
    \label{fig:isocurvature}
\end{figure}

We now present the time evolution of the power spectra.
\begin{figure}[p]
    \centering
    \begin{subfigure}{0.49\linewidth}
        \centering
        \includegraphics[height=0.32\textheight]{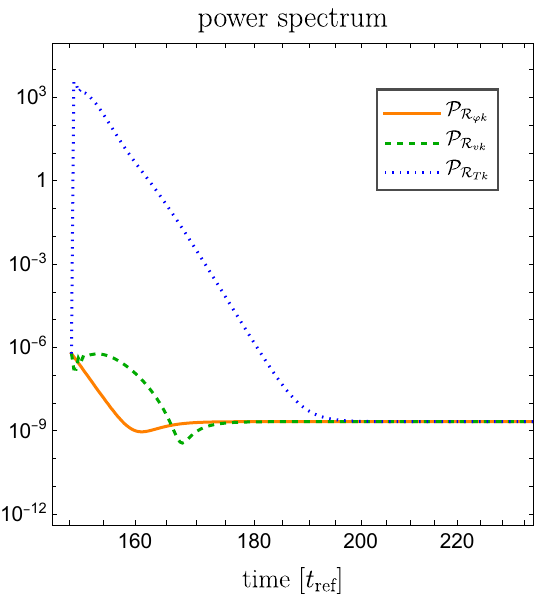}
        \caption{Weak regime: $Q_\star=0.015$,  $T_\star/H_\star=20.4$.}
    \end{subfigure}
     \hfill
    \begin{subfigure}{0.49\linewidth}
        \centering
        \includegraphics[height=0.32\textheight]{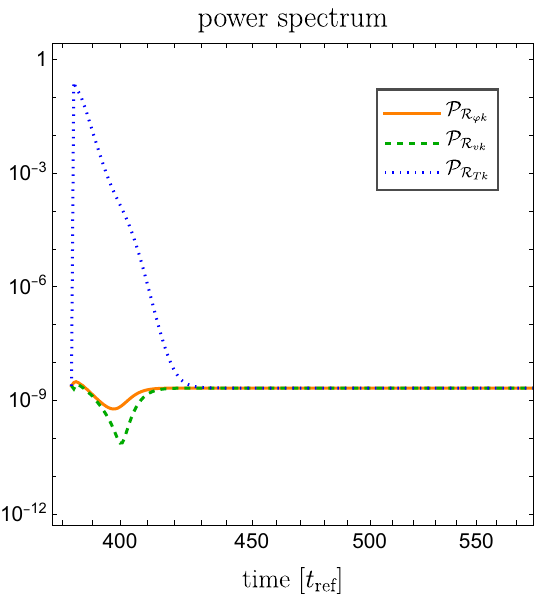}
        \caption{Weak regime:   $Q_\star=0.711$,  $T_\star/H_\star=213$.}
    \end{subfigure}
\par\vspace{1cm}
    \begin{subfigure}{0.49\linewidth}
        \centering
        \includegraphics[height=0.32\textheight]{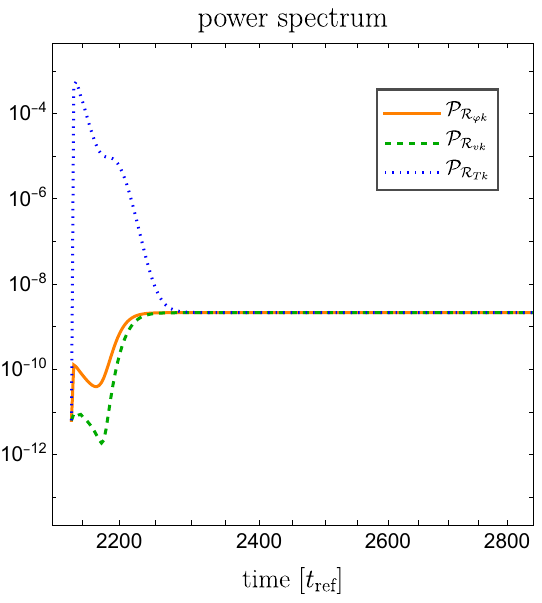}
        \caption{ Strong regime: $Q_\star=4.88$,  $T_\star/H_\star=1531$.}
    \end{subfigure}
    \hfill
    \begin{subfigure}{0.49\linewidth}
        \centering
        \includegraphics[height=0.32\textheight]{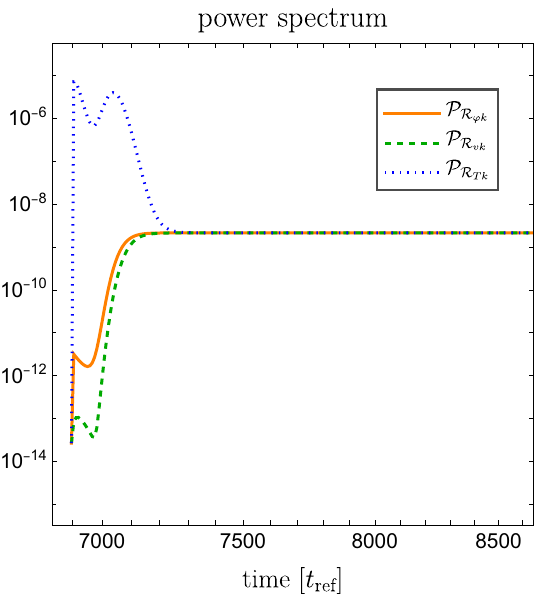}
        \caption{Strong regime: $Q_\star=13.95$,  $T_\star/H_\star=7804$.}
    \end{subfigure}
    \caption{Power spectrum evolution of three curvature perturbations: $\mathcal{R}_{\varphi k}$ (inflaton), $\mathcal{R}_{vk}$, and $\mathcal{R}_{Tk}$ (plasma). We show the evolution from shortly before the freeze-out, when $k/a=100 \,H$, until after all the curvature perturbations have fully frozen-out, when $k/a=10^{-10}\,H$. The notation is like in table \ref{table:sm}.}
    \label{fig:perturb_strong}
\end{figure}
Fig.~\ref{fig:perturb_strong} again shows the evolution for four cases spanning the weak and the strong regimes. In all cases we observe the freeze-out of curvature perturbations. We notice that, especially in the strong regime, $\mathcal{R}_{vk}$ and $\mathcal{R}_{\varphi k}$ are actually growing before the freeze-out. That is because in the strong regime, noise sources more thermal perturbations, but in the end $\mathcal{R}_{Tk}$ must pull the other two to the common value (or vice versa) such that after the horizon crossing, all curvature perturbations obtain the common value. If we take a closer look, however, we see that even in the strong regime, all curvature perturbations are at least initially decreasing. As a check that this is not just an oscillation but a general trend, we show fig.~\ref{fig:kaH10p3} (to be discussed in due detail in Ch.~\ref{ch:inittimeofintegandacoustosc}) --- the same power spectrum as in fig.~\ref{fig:perturb_strong}, just with the evolution starting at an earlier time, when $k/a=10^3\;H$.
%\footnote{\hl{We make a minor comment.} One might notice that the power spectra in fig.~\ref{fig:perturb_strong} are larger in the weak regime, before they settle to a constant value. It is unclear whether that carries any physical significance; one can note, however, that initial conditions for the case in fig.~\ref{fig:perturb_strong}(a), are orders of magnitude larger than those in fig.~\ref{fig:perturb_strong}(d). Similarly, the temperature curvature perturbation is rather large for the choice of parameters in fig.~\ref{fig:perturb_strong}(a); perhaps this could be partially explained by noticing the ratio $H/\dot{T}$, which defines $\mathcal{R}_T$, is about two orders of magnitude larger there than in other cases, largely due to $H$ being significantly larger in the weak regime. } %Additionally, if plasma would not couple to inflaton at all, there would be no a priori reason why all curvature perturbations should obtain the same value. Going deep into the weak regime does not decouple the equations, however, it makes is less strong; this could perhaps lead to longer time needed to obtain the same value. This is, however, just an educated guess, and not a real explanation.
\\\\
Given that the value of $\mathcal{R}_{Tk}$ immediately jumps many orders of magnitude, one might wonder whether it is correct to assume the initial conditions in eqs.~\eqref{ourinitialconditions}. We must recall that, despite the apparent $\mathcal{R}_{Tk}\gg  \{\mathcal{R}_{\varphi k},\, \mathcal{R}_{v k}\} $ at early times, the isocurvature perturbation $\mathcal{S}_{Tk}$ has in its definition a prefactor $e_{r,T} \dot{T}$, which is very small during inflation, see eq.~\eqref{isocurvature_b}. In practice, we indeed find that $\mathcal{S}_{Tk}
\approx \mathcal{S}_{vk}\approx0$ is a good approximation, see fig.~\ref{fig:isocurvature}. It turns out that the magnitudes of isocurvature perturbations are at least 30 orders of magnitude smaller than that of $\mathcal{R}_{\varphi k}$ at all times. Additionally, since eqs.~\eqref{result} are written and solved in terms of $\mathcal{R}_{\varphi k},\;\mathcal{S}_{vk},\;\mathcal{S}_{Tk}$, and not $\mathcal{R}_{Tk}$, there is also no problem regarding the perturbative treatment of perturbations.

%While those initial conditions were derived at very early times, when $k/a$ was by far the largest scale in question, in fig.~\ref{fig:perturb_strong}, we actually implemented initial conditions at much later times, $k=100 \;a\,H$. By that time, the noise is no longer exponentially suppressed, but rather significant enough to source temperature fluctuations, see fig.~\ref{fig:superearlytimeOMEGA}. Therefore, one might predict significant deviation of $\mathcal{R}_T$ from 0, by the time we start our power spectrum evolution. To illustrate this point, see fig.~\ref{fig:superearlytime}. At the beginning, when the noise....

\subsection{Initial time of integration and acoustic oscillations}\label{ch:inittimeofintegandacoustosc}
We make a few comments on the time at which we start integrating eqs.~\eqref{result}. The initial conditions were set in Bunch-Davies vacuum, when the largest scale in eqs.~\eqref{result} is set by $k/a$. Setting the initial conditions too late, when $k/a$ is not anymore the dominant scale in eqs.~\eqref{result}, might result in an inaccurate frozen-out value of the power spectrum, $A_s$. On the other hand, we expect that if we do set the initial conditions at the time when $k/a$ is the dominant scale, $A_s$ should no longer depend on the precise starting moment of integration; rather, the frozen-out power spectrum would ``plateau'' (as a function of the starting time).
\\
\begin{figure}[p]
    \centering
    \begin{subfigure}{0.49\linewidth}
        \centering
        \includegraphics[height=0.32\textheight]{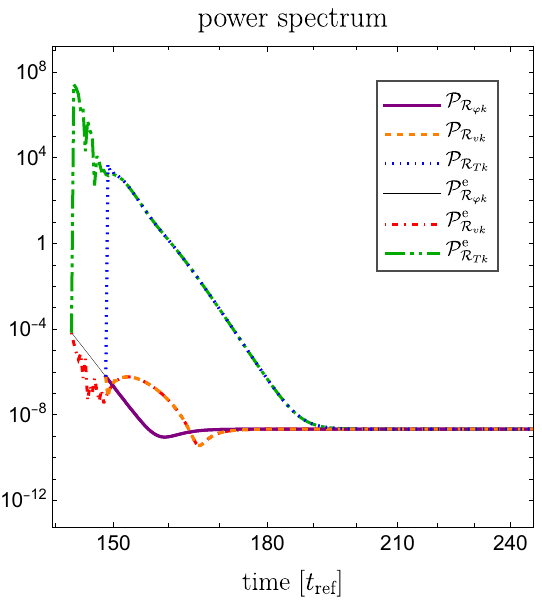}
        \caption{Weak regime: $Q_\star=0.015$,  $T_\star/H_\star=20.4$.}\label{fig:kaH10p3a}
    \end{subfigure}
     \hfill
    \begin{subfigure}{0.49\linewidth}
        \centering
        \includegraphics[height=0.32\textheight]{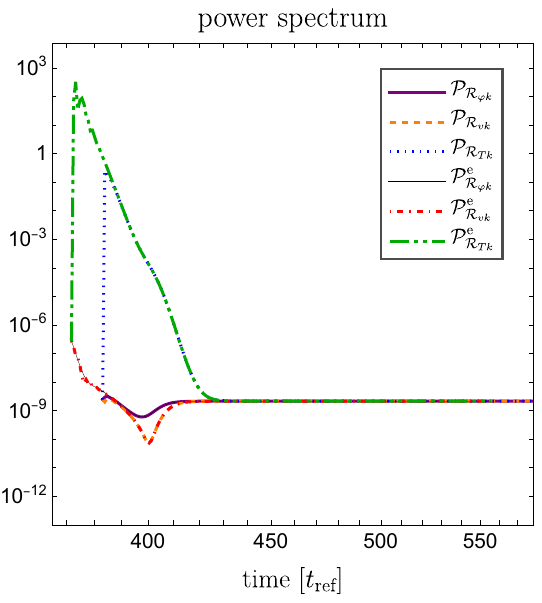}
        \caption{Weak regime:   $Q_\star=0.711$,  $T_\star/H_\star=213$.}\label{fig:kaH10p3b}
    \end{subfigure}
\par\vspace{1cm}
    \begin{subfigure}{0.49\linewidth}
        \centering
        \includegraphics[height=0.32\textheight]{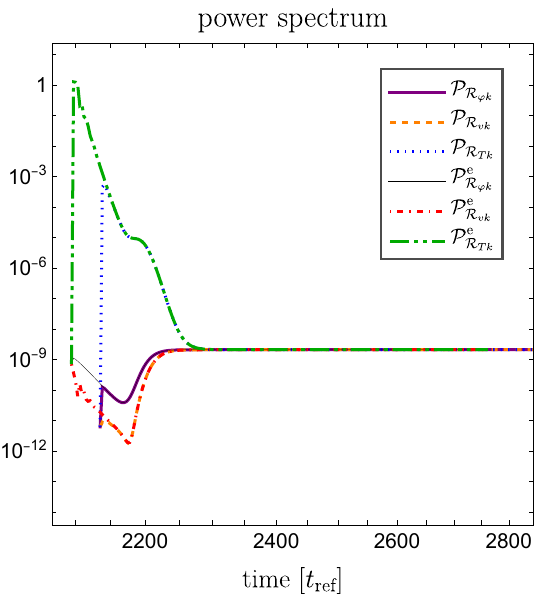}
        \caption{ Strong regime: $Q_\star=4.88$,  $T_\star/H_\star=1531$.}\label{fig:kaH10p3c}
    \end{subfigure}
    \hfill
    \begin{subfigure}{0.49\linewidth}
        \centering
        \includegraphics[height=0.32\textheight]{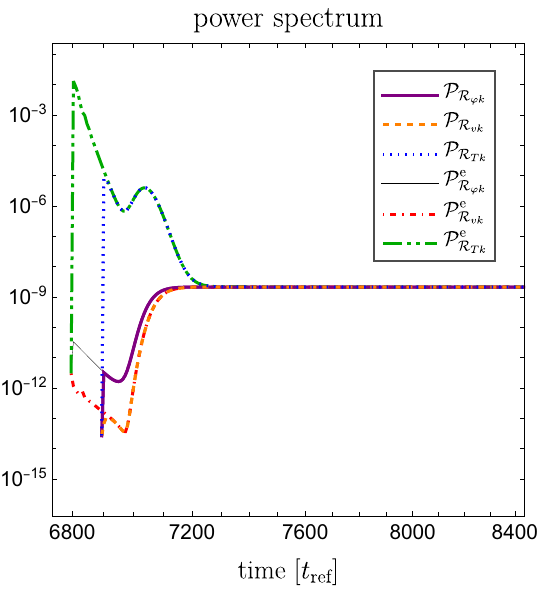}
        \caption{Strong regime: $Q_\star=13.95$,  $T_\star/H_\star=7804$.}\label{fig:kaH10p3d}
    \end{subfigure}
    \caption{Power spectrum evolution of three curvature perturbations, starting at $k/(aH)=10^2$ ($\mathcal{R}_{\varphi k}$, $\mathcal{R}_{vk}$, and $\mathcal{R}_{Tk}$), and starting at en earlier time, when $k/(aH)=10^3$ ($\mathcal{R}_{\varphi k}^\text{e}$, $\mathcal{R}_{vk}^\text{e}$, and $\mathcal{R}_{Tk}^\text{e}$). The notation is like in table \ref{table:sm}.}
    \label{fig:kaH10p3}
\end{figure}
\\
All the results so far were obtained when $k/a=10^2\,H$, assuming this was a large enough ratio to already be in the ``plateau'' region. We shall now investigate whether the power spectra change if one increases the starting ratio to $k/a=10^3\,H$. The results of the study are presented in fig.~\ref{fig:kaH10p3}. Each figure (covering the weak and the strong regimes) plots the new, ``early'' power spectra with $k/a=10^3\,H$ start ($\mathcal{R}_{\varphi k}^\text{e}$, $\mathcal{R}_{v k}^\text{e}$, and $\mathcal{R}_{T k}^\text{e}$), along with the power spectra already presented in fig.~\ref{fig:perturb_strong}, with $k/a=10^2\,H$ start ($\mathcal{R}_{\varphi k}$, $\mathcal{R}_{vk}$, and $\mathcal{R}_{Tk}$).
\\
\\
We observe a remarkable overlap in the evolution of the power spectra, regardless of whether we start the evolution at $k/a=10^2\,H$ or $k/a=10^3\,H$. This suggests that we have already entered the ``plateau'' region, and that the results presented in the previous sections are trustworthy. 
\\\\
We need to make an additional comment about the tolerance, eq.~\eqref{tolerance}. The default value for most results in this thesis was $\texttt{tol}=10^{-3}$. However, when this value was used to obtain fig.~\ref{fig:kaH10p3a}, we initially observed a tiny discrepancy between the ``early'', $k/(a_1 H_1)=10^3$ power spectra, and the $k/(a_1 H_1)=10^2$ power spectra. The difference grew if we further increased the starting ratio to $k/(aH)=10^4$. This was not observed when studying the strong regime at the same starting ratio; see fig.~\ref{fig:ss} (to be further discussed in Ch.~\ref{ch:noiseimp}). It is unlikely the difference should be attributed to physics; after the power spectrum has already reached a ``plateau'', it should not suddenly change again, as no new physics enters with the increasing of the starting ratio $k/(aH)$. One should further note that if anything, the deviation from the Bunch-Davies vacuum would be more pronounced in the strong regime, not in the weak regime, giving further support to attributing the tiny difference in power spectra in the weak regime to the numerics. Indeed, after decreasing \texttt{tol} to $10^{-4}$ the problem disappeared, as shown in fig.~\ref{fig:kaH10p3a}.
\\\\
Going to even earlier times in the weak regime would require us to further decrease the tolerance to obtain accurate results. That is because at very early times, the curvature perturbations undergo acoustic oscillations; they are especially pronounced in the weak regime (there the dissipation does not damp them out), see ``wiggles'' at the top of fig.~\ref{fig:kaH10p3a}. In order to integrate over those rapid oscillations accurately, the time steps should be very small, and so $\texttt{tol}$ from eq.~\eqref{tolerance} needs to be decreased. Otherwise, too large time steps lead to an accumulation of numerical errors and a shift in the power spectrum. However, such tiny time steps as are required for accurate integration over acoustic oscillations quickly become numerically too expensive. Luckily, fig.~\ref{fig:kaH10p3} shows it is not necessary to go to such early times; starting at $k/(a_1 H_1)=10^2$ is sufficient.

\subsection{Importance of noise}\label{ch:noiseimp}
\begin{figure}[!h]
    \centering
    \begin{subfigure}{0.49\linewidth}
        \centering
        \includegraphics[height=0.32\textheight]{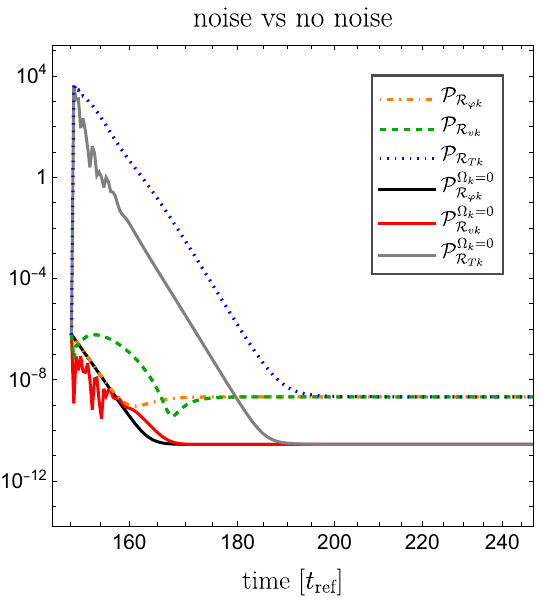}
        \caption{Weak regime: $Q_\star=0.015$,  $T_\star/H_\star=20.4$.}
    \end{subfigure}
     \hfill
    \begin{subfigure}{0.49\linewidth}
        \centering
        \includegraphics[height=0.32\textheight]{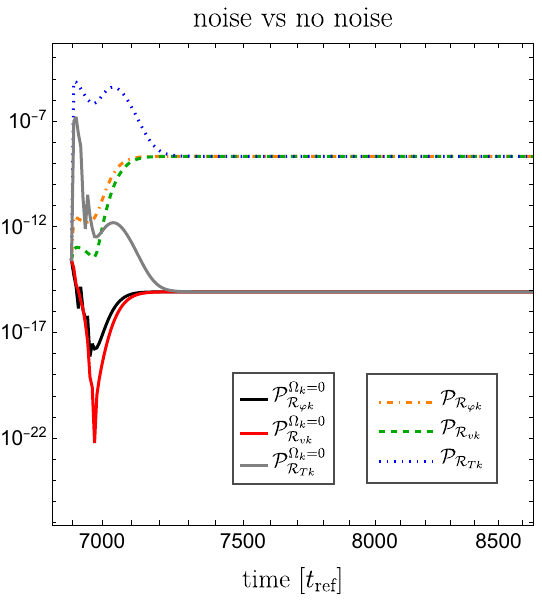}
        \caption{Strong regime: $Q_\star=13.95$,  $T_\star/H_\star=7804$.}
    \end{subfigure}
    \caption{Power spectrum in presence of the noise term (dotted, dashed, dot-dashed lines) and in the absence of it, $\Omega_k\overset{!}{=}0$ (solid lines). The notation is like in table \ref{table:sm}.}
    \label{fig:noiseimportance}
\end{figure}

Here we briefly demonstrate the relevance of the noise term for the power spectrum, with the aid of fig.~\ref{fig:noiseimportance}. The left part of the figure shows the power spectra in the weak regime, both in the presence of the noise term and with the noise autocorrelator set to zero by hand, $\Omega_k=0$. The right part of the figure plots the same quantities, but in the strong regime. We observe that in the strong regime the noise-less power spectra are larger by approximately seven orders of magnitude, as opposed to two orders of magnitude in the weak regime (going deeper into the weak regime would bring the difference between the noisy and the noise-less computation to zero).\footnote{In the latter computation, we ``extrapolated'' $\alpha_s=0.0241$ for $\lambda=10^{-14}$ and tuned $f_a=13.92$ such that $A_s=2.10\times 10^{-9}$; then the noise-less computation is identical to the noisy one. However, for these parameter values we also obtain $n_s\approx 0.34$ and $r\approx 0.068$, both of which are observationally excluded.} %\footnote{\hl{decreasing the inflaton coupling $\lambda$ an order of magnitude more, to $10^{-22}$, and then adjusting $f_a=0.083$ such that, given extrapolated $\alpha_s=0.0275$, $A_s=2.10\times10^{-9}$, would already lead to no difference in power spectra, noisy or noise-less. However, then, for some reason, $Q_\star=21.8$; why? Is $\alpha_s$ wrong? Previously we just kept their $\alpha_s$ values... Also since T is changing, shouldn't $\alpha_s$ in principle be changing also?}}  
\\\\
The general trend is that the more we approach the strong regime, the bigger the ratio between the ``noisy'' and ``noise-less'' power spectra becomes, and hence the bigger the influence of the noise on the power spectra.
\\\\
\begin{figure}[!h]
    \centering
    \begin{subfigure}{0.48\linewidth}
        \centering
        \includegraphics[height=0.32\textheight]{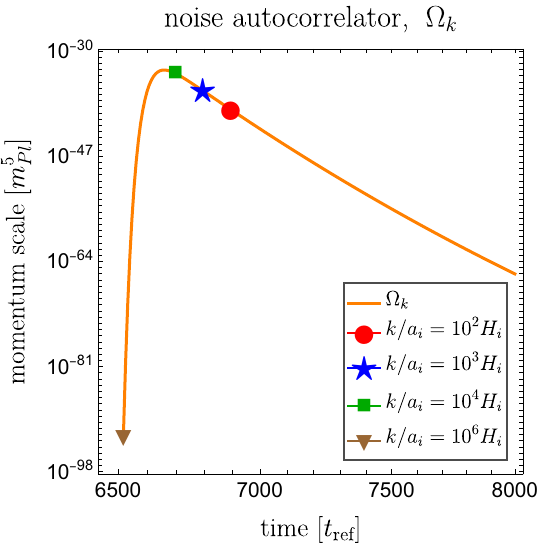}
        \caption{Plot of $\Omega_k$. The red dot, blue star, green square, and brown triangle represent the starting times of integration for different initial ratios of $k/(aH)$.}\label{fig:so}
    \end{subfigure}
     \hfill
    \begin{subfigure}{0.48\linewidth}
        \centering
        \includegraphics[height=0.32\textheight]{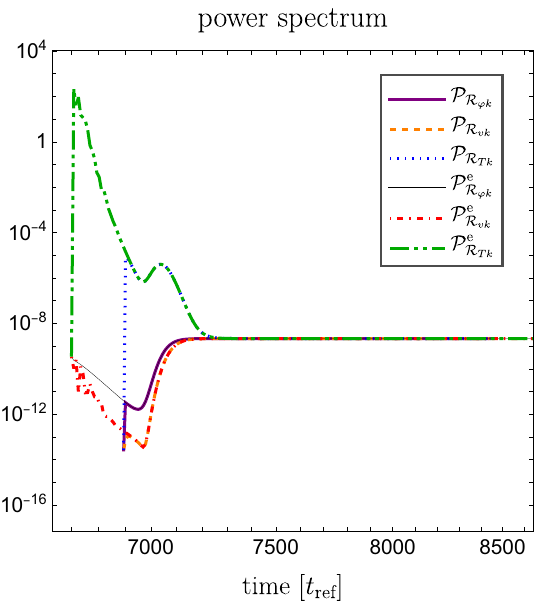}
        \caption{Power spectrum evolution, starting at initial ratio $k/(aH)$ values of $10^2$ and $10^4$ (superscript $e$). They overlap perfectly.}\label{fig:ss}
    \end{subfigure}
    \caption{Independence of the precise starting point of integration, in relation to the noise. Values correspond to the strong regime, $Q_\star=13.95$. The notation is like in table \ref{table:sm}.}
    \label{fig:noiseimportance1}
\end{figure}
\clearpage
\noindent
We make a final comment on the noise, in relation to the starting point of integration. We recall that the initial conditions were fixed in Bunch-Davies vacuum, where the noise exponentially decays. A theoretically clean approach to determining the power spectrum would therefore be to start the evolution of power spectra when $\Omega_k$ is exponentially suppressed (see the brown triangle in fig.~\ref{fig:noiseimportance1}, i.e.~the starting point of integration at $k/(aH)=10^6$, for strong regime, $Q_\star=13.95$). Note that starting this early is a sufficient, but not a necessary condition to fixing the initial time. As seen from fig.~\ref{fig:kaH10p3d}, the ``plateau'' is already reached at $k/(aH)=10^2$. If the power spectra results do not change through the starting ratios $k/(aH)=\{10^2,\,10^3,\,10^4\}$ (as seen on fig.~\ref{fig:ss}), the results should not change for the starting ratio $k/(aH)=10^6$ either. Additionally, as seen from fig.~\ref{fig:so}, starting at $k/(aH)=10^4$ (green square) already captures most of the noise; earlier, the noise is exponentially suppressed and should thus be negligible.

%\footnote{\hl{Nonetheless, we know that in the absence of a plasma, noise should have no influence on the power spectrum; the last line of table} \ref{table:sm} \hl{is not deep enough in the weak regime to see that, as also evidenced by fig.~\ref{fig:noiseimportance}\hl{.But it seems to be that in the weak (not ultra weak) regime noise has a nontrivial influence. Perhaps that is because initial conditions are not strongly washed away by large $\Upsilon$; the noise fluctuations in the weak regime then take longer to start having effect. Perhaps?}}

\chapter{Conclusions and outlook}\label{chapter:conclusion}
The work discussed in this thesis provides a general framework for describing the late stage of inflation and the subsequent period of reheating, assuming the existence of a thermal plasma and a single scalar inflaton field. In particular, we derive gauge-invariant, model-independent evolution equations for curvature perturbations, see eqs.~\eqref{result}. A key property of the equations is to address the question of inflaton equilibration dynamically, rather than assuming it from the start. Furthermore, we suggest a new form of the noise autocorrelator, eqs.~\eqref{correlator} and \eqref{omega}, which smoothly interpolates between quantum and classical domains. 
\\\\
In order to obtain solutions to eqs.~\eqref{result}, we provide a numerical implementation of our framework, see App.~\ref{app:CodeOverview}. We test the code on a particular model, defined in ref.~\cite{ZellSM}, summarized in Ch.~\ref{sec:ModelBuilding}. The solutions exhibit early acoustic oscillations as well as the freeze-out of curvature perturbations shortly after the horizon crossing. By testing the model, we also provide its more precise\footnote{``More'' as compared to the original estimate by ref.~\cite{ZellSM}; see Ch.~\ref{ch:keyresults}.} predictions for the power spectrum, spectral tilt, and tensor-to-scalar ratio. 
The code can accommodate other models as well. 
\\\\
Indeed, there exist other warm inflation codes, written in different programming languages (\texttt{Python}, ref.~\cite{WarmSPy}, \texttt{Mathematica}, ref.~\cite{WI2Easy}, \texttt{C++} and \texttt{Python}, refs.~\cite{swim,dswim}). While some other codes may be more efficient than  our \texttt{Mathematica} implementation, none of them solves the full equations, eqs.~\eqref{result} with eqs.~\eqref{correlator}, \eqref{omega}. The purpose of this work was to present an implementation based on first principles, even if it comes at the cost of more demanding numerics. 
\\\\
Natural next steps are to make the code more robust while simultaneously using it to make predictions for a few more models. More conceptually, it would be interesting to try to distill the predictions for power spectra into some model-independent observations. Furthermore, at the second order in the perturbative expansion, scalar perturbations source gravitational waves. It would be interesting to explore gravitational-wave signatures of e.g.~the model discussed here with our dynamically thermal and gauge-independent framework.

% Appendix______________________________________________________________________
\appendix

\clearpage\phantomsection\addcontentsline{toc}{chapter}{Bibliography}\addcontentsline{toc}{chapter}{Appendix}
\chapter{Code overview}\label{app:CodeOverview}
\SetWatermarkText{}
Here we describe our Mathematica code in detail. It is composed of three parts: \textit{Model-dependent inputs}, which is meant to be modified by a user, \textit{Computing: background and perturbations}, which computes the power spectrum, spectral tilt and tensor-to-scalar ratio, and \textit{Results}, which plots and saves the results of the second part. The last two parts do not need to be modified by the user.

\section{Model-dependent inputs}
The \textit{model-dependent inputs} part of the code is composed of five small subparts: \textit{model parameters, initial values, numerical parameters, parameter scan} and \textit{save results}.
\\\\
At the beginning, we define radiation energy density $ e_r$ and pressure $ p_r$, inflaton potential $V$, friction coefficient $\Upsilon$, along with potential fixed coefficients that enter them, see e.g.~Ch.~\ref{ch:SM}. As an example, the code snippet below defines the radiation energy density as the number of effective degrees of freedom $g_*$ times $T^4$, rescaled by $\frac{\pi^2}{30}$. \\
\begin{center}
\includegraphics[width=1\linewidth]{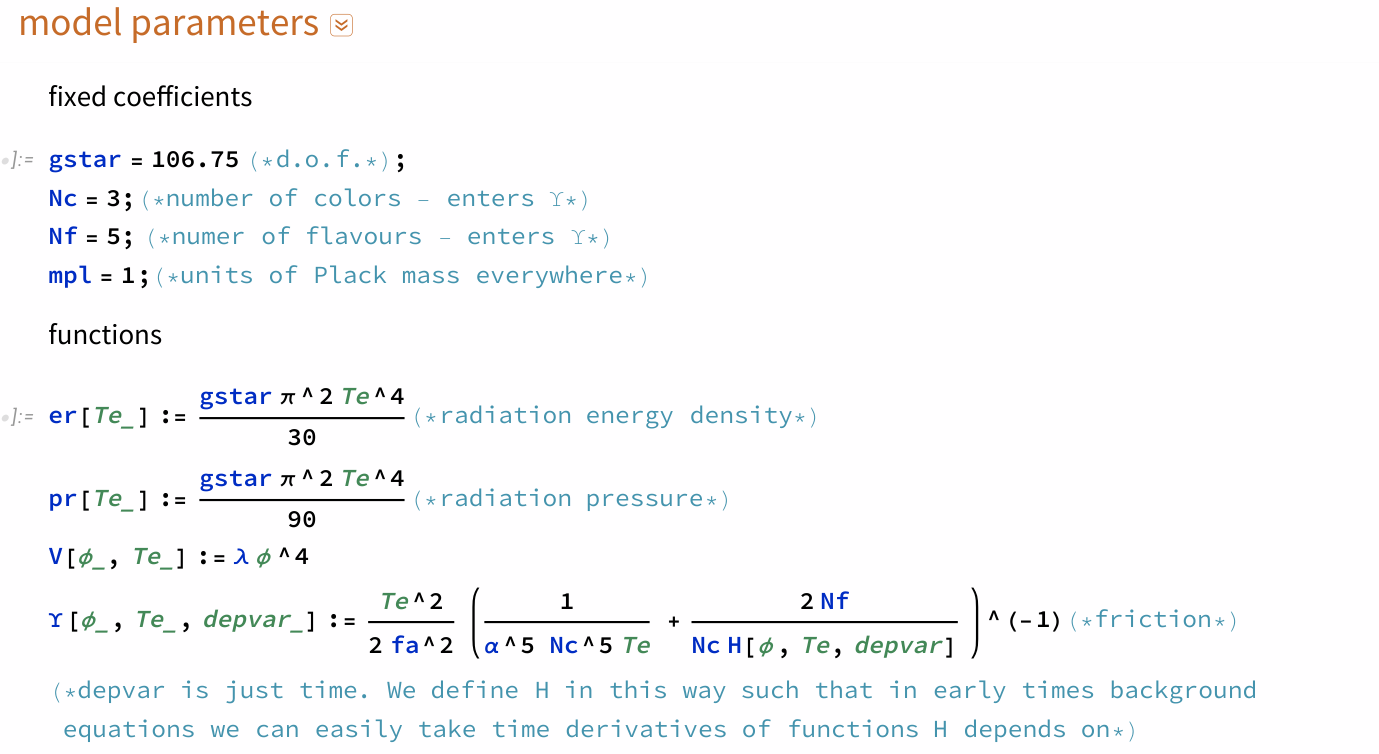}
\end{center}
We continue by defining initial values, which were chosen as explained in Ch.~\ref{ch:initialvalues}.
\begin{center}
    \includegraphics[width=1\linewidth]{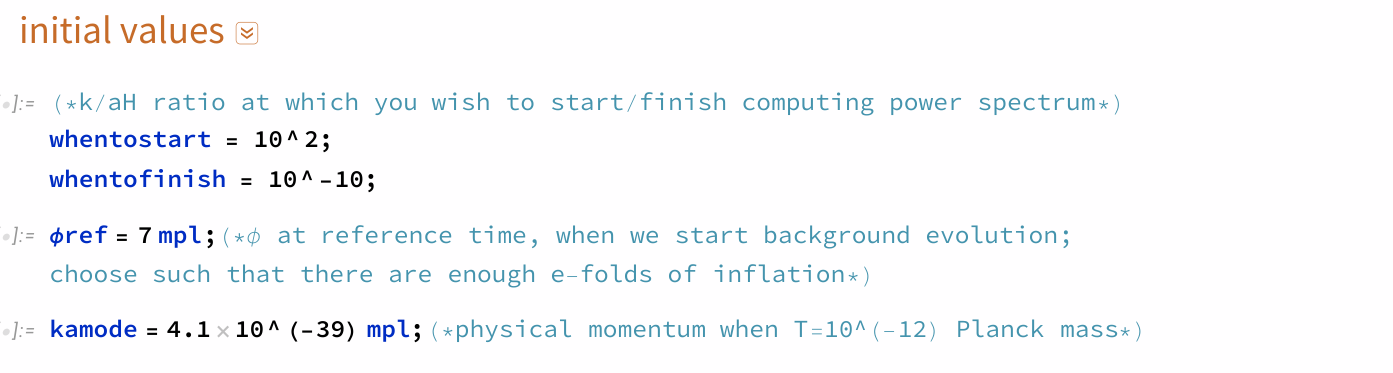}
\end{center}
We further define numerical parameters. Parameter $\delta_t$, see eq.~\eqref{nsapprox}, is denoted as \texttt{tilt}, while the tolerance from eq.~\eqref{tolerance} is $\texttt{tol}=10^{-\texttt{n}}$. The regularization parameter $\delta$ was defined in eq.~\eqref{delta}.\\
\begin{center}
    \includegraphics[width=1\linewidth]{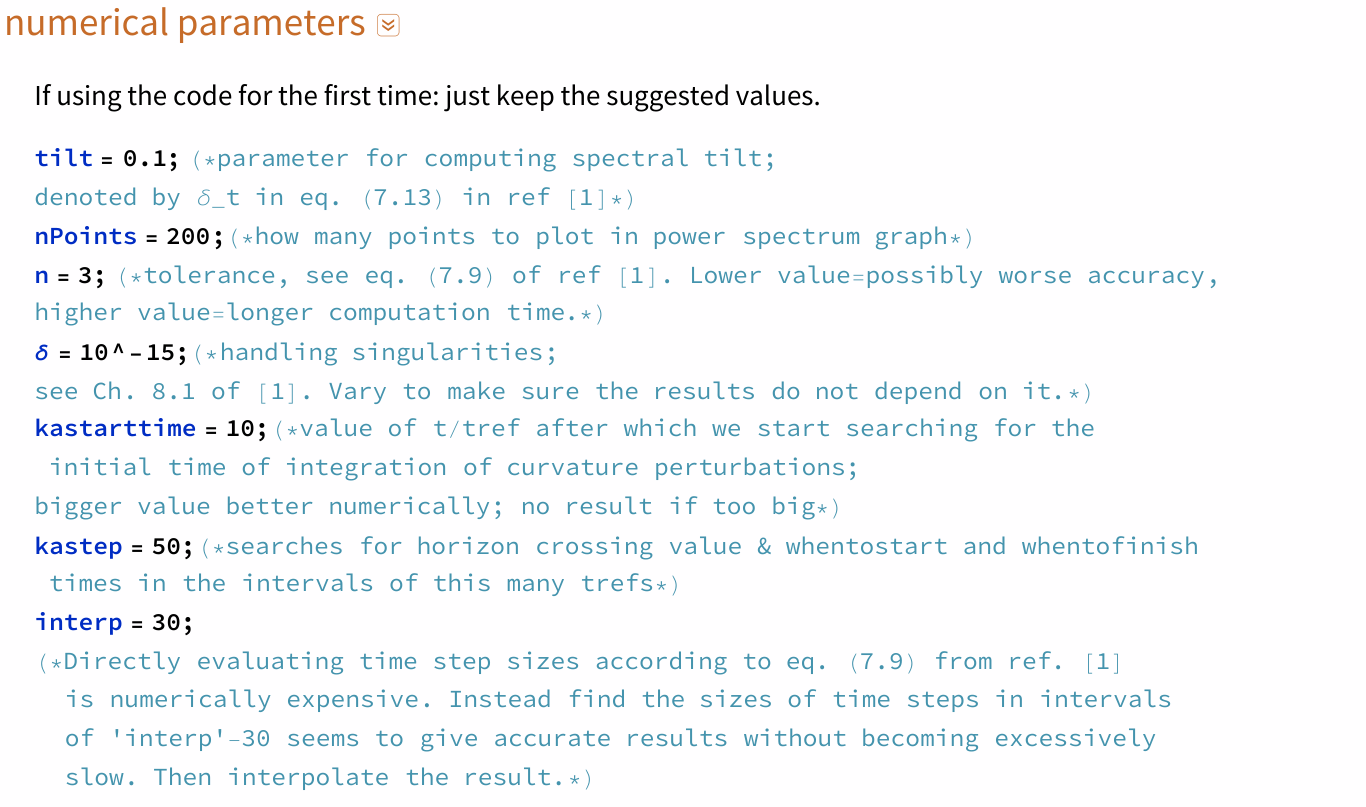}
\end{center}
We continue by sketching how one can obtain solutions for a number of different values of coefficients. If there are just a few sets of coefficients we wish to run our simulation for, we can run this notebook manually a few times, i.e.~by changing \texttt{j} in the code snippet below. If the parameter scan is more substantial, however, it is advisable to write a small control notebook for the parameter scan instead. \\
%one can split the pic in this way; but make sure all text is here fine first, before troubling with formatting...
\begin{center}
    \includegraphics[width=1\linewidth]{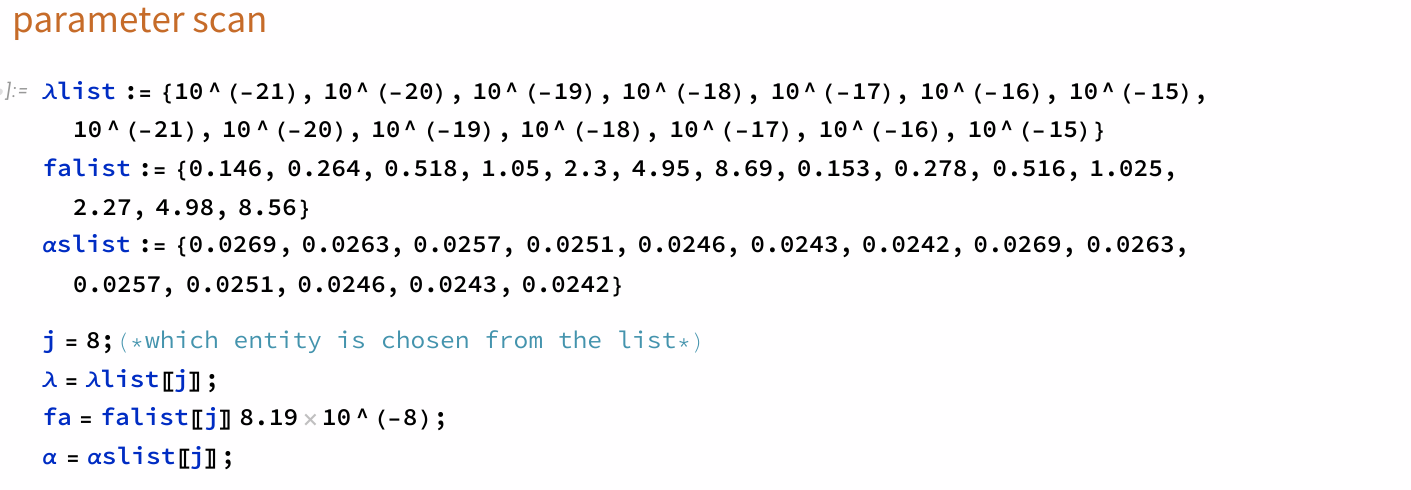}
\end{center}
The code automatically saves selected results locally. Here we provide the folder where we save the results, along with the names to be used for pdf and data files. In the code snippet below, the names of the data files were chosen according to the choice of parameters used to obtain those results.\\
\begin{center}
    \includegraphics[width=1\linewidth]{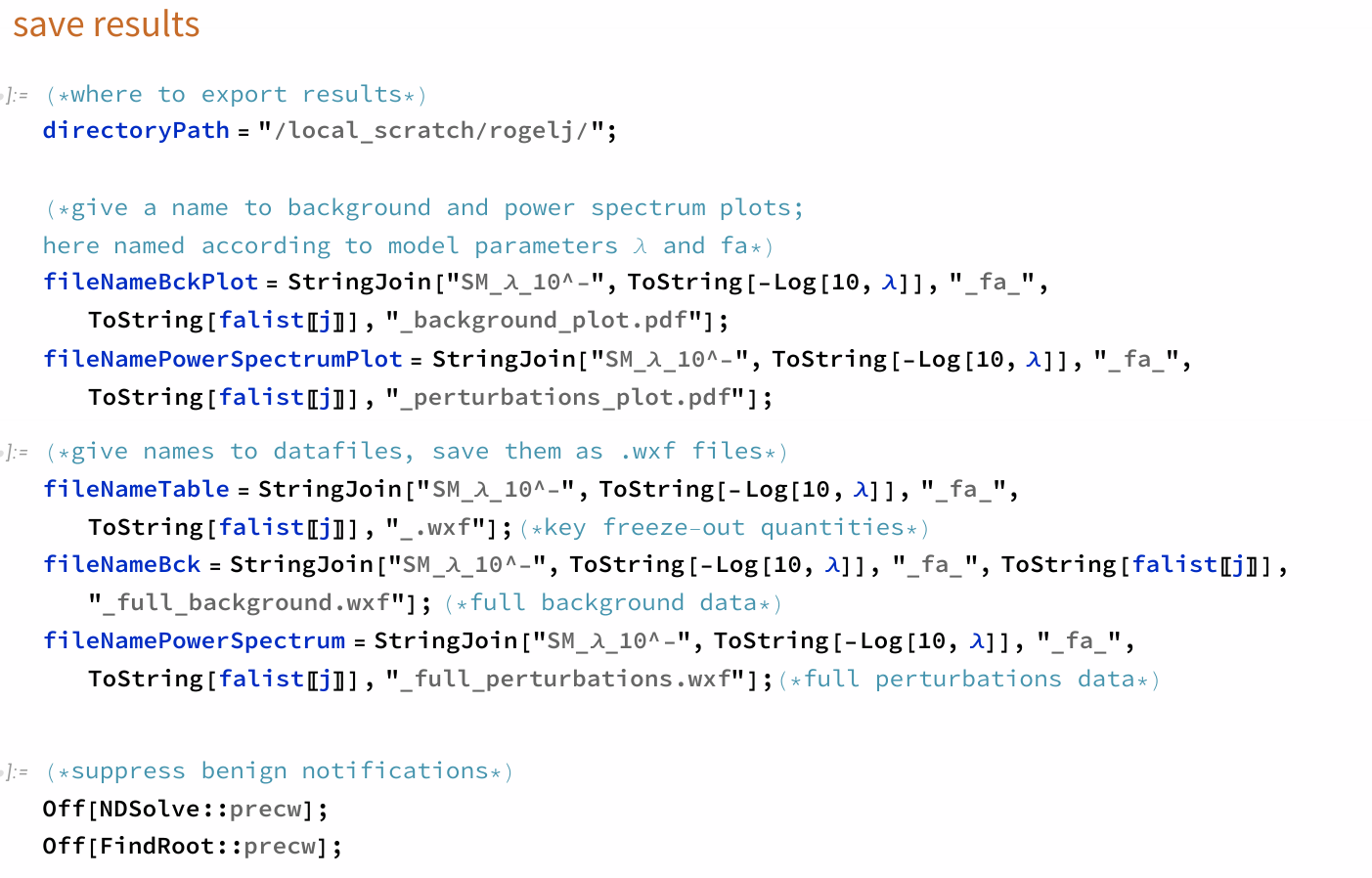}
    \label{fig:saveresults}
\end{center}
The code automatically produces two plots in pdf format: background plot, showing the evolution of $\Upsilon,\;H,\;T,$ and $k/a$, and the power spectrum plot, showing the evolution of $\mathcal{P}_{\mathcal{R}_{\varphi k}}$, $\mathcal{P}_{\mathcal{R}_{v k}}$, and $\mathcal{P}_{\mathcal{R}_{T k}}$. Aside from the plots, three data files are produced. One saves key freeze-out parameters, while the other two tabulate the time evolution of the background quantities and the power spectrum.

\section{Computing: background and perturbations}\label{app:codeCOMPUTING}
This part of the notebook requires no user input. It consists of eight parts: \textit{background evolution at early times}, \textit{background evolution at late times}, \textit{momentum evolution}, \textit{determining time steps}, \textit{tabulating the background}, \textit{computing the power spectrum average}, which  includes the computation of the spectral tilt and tensor-to-scalar ratio, and \textit{tabulating} and \textit{plotting the results}.
\\\\
We begin the background evolution by defining the background energy density and pressure.
\begin{center}
    \includegraphics[width=1\linewidth]{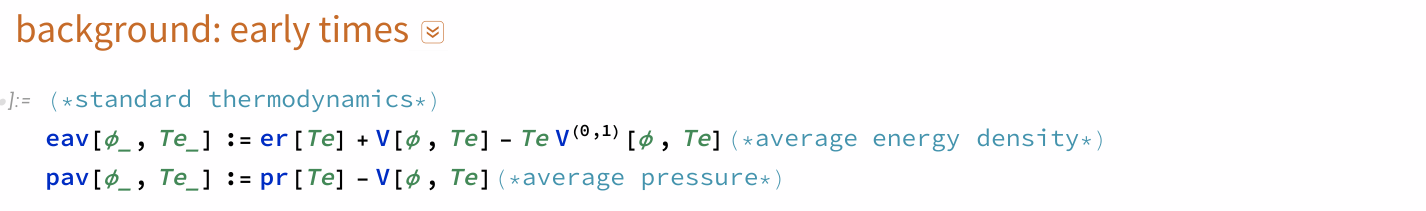}
\end{center}
We continue by deriving initial values for $\dot{\bar{\varphi}}$ and $T$ at the reference time $t_\text{ref}$, as described in Ch.~\ref{ch:initialvalues}. These are only estimates, and in practice the estimates do not even need to be very good, as the solution quickly evolves to the quasi de Sitter expansion anyway.
\begin{center}
    \includegraphics[width=1\linewidth]{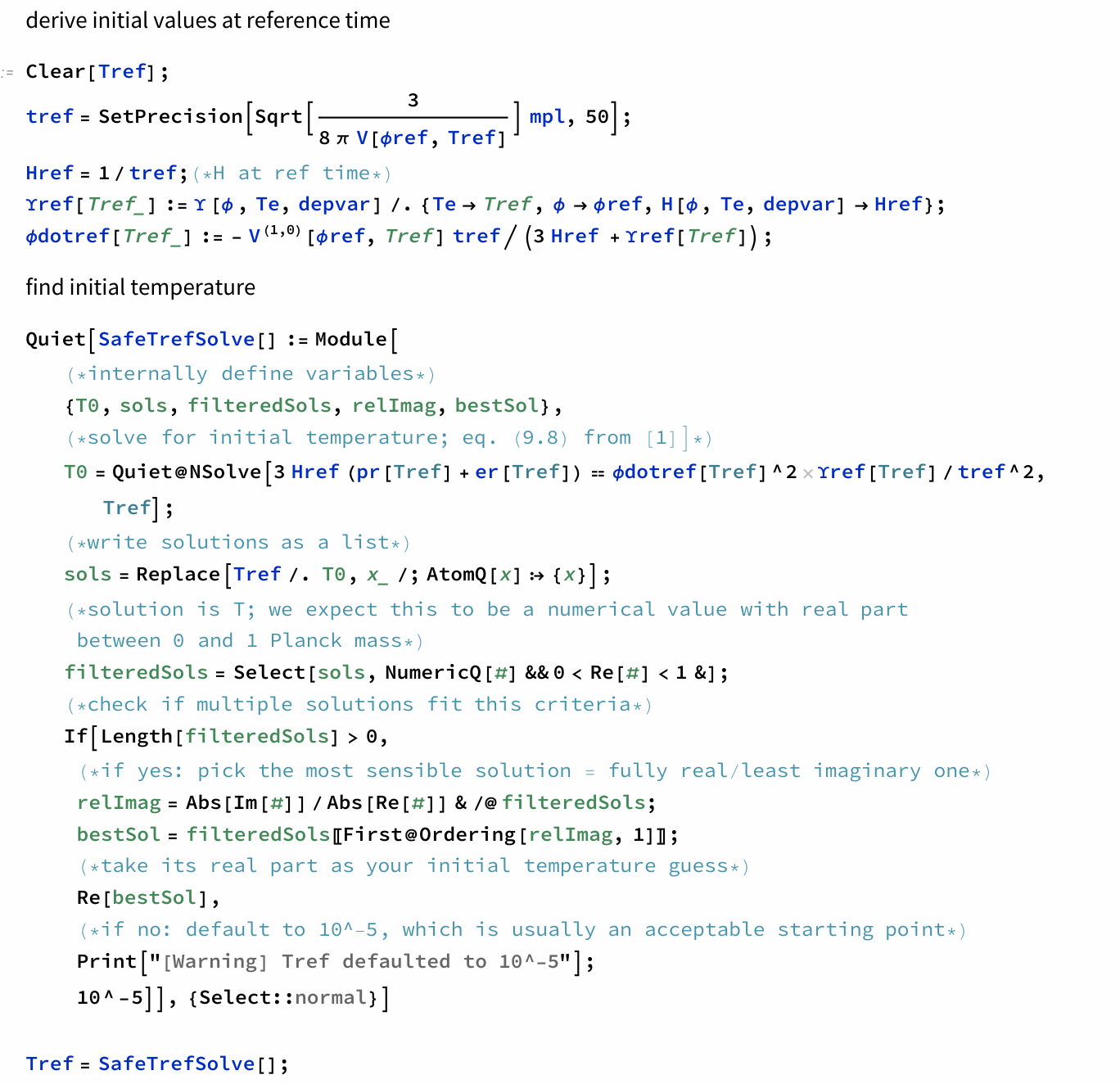}
\end{center}
Then we define the background equations as in eqs.~\eqref{physical}.
\begin{center}
    \includegraphics[width=1\linewidth]{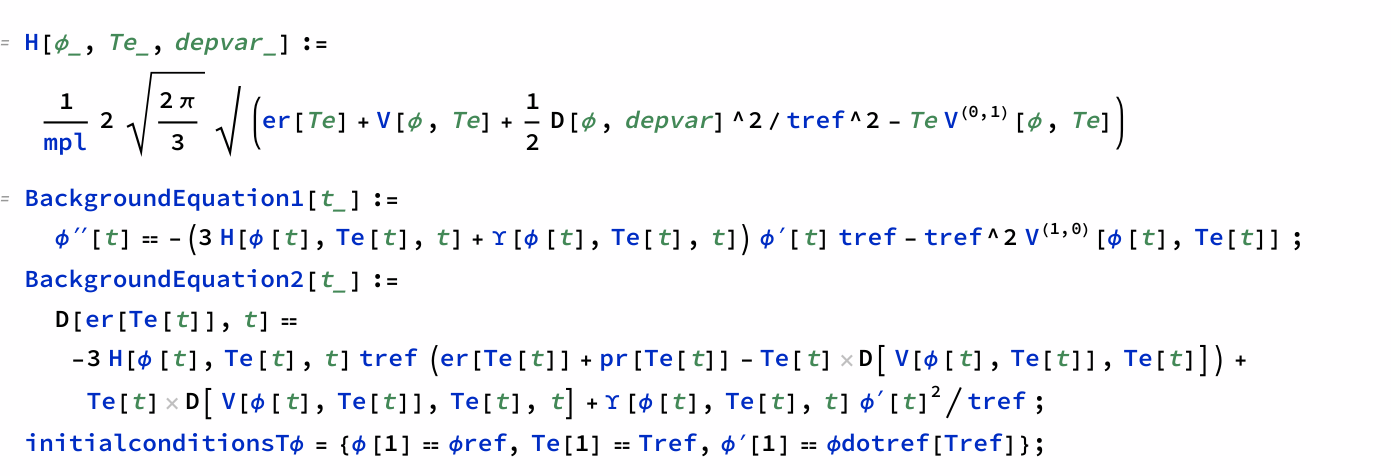}
\end{center}
We proceed to solve them.
\begin{center}
    \includegraphics[width=1\linewidth]{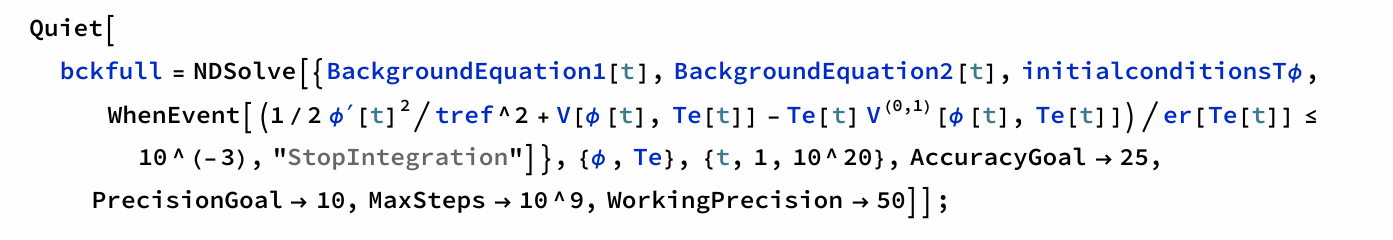}
\end{center}
We save the resulting $\bar{\varphi}$ and $T$ as \texttt{$\phi\text{bck}1[t]$} and \texttt{$T\text{bck}1[t]$} respectively. 
\begin{center}
    \includegraphics[width=1\linewidth]{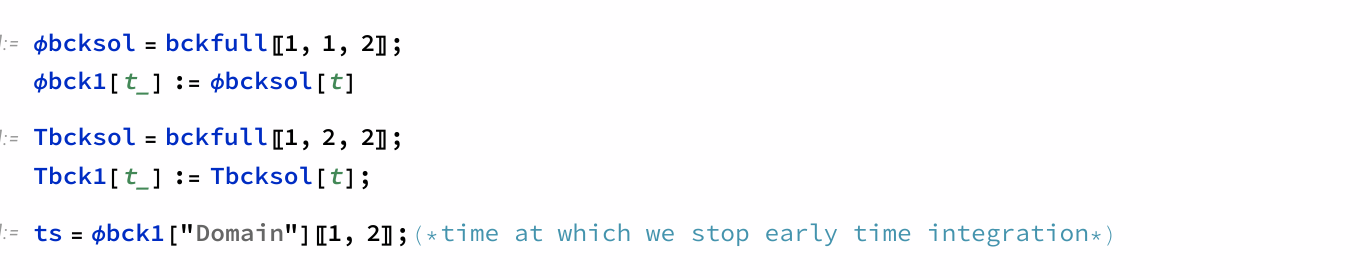}
\end{center}
Note that we only evolve the background until radiation domination (\texttt{WhenEvent} in the code snippet).\footnote{For models with a clear period of matter domination, one may consider stopping the early-time integration after a certain number of oscillations instead; a conservative choice could be 20 oscillations. Then, a different set of late-time evolution equations should be adopted, given by eqs.~\eqref{AVERAGED}. The here implemented eqs.~\eqref{raddombck} can only be used once $e_r\gg e_{\bar{\varphi}}$.} The moment in time when we stop the integration is denoted by \texttt{ts}, for ``t switch''. Afterwards, we go over to the late-time evolution described by the averaged equations from Ch.~\ref{averaging}.
\begin{center}
    \includegraphics[width=1\linewidth]{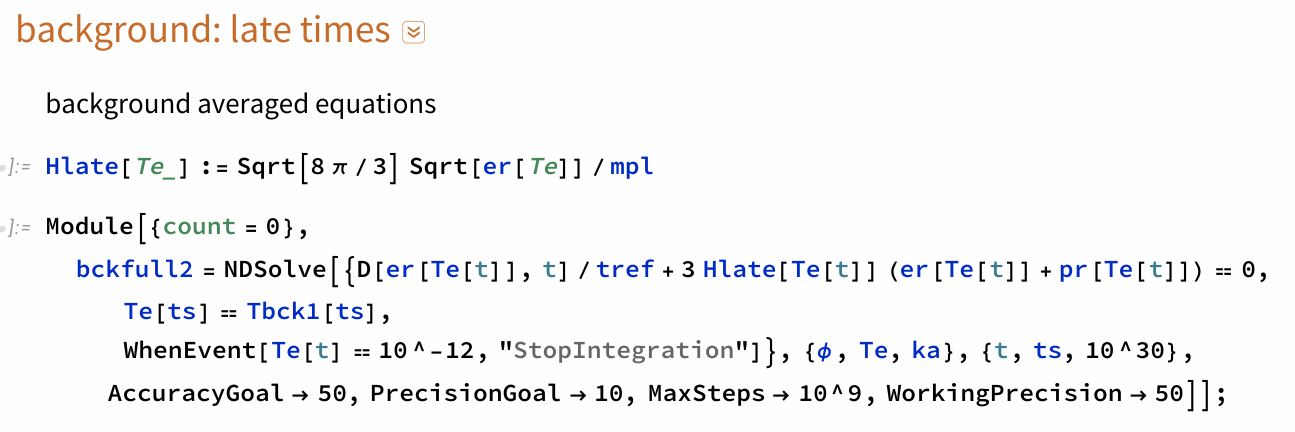}
\end{center}
We stop the integration when the temperature of the universe reaches $10^{-12}\,m_\text{Pl}$. We denote the time at which this happens by \texttt{te}, for ``t end'', and save the obtained temperature as \texttt{$T\text{bck}2[t]$}. 
\begin{center}
    \includegraphics[width=1\linewidth]{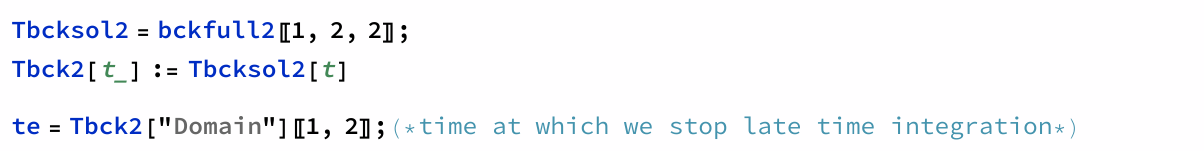}
\end{center}
We continue by constructing one continuous function for the background temperature and the inflaton field.
\begin{center}
    \includegraphics[width=1\linewidth]{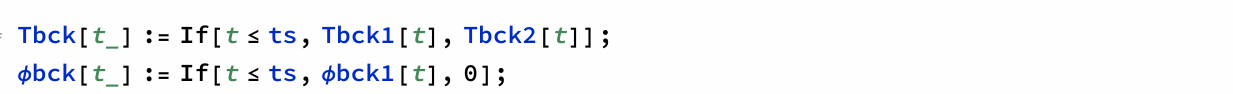}
\end{center}
This allows us to also define other background quantities, which will be used for the determination of the power spectrum evolution. 
\begin{center}
    \includegraphics[width=1\linewidth]{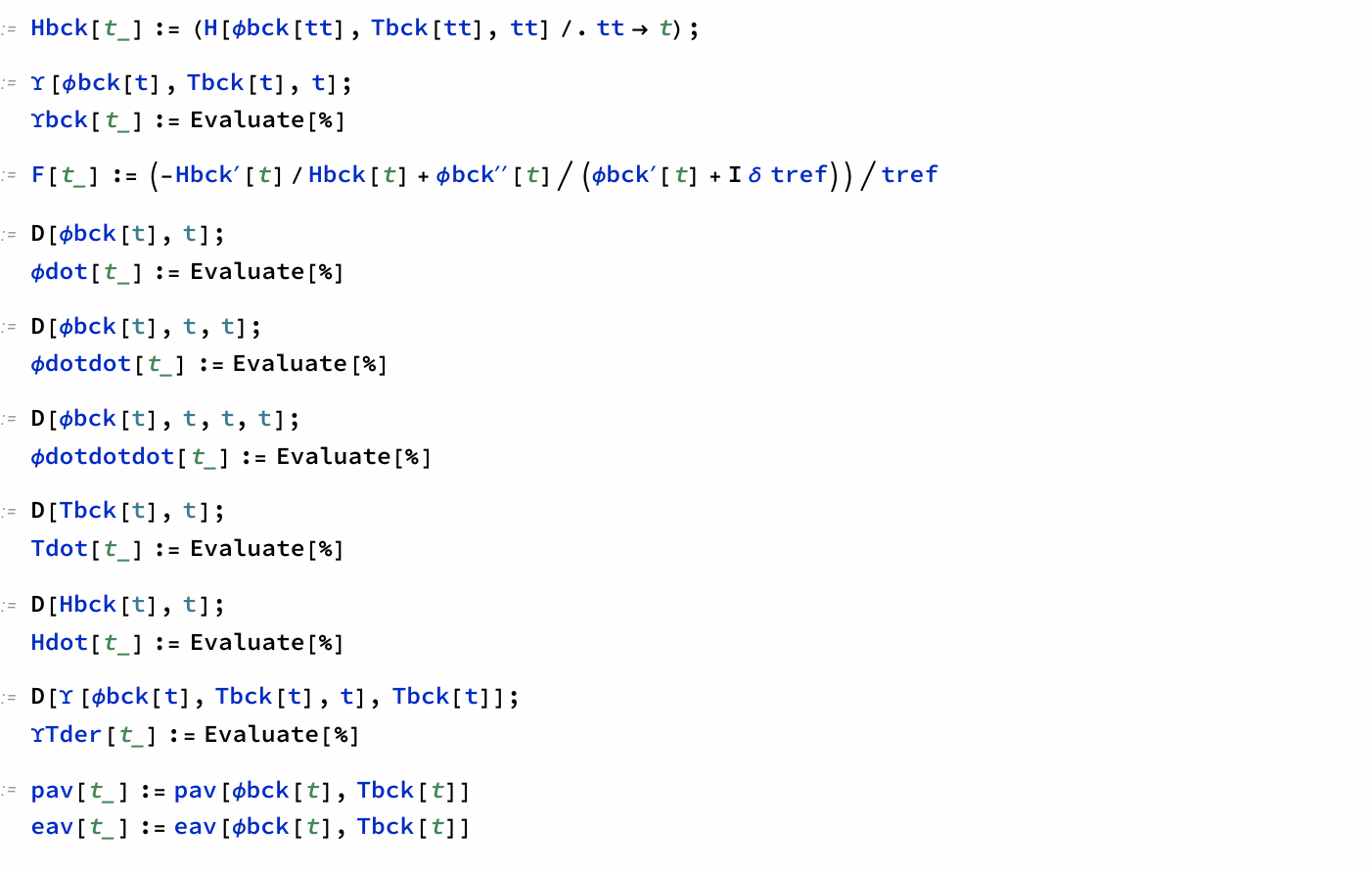}
\end{center}
One more background quantity is needed. While physical momentum, $k/a$, does not appear in the background eqs.~\eqref{physical}, it does enter the coefficients of the perturbation eqs.~\eqref{result}. One can obtain the physical momentum by relating it to the Hubble rate through a differential equation $\frac{d}{dt}\ln{\left(k/a\right)}=- H$. Since $k/a$ becomes very large at early times, we solve for $\texttt{kal}\equiv \ln{\left(k/a\right)}$ and then exponentiate the result.
\begin{center}
    \includegraphics[width=1\linewidth]{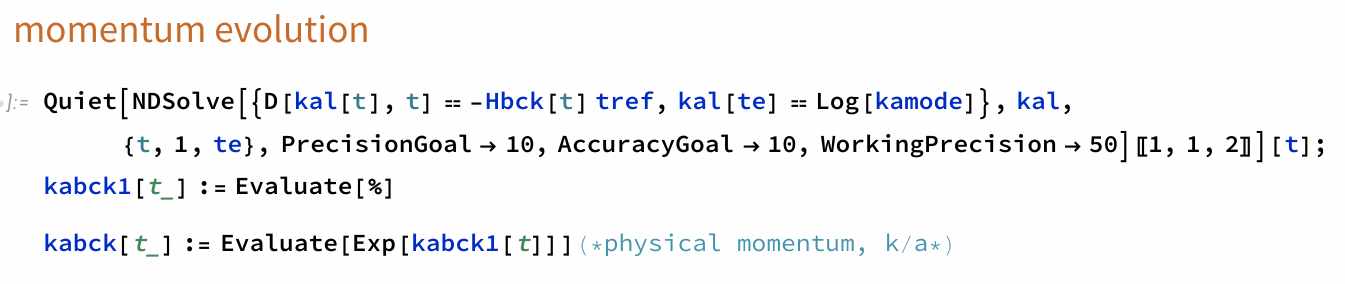}
\end{center}
Having obtained $k/a$, we can use it to define the start ($\texttt{ti}$, ``initial time'') and end point ($\texttt{tf}$, ``final time'') of the integration of curvature perturbations, as well as the point of horizon crossing ($\texttt{tmatch}$), through the ratio $k/(aH)$. This can be done with Mathematica's \texttt{FindRoot} function. Unfortunately, \texttt{FindRoot} is very slow unless we provide it with a narrow interval in which $k/(aH)$ is expected to obtain the desired value. Hence, we begin by defining a function which looks for the interval in which \texttt{funct} $k/(aH)$ obtains a target value.\\
\begin{center}
    \includegraphics[width=1\linewidth]{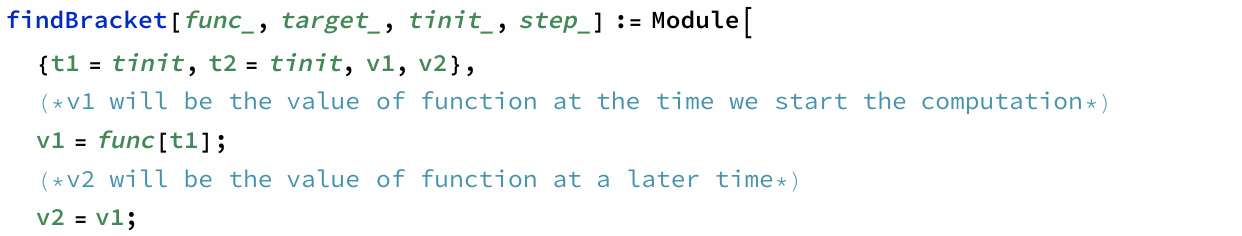}
\end{center}
\begin{center}
    \includegraphics[width=1\linewidth]{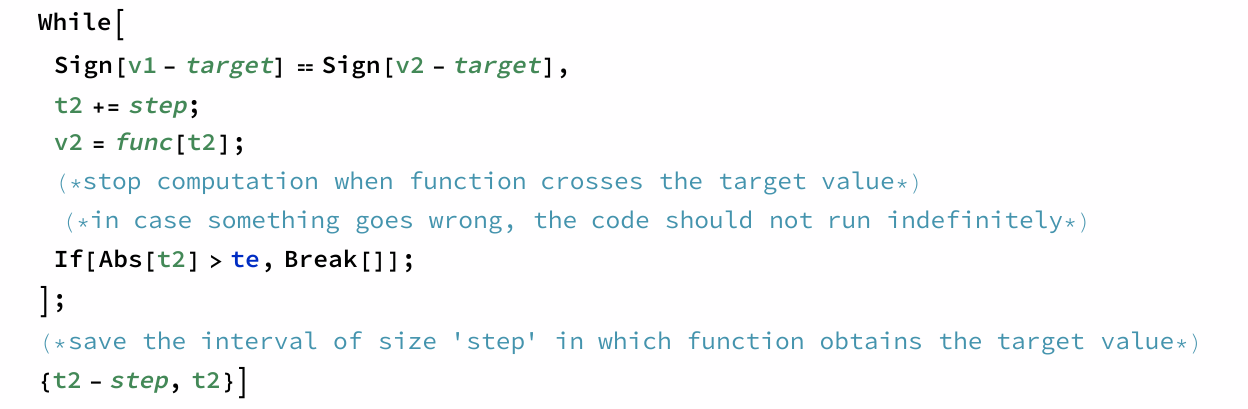}
\end{center}
We proceed to define \texttt{ratio}, $k/(aH)$, which is our function of interest.
\begin{center}
    \includegraphics[width=1\linewidth]{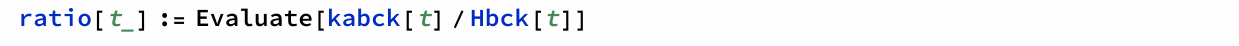}
\end{center}
We can now apply \texttt{findBracket} to the function \texttt{ratio}. The obtained interval serves as an input to \texttt{FindRoot}, which finds the exact values \texttt{ti}, \texttt{tmatch}, and \texttt{tf}.
\begin{center}
    \includegraphics[width=1\linewidth]{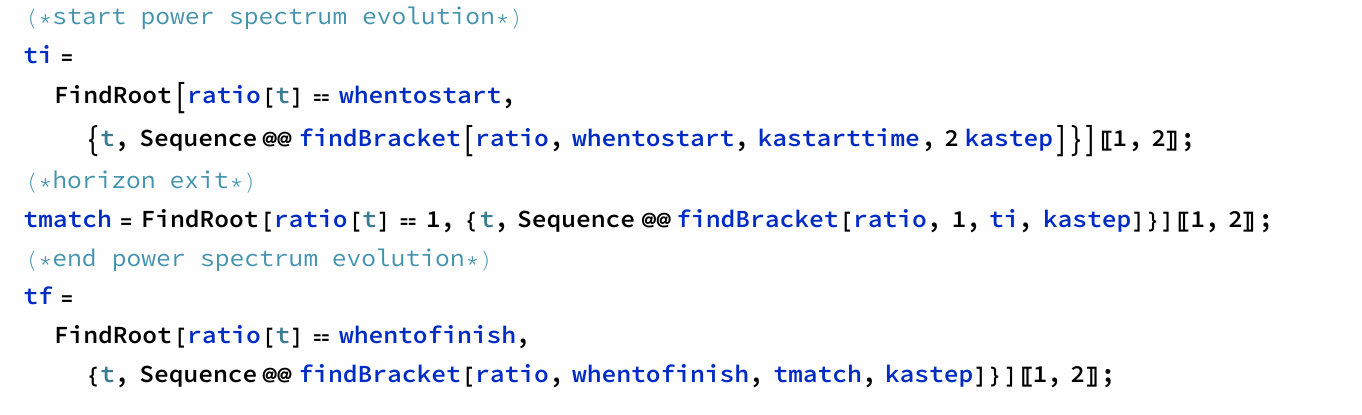}
\end{center}
We start the search for the initial time \texttt{ti} at $t=\texttt{kastarttime}\;(10)$, as earlier the ratio is normally very large and thus more difficult for Mathematica to handle numerically. We choose to conduct the search for the correct time interval in the increments of 2 \texttt{kastep} (100). To find \texttt{tmatch}, we may start at time \texttt{ti}, but we now proceed the search in the increments of \texttt{kastep} (50), as the horizon crossing will normally be relatively close to \texttt{ti}. A similar logic applies to our search for \texttt{tf}, where we start the search from \texttt{tmatch}.
\\\\
Having obtained all the background quantities, we can begin to numerically implement eqs.~\eqref{result} according to the matrix approach, eq.~\eqref{matrixeq}. We begin by defining the matrix $\mathcal{M}$ (\texttt{Matr[t]} in the code) with all of its coefficients.
\begin{center}
    \includegraphics[width=1\linewidth]{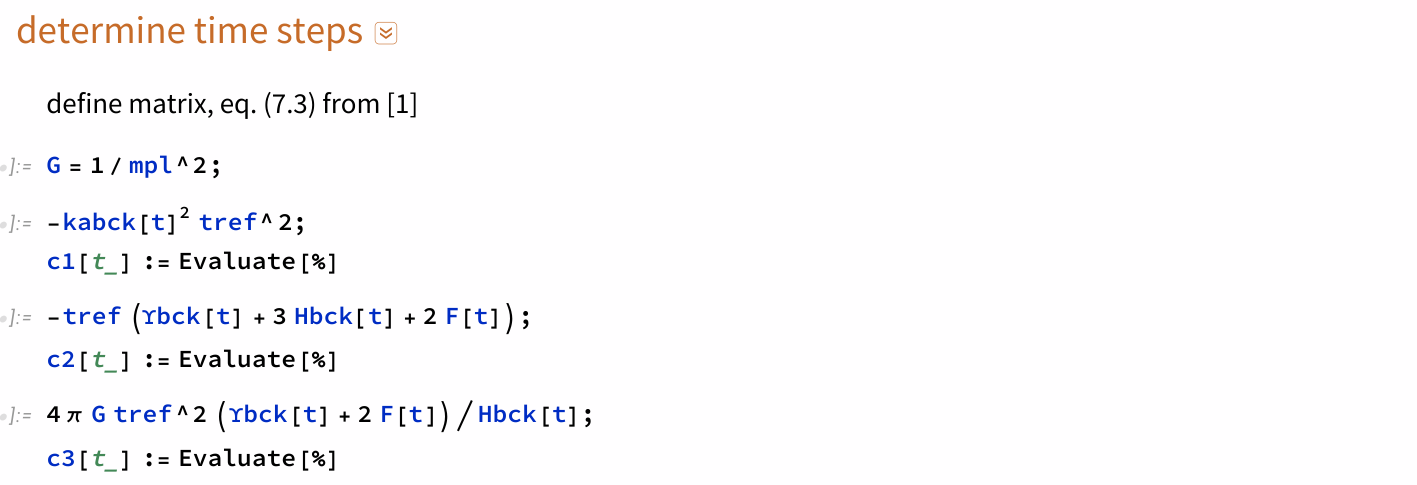}
\end{center}
\begin{center}
    \includegraphics[width=1\linewidth]{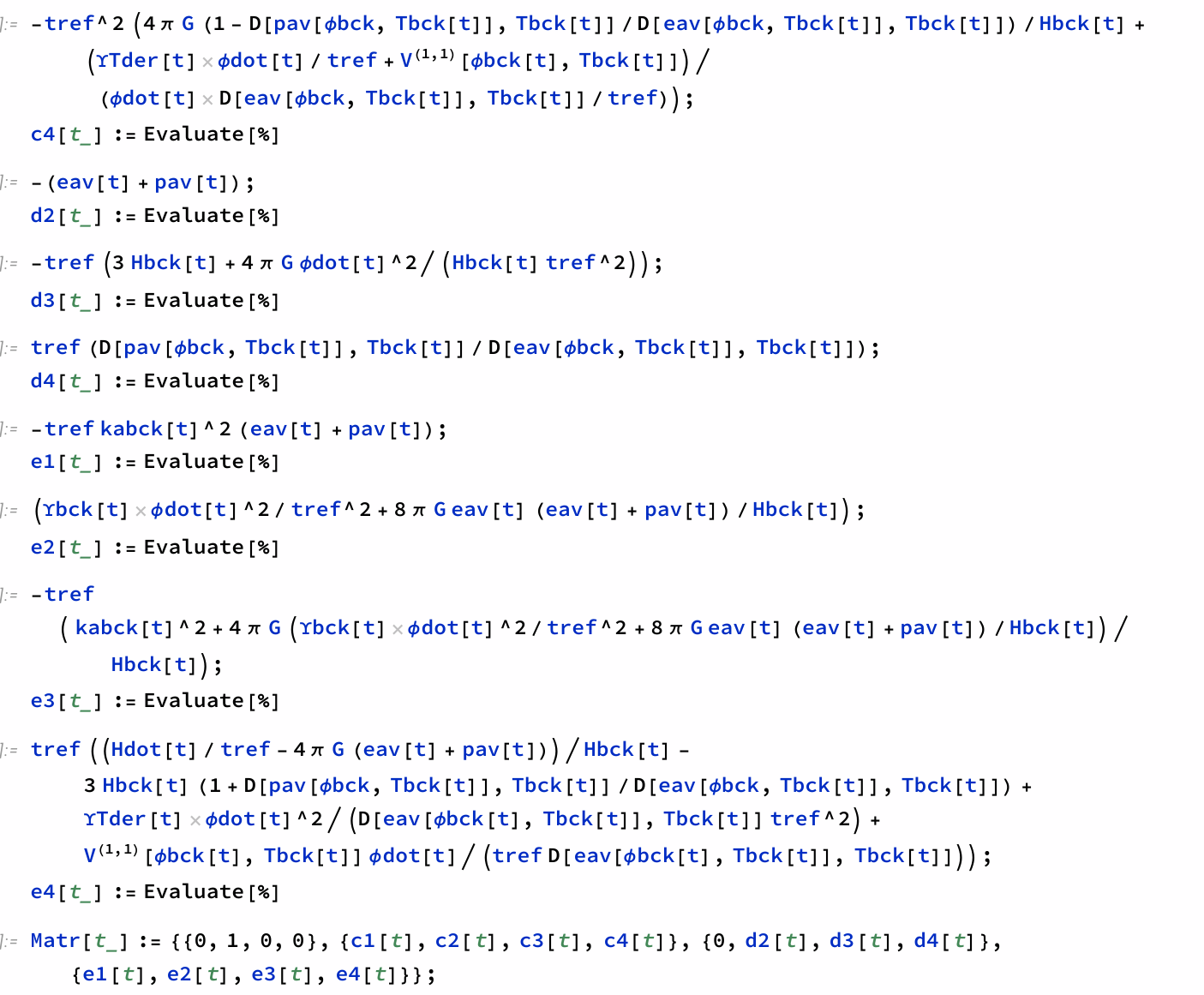}
\end{center}
We use this to compute the step sizes, see eq.~\eqref{tolerance}. The code uses the notation $\varepsilon[$\texttt{t,n}$]$, where \texttt{t} has the role of $i$ in eq.~\eqref{tolerance}, while \texttt{n} is related to the tolerance via $\texttt{tol}=10^{-n}$.
\begin{center}
    \includegraphics[width=1\linewidth]{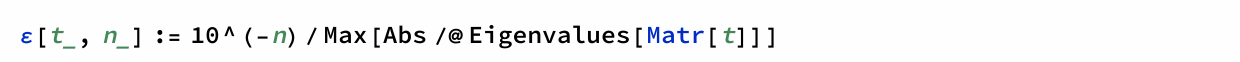}
\end{center}
Directly evaluating $\varepsilon[$\texttt{t,n}$]$ is numerically expensive. Instead, we find the sizes of time steps in the intervals of \texttt{interp} (recommended value 30), and then interpolate the result. We use $\ln{\varepsilon[\texttt{t,n}]}$ as the step sizes are very small at first; additionally, taking a logarithm ensures that $\varepsilon[$\texttt{t,n}$]$ is interpolated to positive values. The final times and time steps are given by \texttt{times} and \texttt{dts} respectively.
\begin{center}
    \includegraphics[width=1\linewidth]{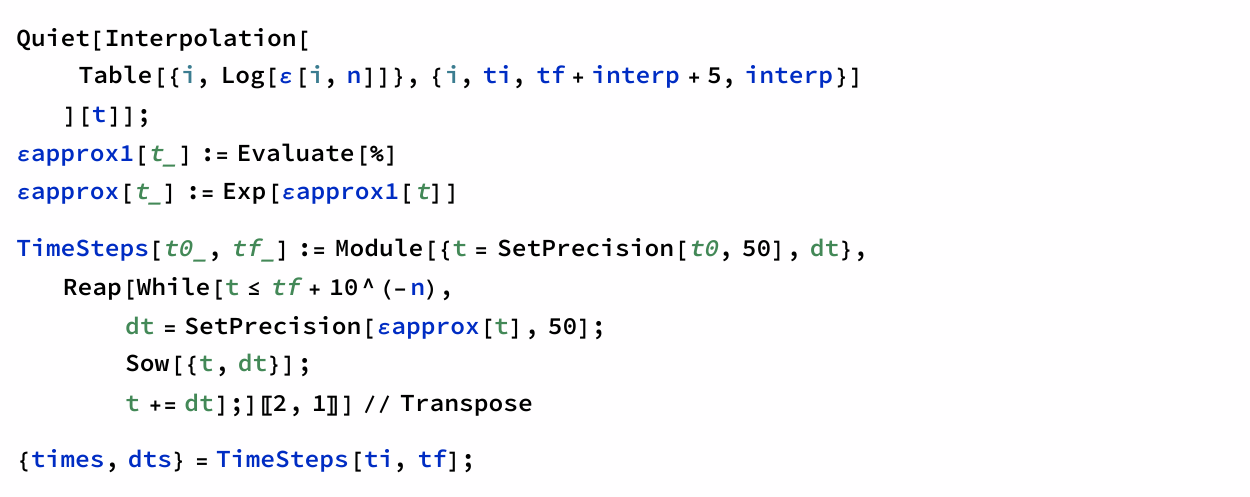}
\end{center}
Having obtained the step sizes, we continue with tabulating the basic quantities, such as $\bar{\varphi}$, $H$, $T$, and their derivatives. When a quantity's name ends with \texttt{dots}, it means derivative with respect to $t/t_\text{ref}$, while \texttt{derT} means derivative with respect to $T$.\\
\begin{center}
    \includegraphics[width=1\linewidth]{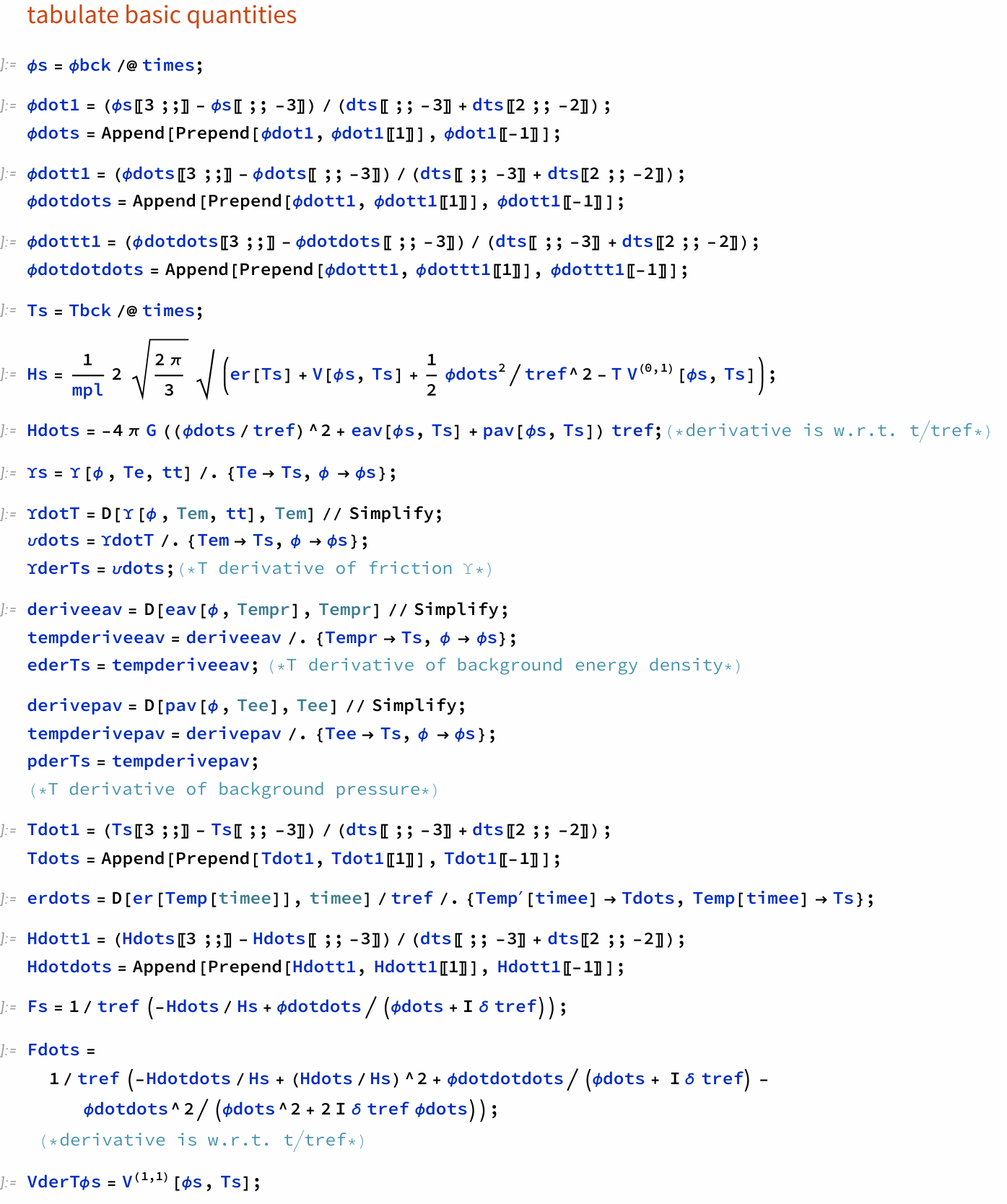}
\end{center}
We proceed to tabulate momentum, as well as the matrix elements \eqref{matrixM}, and the noise autocorrelator \eqref{omega}. As we plan to compute the spectral tilt, it is necessary to compute power spectra not just for the desired momentum mode, but also for a mode just slightly higher and just slightly lower. Hence, the quantities that contain $k/a$ are evaluated three times; the additional two times are in blue cells, and have suffixes -\texttt{p} (higher momentum) and -\texttt{m} (lower momentum).
\begin{center}
\includegraphics[width=1\linewidth]{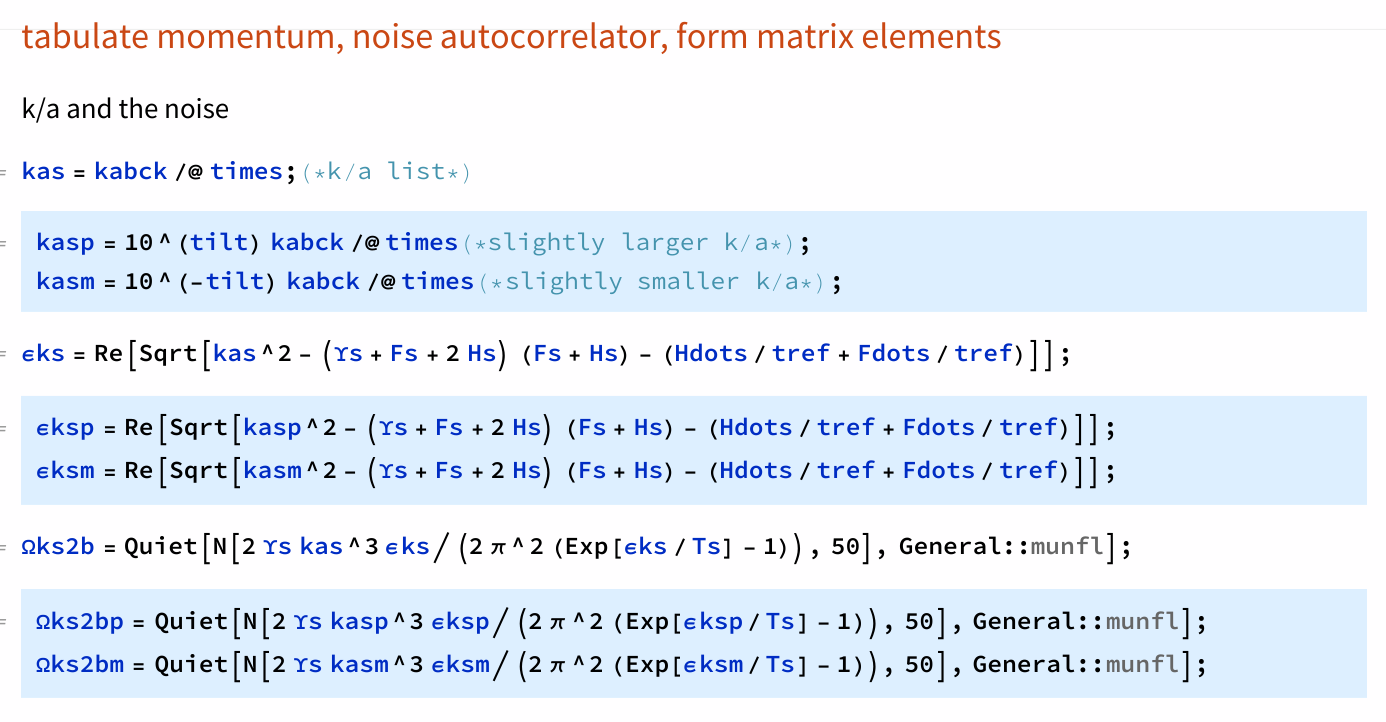}
\end{center}

\begin{center}
\includegraphics[width=1\linewidth]{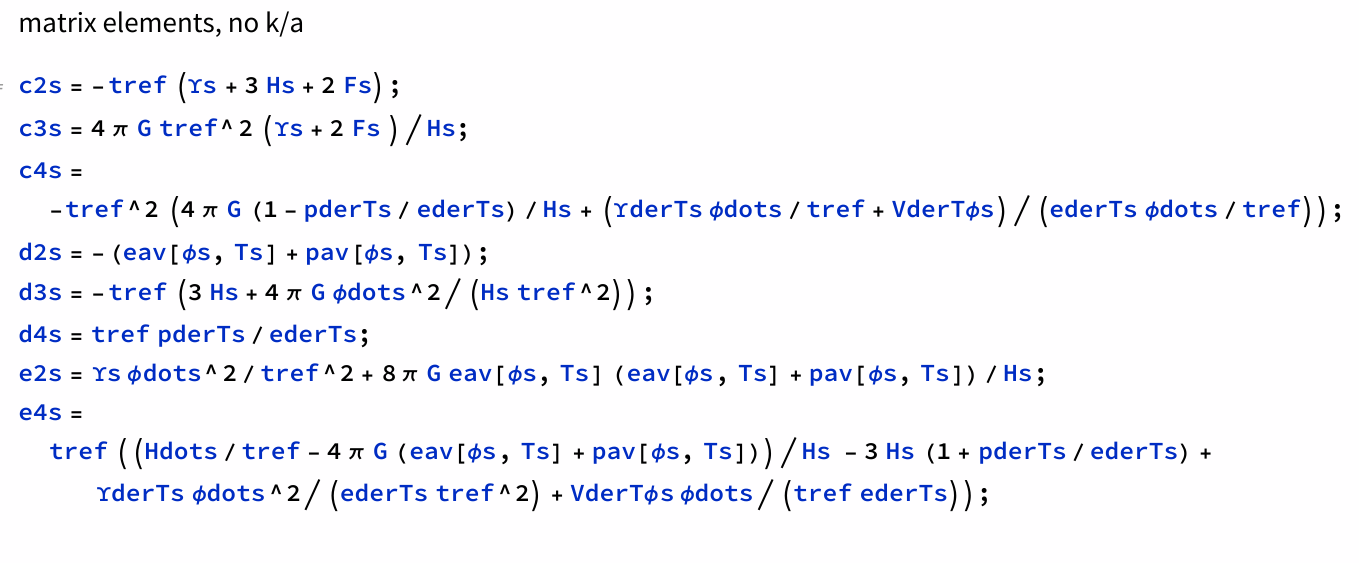}
\end{center}

\begin{center}
\includegraphics[width=1\linewidth]{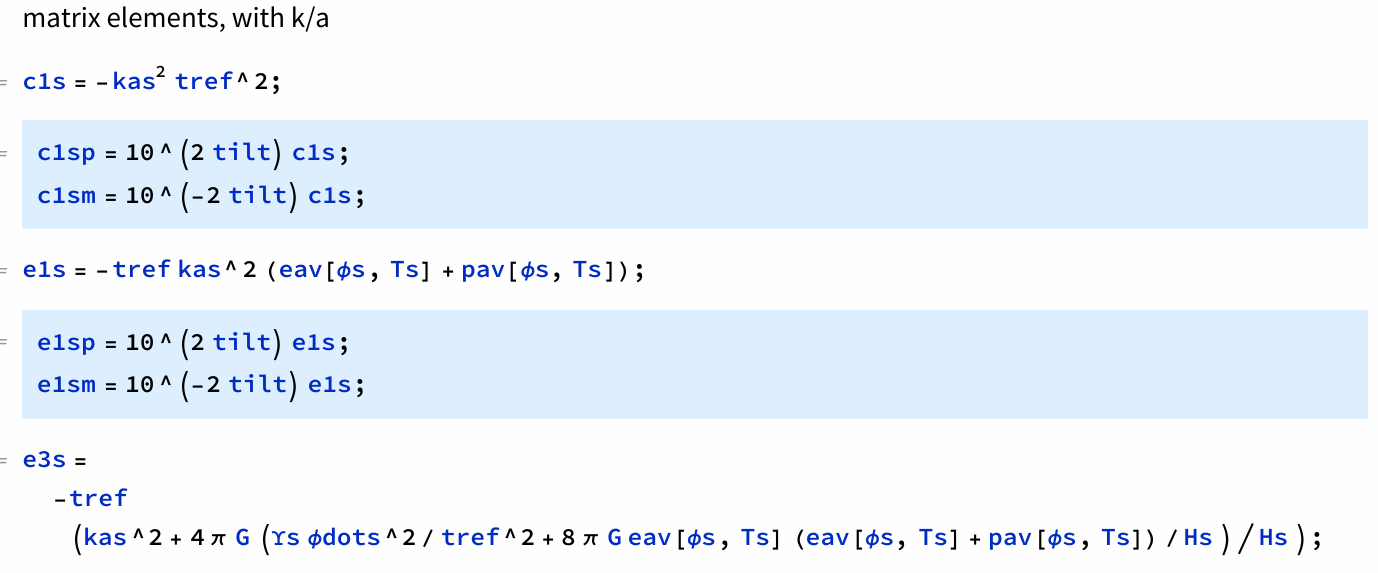}
\end{center}
\begin{center}
\includegraphics[width=1\linewidth]{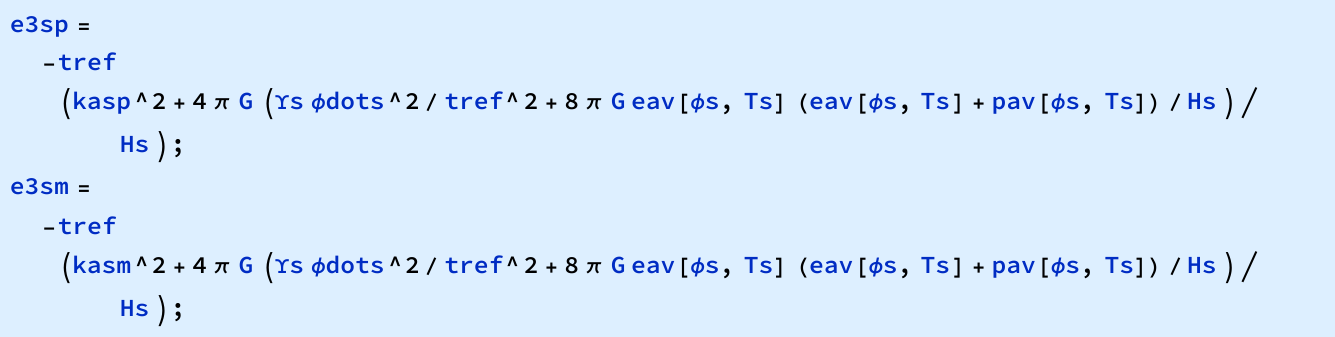}
\end{center}
We move on to compute the power spectrum average, eq.~\eqref{matrixformalismeq}. We start be defining initial conditions according to eq.~\eqref{ourinitialconditions}.
\begin{center}
\includegraphics[width=1\linewidth]{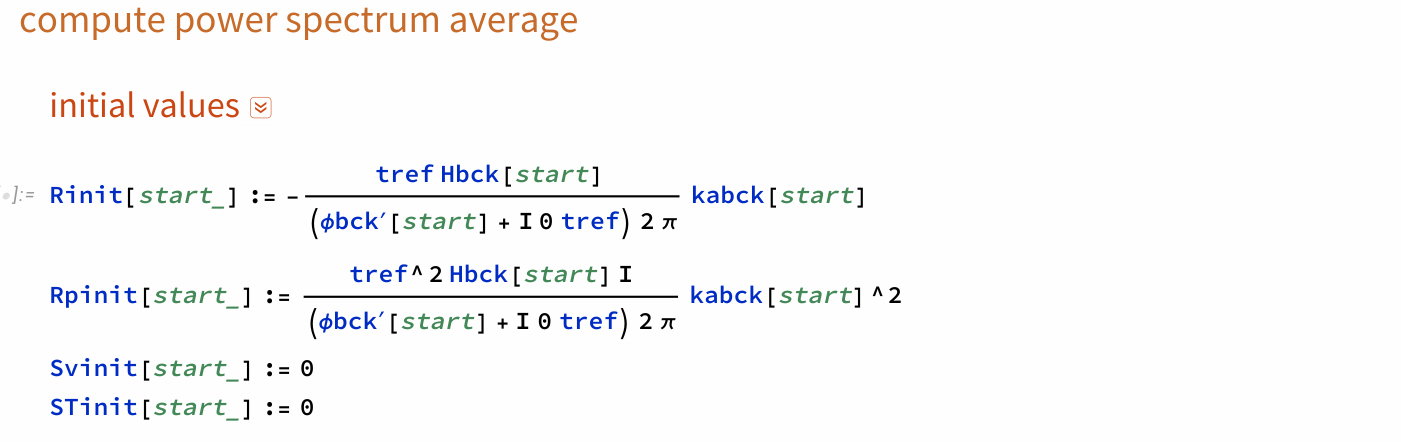}
\end{center}
Then we construct the necessary matrices from eqs.~\eqref{matrixM} and \eqref{noisematrix}.
\begin{center}
\includegraphics[width=1\linewidth]{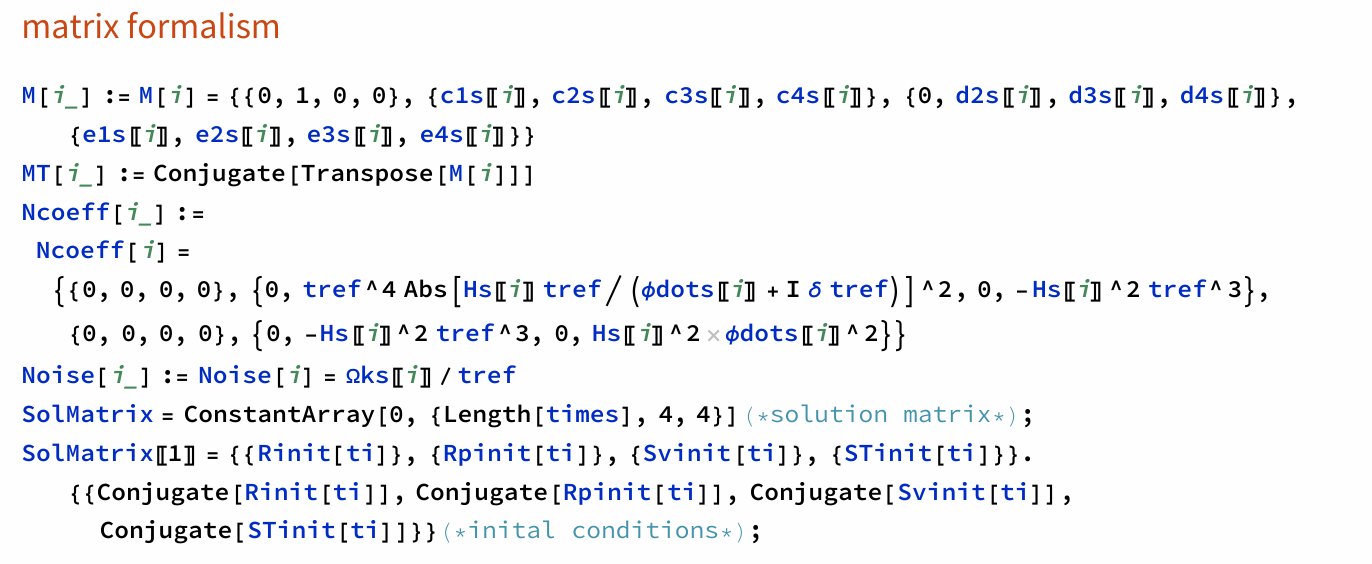}
\end{center}
Subsequently we find the solution matrix.
\begin{center}
\includegraphics[width=1\linewidth]{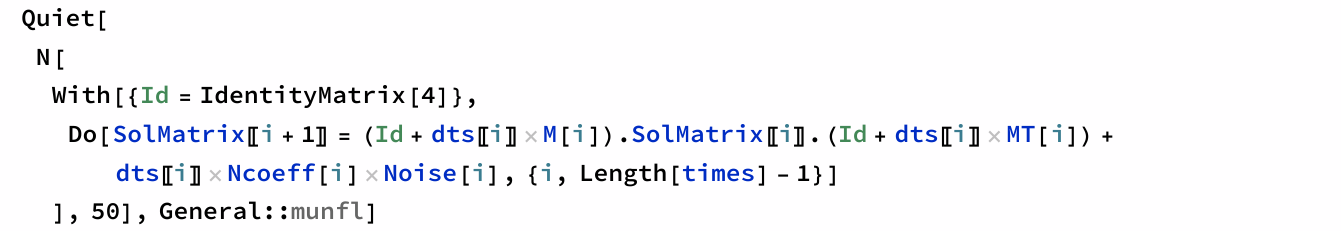}
\end{center}
We later wish to plot the solutions. It is redundant and computationally expensive to plot all time points, so we select a few (\texttt{nPoints}) on a logarithmic scale.
\begin{center}
\includegraphics[width=1\linewidth]{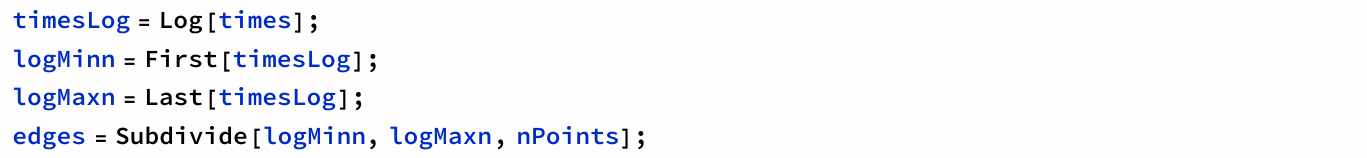}
\end{center}
Our discrete solution may not have been evaluated at exactly those points (\texttt{edges}), so instead we find nearby time points where our solution was actually evaluated. We call this new set of time points to be used for plotting \texttt{inds}.
\begin{center}
\includegraphics[width=1\linewidth]{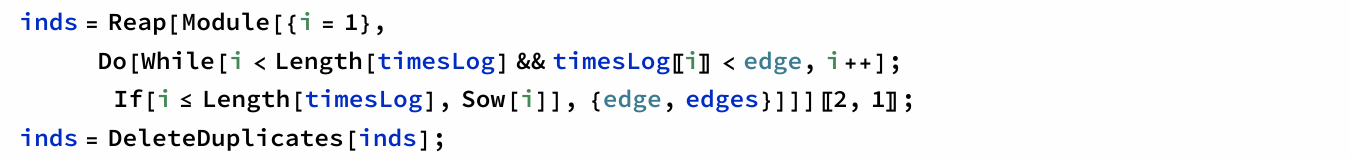}
\end{center}
To switch from the solution matrix to the curvature power spectrum, we need to define background quantities from eqs.~\eqref{isocurvature} at \texttt{inds}.
\begin{center}
\includegraphics[width=1\linewidth]{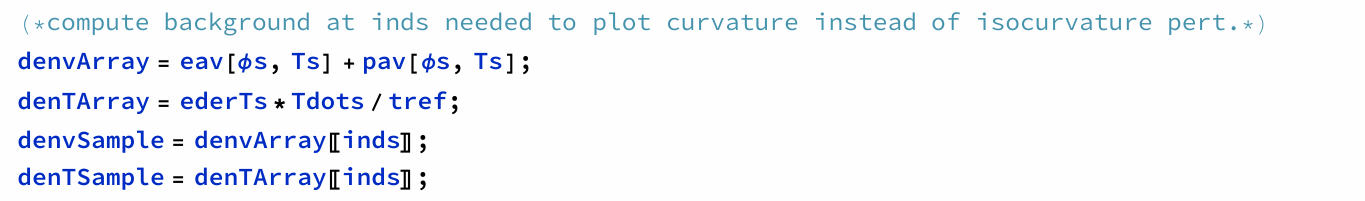}
\end{center}
This allows us to compute the power spectrum, according to eqs.~\eqref{matrixformalismeq}.
\begin{center}
\includegraphics[width=1\linewidth]{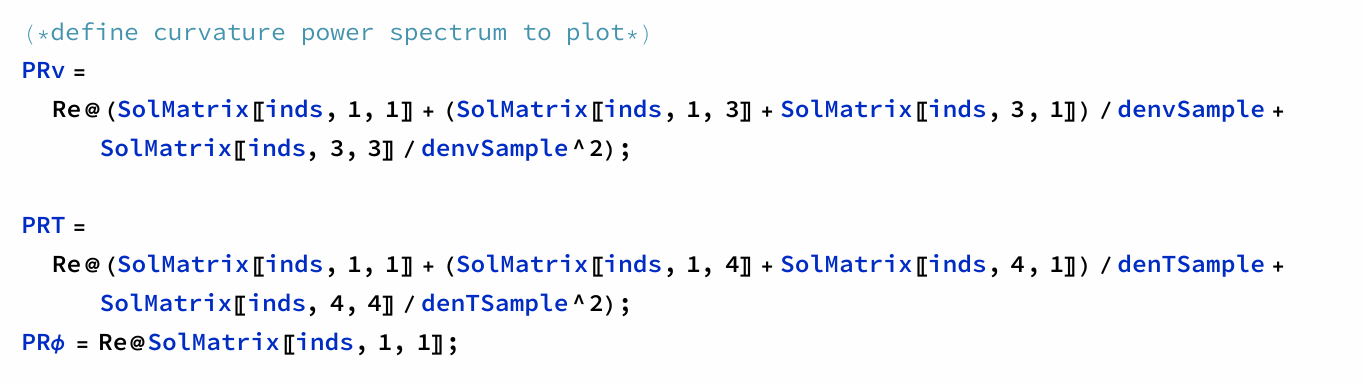}
\end{center}
In order to find the spectral tilt, we need to obtain the power spectrum at a slightly higher and slightly lower momentum. \\
\begin{center}
\includegraphics[width=1\linewidth]{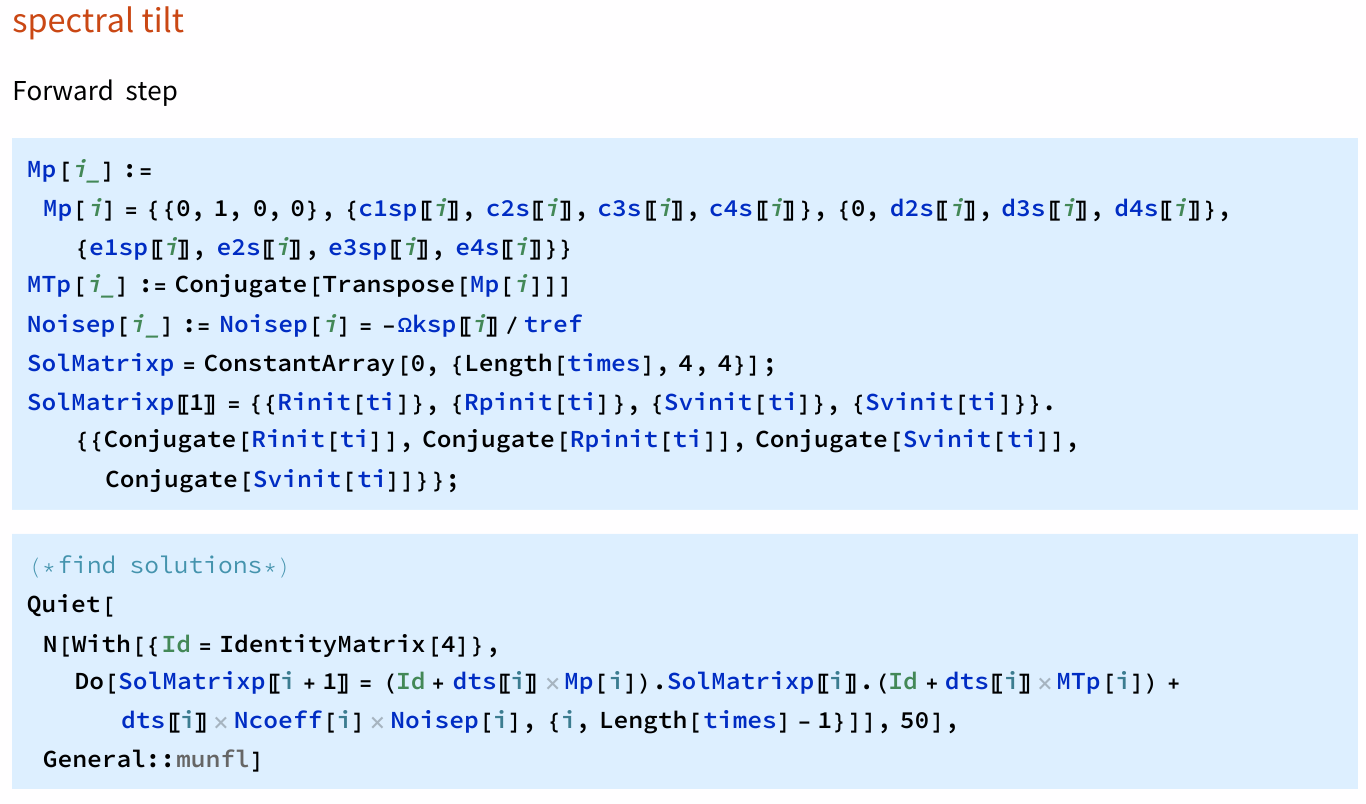}
\end{center}
\begin{center}
\includegraphics[width=1\linewidth]{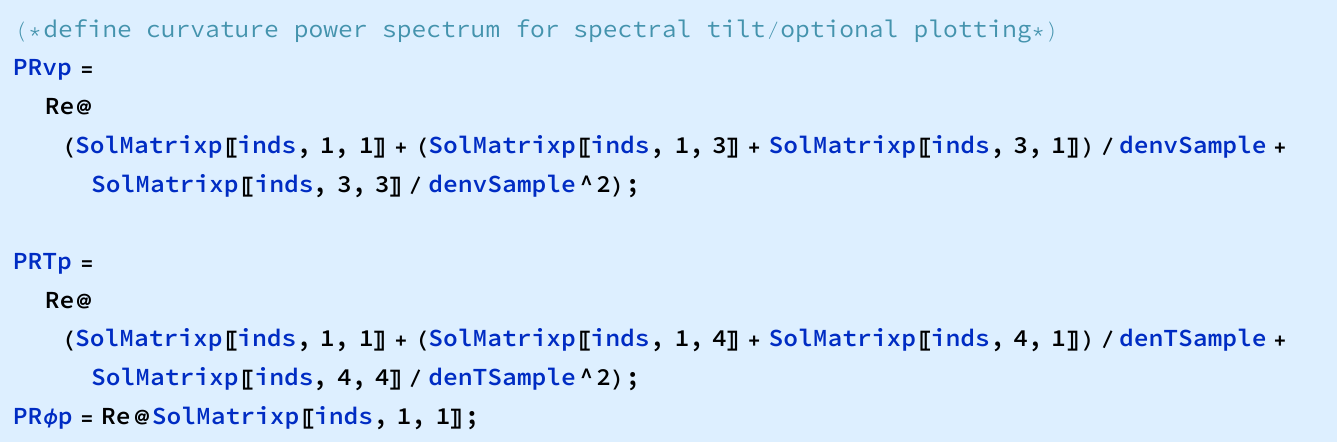}
\end{center}
The ``backward step'', or the power spectrum obtained at slightly lower momentum, is evaluated analogously. Then we can finally obtain the spectral tilt.\vspace{0.2cm}
\begin{center}
\includegraphics[width=1\linewidth]{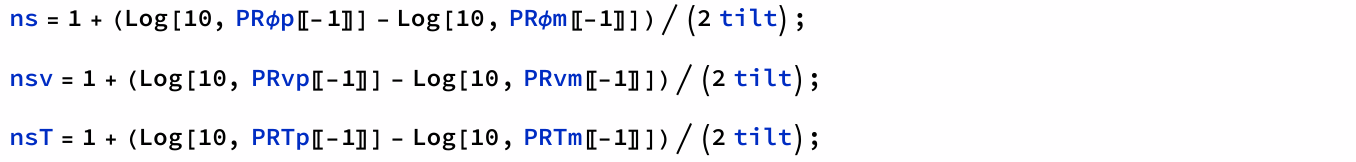}
\end{center}
Lastly, we determine the tensor-to-scalar ratio.
\begin{center}
\includegraphics[width=1\linewidth]{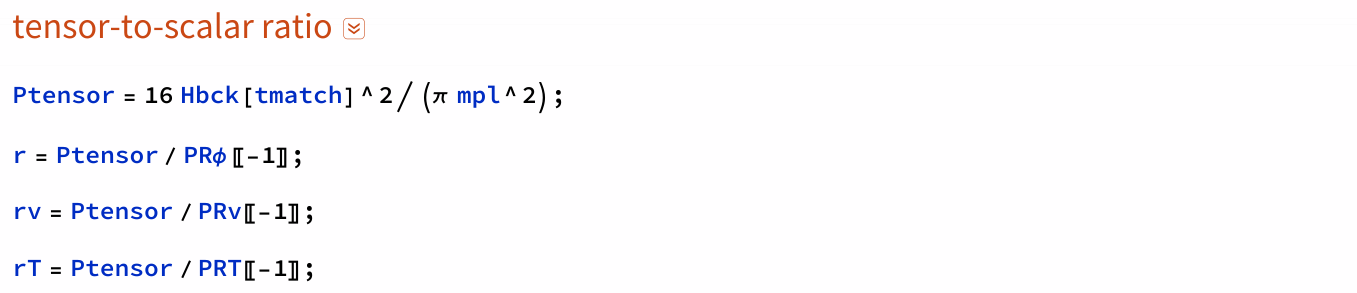}
\end{center}
Having obtained all results, we tabulate the potentially interesting ones. We start by tabulating key freeze-out quantities in a table. Aside from model parameters $\lambda$, $f_a$, $\alpha_s$, we save the initial time \texttt{ti} when we start computing the perturbations, along with $\bar\varphi(\texttt{ti})$. At the time of the horizon exit we save: $T/H$, $\Upsilon/3H$, $\dot{\bar\varphi}/H^2$, $\dot{\bar\varphi}$, and $H$. We additionally save the frozen-out values of the power spectra, spectral tilt, and tensor-to-scalar-ratio. The power spectra should match regardless of whether they describe the perturbations of $\varphi,\;T$ or $v$; if they do not, the value of the ratio $k/(aH)$ should be smaller when we stop evolving the power spectra.
\begin{center}
\includegraphics[width=1\linewidth]{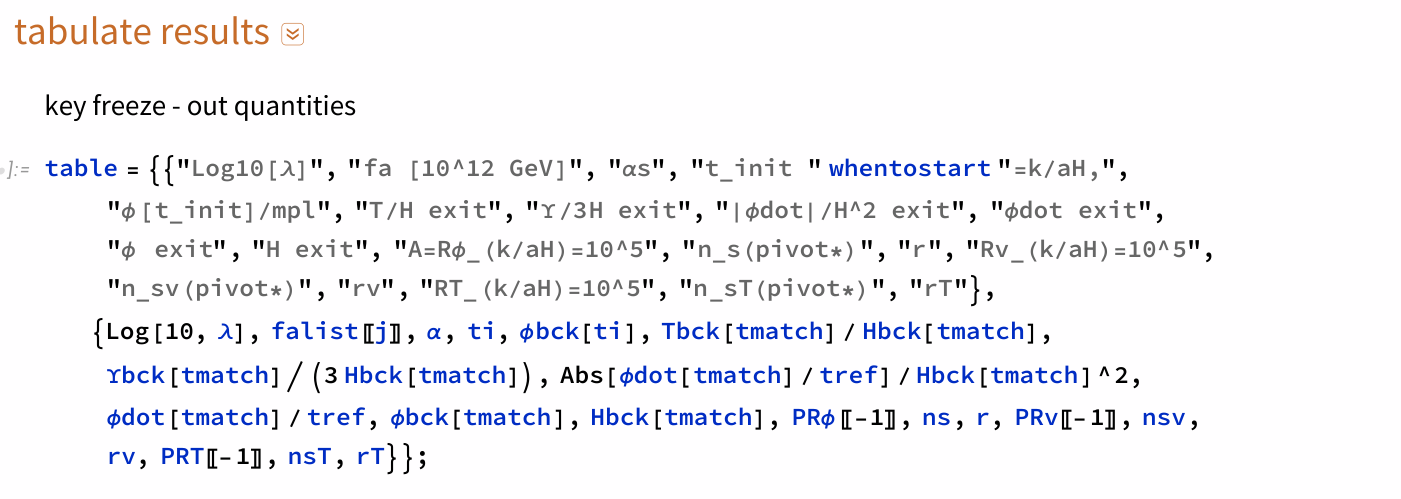}
\end{center}
\vspace{-0.3cm}
For potential future plotting purposes, we also save tabulated data at selected time steps. In case of the background quantities, we tabulate $H$, $T$, $k/a$, $\Upsilon$, $e_r$, $p_r$, $\bar e$, $\bar p$, $V$, $\bar{\varphi}$, $\dot{\bar\varphi}$, $\dot{T}$, and $\dot{H}$. In this way we save background data as \texttt{bckData} table, while the power spectrum evolution is saved as \texttt{powerData} table.
\begin{center}
\includegraphics[width=1\linewidth]{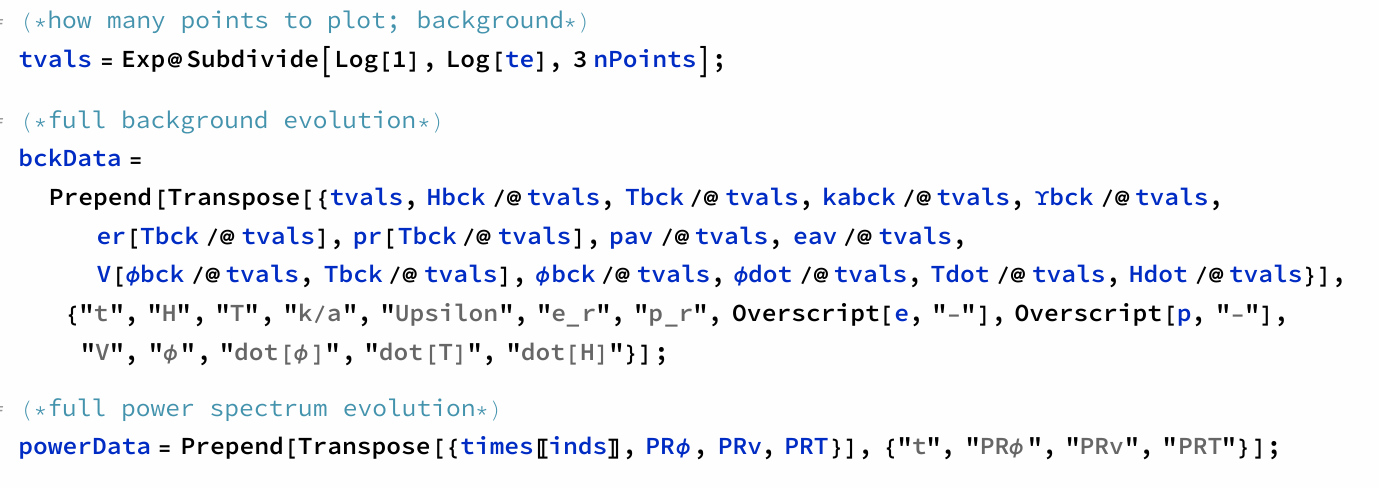}
\end{center}
We can now plot the results. We define multiple background quantities, in case plots of functions other than $H$, $\Upsilon$, $T$, and $k/a$ are desired.
\begin{center}
\includegraphics[width=1\linewidth]{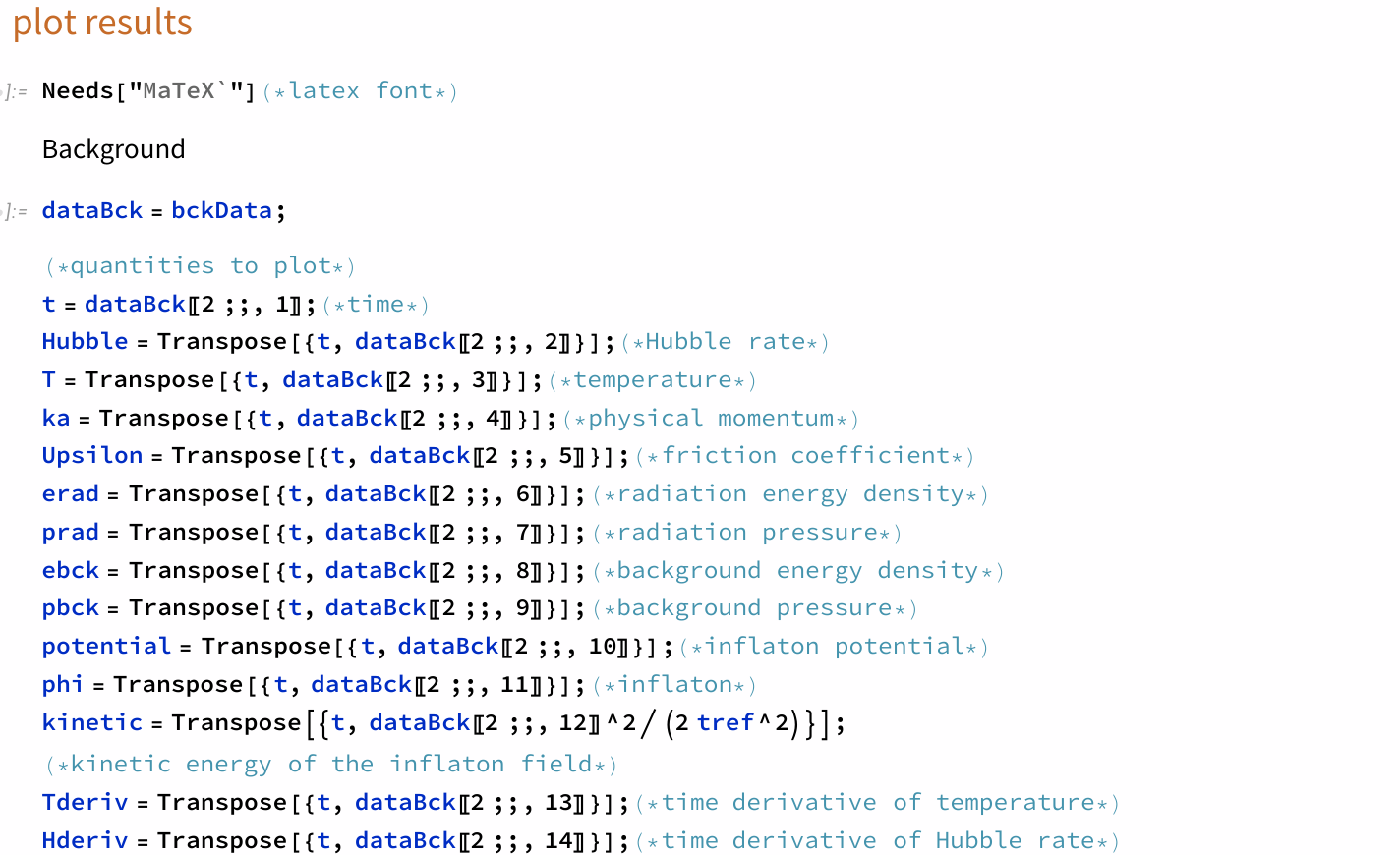}
\end{center}
\vspace{-0.1cm}
We can now plot the selected background data.
\vspace{0.1cm}
\begin{center}
\includegraphics[width=1\linewidth]{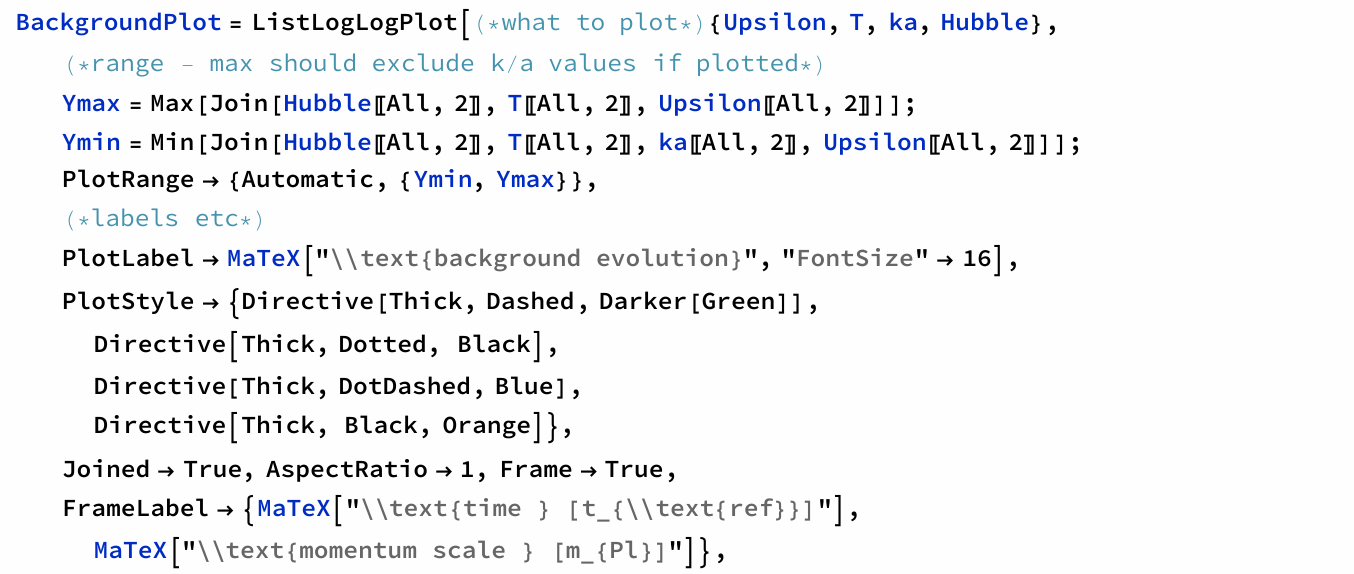}
\end{center}
\begin{center}
\includegraphics[width=1\linewidth]{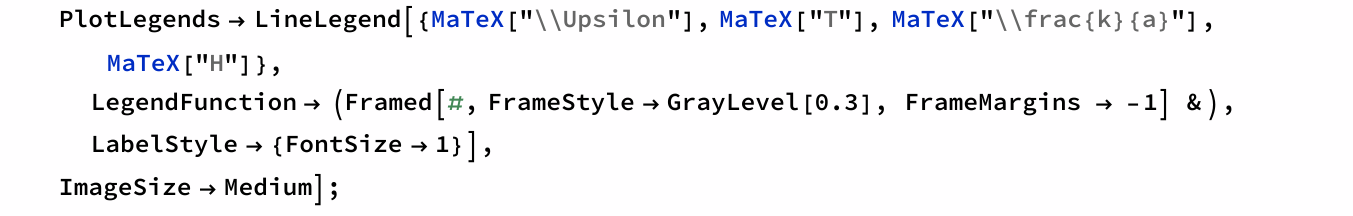}
\end{center}
Similarly, we plot the power spectra.
\vspace{0.2cm}
\begin{center}
\includegraphics[width=1\linewidth]{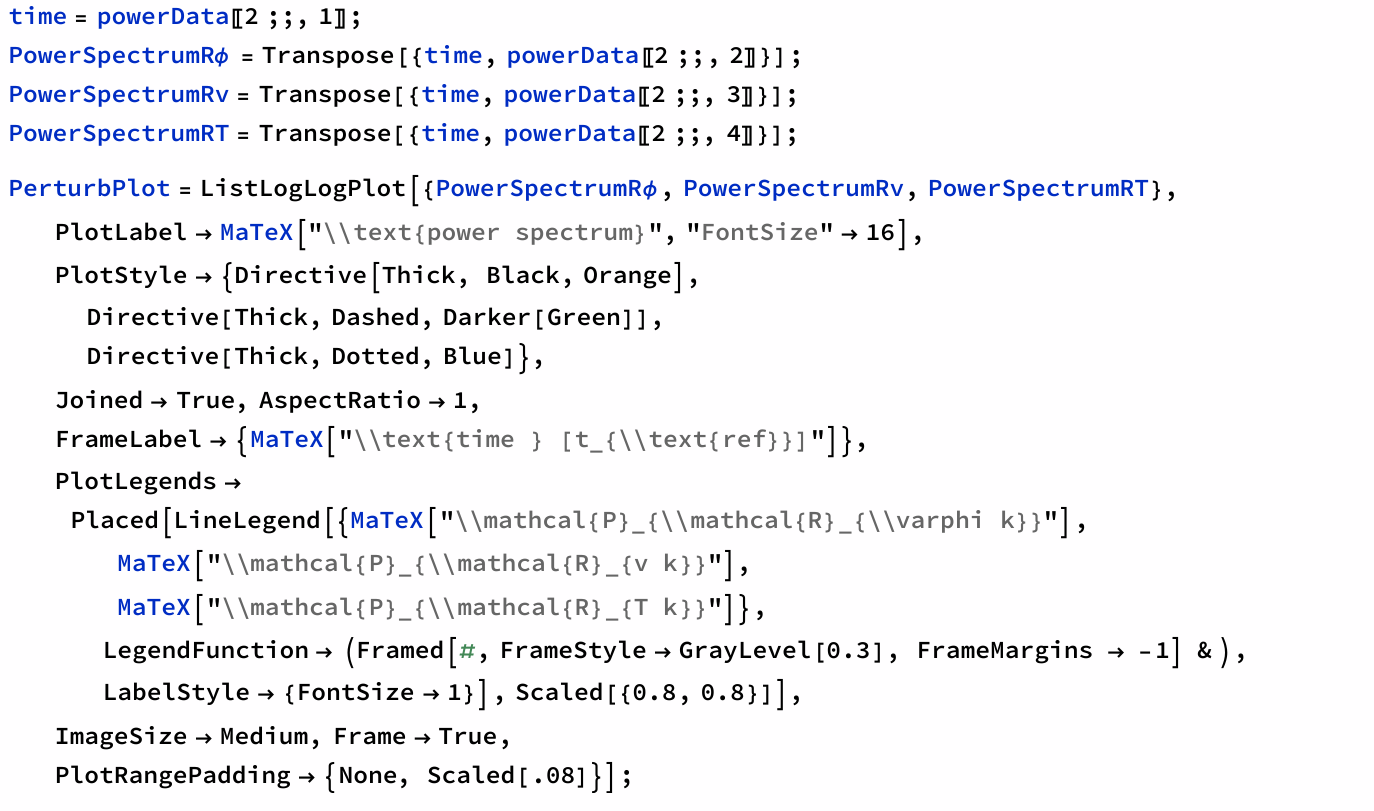}
\end{center}

\section{Results}\label{resultsmathematica}
In this last section, we can see the results: plots and tables.\\
\begin{center}
\includegraphics[width=1\linewidth]{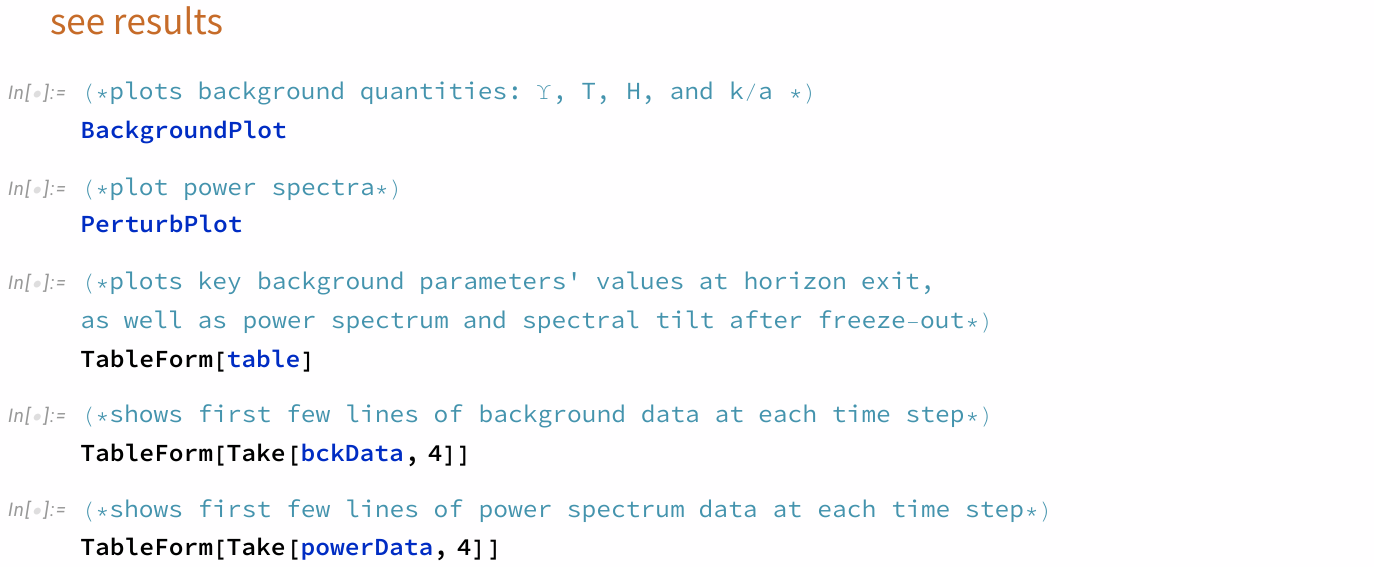}
\end{center}
In addition, we may export the results.
\begin{center}
\includegraphics[width=1\linewidth]{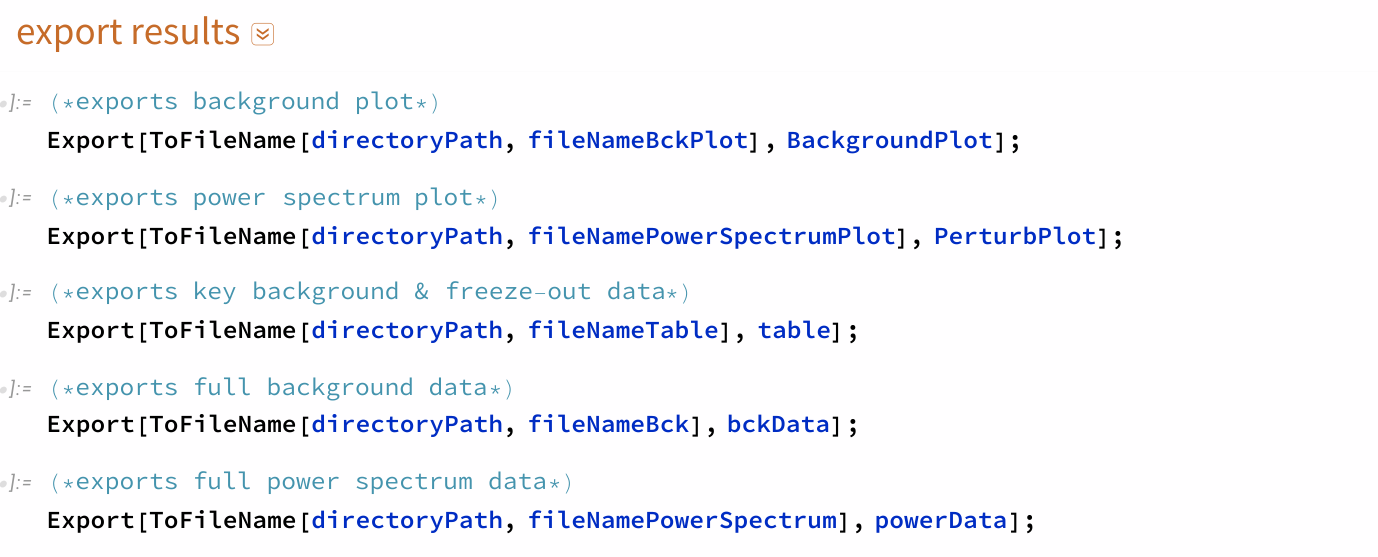}
\end{center}
If we choose to export them, they get saved in the folder specified at the beginning.\\
\begin{center}
\includegraphics[width=1\linewidth]{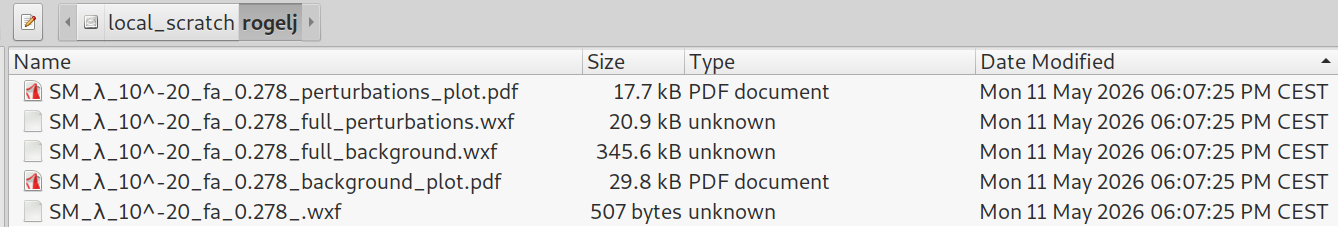}
\end{center}
Five data files are saved, from top to bottom: power spectrum plot, full time evolution of the power spectrum, full background time evolution, background plot, and the selected freeze-out quantities.

\newpage
\SetWatermarkText{}
\chapter{Evolution equations for curvature perturbations}\label{sec:curvature}
Here we explicitly perform the lengthy derivations of the curvature perturbation equations, as summarized in Ch.~\ref{sec:gaugeinveq}.
\section{First equation, $\mathcal{R}_{\varphi}$}
We start by considering eq.~\eqref{LangevinPerturbation}. Aside from $\delta\varphi$, it contains perturbations $h_0,h_D,h$, and $\delta T$. Since our curvature perturbations do not include $h_0$, we seek to eliminate it. We use eq.~\eqref{EFEpertexpanded2} to remove the first $h_0$, while the second appears in the combination $h_0'+3h_D'+\nabla^2 h$. We can remove it using the following equation, which was obtained by subtracting eq.~\eqref{EFEpertexpanded1} from eq.~\eqref{EFEpertexpanded4},
\begin{equation}\begin{split}\label{efeintermid}
     2 \bigl( \HH' + 2 \HH^2 \bigr) h^{ }_0 
 +
 \HH \bigl( h_0' + 3 h_{D}' + \nabla^2 h \bigr)
 +& \bigl(\partial^{2}_\tau + 2 \HH \partial^{ }_\tau - \nabla^2 \bigr)
    \biggl( h^{ }_{D} + \frac{\nabla^2\vartheta}{3} \biggr)\\
  &=  
 4 \pi G a^2_{ } \bigl[\,
 (\bar{p}^{ }_{, T} - \bar{e}^{ }_{,T})\, \delta T 
 + 
 (\bar{p}^{ }_{,\varphi} - \bar{e}^{ }_{,\varphi})\, \delta\varphi
 \,\bigr] \;.
\end{split}
\end{equation}
This introduces a new $h_0$ term, which can be again eliminated by eq.~\eqref{EFEpertexpanded2}, leading to
\begin{equation}\begin{split}\label{im1}
    \delta\varphi''
 + \left(2\mathcal{H} + a \Upsilon\right)  \delta\varphi'
 - \nabla^2 \delta\varphi 
 + \left( a^2 V^{ }_{,\varphi\varphi}  
           + a \bar\varphi' \Upsilon^{ }_{,\varphi}  \right) 
  \, \delta\varphi &\\ \,
 +\frac{ 4\pi G\bar\varphi'}{\mathcal{H}}\, \left[
   2 a^2 V^{ }_{,\varphi} 
 + a \Upsilon \bar\varphi'+\frac{\bar\varphi'}{\HH}\,
 2 \bigl( \HH' + 2 \HH^2 \bigr)  -a^2_{ } 
 (\bar{p}^{ }_{,\varphi} - \bar{e}^{ }_{,\varphi})
 \right]\,\delta\varphi &
 \\
 + 
 \left( a^2 V^{ }_{,\varphi T} + a \bar\varphi' \Upsilon^{ }_{,T}- \frac{\bar\varphi'}{\HH}
 4 \pi G a^2_{ }  (\bar{p}^{ }_{, T} - \bar{e}^{ }_{,T}) \right)
 \, \delta T&\\
 + 
 \left\{ \frac{\bar\varphi'}{\HH}\,
  \bigl(\partial^{2}_\tau - \nabla^2 \bigr)
   - \,
  \left[
   2 a^2 V^{ }_{,\varphi} 
 + a \Upsilon \bar\varphi'+\frac{\bar\varphi'}{\HH}\,
 2 \bigl( \HH' +  \HH^2 \bigr) 
 \right]\,\frac{ \partial_\tau}{\mathcal{H}}\right\} \biggl( h^{ }_{D} + \frac{\nabla^2\vartheta}{3} \biggr)&  \\
 - \,
  \left[
   2 a^2 V^{ }_{,\varphi} 
 + a \Upsilon \bar\varphi'+\frac{\bar\varphi'}{\HH}\,
 2 \bigl( \HH' + 2 \HH^2 \bigr)  
 \right]\,\frac{4\pi G}{\mathcal{H}} a^2(\bar e+\bar p)(h-v)&= 
 a^2_{ }\varrho    \;.
\end{split}
\end{equation}
We now multiply the equation by $\frac{\HH}{\bar\varphi'}$, collect all the perturbations including $-\HH\frac{\delta T}{T'}$, and name the coefficient in front $\mathcal{M}_{1T}$. We then add and subtract $\mathcal{M}_{1T}\HH\frac{\delta\varphi}{\bar\varphi'}$, such that we obtain a gauge-invariant term $\mathcal{M}_{1T}(\mathcal{R}_T-\mathcal{R}_\varphi)$ at the expense of adding several $\delta\varphi$ terms. We do exactly the same with the $\HH(h-v)$ terms, leading to $\mathcal{M}_{1v}(\mathcal{R}_v-\mathcal{R}_\varphi)$ at the expense of additional $\delta\varphi$ terms. Lastly, we collect the $\bigl( h^{ }_{D} + \frac{\nabla^2\vartheta}{3} \bigr)$ terms and construct $\mathcal{M}_{1\varphi}\mathcal{R}_\varphi$ by again adding and subtracting terms proportional to $\delta \varphi$. We then end up with an equation including only gauge-invariant variables and $\delta \varphi$ terms, 
\begin{equation}\label{im2}
    \begin{split}
        a^2\frac{\HH}{\bar\varphi'}\varrho=&-\mathcal{M}_{1\varphi}\mathcal{R}_\varphi-\mathcal{M}_{1v}(\mathcal{R}_v-\mathcal{R}_\varphi)-\mathcal{M}_{1T}(\mathcal{R}_T-\mathcal{R}_\varphi)\\
        &+\frac{\HH}{\bar\varphi'}\left[
    \delta\varphi''
 + \left(2\mathcal{H} + a \Upsilon\right)  \delta\varphi'
 - \nabla^2 \delta\varphi 
 + \left( a^2 V^{ }_{,\varphi\varphi}  
           + a \bar\varphi' \Upsilon^{ }_{,\varphi} \right) 
  \, \delta\varphi\right]  \\
 &-\left[a\Upsilon+2\left(\frac{\bar\varphi''}{\bar\varphi'}-\frac{\HH'}{\HH}\right)\right]\frac{4\pi G}{\bar\varphi'}\left[a^2(\bar e+\bar p)+\bar\varphi'^2 \right]\delta\varphi
 \\
 &-\left\{\partial_\tau^2+\left[2\HH+a\Upsilon+2\left(\frac{\bar\varphi''}{\bar\varphi'}-\frac{\HH'}{\HH}\right)\right]\partial_\tau-\nabla^2 \right\}\left(\frac{\HH}{\bar\varphi'}\delta\varphi\right)
 \\
 &-
 4 \pi G a^2_{ } \,
 (\bar{p}^{ }_{,\varphi} - \bar{e}^{ }_{,\varphi})\, \delta\varphi
 \,  \\
  &-\left[
 \frac{4 \pi G a^2_{ }}{\HH} \,
 (\bar{p}^{ }_{, T} - \bar{e}^{ }_{,T})\, 
 \,  - \frac{1}{\bar\varphi'}
 \left( a^2 V^{ }_{,\varphi T} + a \bar\varphi' \Upsilon^{ }_{,T} \right)\right]T'
 \, \left(\frac{\HH}{\bar\varphi'}\delta \varphi    \right)\;.
    \end{split}
\end{equation}
We used the background Langevin eq.~\eqref{LangevinBackground} in order to simplify the equation above. The coefficients $\mathcal{M}_{1\varphi},\,\mathcal{M}_{1v},\,\mathcal{M}_{1T}$ were defined as follows,
\begin{equation}
    \begin{split}
 \mathcal{M}^{ }_{1\varphi}
 & =  
   \partial_\tau^2 - \nabla^2_{ }
   + 
   \biggl[\, 
    a \Upsilon + 2 \HH + 
    \frac{2\HH}{\bar\varphi'}
    \biggl( \frac{\bar\varphi'}{\HH} \biggr)'_{ }
   \,\biggr]\,\partial^{ }_\tau
  \;, 
  \\
%%%%%
 \mathcal{M}^{ }_{1v}
 & =   -\, 
 \biggl[\, 
   a \Upsilon
   +
   \frac{2 \HH}{\bar\varphi'}
   \biggl( \frac{\bar\varphi'}{\HH} \biggr)'   
 \,\biggr]
    \frac{ 4\pi G a^2_{ }(\bar{e} + \bar{p}) }{\HH }
  \;, 
  \\
%%%%%
 \mathcal{M}^{ }_{1T}
 & =  
 \biggl[\, 
   \frac{4\pi G a^2_{ } ( \bar{e}^{ }_{,T}  - \bar{p}^{ }_{,T} ) }{ \HH}
  +
  \frac{  
    a^2_{ } V^{ }_{,\varphi T}  
  +  a \bar\varphi' \Upsilon^{ }_{,T}  }
  { \bar\varphi' }
 \,\biggr] \, T' 
 \;.
    \end{split}
\end{equation} 
Since all our equations should be gauge-invariant, the coefficients in front of $\delta \varphi$ should vanish. One can check this for completeness. The terms involving the second or first derivatives of $\delta\varphi$ vanish simply by using the chain rule in line 4. On the other hand, it is a bit more involved to show that the terms with no derivatives of $\delta\varphi$ vanish as well. We first notice that eq.~\eqref{im2} contains third time derivatives of $\varphi$, as well as a time derivative of $\Upsilon$ and $V_{,\varphi}$. To eliminate such terms, we add to eq.~\eqref{im2} a time derivative of the background Langevin eq.~\eqref{LangevinBackground},
\begin{equation}\label{timederLangevinbck}
   -\frac{\HH}{\bar\varphi'^2}\left\{ \bar\varphi''' + \bigl[ 2 \HH' + (a\Upsilon)' 
 - (2 \HH + a \Upsilon)^2 \bigr] \bar\varphi' 
 + a^2 \bigl( V^{ }_{,\varphi\varphi}\, \bar\varphi'
            + V^{ }_{,\varphi T}\, T'
            - a \Upsilon V^{ }_{,\varphi} \bigr)\right\} =0\;.
\end{equation}
This helps cancel many terms. We can further eliminate four terms by recognizing a Langevin equation \eqref{LangevinBackground} multiplied by $\frac{a\Upsilon\HH}{\bar\varphi'^2}$. There are still many terms left, including combinations $4 \pi G \bigl[ (\bar\varphi')^2_{ }
+ a^2(\bar{e} + \bar{p})\bigr]$ and $\bar e'-\bar p'$, as well as $\HH^3$- and $\HH''$-like terms. Upon using eq.~\eqref{bck1_a} and eq.~\eqref{bck2_b}, everything but the first line in eq.~\eqref{im2} vanishes,
\begin{equation}\label{e1APP}
     a^2\frac{\HH}{\bar\varphi'}\varrho=-\mathcal{M}_{1\varphi}\mathcal{R}_\varphi-\mathcal{M}_{1v}(\mathcal{R}_v-\mathcal{R}_\varphi)-\mathcal{M}_{1T}(\mathcal{R}_T-\mathcal{R}_\varphi)\;.
\end{equation}

\section{Second equation, $\mathcal{S}_{v}$}
As a starting point for obtaining the second gauge-invariant equation, we consider eq.~\eqref{plasmaconservP1}. As before, we start by eliminating $h_0$ through eq.~\eqref{EFEpertexpanded2}. We multiply the result by $-\frac{\HH}{(\bar e+\bar p)a^2}$ and use eq.~\eqref{bck1_a} to simplify the remaining $-a^2 4\pi G(\bar e+\bar p)(h-v)$ term. We search for the combinations $\mathcal{R}_\varphi$, $\mathcal{R}_v-\mathcal{R}_\varphi$, and $\mathcal{R}_T-\mathcal{R}_\varphi$ by adding and subtracting $\delta \varphi$ terms. We obtain an equation analogous to \eqref{im2},
\begin{equation}\begin{split}
   -\left[\left(3\HH
    + \frac{(\bar e + \bar p)'}{\bar e + \bar p}\right)\frac{\HH}{\bar\varphi'}+\frac{\HH}{(\bar e+\bar p)a^2}\left(\bar\varphi''+2\HH\bar\varphi'-a^2\bar p_{,\varphi}-\frac{\bar p_{,T} T'a^2}{\bar \varphi'}\right)\right]\delta\varphi& \\
    +\mathcal{M}^{ }_{2\varphi}\, \mathcal{R}^{ }_\varphi 
 + \mathcal{M}^{ }_{2v} 
   \bigl(\mathcal{R}^{ }_v - \mathcal{R}^{ }_\varphi\bigr)  
 + \mathcal{M}^{ }_{2T} 
   \bigl(\mathcal{R}^{ }_T - \mathcal{R}^{ }_\varphi\bigr)&=0\;.
\end{split}
\end{equation} 
Upon using the background energy-momentum tensor conservation \eqref{plasmaconservbck}, we quickly end up with
\begin{equation}\begin{split}\label{e2APP}
       0=& \mathcal{M}^{ }_{2\varphi}\, \mathcal{R}^{ }_\varphi 
 + \mathcal{M}^{ }_{2v} 
   \bigl(\mathcal{R}^{ }_v - \mathcal{R}^{ }_\varphi\bigr)  
 + \mathcal{M}^{ }_{2T} 
   \bigl(\mathcal{R}^{ }_T - \mathcal{R}^{ }_\varphi\bigr)\;,
\end{split}
\end{equation}
where the coefficients are given by
\begin{equation}
    \mathcal{M}^{ }_{2\varphi}
  =
 \partial^{ }_\tau
 \;, \quad
%%%%
 \mathcal{M}^{ }_{2v}
 = 
 \partial^{ }_\tau
 + 
    \frac{3 \HH^2_{ }+ 4\pi G (\bar\varphi')^2_{ } }{\HH }
    + \frac{(\bar e + \bar p)'}{\bar e + \bar p}
 \;, \quad
%%%%
 \mathcal{M}^{ }_{2T}
 = 
 - 
 \frac{\bar{p}^{ }_{,T} \, T'}{\bar e + \bar p} 
 \;.
\end{equation}

\section{Third equation, $\mathcal{S}_{T}$}
The third and last curvature perturbation equation is conceptually no more difficult to obtain than the previous two, however, the computation is rather lengthy and error-prone. Hence, we perform the derivation slightly more explicitly. We start with the remaining stress-energy tensor conservation eq.~\eqref{plasmaconservP2}. Firstly, we eliminate the combination $h_0' + 3 h_{D}' + \nabla^2 h $ through eq.~\eqref{efeintermid}, as well as remove the $h_0$ term by employing eq.~\eqref{EFEpertexpanded2}. We express the $\delta e$ and $\delta p$ terms through $\delta T$ and $\delta \varphi$ perturbations. After rewriting $v=(v-h)+h$, and adding and subtracting $h_0'$ in the last line, we end up with
\begin{equation}\label{mmmmm}
\begin{split}
 a^2\bigl[\, 
 (\bar e'_{,T} \delta T+\bar e_{,T} \delta T') + 3\HH \bigl( \bar e_{,T} \delta T + \bar p_{,T} \delta T \bigr)
 \,\bigr]
  - 
 \frac{(\bar\varphi' )^2_{ }}{\HH}
   4 \pi G a^2_{ } \bigl[\,
 (\bar{p}^{ }_{, T} - \bar{e}^{ }_{,T})\, \delta T 
 \,\bigr]&\\
  +\bar\varphi' \, 
 \bigl(\, \delta\varphi'' - \nabla^2 \delta\varphi \,\bigr)
 + 
 \bigl(\, \bar\varphi'' + 4\HH \bar\varphi' \,\bigr)
 \delta\varphi' - 
 \frac{(\bar\varphi' )^2_{ }}{\HH}
   4 \pi G a^2_{ } \bigl[\, (\bar{p}^{ }_{,\varphi} - \bar{e}^{ }_{,\varphi})\, \delta\varphi
 \,\bigr]&
 \\
+
 a^2\bigl[\, 
 (\bar e'_{,\varphi} +\bar e_{,\varphi} \partial_\tau) + 3\HH \bigl( \bar e_{,\varphi} +\bar p_{,\varphi}  \bigr)
 \,\bigr]\delta \varphi+\frac{(\bar\varphi' )^2_{ }}{\HH}\bigl(\,
    \partial^{2}_\tau + 2 \HH \partial^{ }_\tau - \nabla^2 \bigr)
    \biggl( h^{ }_{D} + \frac{\nabla^2\vartheta}{3} \biggr)&
 \\
  + 
 \frac{(\bar\varphi' )^2_{ }}{\HH}
    2 \bigl( \HH' + 2 \HH^2 \bigr) h^{ }_0
 - 2 h^{ }_0\, \bar\varphi' 
 \bigl(\, \bar\varphi'' + 2 \HH \bar\varphi' \,\bigr)-
 a^2
  \bigl( \bar{e} + \bar{p} \bigr)
   \nabla^2 (v-h) &
\\
 -
 a^2
  \bigl( \bar{e} + \bar{p} \bigr)
   \bigl(\,
   \nabla^2 h + 3 h_{D}'+h_0'  
   \,\bigr)
 \,
 +
 a^2 
  \bigl( \bar{e} + \bar{p} \bigr)
   h_0' 
 \; &= \; 
 0 \;.
\end{split}
\end{equation}
To remove $h_0'$ that does not appear in the final combination $h_0' + 3 h_{D}' + \nabla^2 h $, we take a derivative of \eqref{EFEpertexpanded2},
\begin{equation}\begin{split}
    \mathcal{H}h_0'=&- \mathcal{H}'h_0-\left(h_D+\frac{\nabla^2\vartheta}{3}\right)''+4\pi G\left[\bar\varphi''\delta\varphi+\bar\varphi'\delta\varphi'\right]\\&-4\pi G\left[2a^2\HH(\bar e+\bar p)(h-v)+a^2(\bar e'+\bar p')(h-v)+a^2(\bar e+\bar p)(h-v)'\right]\;.\end{split}
\end{equation}
We substitute the expression above into eq.~\eqref{mmmmm}, giving
\begin{equation}
\begin{split}
    \bar\varphi' \, 
 \bigl(\, \delta\varphi'' - \nabla^2 \delta\varphi \,\bigr)
 + 
 \bigl(\, \bar\varphi'' + 4\HH \bar\varphi' \,\bigr)
 \delta\varphi' +a^2 
  \bigl( \bar{e} + \bar{p} \bigr)
   \frac{4\pi G}{\HH}\left(\bar\varphi''\delta\varphi+\bar\varphi'\delta\varphi'\right)&
 \\
   +
 a^2\bigl[\, 
 (\bar e'_{,\varphi} \delta \varphi+\bar e_{,\varphi} \delta \varphi') + 3\HH \bigl( \bar e_{,\varphi} \delta \varphi +\bar p_{,\varphi} \delta \varphi \bigr)
 \,\bigr]- 
 \frac{(\bar\varphi' )^2_{ }}{\HH}
   4 \pi G a^2_{ } \bigl[\,
 (\bar{p}^{ }_{,\varphi} - \bar{e}^{ }_{,\varphi})\, \delta\varphi
 \,\bigr]&
 \\
   +a^2\left[\,
 (\bar e'_{,T} +\bar e_{,T} \partial_\tau) + 3\HH \bigl( \bar e_{,T}  + \bar p_{,T}  \bigr)
- 
 \frac{(\bar\varphi' )^2_{ }}{\HH}
   4 \pi G 
 (\bar{p}^{ }_{, T} - \bar{e}^{ }_{,T})\,\,\right] \delta T&
 \\
 +
 \left[-a^2 
  \bigl( \bar{e} + \bar{p} \bigr)
   \frac{1}{\HH}\partial_\tau^2+\frac{(\bar\varphi' )^2_{ }}{\HH}\bigl(\,
    \partial^{2}_\tau + 2 \HH \partial^{ }_\tau - \nabla^2 \bigr)\right]
    \biggl( h^{ }_{D} + \frac{\nabla^2\vartheta}{3} \biggr)&
   \\
   +
 a^2 
  \bigl( \bar{e} + \bar{p} \bigr)\left\{a^2
   \frac{4\pi G}{\HH}\left[-2\HH(\bar e+\bar p)-(\bar e'+\bar p')-(\bar e+\bar p)\partial_\tau\right]+
   \nabla^2  \right\}(h-v)&\\
  + 
\left[ \frac{(\bar\varphi' )^2_{ }}{\HH^2}
    2 \bigl( \HH' + 2 \HH^2 \bigr)-
 a^2 
  \bigl( \bar{e} + \bar{p} \bigr)
   \frac{\HH'}{\HH^2}- 2 \frac{1}{\HH}\, \bar\varphi' 
 \bigl(\, \bar\varphi'' + 2 \HH \bar\varphi' \,\bigr)\right]\mathcal{H}h_0&
\\
 -
 a^2
  \bigl( \bar{e} + \bar{p} \bigr)
   \bigl(\,
   \nabla^2 h + 3 h_{D}'+h_0'  
   \,\bigr)
 \,
 \;& = \; 
 0 \;.
\end{split}
\end{equation}
Once again we employ equations \eqref{efeintermid} and \eqref{EFEpertexpanded2} on the remaining $h_0' + 3 h_{D}' + \nabla^2 h $ and $h_0$,
\begin{equation}
\begin{split}
    \bar\varphi' \, 
 \bigl(\, \delta\varphi'' - \nabla^2 \delta\varphi \,\bigr)
 + 
 \bigl(\, \bar\varphi'' + 4\HH \bar\varphi' \,\bigr)
 \delta\varphi' +a^2 
  \bigl( \bar{e} + \bar{p} \bigr)
   \frac{4\pi G}{\HH}\left(\bar\varphi''\delta\varphi+\bar\varphi'\delta\varphi'\right)&\\ -a^2\frac{1}{\HH}\bigl( \bar{e} + \bar{p} \bigr)
 4 \pi G a^2_{ } \bigl[\, 
 (\bar{p}^{ }_{,\varphi} - \bar{e}^{ }_{,\varphi})\, \delta\varphi
 \,\bigr]+ 4\pi G \bar\varphi'
\left[\frac{a^2}{\HH^2}\bigl( \bar{e} + \bar{p} \bigr)2 \bigl( \HH' + 2 \HH^2 \bigr)\right]\delta\varphi&
 \\
   +
 a^2\bigl[\, 
 (\bar e'_{,\varphi} \delta \varphi+\bar e_{,\varphi} \delta \varphi') + 3\HH \bigl( \bar e_{,\varphi} \delta \varphi +\bar p_{,\varphi} \delta \varphi \bigr)
 \,\bigr]- 
 \frac{(\bar\varphi' )^2_{ }}{\HH}
   4 \pi G a^2_{ } \bigl[\,
 (\bar{p}^{ }_{,\varphi} - \bar{e}^{ }_{,\varphi})\, \delta\varphi
 \,\bigr]&
 \\
  + 4\pi G \bar\varphi'
\left[ \frac{(\bar\varphi' )^2_{ }}{\HH^2}
    2 \bigl( \HH' + 2 \HH^2 \bigr)-
 a^2 
  \bigl( \bar{e} + \bar{p} \bigr)
   \frac{\HH'}{\HH^2}- 2 \frac{1}{\HH}\, \bar\varphi' 
 \bigl(\, \bar\varphi'' + 2 \HH \bar\varphi' \,\bigr)\right]\delta\varphi&
 \\
   +a^2\left\{\,
 (\bar e'_{,T} +\bar e_{,T} \partial_\tau) + 3\HH \bigl( \bar e_{,T}  + \bar p_{,T}  \bigr)
- 
 \frac{4 \pi G }{\HH}
 (\bar{p}^{ }_{, T} - \bar{e}^{ }_{,T})\,\left[(\bar\varphi' )^2_{ } +\bigl( \bar{e} + \bar{p} \bigr)
  a^2_{ } 
 \right]\,\right\} \delta T&
 \\
 +
 \left[-a^2 
  \bigl( \bar{e} + \bar{p} \bigr)
   \frac{1}{\HH}\partial_\tau^2+\frac{(\bar\varphi' )^2_{ }+a^2(\bar e+\bar p)}{\HH}\bigl(\,
    \partial^{2}_\tau + 2 \HH \partial^{ }_\tau - \nabla^2 \bigr)\right]
    \biggl( h^{ }_{D} + \frac{\nabla^2\vartheta}{3} \biggr)&\\
    -
\left[ \frac{(\bar\varphi' )^2_{ }}{\HH^2}
    2 \bigl( \HH' -\frac{\bar{\varphi}''}{\bar{\varphi}'}\HH \bigr)+\frac{a^2}{\HH^2}\bigl( \bar{e} + \bar{p} \bigr) \bigl(\HH' + 4 \HH^2 \bigr)\right]\left(h_D+\frac{\nabla^2\vartheta}{3}\right)'&\\
   +
 a^2 
  \bigl( \bar{e} + \bar{p} \bigr)\left[\nabla^2-a^2
   \frac{4\pi G}{\HH}(\bar e+\bar p)\left(2\HH+\frac{(\bar e'+\bar p')}{(\bar e+\bar p)}+\partial_\tau\right)  \right](h-v)&
 \\
  -4\pi G a^2(\bar e+\bar p)
\left[ \frac{(\bar\varphi' )^2_{ }}{\HH^2}
    2 \bigl( \HH' -\frac{\bar{\varphi}''}{\bar{\varphi}'}\HH \bigr)+a^2\frac{1}{\HH^2}\bigl( \bar{e} + \bar{p} \bigr)\bigl( \HH' + 4 \HH^2 \bigr)\right](h-v)
 \; &= \; 
 0 \;.
\end{split}
\end{equation}
We have now eliminated all gauge-dependent variables that do not appear in the curvature perturbations. Hence, we are in a good position to rewrite our equation in terms of the combinations $\mathcal{R}_\varphi$, $\mathcal{R}_v-\mathcal{R}_\varphi$, and $\mathcal{R}_T-\mathcal{R}_\varphi$, 
\begin{equation}
    \begin{split}
        \left(h_D+\frac{\nabla^2\vartheta}{3}\right)&=-\left(\mathcal{R}_{\varphi}+\frac{\HH}{\bar{\varphi}'}\delta\varphi\right)\;,\\
        (h-v)&=\frac{1}{\HH}\left(\mathcal{R}_{v}-\mathcal{R}_{\varphi}-\frac{\HH}{\bar{\varphi}'}\delta\varphi\right)\;,\\
        \delta T&=-\frac{T'}{\HH}\left(\mathcal{R}_{T}-\mathcal{R}_{\varphi}-\frac{\HH}{\bar{\varphi}'}\delta\varphi\right)\;.
        \end{split}
\end{equation}
This procedure evidently produces additional $\delta\varphi$ terms which are not gauge-invariant. But since the terms proportional to $\delta\varphi$ are the only gauge dependent terms, we will later show they vanish by using various background identities. Making use of the identities above and using the chain rule when necessary, as well as multiplying by $\frac{\HH}{a^2}$, we arrive at
\begin{align}
    \frac{\HH}{a^2}\bar\varphi' \, 
 \bigl(\, \delta\varphi'' - \nabla^2 \delta\varphi \,\bigr)
 + 
 \frac{\HH}{a^2}\bigl(\, \bar\varphi'' + 4\HH \bar\varphi' \,\bigr)
 \delta\varphi' + 
  \bigl( \bar{e} + \bar{p} \bigr)
   4\pi G\left(\bar\varphi''\delta\varphi+\bar\varphi'\delta\varphi'\right)&
  \notag \\ 
    + 4\pi G \bar\varphi'\frac{1}{\HH}\bigl( \bar{e} + \bar{p} \bigr)2 \bigl( \HH' + 2 \HH^2 \bigr)-\bigl( \bar{e} + \bar{p} \bigr)
 4 \pi G a^2_{ } \bigl[\, 
 (\bar{p}^{ }_{,\varphi} - \bar{e}^{ }_{,\varphi})\, \delta\varphi
 \,\bigr]&
\notag  \\
   +\HH 
 \bigl[\, 
 (\bar e'_{,\varphi} \delta \varphi+\bar e_{,\varphi} \delta \varphi') + 3\HH \bigl( \bar e_{,\varphi} \delta \varphi +\bar p_{,\varphi} \delta \varphi \bigr)
 \,\bigr]- 
 (\bar\varphi' )^2_{ }
   4 \pi G  \bigl[\,
 (\bar{p}^{ }_{,\varphi} - \bar{e}^{ }_{,\varphi})\, \delta\varphi
 \,\bigr]&
 \notag \\
  + 4\pi G \bar\varphi'
\left[ \frac{(\bar\varphi' )^2_{ }}{\HH a^2}
    2 \bigl( \HH' + 2 \HH^2 \bigr)-
  \bigl( \bar{e} + \bar{p} \bigr)
   \frac{\HH'}{\HH}- 2 \frac{1}{a^2}\, \bar\varphi' 
 \bigl(\, \bar\varphi'' + 2 \HH \bar\varphi' \,\bigr)\right]\delta\varphi&
 \notag  \\
  -T'\left\{\,
 \bar e'_{,T}  + 3\HH \bigl( \bar e_{,T}  + \bar p_{,T}  \bigr)
- \frac{4 \pi G }{\HH}(\bar{p}^{ }_{, T} - \bar{e}^{ }_{,T})\left[
 (\bar\varphi' )^2_{ }
 \, +\bigl( \bar{e} + \bar{p} \bigr)
  a^2_{ } 
 \right]\,\right\}\left( -\frac{\HH}{\bar\varphi'}\delta\varphi\right)&
 \notag \\
    +
 \bar e_{,T}T'\left(\, -\partial_\tau+\frac{\HH'}{\HH}-\frac{T''}{T'} \right) \left(-\frac{\HH}{\bar\varphi'}\delta\varphi\right)&
 \notag  \\
 +\left[- 
  \bigl( \bar{e} + \bar{p} \bigr)
   \partial_\tau^2+\frac{(\bar\varphi' )^2_{ }+a^2(\bar e+\bar p)}{a^2}\bigl(\,
    \partial^{2}_\tau + 2 \HH \partial^{ }_\tau - \nabla^2 \bigr)\right]\left(-\frac{\HH}{\bar\varphi'}\delta\varphi\right)&
 \notag \\
    +
\left[ \frac{(\bar\varphi' )^2_{ }}{\HH a^2}
    2 \bigl( \HH' -\frac{\bar\varphi''}{\bar\varphi'} \HH \bigr)+\frac{1}{\HH}\bigl( \bar{e} + \bar{p} \bigr) \bigl( \HH' + 4 \HH^2 \bigr)\right]\partial_\tau\left(\frac{\HH}{\bar\varphi'}\delta\varphi\right)&
  \notag  \\
   +\bigl( \bar{e} + \bar{p} \bigr)\left\{
   \nabla^2-a^2
   \frac{4\pi G}{\HH}(\bar e+\bar p)\left[2\HH+\frac{(\bar e'+\bar p')}{(\bar e+\bar p)}+\left(\partial_\tau-\frac{\HH'}{\HH} \right)\right]  \right\}\left(-\frac{\HH}{\bar\varphi'}\delta\varphi\right)&
  \notag  \\
   -4\pi G (\bar e+\bar p)
\left[ \frac{(\bar\varphi' )^2_{ }}{\HH^2}
    2 \bigl( \HH' -\frac{\bar\varphi''}{\bar\varphi'} \HH  \bigr)+a^2\frac{1}{\HH^2}\bigl( \bar{e} + \bar{p} \bigr) \bigl( \HH' + 4 \HH^2 \bigr)\right]\left(-\frac{\HH}{\bar\varphi'}\delta\varphi\right)&
      \\\notag \\
    -T'\left\{\,
 \bar e'_{,T}  + 3\HH \bigl( \bar e_{,T}  + \bar p_{,T}  \bigr)
- \frac{4 \pi G }{\HH}(\bar{p}^{ }_{, T} - \bar{e}^{ }_{,T})\left[
 (\bar\varphi' )^2_{ }
 \, +\bigl( \bar{e} + \bar{p} \bigr)
  a^2_{ } 
 \right]\,\right\}\left( \mathcal{R}_T-\mathcal{R}_\varphi\right)&
\notag  \\
  +
 \bar e_{,T}T'\left(\, -\partial_\tau+\frac{\HH'}{\HH}-\frac{T''}{T'} \right)\left(\mathcal{R}_T-\mathcal{R}_\varphi\right)&
\notag  \\\notag 
 \\
 -
 \left[- 
  \bigl( \bar{e} + \bar{p} \bigr)
   \partial_\tau^2+\frac{(\bar\varphi' )^2_{ }+a^2(\bar e+\bar p)}{a^2}\bigl(\,
    \partial^{2}_\tau + 2 \HH \partial^{ }_\tau - \nabla^2 \bigr)\right]\mathcal{R}_\varphi&
\notag  \\
    +
\left[ \frac{(\bar\varphi' )^2_{ }}{\HH a^2}
    2 \bigl( \HH' -\frac{\bar\varphi''}{\bar\varphi'} \HH \bigr)+\frac{1}{\HH}\bigl( \bar{e} + \bar{p} \bigr) \bigl( \HH' + 4 \HH^2 \bigr)\right]\partial_\tau\mathcal{R}_\varphi&
 \notag \\\notag \\
    -4\pi G (\bar e+\bar p)
\left[ \frac{(\bar\varphi' )^2_{ }}{\HH^2}
    2 \bigl( \HH' -\frac{\bar\varphi''}{\bar\varphi'} \HH \bigr)+a^2\frac{1}{\HH^2}\bigl( \bar{e} + \bar{p} \bigr) \bigl( \HH' + 4 \HH^2 \bigr)\right]\left(\mathcal{R}_v-\mathcal{R}_\varphi\right)&
\notag  \\
   +
 \bigl( \bar{e} + \bar{p} \bigr)\left\{
   \nabla^2-a^2
   \frac{4\pi G}{\HH}(\bar e+\bar p)\left[2\HH+\frac{(\bar e'+\bar p')}{(\bar e+\bar p)}+\left(\partial_\tau-\frac{\HH'}{\HH} \right)\right]  \right\}\left(\mathcal{R}_v-\mathcal{R}_\varphi\right)
 \;& = \; 
 0 \;.\notag 
\end{align}
The first four lines are the same $\delta\varphi$ terms as in the preceding equation, the next two lines come from the $\delta T$ terms, the seventh and eight line come from $\left(h_D+\frac{\nabla^2\vartheta}{3}\right)$, while the last two lines with $\delta\varphi$ come from the $(h-v)$ terms. Similarly, the eleventh and twelfth line (which include $\mathcal{R}_{T}-\mathcal{R}_\varphi$) originate from $\delta T$, the next two lines come from $\left(h_D+\frac{\nabla^2\vartheta}{3}\right)$, while the last two lines come from the $(h-v)$ terms. We now define the $\mathcal{M}_{3...}$ coefficients, 
\begin{equation}\label{M3}
    \begin{split}
         \mathcal{M}^{ }_{3\varphi}
  =&  -
 \frac{(\bar\varphi')^2_{ }}{a^2_{ }}
 \biggl\{\, 
   \partial_\tau^2 - \nabla^2_{ }
   + 
   2 \biggl[\, 
    \HH + 
    \frac{\HH}{\bar\varphi'}
    \biggl( \frac{\bar\varphi'}{\HH} \biggr)'_{ }
   \,\biggr]\,\partial^{ }_\tau
  \,\biggr\}
\\
  &+
 \bigl( \bar{e} + \bar{p} \bigr) 
 \,\biggl[\,
   \biggl( 2\HH + \frac{\HH'}{\HH} \biggr)
   \,\partial^{ }_\tau + \nabla^2_{ }
 \,\biggr]
 \;, \\
 \mathcal{M}^{ }_{3v}
  =& 
 + \frac{(\bar\varphi')^2_{ }}{a^2_{ }}
 \biggl[\, 
    \frac{2 \HH}{\bar\varphi'}
    \biggl( \frac{\bar\varphi'}{\HH} \biggr)'_{ }
    \frac{4\pi G a^2(\bar{e} + \bar{p})}{\HH}
 \biggr]
\\
 &- 
 \bigl( \bar{e} + \bar{p} \bigr) 
 \biggl\{\, 
    \frac{4\pi G a^2(\bar{e} + \bar{p})}{\HH}
 \biggl[\,
    \partial^{ }_\tau
    + 6 \HH 
    + \frac{(\bar e + \bar p)'}{\bar e + \bar p}   
 \,\biggr]
 -\nabla^2_{ }
 \,\biggr\} 
 \;, 
\\
 \mathcal{M}^{ }_{3T}
 = &
 - \bar{e}^{ }_{,T}  \, T'
 \biggl\{\,
   \partial^{ }_\tau 
  + 2 \, \biggl(2\HH - \frac{\HH'}{\HH} \biggr)
  + \frac{(\bar{e}^{ }_{,T} T')'}
         {\bar{e}^{ }_{,T} T'}
  + \biggl(2\HH + \frac{\HH'}{\HH} \biggr)
    \frac{\bar{p}^{ }_{,T}}{\bar{e}^{ }_{,T}}
 \,\biggr\}
 \;. 
\end{split}
\end{equation}
In addition, we define a $\delta \mathcal{M}_\varphi$ term,
\begin{align}
  \delta \mathcal{M}_\varphi = &    \frac{\HH}{a^2}\bar\varphi' \, 
 \bigl(\, \delta\varphi'' - \nabla^2 \delta\varphi \,\bigr)
 + 
 \frac{\HH}{a^2}\bigl(\, \bar\varphi'' + 4\HH \bar\varphi' \,\bigr)
 \delta\varphi' + 
  \bigl( \bar{e} + \bar{p} \bigr)
   4\pi G\left(\bar\varphi''\delta\varphi+\bar\varphi'\delta\varphi'\right) \notag 
   \\ 
    &+ 4\pi G \bar\varphi'\frac{1}{\HH}\bigl( \bar{e} + \bar{p} \bigr)2 \bigl( \HH' + 2 \HH^2 \bigr)-\bigl( \bar{e} + \bar{p} \bigr)
 4 \pi G a^2_{ } \bigl[\, 
 (\bar{p}^{ }_{,\varphi} - \bar{e}^{ }_{,\varphi})\, \delta\varphi
 \,\bigr]\notag 
 \\
  & +\HH 
 \bigl[\, 
 (\bar e'_{,\varphi} \delta \varphi+\bar e_{,\varphi} \delta \varphi') + 3\HH \bigl( \bar e_{,\varphi} \delta \varphi +\bar p_{,\varphi} \delta \varphi \bigr)
 \,\bigr]- 
 (\bar\varphi' )^2_{ }
   4 \pi G  \bigl[\,
 (\bar{p}^{ }_{,\varphi} - \bar{e}^{ }_{,\varphi})\, \delta\varphi
 \,\bigr]\notag 
 \\
  &+ 4\pi G \bar\varphi'
\left[ \frac{(\bar\varphi' )^2_{ }}{\HH a^2}
    2 \bigl( \HH' + 2 \HH^2 \bigr)-
  \bigl( \bar{e} + \bar{p} \bigr)
   \frac{\HH'}{\HH}- 2 \frac{1}{a^2}\, \bar\varphi' 
 \bigl(\, \bar\varphi'' + 2 \HH \bar\varphi' \,\bigr)\right]\delta\varphi \notag 
  \\
  &+T'\left\{\,
 \bar e'_{,T}  + 3\HH \bigl( \bar e_{,T}  + \bar p_{,T}  \bigr)
- \frac{4 \pi G }{\HH}(\bar{p}^{ }_{, T} - \bar{e}^{ }_{,T})\left[
 (\bar\varphi' )^2_{ }
 \, +\bigl( \bar{e} + \bar{p} \bigr)
  a^2_{ } 
 \right]\,\right\}\left( \frac{\HH}{\bar\varphi'}\delta\varphi\right)\notag 
 \\
    &+
 \bar e_{,T}T'\left(\, -\partial_\tau+\frac{\HH'}{\HH}-\frac{T''}{T'} \right) \left(-\frac{\HH}{\bar\varphi'}\delta\varphi\right)  
  \\
   &-
 \left[- 
  \bigl( \bar{e} + \bar{p} \bigr)
   \partial_\tau^2+\frac{(\bar\varphi' )^2_{ }+a^2(\bar e+\bar p)}{a^2}\bigl(\,
    \partial^{2}_\tau + 2 \HH \partial^{ }_\tau - \nabla^2 \bigr)\right]\left(\frac{\HH}{\bar\varphi'}\delta\varphi\right) \notag 
 \\
    &+
\left[ \frac{(\bar\varphi' )^2_{ }}{\HH a^2}
    2 \bigl( \HH' -\frac{\bar\varphi''}{\bar\varphi'} \HH \bigr)+\frac{1}{\HH}\bigl( \bar{e} + \bar{p} \bigr) \bigl( \HH' + 4 \HH^2 \bigr)\right]\partial_\tau\left(\frac{\HH}{\bar\varphi'}\delta\varphi\right) \notag 
   \\
   &+\bigl( \bar{e} + \bar{p} \bigr)\left\{
   \nabla^2-a^2
   \frac{4\pi G}{\HH}(\bar e+\bar p)\left[2\HH+\frac{(\bar e'+\bar p')}{(\bar e+\bar p)}+\left(\partial_\tau-\frac{\HH'}{\HH} \right)\right]  \right\}\left(-\frac{\HH}{\bar\varphi'}\delta\varphi\right) \notag 
   \\
   &-4\pi G (\bar e+\bar p)
\left[ \frac{(\bar\varphi' )^2_{ }}{\HH^2}
    2 \bigl( \HH' -\frac{\bar\varphi''}{\bar\varphi'} \HH  \bigr)+a^2\frac{1}{\HH^2}\bigl( \bar{e} + \bar{p} \bigr) \bigl( \HH' + 4 \HH^2 \bigr)\right]\left(-\frac{\HH}{\bar\varphi'}\delta\varphi\right)\;.\notag 
\end{align}
With these definitions, we can rewrite our third equation simply as
\begin{equation}
\begin{split}
   \delta \mathcal{M}_\varphi+\mathcal{M}_{3\varphi}\mathcal{R}_\varphi+\mathcal{M}_{3v}\left[\mathcal{R}_v-\mathcal{R}_\varphi\right]+\mathcal{M}_{3T} \left[\mathcal{R}_T-\mathcal{R}_\varphi\right]
 \; = \; 
 0 \;.
\end{split}
\end{equation}
To simplify $\mathcal{M}_{3T}$ we have used the background identity \eqref{bck1}. What remains to show is that all the terms with $\delta\varphi$ vanish, i.e.~that $\delta\mathcal{M}_\varphi=0$. Using the chain rule, we rewrite $\delta \varphi$ terms by grouping those with the same number of derivatives of $\delta\varphi$ together:
\begin{align}
    \delta\mathcal{M}_\varphi=&
  \bigl( \bar{e} + \bar{p} \bigr)
   4\pi G\left[\bar\varphi''\delta\varphi\right]-\bigl( \bar{e} + \bar{p} \bigr)
 4 \pi G a^2_{ } \bigl[\, 
 (\bar{p}^{ }_{,\varphi} - \bar{e}^{ }_{,\varphi})\, \delta\varphi
 \,\bigr]\notag 
 \\
   &+\HH 
 \bigl[\, 
 \bar e'_{,\varphi} \delta \varphi+ 3\HH \bigl( \bar e_{,\varphi} \delta \varphi +\bar p_{,\varphi} \delta \varphi \bigr)
 \,\bigr]- 
 (\bar\varphi' )^2_{ }
   4 \pi G  \bigl[\,
 (\bar{p}^{ }_{,\varphi} - \bar{e}^{ }_{,\varphi})\, \delta\varphi
 \,\bigr]\notag 
 \\
  &+ 4\pi G \bar\varphi'
\left[ \frac{(\bar\varphi' )^2_{ }}{\HH a^2}
    2 \bigl( \HH' + 2 \HH^2 \bigr)- 2 \frac{1}{a^2}\, \bar\varphi' 
 \bigl(\, \bar\varphi'' + 2 \HH \bar\varphi' \,\bigr)+\frac{1}{\HH}\bigl( \bar{e} + \bar{p} \bigr) \bigl( \HH' + 4 \HH^2 \bigr)\right]\delta\varphi\notag 
  \\
  &-4\pi G (\bar e+\bar p)
\left[ \frac{(\bar\varphi' )^2_{ }}{\HH^2}
    2 \HH'\right]\left(-\frac{\HH}{\bar\varphi'}\delta\varphi\right)+
 \bar e_{,T}T'\left(\,\frac{\HH'}{\HH}-\frac{T''}{T'} \right) \left(-\frac{\HH}{\bar\varphi'}\delta\varphi\right)\notag 
 \\
   &-T'\left\{\,
 \bar e'_{,T}  + 3\HH \bigl( \bar e_{,T}  + \bar p_{,T}  \bigr)
- \frac{4 \pi G }{\HH}(\bar{p}^{ }_{, T} - \bar{e}^{ }_{,T})\left[
 (\bar\varphi' )^2_{ }
 \, +\bigl( \bar{e} + \bar{p} \bigr)
  a^2_{ } 
 \right]\,\right\}\left( -\frac{\HH}{\bar\varphi'}\delta\varphi\right)\notag 
 \\
   &+\bigl( \bar{e} + \bar{p} \bigr)\left\{a^2
   \frac{4\pi G}{\HH}\left[-6\HH(\bar e+\bar p)-(\bar e'+\bar p')\right] +  \frac{8\pi G}{\HH}\, \bar\varphi' 
 \, \bar\varphi''  \right\}\left(-\frac{\HH}{\bar\varphi'}\delta\varphi\right)\notag 
 \\
    &+\left[
  \frac{4\pi G}{\HH} a^2 
  \bigl( \bar{e} + \bar{p} \bigr)^2
  +  \bar e_{,T}T' \right]\frac{\HH}{\bar\varphi'}\left[\left(\frac{\HH'}{\HH}\right)\delta\varphi\right]\notag 
   \\
    &+\left[\frac{4\pi G}{\HH}
 a^2 
  \bigl( \bar{e} + \bar{p} \bigr)^2+  \bar e_{,T}T' \right]\frac{\HH}{\bar\varphi'}\left[\left(-\frac{\bar\varphi''}{\bar\varphi'}\right)\delta\varphi\right]
      \\  
    &+
\left[ \frac{(\bar\varphi' )^2_{ }}{\HH a^2}
    2 \bigl( \HH' - \HH^2 \bigr)+\frac{1}{\HH}\bigl( \bar{e} + \bar{p} \bigr) \bigl( \HH' + 2 \HH^2 \bigr)- \frac{2}{a^2}\, \bar\varphi' \, \bar\varphi''  \right]\frac{\HH}{\bar\varphi'}\left[\left(-\frac{\bar\varphi''}{\bar\varphi'}\right)\delta\varphi\right]\notag 
    \\  
    &+
\left[ \frac{(\bar\varphi' )^2_{ }}{\HH a^2}
    2 \bigl( \HH' - \HH^2 \bigr)+\frac{1}{\HH}\bigl( \bar{e} + \bar{p} \bigr) \bigl( \HH' + 2 \HH^2 \bigr)- \frac{2}{a^2}\, \bar\varphi' 
 \, \bar\varphi''  \right]\frac{\HH}{\bar\varphi'}\left[\left(\frac{\HH'}{\HH}\right)\delta\varphi\right]\notag 
 \\
   &-
 \left[\frac{(\bar\varphi' )^2_{ }}{a^2}\right]\frac{\HH}{\bar\varphi'}\left[\frac{\HH''}{\HH}-2\frac{\HH'}{\HH}\frac{\bar\varphi''}{\bar\varphi'}-\frac{\bar\varphi'''}{\bar\varphi'}+2\frac{(\bar\varphi'')^2}{(\bar\varphi')^2}\right]\delta\varphi\notag 
\\
    & + 
 \frac{\HH}{a^2}\bigl(\, \bar\varphi'' + 4\HH \bar\varphi' \,\bigr)
 \delta\varphi' + 
  \bigl( \bar{e} + \bar{p} \bigr)
   4\pi G\,\bar\varphi'\delta\varphi'+\HH \, 
 \bar e_{,\varphi} \,\delta \varphi' 
 \notag 
  \\
   &+\left[\frac{4\pi G}{\HH}
 a^2 
  \bigl( \bar{e} + \bar{p} \bigr)^2
   +  \bar e_{,T}T' \right]\frac{\HH}{\bar\varphi'}\,\partial_\tau \delta\varphi\notag 
 \\  
    &+
\left[ \frac{(\bar\varphi' )^2_{ }}{\HH a^2}
    2 \bigl( \HH' - \HH^2 \bigr)+\frac{1}{\HH}\bigl( \bar{e} + \bar{p} \bigr) \bigl( \HH' + 2 \HH^2 \bigr)- \frac{2}{a^2}\, \bar\varphi' 
  \bar\varphi'' \right]\frac{\HH}{\bar\varphi'}\,\partial_\tau \delta\varphi\notag 
     \\
   &-
 \frac{(\bar\varphi' )^2_{ }}{a^2}\frac{\HH}{\bar\varphi'}\left(2\frac{\HH'}{\HH}-2\frac{\bar\varphi''}{\bar\varphi'}\right)\partial_\tau\delta\varphi\;.\notag 
    \end{align}
Since there were only two terms with second derivatives, $\delta \varphi''$, we have already canceled them. The same goes for the terms with the Laplacian. Upon using eq.~\eqref{bck1_a}, the coefficients of the first derivatives add up to $\frac{\HH}{a^2\bar\varphi'} $ times eq.~\eqref{plasmaconservbck}, hence the $\delta\varphi'$ terms vanish as well. Lastly, we need to show that the terms with no derivatives add up to 0 as well. To do so, we first focus our attention on terms containing derivatives of $\bar{e}$ and $\bar{p}$. Upon using eq.~\eqref{bck1_a}, and adding and subtracting $\frac{\HH}{\bar\varphi'} \bar e_{,\varphi}\bar\varphi''$, we turn all $\varphi$ and $T$ derivatives in derivatives with respect to conformal time $\tau$. 
We further use eq.~\eqref{bck1_a} to eliminate four terms proportional to $\bar\varphi''$, which simplifies $\delta\mathcal{M}_\varphi$,
\begin{equation}\label{intermediatemonster}
    \begin{split}
        \delta\mathcal{M}_\varphi=
        &(\bar e'-\bar p')\frac{\HH}{\bar\varphi'}\left(\HH-\frac{\HH'}{\HH}\right)\delta\varphi+3(\bar e'+\bar p')\frac{\HH^2}{\bar\varphi'}\delta\varphi+\frac{\HH}{\bar\varphi'}\bar{e}''\delta\varphi-\HH\frac{\bar\varphi''}{\bar\varphi'^2}\bar{e}'\delta\varphi\\
  &+ 4\pi G \bar\varphi'
\left[ \frac{(\bar\varphi' )^2_{ }}{\HH a^2}
    2 \HH' +\frac{1}{\HH}\bigl( \bar{e} + \bar{p} \bigr) \bigl(3 \HH' + 4 \HH^2 \bigr)\right]\delta\varphi
   \\
   &+\bigl( \bar{e} + \bar{p} \bigr)\,4\pi G\,a^2
   (\bar e'+\bar p') \left(\frac{\delta\varphi}{\bar\varphi'}\right)-
 \frac{\bar\varphi' _{ }}{a^2}\HH\left[\frac{\HH''}{\HH}-\frac{\bar\varphi'''}{\bar\varphi'}\right]\delta\varphi
 \\
    &+\left[ \frac{4\pi G}{\HH}
 a^2 
  \bigl( \bar{e} + \bar{p} \bigr)^2\right]\frac{\HH}{\bar\varphi'}\left[\left(\frac{\HH'}{\HH}-\frac{\bar\varphi''}{\bar\varphi'}+6\HH\right)\delta\varphi\right]
      \\  
    &+
\left[\frac{1}{\HH}\bigl( \bar{e} + \bar{p} \bigr) \bigl( \HH' + 2 \HH^2 \bigr)\right]\frac{\HH}{\bar\varphi'}\left[\left(-\frac{\bar\varphi''}{\bar\varphi'}\right)\delta\varphi\right]+ 
  \bigl( \bar{e} + \bar{p} \bigr)
   4\pi G\,\bar\varphi''\delta\varphi
    \\  
    &+
\left[ \frac{(\bar\varphi' )^2_{ }}{\HH a^2}
    2 \bigl( \HH' - \HH^2 \bigr)+\frac{1}{\HH}\bigl( \bar{e} + \bar{p} \bigr) \bigl( \HH' + 2 \HH^2 \bigr)\right]\frac{\HH'}{\bar\varphi'}\delta\varphi\;.
    \end{split}
\end{equation}
In order to eliminate $\bar{e}''$, we may want to take a derivative of the stress-tensor-conservation eq.~\eqref{plasmaconservbck} and multiply it with $\frac{\HH}{a^2 \bar\varphi'}\delta\varphi$. We can add the result,
\begin{equation}
    \begin{split}
        \left[\frac{\HH}{a^2 \bar\varphi'} (\bar\varphi'')^2+\frac{\HH}{a^2 }\bar\varphi'''+2\bar\varphi'\HH' \frac{\HH}{a^2 } +4\bar\varphi'' \frac{\HH^2}{a^2 } +6\HH^3\frac{(\bar e+\bar p)}{\bar\varphi'}+3\HH' \frac{\HH}{ \bar\varphi'} (\bar e+\bar p)\right]\delta\varphi &\\+\left[3\HH^2\frac{(\bar e'+\bar p')}{\bar\varphi'}+2\HH^2\frac{\bar e'}{\bar\varphi'}+\frac{\HH}{\bar\varphi'}\bar e''\right]\delta\varphi&=0\;,
    \end{split}
\end{equation} 
to eq.~\eqref{intermediatemonster}, helping us cancel a number of terms. We must further remove the ``difficult'' terms, such as $\HH''$ or $(\bar\varphi'')^2$, which do not appear in the simple background equations \eqref{plasmaconservbck}, \eqref{EFEbckexpanded} and \eqref{bck1}. We do so by adding to eq.~\eqref{intermediatemonster} two more equations: the stress tensor conservation eq.~\eqref{plasmaconservbck}, multiplied by $\left(\frac{\HH \bar\varphi''}{a^2(\bar\varphi')^2}\delta\varphi+2\frac{\HH^2}{a^2\bar\varphi'}\delta\varphi\right)$, and eq.~\eqref{bck2_b}, rescaled by $\frac{\bar\varphi' }{a^2}\delta\varphi$. This leads to
\begin{equation}
    \begin{split}
              \delta\mathcal{M}_\varphi=
               & \frac{(\bar e'-\bar p')}{\bar\varphi'}\left[\HH^2-\HH'-4\pi G(\bar\varphi')^2\right]\delta\varphi
  \\
  &+ 4\pi G \bar\varphi'
\left[ \frac{(\bar\varphi' )^2_{ }}{\HH a^2}
    2 \HH' +\frac{1}{\HH}\bigl( \bar{e} + \bar{p} \bigr) \bigl( 3\HH' + 4 \HH^2 \bigr)\right]\delta\varphi
   \\
   &+\bigl( \bar{e} + \bar{p} \bigr)4\pi G\left[a^2
   (\bar e'+\bar p') \right]\left(\frac{\delta\varphi}{\bar\varphi'}\right)+ 
  \bigl( \bar{e} + \bar{p} \bigr)
   4\pi G\,\bar\varphi''\delta\varphi
 \\
    &+\left[\frac{4\pi G}{\HH}
 a^2 
  \bigl( \bar{e} + \bar{p} \bigr)^2 \right]\frac{\HH}{\bar\varphi'}\left[\left(\frac{\HH'}{\HH}+6\HH\right)\delta\varphi\right]
      \\  
    &+
\left[ \bigl( \HH' -\HH^2 \bigr)+ a^2 
  4\pi G(\bar e+\bar p)\right]\frac{\bigl( \bar{e} + \bar{p} \bigr)}{\bar\varphi'}\left[\left(-\frac{\bar\varphi''}{\bar\varphi'}\right)\delta\varphi\right]
    \\  
    &+
\left[ \frac{(\bar\varphi' )^2_{ }}{\HH a^2}
    2 \bigl( \HH' - \HH^2 \bigr)\frac{\HH'}{\bar\varphi'}+\frac{1}{\HH}\bigl( \bar{e} + \bar{p} \bigr) \bigl( \HH' - \HH^2 \bigr)\frac{\HH'}{\bar\varphi'}\right]\delta\varphi\;.
    \end{split}
\end{equation}
In the first and fifth line we now use eq.~\eqref{bck1_a}. We notice the remaining $p'$ cancel, while the $e'$ terms add up to $\frac{\bar e'}{\bar\varphi'}8\pi G a^2 (\bar e+\bar p)$. Once again we use eq.~\eqref{plasmaconservbck} to eliminate this last $e'$ term,
\begin{equation}
    \begin{split}
               \delta\mathcal{M}_\varphi= &\, (\bar e+\bar p)\frac{\HH'}{\HH \bar\varphi'}\left[- \HH^2 + 4\pi G \bar\varphi'^2+
   4\pi G a^2 (\bar e+\bar p)+
\HH'\right]\delta\varphi
   \\
    &+2\frac{\HH' \bar\varphi'}{\HH a^2}\left\{ 4\pi G
\left[ \bar\varphi'^2
     +a^2\bigl( \bar{e} + \bar{p} \bigr) \right]+
\HH' - \HH^2 \right\}\delta\varphi\;.
    \end{split}
\end{equation}
Both lines are separately zero, as we recognize eq.~\eqref{bck1_a}. Then, the last equation becomes
\begin{equation}\label{e3APP}
    \mathcal{M}_{3\varphi}\mathcal{R}_\varphi+\mathcal{M}_{3v}\left(\mathcal{R}_v-\mathcal{R}_\varphi\right)+\mathcal{M}_{3T} \left(\mathcal{R}_T-\mathcal{R}_\varphi\right)
 \; = \; 
 0 \;,
\end{equation}
with coefficients defined in eq.~\eqref{M3}. 
\section*{Acknowledgements}
I would like to warmly thank Mikko Laine and Simona Procacci, with whom I have worked on projects leading up to this PhD thesis.

% Bibliography__________________________________________________________________
%
%%%%%%%%%%%%%%%%%%%%%%%%%%%%%%%%%%%%%%%%%%%%%%%%%%%%%%%%%%%%%%%%%%%%%%%%%%%%%%%%%%%%%%%%%%%
%
% FOR THE PRINTED VERSION AN EMPTY PAGE BETWEEN END OF THE CHAPTERS AND BIBLIOGRAPHY IS NEEDED
%\clearpage

\clearpage\phantomsection\addcontentsline{toc}{chapter}{Bibliography}
\bibliographystyle{JHEP}
\bibliography{bibliography}

%========================Selbstständigkeitserklärung==================================

\end{document}